\documentclass[twocolumn,aps,superscriptaddress,showpacs,nofootinbib,floatfix]{revtex4}
\usepackage{bm,feynmf}
\usepackage{graphicx}
\usepackage{amsmath}
\usepackage{dcolumn}
\usepackage{xcolor}

\begin{document}


\title{In-medium effects in strangeness production  in heavy-ion collisions at 
(sub-)threshold energies}


\author{Taesoo Song}\email{t.song@gsi.de}
\affiliation{GSI Helmholtzzentrum f\"{u}r Schwerionenforschung GmbH, Planckstrasse 1, 64291 Darmstadt, Germany}

\author{Laura Tolos}\email{tolos@ice.csic.es}
\affiliation{Institute of Space Sciences (ICE, CSIC), Campus Universitat Autonoma de Barcelona,
Carrer de Can Magrans, 08193 Bellaterra, Spain}
\affiliation{Institut d'Estudis Espacials de Catalunya (IEEC), 08034 Barcelona, Spain}
\affiliation{Frankfurt Institute for Advanced Studies, Ruth-Moufang-Strasse 1, 60438 Frankfurt am Main, Germany}

\author{Joana Wirth}\email{joana.wirth@tum.de}
\affiliation{Excellence Cluster 'Origin and Structure of the Universe', 85748 Garching, Germany}
\affiliation{Physik Department E62, Technische Universit\"{a}t M\"{u}nchen, 85748 Garching, Germany}

\author{Joerg Aichelin}\email{aichelin@subatech.in2p3.fr}
\affiliation{SUBATECH UMR 6457 (IMT Atlantique, Universit\'{e} de Nantes, IN2P3/CNRS), 4 Rue Alfred Kastler, F-44307 Nantes, France}
\affiliation{Frankfurt Institute for Advanced Studies, Ruth-Moufang-Strasse 1, 60438 Frankfurt am Main, Germany}

\author{Elena Bratkovskaya}\email{E.Bratkovskaya@gsi.de}
\affiliation{GSI Helmholtzzentrum f\"{u}r Schwerionenforschung GmbH, Planckstrasse 1, 64291 Darmstadt, Germany}
\affiliation{Institute for Theoretical Physics, Johann Wolfgang Goethe Universit\"{a}t, Frankfurt am Main, Germany}


\begin{abstract}
We study the in-medium effects in strangeness production 
in heavy-ion collisions at (sub-)\-threshold energies of 1 - 2 A GeV  based on
the microscopic Parton-Hadron-String Dynamics (PHSD) transport approach. 
The in-medium modifications of the antikaon $(\bar K = K^-, \bar K^0)$ properties 
are described via the self-consistent coupled-channel unitarized scheme 
based on a SU(3) chiral Lagrangian which incorporates explicitly the $s-$ and 
$p-$ waves of the kaon-nucleon interaction. This scheme provides the antikaon
potential, spectral functions and reaction cross sections as well as their dependence 
on baryon density, temperature and antikaon momentum in the nuclear medium, 
which are incorporated in the off-shell dynamics of the PHSD.
The in-medium modification of kaons $(K = K^+, K^0)$ are accounted via
the kaon-nuclear potential, which is assumed to be proportional to
the local baryon density.
The manifestation of the medium effects in observables is investigated for 
the $K$ and ${\bar K}$ rapidity distributions, $p_T$-spectra as well as the polar and
azimuthal angular distributions, directed ($v_1$) and elliptic ($v_2$) flow  
in C+C, Ni+Ni, and Au+Au collisions.
We find - by comparison to experimental data from the KaoS, FOPI and HADES 
Collaborations - that the  modifications of (anti)kaon properties 
in nuclear matter are necessary to explain 
the data  in a consistent manner. Moreover, we demonstrate the sensitivity of 
kaon observables to the equation-of-state of nuclear matter.
\end{abstract}


\maketitle

\section{Introduction}

The modification of hadron properties at finite temperatures and
densities due to the onset of chiral symmetry restoration 
is an open problem in present strong interaction physics.
This question is of primary importance for the field of heavy-ion collisions (HICs) 
and it has also implications for astrophysics, in particular for the evolution of the 
early universe and neutron stars. As suggested early by Kaplan and Nelson \cite{kaplan},
the attractive kaon-nucleon interaction might lead to kaon condensation in the 
interior of neutron stars. 
Experimental data
\cite{FOPI,Laue99,Menzel:2000vv,Sturm01,Forster02,Zinyuk:2014zor,Agakishiev:2010rs,Adamczewski-Musch:2018xwg} and theoretical studies
\cite{Aichelin:1986ss,Ko1,Ko4,lix,cmko,Li2001,Cass97,brat97,CB99,laura03,Effenber00,Aichelin,Fuchs,Mishra:2004te,Hartnack:2011cn,Kolomeitsev:2004np} 
on $K^\pm$ production from A+A collisions at SIS energies of 1-2
A$\cdot$GeV have shown that in-medium modifications of the properties of (anti)kaons have been
seen in the collective flow pattern of $K^+$ mesons \cite{Ko2} as well
as in the abundance and spectra of kaons and antikaons \cite{brat97,laura03,Ko2,Hartnack:2011cn}.
Moreover, as advocated early in Ref. \cite{Aichelin:1986ss}, 
the strangeness production at threshold energies is sensitive to the equation-of-state  (EoS)
of nuclear matter and thus can provide a comprehensive information
about the compressibility of nuclear matter  \cite{Fuchs,Hartnack:2011cn}.

The theoretical study of in-medium properties of hadrons has been launched  
in part by the early suggestion of Brown and Rho
\cite{brown}, that the modifications of hadron masses should scale with
the scalar quark condensate $\langle \bar{q}q\rangle$ at finite baryon
density and temperature. In this scenario the kaon masses 'grow' with the density 
while antikaon masses "drop". 
The mass modification at normal nuclear matter density $\rho_0$ could be attributed 
to the (anti)kaon-nuclear potential.
The first theoretical attempts on the extraction of the antikaon-nucleus potential
from the analysis of kaonic-atom data were in favour of very strong
attractive potentials of the order of -150 to -200 MeV at normal
nuclear matter density $\rho_0$ \cite{FGB94,Gal,Friedman:2007zza}.  
However, later self-consistent calculations based on a chiral Lagrangian
\cite{Lutz:1997wt,Waas:1996xh,Waas:1996fy,Lutz021,Lutz02,Ramos:1999ku,Mishra:2003tr} 
or the coupled-channel unitarized scheme (or G-matrix approach)
using meson-exchange potentials \cite{Tolos:2000fj,Tolos:2002ud}  predicted
a moderate attractive depths of -40 to -60 MeV at density $\rho_0$.

The complexity in the determination of the antikaon potential at finite baryon 
density is related to the fact that the antikaon-nucleon amplitude in the 
isospin channel $I=0$  is  dominated by the $\Lambda(1405)$ resonant 
structure \cite{Koch:1994mj}, which in free space is only $\simeq$ 27 MeV below 
the ${\bar K}N$ threshold.  In the vacuum the antikaon-nucleon interaction 
is repulsive at low energies, however, in the medium it becomes attractive. 
The onset of an attractive $\bar K N$ interaction at low densities is a
consequence of an up-shift of the $\Lambda(1405)$ resonance mass induced
by Pauli blocking on the intermediate nucleon states
\cite{Koch:1994mj,Waas:1996xh,Waas:1996fy,Lutz:1997wt}. Additional
medium effects such as the self-energy of mesons in related coupled
channels and the binding of hyperons in the nuclear environment bring a
smoothed $\Lambda (1405)$ back to its vacuum position
\cite{Ramos:1999ku}, while keeping the attractive character of the
$\bar K N$ interaction in the medium. We refer the reader to 
Ref.~\cite{Tolos:2020aln} for a recent review on strangeness.

Another problem related to the complexity of the description of strangeness dynamics in
relativistic heavy-ion reactions is that 
their in-medium complex self-energies and scattering amplitudes, partly far from the mass shell,
have to be considered \cite{Tolos:2002ud,laura03},  because the antikaon couples strongly to the baryons 
and achieves a nontrivial spectral shape in the medium. The propagation
of such broad states which change their properties dynamically 
depending on the local environment, i.e. a temperature $T$ and a baryon density $\rho_B$ 
(or baryon chemical potential $\mu_B$), requires to go beyond the standard on-shell BUU 
type of approaches and to employ a transport theory for strongly interacting systems.
This became possible after a sizeable progress in the understanding of the
propagation of strongly interacting off-shell particles in phase-space configurations 
has been obtained in the end of the last millennium:  
in Refs. \cite{Cassing:1999mh,Cassing:1999wx}
an off-shell transport approach has been developed by Cassing and Juchem on the basis 
of the Kadanoff-Baym equations that include the propagation of particles with 
dynamical spectral functions (cf. also the review \cite{Cassing:2008nn}). 
They derived relativistic off-shell generalized transport equations 
from the Wigner transformed Kadanoff-Baym equations in the limit of first order 
gradients in phase space and realized them - within an extended  test-particle representation -
in the Hadron-String Dynamics (HSD) transport approach \cite{Ehehalt:1996uq,CB99}
extended later to the Parton-Hadron-String Dynamics (PHSD) approach 
~\cite{Cassing:2008sv,Cassing:2008nn,Cassing:2009vt,Bratkovskaya:2011wp,Linnyk:2015rco,Moreau:2019vhw}.
The PHSD incorporates also partonic degrees-of-freedom in terms of strongly 
interacting dynamical quasiparticles, whose properties are evaluated within the 
Dynamical Quasi-Particle Model (DQPM) \cite{Peshier:2005pp,Cassing:2007nb,Cassing:2007yg}
in line with the lattice QCD EoS. 

The formulation of off-shell transport found an immediate application for the description of 
the strangeness dynamics: whereas in the early calculations \cite{lix,cmko,Li2001,Cass97,brat97,CB99} 
the in-medium cross sections have been simply extrapolated from on-shell cross
sections in vacuum, in Ref.~\cite{laura03} the full off-shell dynamics - based on 
coupled channel G-matrix calculations \cite{Tolos:2002ud} - has been incorporated for the first time 
in the HSD transport approach for the description of strangeness
production at SIS energies. It has been shown  that the antikaon observables (multiplicities, 
rapidity and $p_T$-spectra, angular distributions, flow harmonics 
$v_1, v_2$ etc.) are very sensitive to the in-medium interactions, i.e. to the details 
of the G-matrix approach. A summary of the research on strangeness dynamics 
at SIS energy can be found in the review \cite{Hartnack:2011cn}.

The results obtained with the G-matrix approach in Ref. \cite{laura03} are based on the J\"ulich 
meson-exchange model \cite{Tolos:2000fj,Tolos:2002ud} as the effective $\bar KN$
interaction in matter.  During the last decade the understanding of the $\bar KN$ interaction has been further improved:
in Ref. \cite{Tolos:2006ny} a chiral unitarity approach in coupled channels 
has been developed by incorporating $s$- and $p$- waves of the kaon-nucleon 
interaction at finite density and zero temperature in a self-consistent manner. 
In Ref. \cite{Tolos:2008di}  finite temperature effects have been also 
implemented, however, a full self-consistent solution has been only  achieved 
for the $s$- wave effective $\bar KN$ interaction, while
the $p$-wave contribution was treated by means of hyperon-nucleon
insertions. 
Later on, the chiral effective scheme  in dense (and hot) matter developed in
Refs.~\cite{Tolos:2006ny,Tolos:2008di} has been substantially improved 
in Ref.~\cite{Cabrera:2014lca}. There the full
self-consistency in $s$- and $p$-waves at finite density and
temperature has been achieved. In this way, it became possible
to generate in-medium antistrange meson-baryon cross sections (amplitudes) 
 at finite  density and temperature as well as to determine the 
single-particle properties of hyperons, 
such as the $\Lambda(1115)$, $\Sigma(1195)$ and $\Sigma^*(1385)$, at finite
momentum with respect to the medium at rest, and finite density and temperature. 
The latter are important for calculations of the in-medium antikaon 
scattering cross sections.

While antikaons follow strongly attractive interactions in the medium, which broadens 
their spectral function substantially, the kaon spectral function stays narrow 
since there are no baryonic resonances that couple to kaons. 
Nevertheless, they are affected by the repulsive interaction with nucleons, 
which can be approximated by the kaon-nuclear potential proportional to 
the local baryon density.
This scenario has been widely explored in different transport approaches
\cite{Ko1,Ko4,lix,cmko,Li2001,Cass97,brat97,CB99,laura03,Effenber00,Aichelin,Fuchs,Mishra:2004te} 
(although there are models which includes the (anti)kaon on-shell production 
via the coupling to heavy baryonic resonances with vacuum properties
\cite{UrQMDstrange,Steinberg:2018jvv}). 
It has been found that the kaon repulsive potential of +20-30 MeV at normal nuclear density $\rho_0$
is mostly consistent with a variety of experimental data on
$p+A$, $\pi+A$ and $A+A$ collisions. We refer the reader to the review
\cite{Hartnack:2011cn} and references therein.

The goal of this work  is to study the in-medium effects in the strangeness 
production in heavy-ion collisions  at (sub-)threshold energies employing 
the microscopic transport approach PHSD  which incorporates the in-medium 
description of the antikaon-nucleon interactions based on 'state of the art' 
many-body theory realized by the G-matrix formalism \cite{Cabrera:2014lca}. 
This chiral unitary approach in coupled channels
incorporates the $s$- and $p$-waves of the kaon-nucleon interaction, 
its modification in the hot and dense medium to account for Pauli blocking effects, 
mean-field binding for baryons, as well as pion and kaon self-energies. Moreover, 
it implements unitarization and self-consistency for both 
the $s$- and $p$-wave interactions at finite temperature and density. 
This provides access to in-medium amplitudes in several elastic and inelastic coupled 
channels with strangeness content $S=-1$.
For the in-medium scenario of the kaon-nucleon interaction we will adopt here a
repulsive potential linear in the baryon density.
As an other novel development, we mention also the implementation of 
detailed balance on the level of  $2\leftrightarrow 3$ reactions 
for the main channels for strangeness production/absorption 
by baryons ($B=N, \Delta$) and pions: 
$B+B \leftrightarrow N+Y+K$ and  $B+\pi \leftrightarrow N+N+\bar K$,
as well as for the non-resonant reactions $ N+N \leftrightarrow N+N+\pi$ and
$\pi+N \leftrightarrow N+\pi+\pi$.
We will confront our theoretical results with available experimental data 
on (anti)kaon production at SIS energies and analyse the consequences of 
chiral symmetry restoration in strangeness observables.


The outline of the paper is as follows: In section \ref{gmatrix} we shall briefly
recall the basic concepts of the chiral SU(3) model and G-matrix method used 
in the present investigation and describe the medium modifications of the
$K(\bar K$) mesons in this effective many-body model. In Section \ref{PHSDbasic}
we remind the basic ideas of the PHSD approach.
Sections \ref{vacuum} and \ref{medium} are, respectively, dedicated to 
the production and absorption of $K$ and $\bar K$ mesons and their modifications 
in the nuclear medium. In section \ref{heavy-ion} we investigate the properties of 
$K(\bar K$) mesons in heavy-ion collisions and the results are compared to
experimental data from different collaborations in section~\ref{compare}. 
Section \ref{summary} summarizes the findings of the present investigation 
and discusses future extensions.

\section{Chiral unitarized model for $\bar KN$ in hot nuclear matter}\label{gmatrix}

Here we summarize the main features of the self-consistent unitarized coupled-channel model 
in dense and hot matter (or G-matrix approach) based on the $SU(3)$ meson-baryon 
chiral Lagrangian, which incorporates the $s$- and  $p$- waves of the antikaon-nucleon 
interaction, as developed in Ref.~\cite{Cabrera:2014lca}.

The lowest-order chiral Lagrangian which couples the octet of light pseudoscalar
mesons to the octet of $1/2^+$ baryons is given by
\begin{eqnarray}
L=\langle\bar{B}i\gamma^\mu \nabla_{\mu}B\rangle - M \langle\bar{B}B\rangle~~~~~~~~~~~~~~~~~~~\nonumber\\
+\frac{1}{2}D\langle\bar{B}\gamma^\mu \gamma_5\{u_\mu,B\}\rangle+\frac{1}{2}F\langle\bar{B}\gamma^\mu\gamma_5[u_\mu,B]\rangle,
\label{chiralL}
\end{eqnarray}
where $\langle .. \rangle$ denotes the trace over SU(3) flavor matrices, $M$ is the baryon mass, $D$ and $F$ stand for the vector and axial-vector coupling constants,  
and
$\nabla_{\mu}$ is the covariant derivative that couples the baryon fields to the pseudoscalar meson axial vector current $\Gamma_{\mu}$,
\begin{eqnarray}
  \nabla_{\mu} B &=& \partial_{\mu} B + [\Gamma_{\mu}, B] \nonumber \ ,\\
  \Gamma_{\mu} &=& \frac{1}{2} (u^\dagger \partial_{\mu} u + u\, \partial_{\mu}
      u^\dagger) \nonumber \ , \\
  U &=& u^2 = {\rm exp} (i \sqrt{2} \Phi / f) \ , \\
  u_{\mu} &=& i u ^\dagger \partial_{\mu} U u^\dagger \nonumber \ ,
\end{eqnarray}
with $f$ the meson weak decay constant, and $\Phi$ and $B$ the standard SU(3) meson and baryon field matrices.

The $s$-wave meson-baryon interaction results from the covariant derivative term in
Eq.~(\ref{chiralL}). Keeping terms up to two meson fields, one gets
\begin{eqnarray}
L\sim\frac{i}{4f^2}\bigg\langle\bar{B}\gamma^\mu[\Phi\partial_\mu \Phi-\partial_\mu \Phi\Phi,B]\bigg\rangle. \label{lagrred}
\end{eqnarray}
from which one derives the meson-baryon (tree-level) $s$- amplitudes 
\begin{eqnarray}
 V_{ij}^s = - C_{ij} {1 \over 4 f^2} \bar{u}(p^\prime) \gamma^\mu u(p) 
   (k_\mu + k^\prime_\mu) ,
\label{swave}
\end{eqnarray}
where $k,k^\prime~(p,p^\prime)$ are the initial and final meson (baryon) momenta, respectively, and the
coefficients $C_{ij}$, with
$i$, $j$ indicating the particular meson-baryon channel, can be found explicitly in
Ref.~\cite{Oset:1997it}. For low-energy scattering, the $s$-wave meson-baryon interaction can be written as
\begin{eqnarray}
V_{i j}^s &=& - C_{i j} \, \frac{1}{4 f^2} \, (2 \, \sqrt{s}-M_{B_i}-M_{B_j}) \nonumber \\
&\times&\left( \frac{M_{B_i}+E_i}{2 \, M_{B_i}} \right)^{1/2} \, \left( \frac{M_{B_j}+E_j}{2 \, M_{B_j}} \right)^{1/2} \nonumber \\
&\simeq& - C_{i j} \, {1 \over 4 f^2} (k^0_i + k^0_j)
\ ,
\label{swa}
\end{eqnarray}
where $\sqrt{s}$ is the center-of-mass (c.m.) energy, $M_{B_{i(j)}}$ and $E_{i(j)}$ are the mass and energy of the baryon in the $i(j)$ channel, respectively. The second equation is satisfied to a good approximation for practical purposes. In this study, we consider the following channels which couple to $K^-p$ : 
\begin{eqnarray}
  & K^-p,~ \bar{K}^0n,~ \pi^0\Lambda,~ \pi^0\Sigma^0,~ \eta\Lambda,~ \eta\Sigma^0, \hfill  \phantom{Aa}\label{i30} \nonumber \\
 &  \pi^+\Sigma^-,~ \pi^-\Sigma^+, K^+\Xi^-,~ K^0\Xi^0~,  
 \label{channel1}
\end{eqnarray}
and  the following channels which couple to $K^-n$ :
 \begin{eqnarray}
& K^-n,~ \pi^0\Sigma^-,~ \pi^-\Sigma^0,~ \pi^-\Lambda,~ \eta\Sigma^-,
 K^0\Xi^-.\hfill \label{i3-1} 
 \label{channel2}
\end{eqnarray}

The main contribution to the $p$-wave meson-baryon interactions comes from the $\Lambda$ and $\Sigma$ pole terms, which result from the $D$ and $F$ terms of the lowest-order meson-baryon chiral Lagrangian of Eq.~(\ref{chiralL}). The $\Sigma^*(1385)$ is also taken into account explicitly, as done in Ref.~\cite{Oset:2000eg}. Moreover,  the  Lagrangian in Eq.~(\ref{lagrred}) also provides a small part of the $p$-wave.

Following Refs.~\cite{Tolos:2006ny,Cabrera:2014lca}, the $p$-wave meson-baryon interaction is given by
\begin{eqnarray}
\tilde{V}_{ij}^p= 3 \, \lbrack f_{ij}(\sqrt{s})\hat{q}'\cdot \hat{q}-ig_{ij}(\sqrt{s})(\hat{q}'\times \hat{q})\cdot \vec{\sigma} \rbrack \label{pwamp} \ ,
\end{eqnarray}
where $\vec{q}\,(\vec{q}\,')$ are the on-shell c.m. three-momentum of incoming (outgoing) mesons, and $f_{ij}(\sqrt{s})$ and $g_{ij}(\sqrt{s})$ correspond to the spin-non-flip and spin-flip amplitudes, respectively, that read
\begin{eqnarray}
    f_{ij}(\sqrt{s}) &=& {1 \over 3} \left\{ - C_{ij} {1 \over
4 f^2}\, a_i\,
    a_j \left({1 \over b_i} + {1 \over b_j} \right)  \right. \nonumber \\
    &&+ \left. { D^{\Lambda}_i D^{\Lambda}_j \left(1+{q_i^0 \over M_i} \right)
    \left(1+{q_j^0 \over M_j} \right) \over \sqrt{s} - \tilde M_\Lambda}
    \right.  \nonumber \\
    && \left.  + { D^{\Sigma}_i D^{\Sigma}_j \left(1+{q_i^0 \over M_i}
    \right) \left(1+{q_j^0 \over M_j} \right) \over \sqrt{s} - \tilde
    M_\Sigma} \right. \nonumber \\
    && \left. + {2 \over 3} {D^{\Sigma^{*}}_i D^{\Sigma^{*}}_j \over
    \sqrt{s} - \tilde M_\Sigma^{*}} \right\} q_{i} q_{j}
    \label{f1}\\
    g_{ij}(\sqrt{s}) &=& {1 \over 3} \left\{  C_{ij} {1 \over 4
f^2}\, a_i\,
    a_j \left({1 \over b_i} + {1 \over b_j} \right) \right. \nonumber \\
   && \left. - { D^{\Lambda}_i D^{\Lambda}_j \left(1+{q_i^0 \over M_i} \right)
    \left(1+{q_j^0 \over M_j} \right) \over \sqrt{s} - \tilde M_\Lambda}
    \right.   \nonumber \\
    && \left.  - { D^{\Sigma}_i D^{\Sigma}_j \left(1+{q_i^0 \over M_i}
    \right) \left(1+{q_j^0 \over M_j} \right) \over \sqrt{s} - \tilde
    M_\Sigma}  \right. \nonumber \\
    && \left. + {1 \over 3} {D^{\Sigma^{*}}_i D^{\Sigma^{*}}_j \over
    \sqrt{s} - \tilde M_\Sigma^{*}} \right\} q_{i} q_{j} \label{g1} \ ,
\end{eqnarray}
where  $q_{i(j)}\equiv|\vec{q}_{i(j)}|$. The first term in both, $f_{ij}$ and $g_{ij}$, comes from the small $p$-wave component in the meson-baryon amplitudes from the lowest order chiral Lagrangian in Eq.~(\ref{chiralL}) \cite{Tolos:2006ny,Cabrera:2014lca}, with
\begin{equation}
   a_i = \sqrt{E_i + M_i \over 2 M_i}\ , \hspace{0.7cm} b_i = E_i +
   M_i\ , \hspace{0.7cm} E_i = \sqrt{M_i^{\, 2} + \vec q_i\,^{ 2}} \  ,
\end{equation}
given in the c.m. frame. Moreover, $D^Y_i$ are the couplings of $\Lambda$, $\Sigma$ and $\Sigma^*$ to a given meson-baryon pair:
\begin{eqnarray}
   D^\Lambda_i &=& c_i^{D,\Lambda} \sqrt{20 \over 3} {D \over 2 f} -
   c_i^{F,\Lambda} \sqrt{12} { F \over 2 f} \nonumber \ , \\
   D^\Sigma_i &=& c_i^{D,\Sigma} \sqrt{20 \over 3} {D \over 2 f} -
   c_i^{F,\Sigma} \sqrt{12} { F \over 2 f} \ , \\
   D^{\Sigma^*}_i &=& c_i^{S,\Sigma^*} {12 \over 5} {D + F\over 2 f}
   \nonumber \ .
\end{eqnarray}
The constants $c^D$, $c^F$, $c^S$ are the $SU(3)$ Clebsch-Gordan coefficients (see Table~I of Ref.~\cite{Jido:2002zk}), whereas the couplings $D$ and $F$ are chosen as $D=0.85$ and $F=0.52$. The masses $\tilde M_\Lambda$,
$\tilde M_\Sigma$, $\tilde M_{\Sigma^*}$ are bare masses of the
hyperons ($\tilde M_\Lambda$$=$1030 MeV, $\tilde M_\Sigma$$=$1120 MeV, $\tilde M_{\Sigma^*}$$=$1371 MeV),  which will turn into physical masses upon the unitarization procedure described below.

Due to its spin structure, the $p$-wave contribution mixes different total angular momenta ($J=1/2,3/2$). For unitarization, it is convenient to rewrite the $p$-wave amplitudes according to the total angular momentum $J$, since these amplitudes can be unitarized independently. Thus, one can then define the $p$-wave tree-level amplitudes according to the total angular momentum, $V^p_{ij-}$ ($L=1$, $J=1/2$) and
$V^p_{ij+}$ ($L=1$, $J=3/2$), as
\begin{eqnarray}
    V^p_{ij-} &=& f_{ij}-2g_{ij} \nonumber \\
    V^p_{ij+} &=& f_{ij}+g_{ij} .
\end{eqnarray}
Note that the  $\Sigma^*$ pole 
is contained in the  $V^p_{+}$ amplitude, whereas  the $V^p_{-}$ amplitude includes the $\Lambda$ and $\Sigma$ poles.

The transition amplitudes can be obtained by means of a unitarization coupled-channel procedure based on solving the Bethe-Salpeter equation by using the tree level contributions to the $s$- and $p$-wave meson-baryon scattering as the kernel of the equation. Within the on-shell factorization \cite{Oset:1997it,Oller:2000fj}, the Bethe-Salpeter equation is  given in matrix notation as
\begin{eqnarray}
 T &=& V + V G T, \nonumber \\
T &=& [1-VG]^{-1} V \label{eq:BS-matrix} \ , 
\end{eqnarray}
where $V$ is the kernel (potential), and the $G$ is a diagonal matrix 
accounting for the loop function of a meson-baryon $G_l$ propagator, 
\begin{eqnarray}
G_{l}(\sqrt{s})&=&i\int\frac{d^4q}{(2\pi)^4}\frac{M_l}{E_l(\vec{P}-\vec{q})}\frac{1}{\sqrt{s}-q_0-E_l(\vec{P}-\vec{q})+i\varepsilon} \nonumber \\
&\times&\frac{1}{q_0^2-\vec{q}\,^2-m_l^2+i\varepsilon},
\end{eqnarray}
with $(P_0,\vec{P})$ being the total four-momentum of the meson-baryon pair and $s=P_0^2-\vec{P}\,^2$. The loop is divergent 
and  needs to be
regularized. This can be done by adopting either a cutoff method or dimensional regularization. We will adopt the cut-off method as it is easier and more transparent when dealing with particles in the medium. Within this method, and taking advantage of Lorentz invariance to calculate in the c.m. frame, the loop function reads
\begin{eqnarray}
G_{l}(\sqrt{s})&=& \int_{\mid {\vec q}\, \mid < q_{\rm max}} \, \frac{d^3 q}{(2 \pi)^3} \,
\frac{1}{2 \omega_l (\vec q\,)} \frac{M_l}{E_l (-\vec{q}\,)} \nonumber \\
&\times&\frac{1}{\sqrt{s}- \omega_l (\vec{q}\,) - E_l (-\vec{q}\,) + i \varepsilon} \, ,
\label{eq:gprop}
\end{eqnarray}
with $\omega_l$ ($E_l$) being the energy of the meson (baryon) in the intermediate state in the c.m. frame, respectively, and $q_{\rm max}=630$~MeV, which has been fixed in this scheme to reproduce the $\Lambda(1405)$ properties and several threshold branching ratios \cite{Oset:1997it}.

Whereas the $s$-wave amplitude of Eq.~(\ref{swa}) can be used directly to solve Eq.~(\ref{eq:BS-matrix}) to obtain the transition amplitude in the $s$-wave ($T^s$), for the $p$-wave one proceeds as
\begin{eqnarray}
    \label{eq:pw}
    T^p_{+} &=& [1-V^p_{+} G ]^{-1} V^p_{+}  \ , \nonumber \\ 
    T^p_{-} &=& [1-V^p_{-} G ]^{-1} V^p_{-}\ .  
\end{eqnarray}

Nuclear matter effects at  finite temperature are introduced through the
modification of the meson-baryon propagators, as described in Refs.~\cite{Tolos:2008di,Cabrera:2014lca}. On the one hand, one of the main sources of medium modifications comes from the Pauli principle. On the other hand, all mesons and baryons in the intermediate loops interact with the nucleons of the Fermi sea, and their properties are modified with respect to those in vacuum. All these changes are straightforward implemented within the Imaginary Time Formalism, as shown in Refs.~\cite{Tolos:2008di,Cabrera:2014lca}. Here we summarize the most important findings.

As found in Ref.~\cite{Cabrera:2014lca}, the in-medium loop function $\bar KN$ states reads
\begin{eqnarray}
\label{eq:gmed}
&&{G}_{\bar KN}(P_0,\vec{P};T)
=\int \frac{d^3 q}{(2 \pi)^3}
\frac{M_N}{E_N (\vec{P}-\vec{q},T)} \nonumber \\
&\times  &\left[ \int_0^\infty d\omega
 S_{\bar K}(\omega,{\vec q};T)
\frac{1-n_N(\vec{P}-\vec{q},T)}{P_0 + {\rm i} \varepsilon - \omega
- E_N
(\vec{P}-\vec{q},T) } \right. \nonumber \\
&+& \left. \int_0^\infty d\omega
 S_{K}(\omega,{\vec q};T)
\frac{n_N(\vec{P}-\vec{q},T)} {P_0 +{\rm i} \varepsilon + \omega -
E_N(\vec{P}-\vec{q},T)} \right] .  \ \ \ \ \ 
\end{eqnarray}
The Pauli blocking of nucleons is given by the 1-$n_N(\vec{P}-\vec{q},T)$ term, where $n_N(\vec{P}-\vec{q},T)=[1+\exp(E_N(\vec{p},T)-\mu_B)/T)]^{-1}$ is the nucleon Fermi-Dirac distribution. The medium modification on kaons and antikaons appears through the corresponding spectral 
functions, defined as
\begin{eqnarray}
S_M(\omega,{\vec q}; T)
= -\frac{1}{\pi}\frac{{\rm Im}\, \Sigma_M(\omega,\vec{q};T)}{\mid
\omega^2-\vec{q}\,^2-m_{M}^2- \Sigma_{M}(\omega,\vec{q};T) \mid^2 } \ , \ \
\label{eq:spec1}
\end{eqnarray}
where $\Sigma_{M}$ is the meson self-energy that will be described in more detail below for the $\bar K$ case. The $K$ spectral function, $S_K$, can be replaced by a free-space delta function, as the interaction of kaons with matter is rather weak \cite{Tolos:2008di}.  Note also that the second term in the $\bar KN$ loop function typically provides a small, real contribution for the studied energy range in $P_0$. 

In the case of $\pi Y(=\Lambda, \Sigma)$  one finds
\begin{eqnarray}
\label{eq:gmed_piY}
&&{G}_{\pi Y}(P_0,\vec{P}; T)
=  \int \frac{d^3 q}{(2 \pi)^3} \frac{M_{Y}}{E_{Y}
(\vec{P}-\vec{q},T)} \nonumber \\
& \times &
\int_0^\infty d\omega
 S_\pi(\omega,{\vec q},T)
\left[
\frac{1+f(\omega,T)}
{P_0 + {\rm i} \varepsilon - \omega - E_{Y}
(\vec{P}-\vec{q},T) }   \right.
\nonumber \\
& + &
\left.
\frac{f(\omega,T)}
{P_0 + {\rm i} \varepsilon + \omega - E_{Y}
(\vec{P}-\vec{q},T) } \right] \ ,
\end{eqnarray}
where the $\pi Y$ loop function incorporates the pion spectral function $S_{\pi}$. The first term is the dominant one for the studied energy range in $P_0$ and carries the $1+f(\omega ,T)$
enhancement factor, which accounts for the  contribution from thermal pions at
finite temperature. For hyperons, we neglect the fermion distribution given the low population 
for the studied densities and temperatures.

As for $r=\eta \Lambda$, $\eta \Sigma$ and $K \Xi$ intermediate states,
we simply consider
the meson propagator in vacuum and include only the effective baryon
energies modified by the mean-field binding potential for $\Lambda$ and $\Sigma$ hyperons \cite{Cabrera:2014lca}, i.e.
\begin{eqnarray}
G_{r}(P_0,\vec{P};T)&=& \int \frac{d^3 q}{(2 \pi)^3} 
\frac{1}{2 \omega_r (\vec q\,)} \, 
  \frac{M_r}{E_r (\vec{P}-\vec{q},T)} \nonumber \\
&\times&\frac{1}{P_0 +
{\rm i} \varepsilon - \omega_r (\vec{q}\,) - E_r (\vec{P}-\vec{q},T) } \, . \ \
\label{eq:gmed-etaY-KXi}
\end{eqnarray}

The in-medium $s$- and $p$-wave transition amplitudes for $\bar K N$ at finite density and temperature are obtained solving Eq.~(\ref{eq:BS-matrix}) in matter. The on-shell factorization of the amplitudes in the Bethe-Salpeter equation can be maintained in the case of the in-medium calculation for $s$-wave scattering \cite{Tolos:2006ny}. The solution of the Bethe-Salpeter equation with on-shell amplitudes can be kept for the $p$-waves with a simple modification of the meson-baryon loop function, as shown in Ref.~\cite{Tolos:2006ny}. If we denote by $G_l^L(P_0,\vec{P};T)$ the in-medium meson-baryon propagator for $s$- ($L=0$) and $p$-wave ($L=1$), one has
\begin{eqnarray}
\label{eq:Gsummary}
G_l^{s}(P_0,\vec{P};T) &=& G_{l}(P_0,\vec{P}; T) \ , \nonumber \\
G_l^{p}(P_0,\vec{P};T) &=& G_l(s) + \frac{1}{\vec{q}\,^2_{\rm on}} [ \tilde{G}_{l}(P_0,\vec{P}; T) - \tilde{G}_l(s) ] , \ \ \ \ \
\end{eqnarray}
where the $\tilde{G}$ functions carry an extra $\vec{q}\,^2$ factor in the integrand, corresponding to the  off-shell $p$-wave vertex. Moreover, nuclear short-range correlations have to be taken into account, and the $p$-wave amplitudes need to be corrected to incorporate, in the external states,  the proper off-shell momentum  \cite{Tolos:2006ny}.

\begin{figure}[th!]
\centerline{
\includegraphics[width=9. cm]{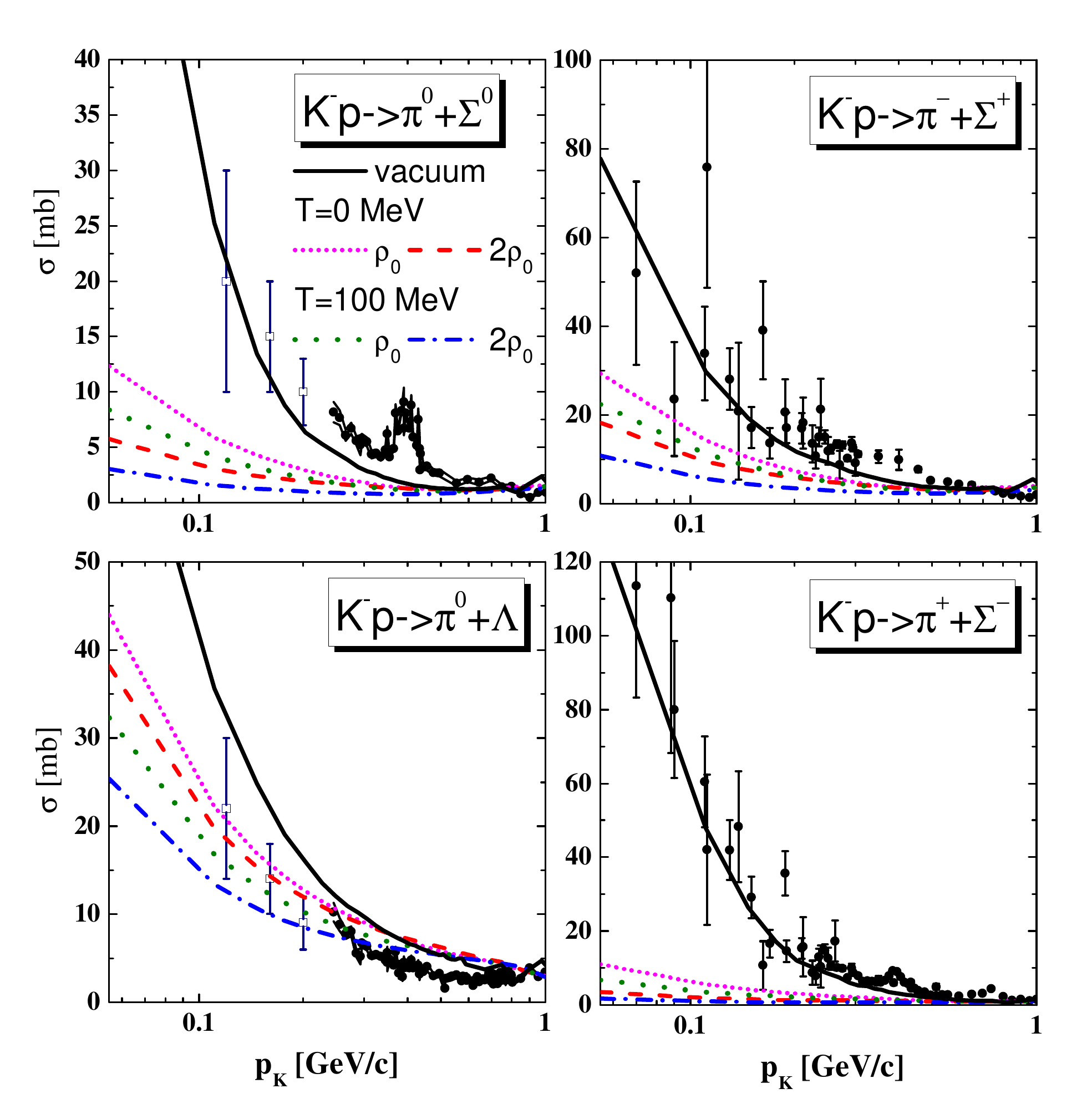}}
\caption{(Color online) The $K^-p$ cross sections to the final states  
$\pi^0\Sigma^0,~\pi^-\Sigma^+,~\pi^+\Sigma^-,$ and $\pi^0\Lambda$ according 
to the G-matrix calculations at $\rho=\rho_0,~2\rho_0$ and $T=0,~100$ MeV
in comparison with the experimental data~\cite{62}. }
\label{check}
\end{figure}

Once the transition amplitudes in free space and in matter are evaluated, the corresponding cross sections can be obtained by
\begin{eqnarray}
\label{eq:diff-cross-sec}
\frac{d\sigma_{ij}}{d\Omega}(\sqrt{s}) &=& \frac{1}{16 \pi^2} \frac{M_i M_j}{s} \frac{q_j}{q_i} \nonumber \\
&\times& \lbrace
| T^{s}_{ij} + (2T^p_{ij+}+T^p_{ij-}) \cos\theta |^2 \nonumber \\
&+&
| T^p_{ij+} - T^p_{ij-} |^2 \sin^2\theta
\rbrace \ ,
\end{eqnarray}
where  $q_i(q_j)$ is  the modulus of the on-shell c.m. three-momentum of the incoming (outgoing) meson-baryon pair, and $\theta$ is the scattering angle in that frame.

Figure~\ref{check} shows the $K^-p$ cross sections to the final states  
$\pi^0\Sigma^0,~\pi^-\Sigma^+,~\pi^+\Sigma^-,$ and $\pi^0\Lambda$  
at $T=0$ and  $T=100$ MeV for several baryon densities. The total three-momentum of $K^-p$ is taken to be zero, while $k_{K^-}$ is calculated in the proton rest frame. One can see that the cross section in each channel is close to the experimental data at low 
densities, while it decreases with increasing baryon density and temperature. As density and temperature increase, the rapid fall of the cross section close to threshold is softened and the strength is distributed over a wide range of energies, as the $\Lambda(1405)$ melts in matter.

\begin{figure*}[th!]
\centerline{
\includegraphics[width=8. cm]{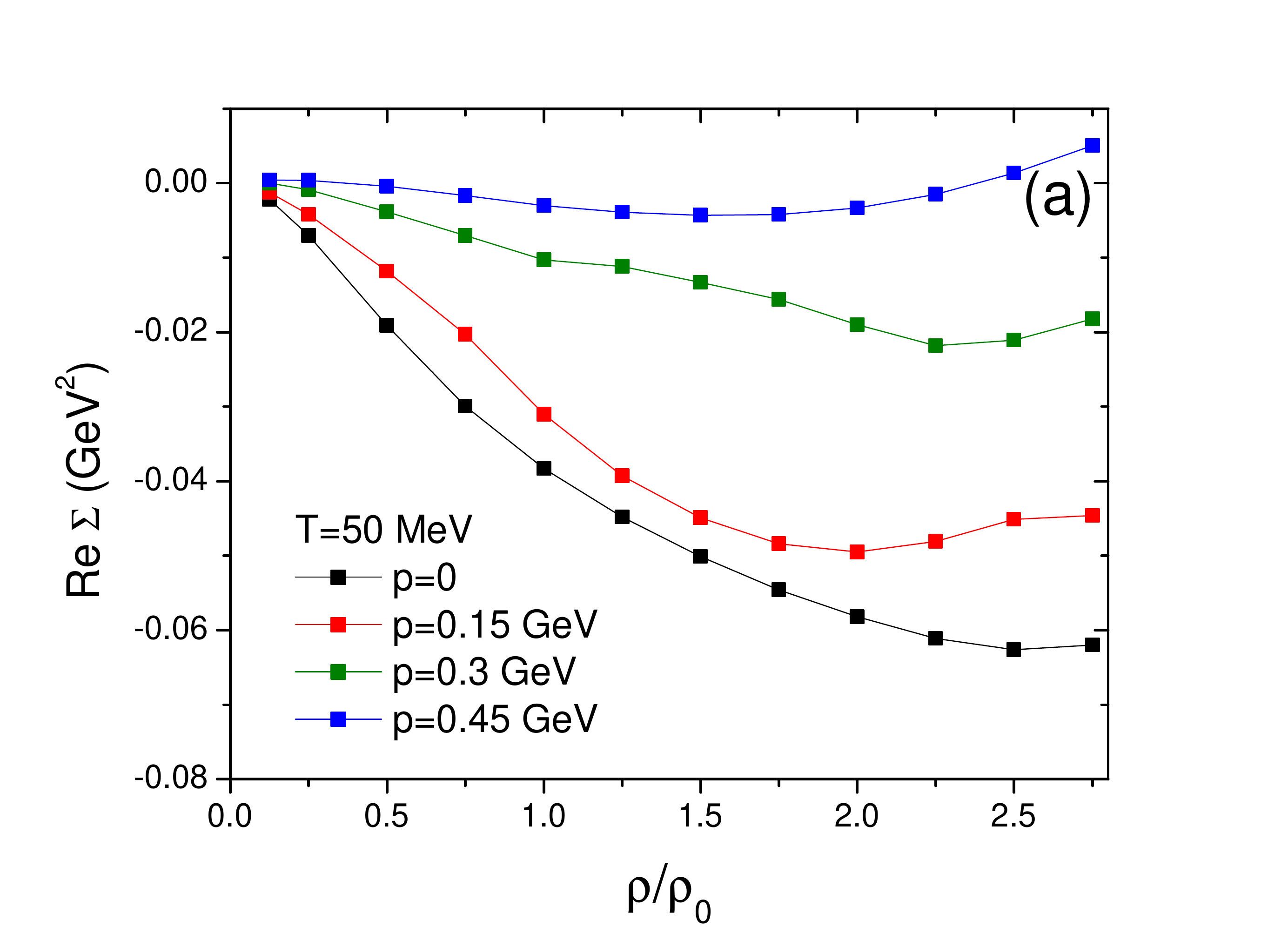}
\includegraphics[width=8. cm]{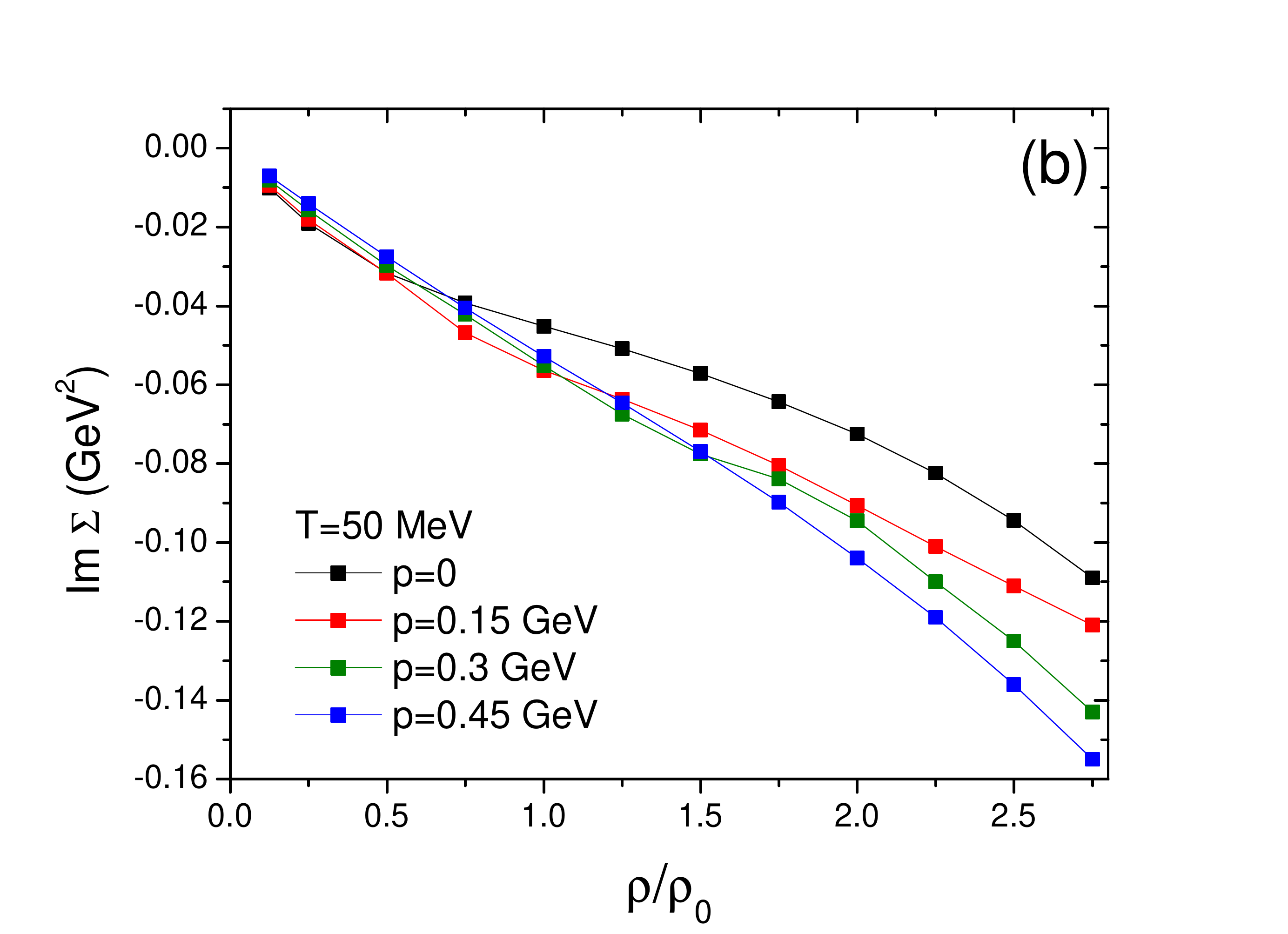}}
\centerline{
\includegraphics[width=8. cm]{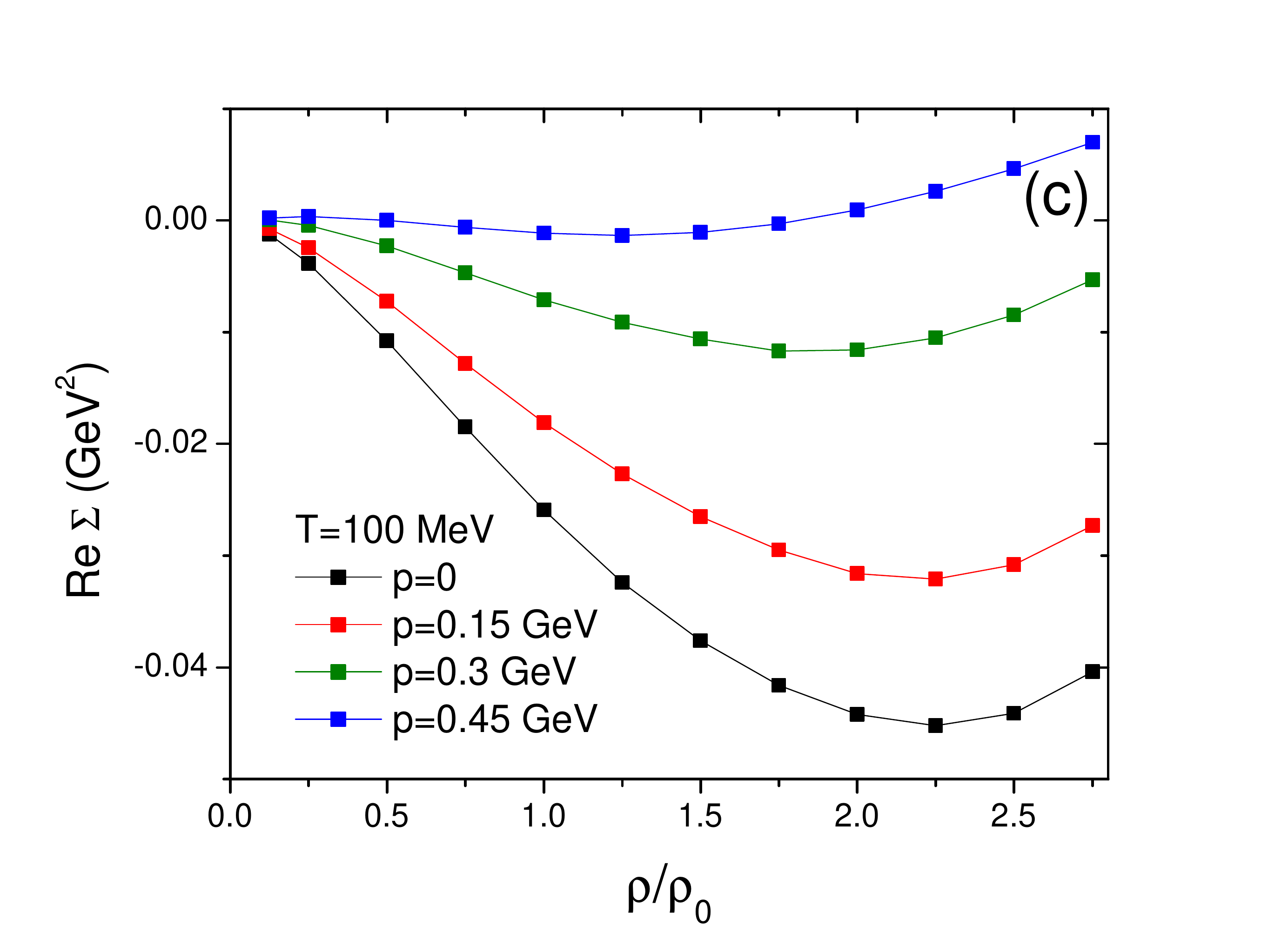}
\includegraphics[width=8. cm]{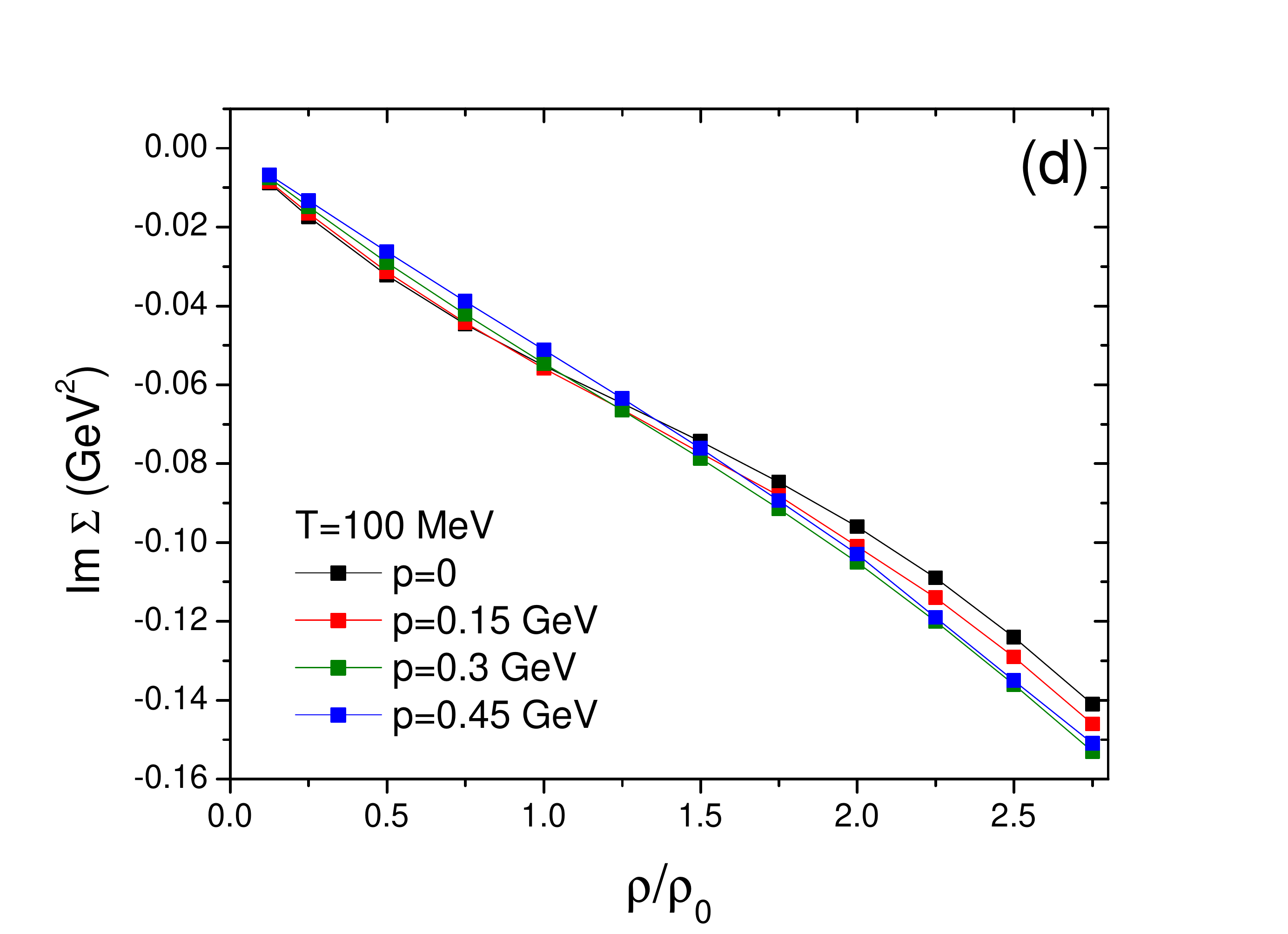}}
\caption{(Color online) (left) The real part and (right) the imaginary part of 
antikaon self-energy as a function of the baryon density for antikaon momenta
of 0, 150, 300, and 450 MeV in the rest frame of nuclear matter at $T=$ 50 MeV 
(upper) and 100 MeV (lower).}
\label{self-energy}
\end{figure*}

Once the transition amplitudes are computed, the self-energy of the antikaon in either $s$- or $p$-wave are obtained by integrating over the nucleon Fermi distribution at a given temperature,
\begin{eqnarray}
\Sigma_{\bar{K}}^L(k_0,\vec{k};T)=4\int\frac{d^3p}{(2\pi)^3}n_N(\vec{p},T)\bar{T}^L_{\bar{K}N\rightarrow \bar{K}N}(P_0,\vec{P};T),\nonumber\\
\end{eqnarray}
where $P_0=k_0+E_N(\vec{p},T)$ and $\vec{P}=\vec{k}+\vec{p}$ are
the total energy and momentum of the $\bar KN$ pair in the
nuclear medium rest frame, $k$ stands for the
momentum of the $\bar K$ meson also in this frame, and $\bar{T}^L$ indicates the spin and isospin averaged scattering amplitude for a given partial wave ($L=0$ or $L=1$). For example, for $K^-$ we have
\begin{eqnarray}
\Sigma_{K^-}^L(k_0,\vec{k};T)=2\int\frac{d^3p}{(2\pi)^3}~~~~~~~~~~\nonumber\\
\times\bigg[n_p(\vec{p},T)\bar{T}^L_{K^-p\rightarrow K^-p}(P_0,\vec{P};T)~~~\nonumber\\
+n_n(\vec{p},T)\bar{T}^L_{K^-n\rightarrow K^-n}(P_0,\vec{P};T)\bigg] ,
\end{eqnarray}
where the $L=1$ transition amplitude reads $T^{p}=3\, \lbrack T^p_- + 2 T^p_+ \rbrack$, with $T^p_-$ and $T^p_+$ given in Eqs.~(\ref{eq:pw}).
The calculation of the antikaon self-energy is a self-consistent process, since the self-energy is
obtained from the in-medium transition amplitude, $T_{\bar{K}N\rightarrow \bar{K}N}$, which
requires the evaluation of the $\bar KN$ loop function,
$G_{\bar{K}N}$, and the latter itself is a function of
$\Sigma_{\bar K}(k_0, \vec k; T)$ through the antikaon spectral function $S_{\bar{K}}(k_0, \vec k; T)$,
\begin{equation}
S_{\bar K}(k_0,{\vec k}; T)
= -\frac{1}{\pi}\frac{{\rm Im}\, \Sigma_{\bar K}(k_0,\vec{k};T)}{\mid
k_0^2-\vec{k}\,^2-m_{\bar{K}}^2- \Sigma_{\bar K}(k_0,\vec{k};T) \mid^2 } \ .
\label{eq:spec}
\end{equation}

Figure~\ref{self-energy} shows the real and imaginary parts of antikaon self-energy as 
a function of the baryon density for a couple of antikaon momenta in the rest frame of 
nuclear matter at $T=$ 50 and 100 MeV.
As the baryon density increases, the imaginary part of the self-energy decreases continuously due to the increase of phase space, whereas the real part decreases up to twice saturation density for momenta up to 0.3 GeV. With increasing antikaon momentum, the changes in the real part of the self-energy become smaller, while the imaginary part is much less sensitive to the momentum.

\begin{figure*}[tbh!]
\includegraphics[width=9. cm]{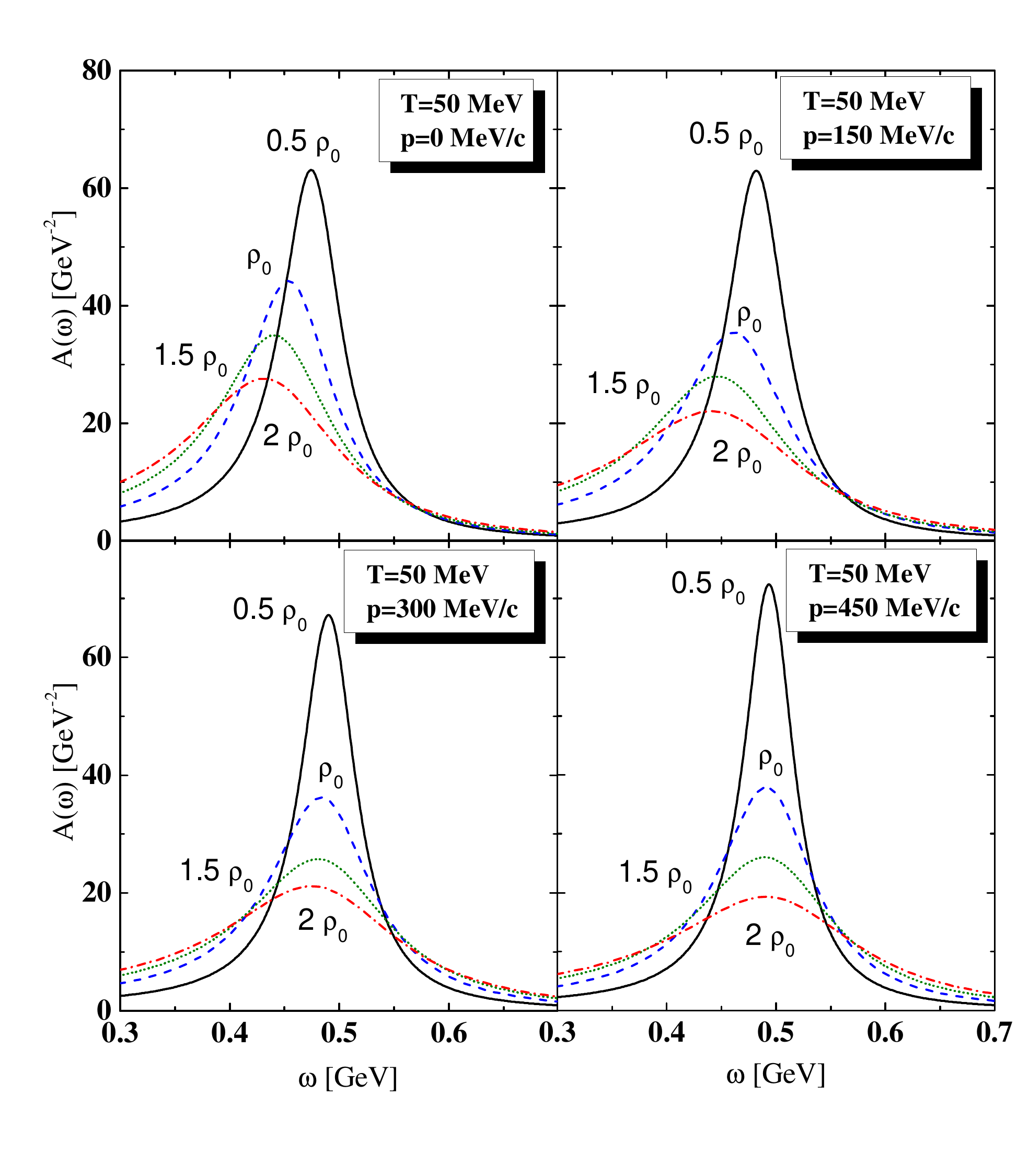}
\includegraphics[width=9. cm]{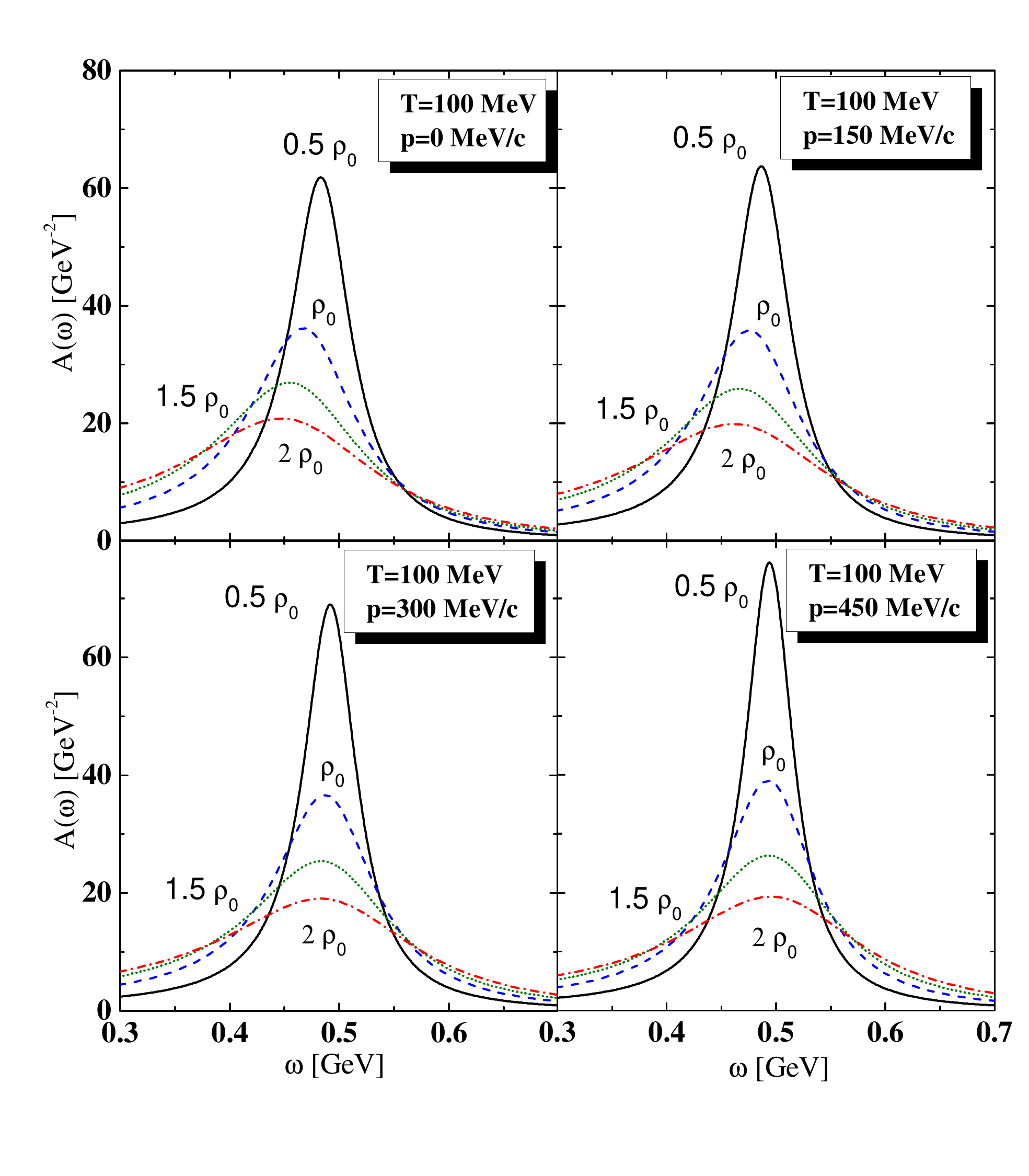} 
\caption{(Color online) Spectral function of $\bar{K}$ as a function of invariant mass at $\rho=0.5\rho_0$, $\rho_0$, $1.5\rho_0$ and $2\rho_0$ with $\rho_0$ being saturation density for antikaon momentum $p=$ 0, 150, 300 and 450 MeV in the rest frame of nuclear matter at $T=$ 50 MeV (left) and 100~MeV (right).}
\label{spectra}
\end{figure*}

 The spectral function can be then reconstructed as
\begin{eqnarray}
A_{\bar K}(\omega,{\bf k})=\frac{-2~{\rm Im}\Sigma_{\bar K}}{(\omega^2-{\bf k}^2-m_{\bar{K}}^2-{\rm Re}\Sigma_{\bar K})^2+({\rm Im}\Sigma_{\bar K})^2},
\end{eqnarray}
where $m_{\bar{K}}$ is the antikaon mass in vacuum and the spectral function is normalized as
\begin{eqnarray}
\int_0^{\infty}\frac{d\omega}{2\pi}\omega A_{\bar K}(\omega,{\bf k})=\frac{1}{2}.
\label{normalization}
\end{eqnarray}
From now on we denote $A_{\bar K}$ as the spectral function, as shown in the literature \cite{laura03}, which is related to $S_{\bar K}$ in Eq.~(\ref{eq:spec}) by $S_{\bar K}=A_{\bar K}/(2\pi)$.

One can change the integral variable from the energy $\omega$ to the invariant mass of the $\bar{K}$,
\begin{eqnarray}
A_{\bar K} (m^2)=\frac{-2~{\rm Im}\Sigma_{\bar K}}{(m^2-m_{\bar{K}}^2-{\rm Re}\Sigma_{\bar K})^2+({\rm Im}\Sigma_{\bar K})^2},
\label{spectralFt}
\end{eqnarray}
where $m^2=\omega^2-k_{\bar K}^2$ and
\begin{eqnarray}
\int_0^{\infty}\frac{dm^2}{2\pi}A_{\bar K}(m^2)=1.
\end{eqnarray}

Figure~\ref{spectra} shows the spectral function of the $\bar{K}$ as a function 
of invariant mass at four baryon densities for antikaon momenta
$p=$ 0, 150, 300, and 450 MeV in the rest frame of nuclear matter at $T=$50 MeV (left panels) and $T=$ 100 MeV (right panels).
The quasiparticle peak appears at $m^2-m_{\bar{K}}^2-{\rm Re}\Sigma_{\bar K}=0$, being its 
height proportional to $-1/{\rm Im}\Sigma_{\bar K}$.

\section{Basic concepts of the PHSD}\label{PHSDbasic}

We start with a reminder of the basic ideas of the PHSD transport approach. 
The~Parton--Hadron--String Dynamics  transport approach 
\cite{Cassing:2008sv,Cassing:2008nn,Cassing:2009vt,Bratkovskaya:2011wp,Linnyk:2015rco}
is a microscopic off-shell transport approach for the description of strongly interacting 
hadronic and partonic matter in and out-of equilibrium. It is based on the solution of
Kadanoff--Baym equations in first-order gradient expansion in phase space 
\cite{Cassing:2008nn}. 
 The approach consistently describes the full evolution of a relativistic
heavy-ion collision from the initial hard scatterings and string
formation through the dynamical deconfinement phase transition to
the strongly-interacting quark-gluon plasma (sQGP) as well as
hadronization and the subsequent interactions in the expanding
hadronic phase as in the Hadron-String-Dynamics (HSD) transport
approach \cite{Ehehalt:1996uq,CB99}. 
The description of partonic degrees-of-freedom and their interactions 
in the PHSD is based on the Dynamical Quasi-Particle
Model (DQPM) \cite{Peshier:2005pp,Cassing:2007nb,Cassing:2007yg}
that is constructed to reproduce lQCD results for a quark-gluon plasma in thermodynamic
equilibrium~\cite{Cassing:2008nn} at finite temperature $T$ and 
baryon (or quark) chemical potential $\mu_q$ on the basis of effective
propagators for quarks and gluons. Correspondingly, the EoS in the PHSD is the lQCD EoS
of crossover type.

The dynamics of heavy-ion collisions within PHSD proceed as follows:\\
$\bullet$ Initially, the penetrating nuclei are initialized 
according to  Wood-Saxon density distributions in coordinate space 
and  by local Thomas-Fermi distributions in momentum space.  
The initial phase of nucleus-nucleus collisions starts with primary nucleon-nucleon scatterings  
from the impinging nuclei. The further  dynamics strongly depends on the collision energy.
For low energy heavy-ion collisions the dynamics is driven by hadronic degrees 
of freedom with binary collisions, possible resonance excitations and decays.
However,  with increasing bombarding energy   
the color-neutral strings are produced which decay to multi-particle states. 
In the PHSD the description of multi-particle production in elementary baryon-baryon ($BB$),
meson-baryon ($mB$) and meson-meson ($mm$) reactions is realized 
within the Lund model \cite{LUND} in terms of the
event generators FRITIOF 7.02 \cite{FRITIOF} and PYTHIA 6.4 \cite{Sjostrand:2006za}. 
We note that in the PHSD the Lund event generators (FRITIOF 7.02 and PYTHIA 6.4)
are "tuned", i.e. adjusted, to get a better agreement with experimental data 
on elementary $p+p$ collisions, especially at low energies (cf. Ref. \cite{Kireyeu:2020wou}). 

$\bullet$ If the local energy density is below the critical one for deconfinement 
($\epsilon_c\simeq 0.5$ GeV/fm$^3$)
the excited strings are dissolved into 'pre-hadrons' (the string decay products)
with a formation time of $\sim$ 0.8 fm/c in their rest frame, except for 
the 'leading hadrons', i.e. the fastest residues of the string ends 
(quarks $q$ or diquarks $qq$), which can re-interact (practically instantly) 
with hadrons with a reduced cross sections in line with the constituent
quark contents. 

We note that in Refs. \cite{PHSD_CSR,Alessia} the chiral symmetry restoration effect 
has been incorporated for the string decay in the dense hadronic medium that is created 
in the early stage of the reaction during the penetration of the colliding nuclei. 
The restoration of chiral symmetry is reflected in the dropping of the scalar 
quark condensate $<\bar q q>$, which can be evaluated in the transport approach
in each local cell within the non-linear $\sigma-\omega$ model from the  
local scalar baryon and meson density according to the Hellman-Feynmann theorem.  
The dropping of the scalar quark condensate leads to a modification of the constituent 
quark masses for light and strange quarks bound in strings and thus affects 
the "chemistry" of decaying strings via the Schwinger mechanism. This leads 
to an enhancement of strangeness production in the dense baryonic medium 
before the deconfined phase may set in \cite{PHSD_CSR,Alessia}.

$\bullet$ If the local energy density in the cell is above the critical value 
of $\epsilon_c\sim 0.5\,$GeV/fm$^3$ \cite{LQCDx}, the 'deconfinement' 
(i.e. a transition of hadronic to partonic degrees-of-freedom) is implemented by dissolving the  
'pre-hadrons' (in this cell) into the massive colored quarks/antiquarks 
and mean-field energy, keeping the 'leading hadrons' out of dissolution
(cf. Refs. \cite{Cassing:2008sv,Cassing:2009vt} for details). This procedure allows
to keep the microscopic description of a changing degrees-of-freedom by conserving 
energy-momentum, charge, flavour etc.
The propagation of the off-shell partons in the hot and dense environment in the 
self-generated repulsive mean-field potential is realized by solving  the
Cassing-Juchem off-shell transport equations for testparticles \cite{Cassing:2008nn}.

$\bullet$ The cross sections for the elastic and inelastic partonic interactions 
in the QGP phase are determined in the DQPM model as well as the covariant transition rates 
for the hadronization to colorless off-shell hadrons when the local energy density 
decreases and become close to or lower than $\epsilon_c$ due to the expansion of the system.
In PHSD 4.0 the partonic cross sections, evaluated from the widths in the quark/gluon propagators,  
depend only on $T$  \cite{Ozvenchuk:2012fn},
while in the recent version PHSD 5.0 the cross sections are evaluated 
from the leading order scattering diagrams using the 'resummed' propagators of the DQPM,
and they have an explicit $(T, \mu_B)$- dependence \cite{Moreau:2019vhw}.  
However, for the present study we will use the PHSD 4.0 since we concentrate on 
low energies where the partonic phase is not the relevance.

$\bullet$ The final stage of a heavy-ion collision at relativistic energies is dominated 
by hadronic interactions. On the other hand, if the initial bombarding energy 
is too low for the formation of the QGP, as in the present study, 
the whole dynamics of the colliding system proceeds  by interactions 
of hadronic degrees-of-freedom. 

The hadronic degrees-of-freedom in the PHSD are the baryon octet and decouplet, 
the ${0}^{-}$ and ${1}^{-}$ meson nonets and higher resonances.
The hadronic interactions include elastic and inelastic collisions between baryons,
mesons and resonances (including the backward reactions through detailed balance)  in line with the HSD approach \cite{Ehehalt:1996uq,CB99}. 
We note that in the PHSD the multi-meson fusion reactions to baryon-antibaryon pairs and backward reactions ($n \ {\rm mesons} \leftrightarrow B+\bar B$) 
are included, too \cite{Cassing:2001ds,Seifert:2017oyb}. 
Following the formulation of the collision term for $n\leftrightarrow m$ reactions in
\cite{Cassing:2001ds,Seifert:2017oyb}, some other  $3\leftrightarrow 2$ reactions
have been implemented recently in the PHSD as the non-resonant reactions 
$N+N\leftrightarrow N+N+\pi$ and $\pi+ N\leftrightarrow N+\pi +\pi$ 
and for the dominant channel in strangeness production/absortion at the threshold
by baryons ($B=N, \Delta$) and pions:
$B+B \leftrightarrow N+Y+K$ and $\pi+B \leftrightarrow K+\bar K + N$
(as will be discussed in the next Section).

We mention that at low energies a density dependent Skyrme nucleon-nucleon 
potential is used for the mean-field propagation of nucleons, which corresponds to 
a 'middle'-soft EoS (default in the PHSD) with the compression modulus 
$K\simeq 300$ MeV (cf. Section VII.E).
The non-strange baryon resonances (such as $\Delta$'s) are
propagated in the same manner as nucleons with  the same $NN$ potential 
while strange baryon resonances (such as $\Lambda$'s, $\Sigma$'s) feel 
only 2/3 of the nucleon-nucleon potential.

The PHSD  incorporates also in-medium effects related to the changes of 
hadronic properties in the dense and hot environment during the time evolution 
of the system, such as a collisional broadening of the spectral functions of 
the light vector mesons ($\rho, \omega, \phi, a_1$) \cite{Bratkovskaya:2007jk}
and strange vector mesons $K^*, \bar K^*$ \cite{Ilner:2016xqr}. 

As mentioned in the Introduction, in Ref. \cite{laura03} the in-medium modification 
of the strange mesons $K, \bar K$ has been implemented in HSD 1.5 using 
an early G-matrix approach. The propagation of broad states - described by the complex 
self-energies and medium-dependent cross sections - is only possible in a consistent way 
by using the off-shell transport theory as reviewed in \cite{Cassing:2008nn}.
This provides a solid ground for the implementation of the novel
G-matrix described in Section \ref{gmatrix} in the PHSD approach.
The  PHSD version, which incorporates the novel G-matrix \cite{Cabrera:2014lca},
we will denote by PHSD 4.5.

\section{On-shell kaon, antikaon production/absorption}\label{vacuum}

We start with the description of the on-shell strangeness production channels in
elementary baryon-baryon ($BB$), meson-baryon ($mB$) and meson-meson ($mm$) reactions
(where $B=p, n, \Delta, ...$ and $m = \pi, \rho, \omega, ... $)
as realized in the PHSD 4.5. Due to the strangeness conservation
$s, \bar s$ quarks can be produced only in  pairs, which implies
the production of kaon-antikaon $K\bar K$ or kaon-hyperon pairs $K Y$, 
where $Y=(\Lambda, \Sigma)$. 
We mention that the production mechanism strongly depends on the bombarding
energy of the heavy-ion collision. At low energies only one or two hadrons
are associated to the production of the strange pair, while with increasing energy 
multi-hadron states are energetically possible. The latter are realized within
string excitation and decay.

\subsection{Baryon+baryon scattering}

There are only two channels for strangeness production in baryon+baryon scattering
at low energy:\\
$\bullet$ The channel with the lowest threshold energy is $N+N\rightarrow N+Y+K$ 
with nucleon $N$, hyperon $Y$ and kaon $K$ in the final state 
($\sqrt{s_{0}}=m_N+m_\Lambda+m_K = 2.546$ GeV).
We take into account the inverse reaction $N+Y+K\rightarrow N+N$ as well, 
which is relevant in a dense baryonic matter and required by detailed balance. 
The implementation of the $N+Y+K\rightarrow N+N$ channel is realized by a calculation 
of the transition rate following Refs. \cite{Cassing:2001ds,Seifert:2017oyb}. 
The details of the numerical realization of $3\rightarrow 2$ reactions in the PHSD
are presented in Appendix \ref{app1}. \\
$\bullet$ With increasing energy the strange pair can be produced in
the channel $N+N\rightarrow N+N+K+\bar{K}$
($\sqrt{s_{0}}=2(m_N+m_K)=2.862$ GeV).

The interactions of nucleons with $\Delta$'s (as well as $\Delta +\Delta$ interactions) 
are treated in the same way as $N+N$ interactions.
We mention that  the contribution of $\Delta N$ reactions (especially $\Delta +N \to N+Y+K$) 
is very important for strangeness production at (sub-)threshold energies since 
$\Delta N$ channel has a larger $\sqrt{s}$  compared to the scattering of two nucleons 
with the same kinetic energy as $\Delta$ and $N$.  
Moreover, since the density of $\Delta$ at SIS energies is high, the $\Delta N$ channels 
dominate $NN$ channels. The quantitative results on this issue will be presented in Section VI.

The cross sections for both channels and the inverse reactions of the first channel 
$N+Y+K\rightarrow N+N$ are, respectively, described in Appendices~\ref{cross-sections} and \ref{app1}. \\
$\bullet$ With further increasing collision energy the string formation and 
fragmentation becomes the main source of strangeness production. The threshold for the 
formation of baryon-baryon strings is taken in the PHSD as $\sqrt{s_{th}}(BB \ str)=$2.65~GeV.

\subsection{Meson+baryon scattering}

$\bullet$ Dominant channels for kaon production in meson+baryon scattering are 
$\pi+N\rightarrow K+Y$ and $\pi+N\rightarrow K+\bar{K}+N$ with cross sections 
given in Appendix~\ref{cross-sections}. 
We also account for the inverse process $N+K+\bar K \to N +\pi$ by detailed balance
(see Appendix \ref{app1}).
\\

$\bullet$ As has been found first in Ref. \cite{Barz:1985xc} and supported by 
other transport calculations \cite{Cass97,Hartnack:2011cn},
the most important channel for antikaon production in nucleus-nucleus collisions is 
$\pi+Y\rightarrow N+\bar{K}$ which involves two secondary particles since the hyperons 
and pions have to be  first produced by initial $NN$ collisions.
The $\pi+Y\rightarrow N+\bar{K}$ cross section is related to  the inverse 
reaction $N+\bar{K}\rightarrow \pi+Y$ by detailed balance:
\begin{eqnarray}
\sigma_{\pi Y\rightarrow \bar{K}N}(s)=\frac{D_{\bar{K}}D_N}{D_\pi D_Y}\bigg(\frac{p_{\bar{K}}}{p_\pi}\bigg)^2\sigma_{\bar{K}N \rightarrow \pi Y}(s)~~~~~~~~\nonumber\\
=\frac{D_{\bar{K}}D_N}{D_\pi D_Y}\frac{\{s-(m_{\bar{K}}+m_N)^2\}\{s-(m_{\bar{K}}-m_N)^2\}}{\{s-(m_\pi+m_Y)^2\}\{s-(m_\pi-m_Y)^2\}}\nonumber\\
\times \sigma_{\bar{K}N \rightarrow \pi Y}(s),~~~
\label{crok3}
\end{eqnarray}
where $D_i$ is the degeneracy factor of particle $i$, and $p_\pi$ and $p_{\bar{K}}$ are, 
respectively, the momenta of pion and antikaon in the c.m. frame.
The cross section for $N+\bar{K}\rightarrow \pi+Y$ is provided by the G-matrix approach
described in Section \ref{gmatrix}, if the initial and final states are included in 
Eqs.~(\ref{channel1}) and (\ref{channel2}).\\
The reactions of mesons with $\Delta$'s are taken into account, too.
For example, for the interactions with $\Delta$'s 
such as $\Delta+\bar{K}\rightarrow \pi+Y$, 
the cross sections are given in Appendix~\ref{cross-sections}. \\
$\bullet$ At energies above the threshold for the string formation in baryon+meson
collisions, which is taken to be $\sqrt{s_{th}}(mB \ str)=$2.4 GeV in the PHSD, 
strangeness can be produced in an associated multi-hadron environment.

\subsection{Meson+meson scattering}

$\bullet$ The most dominant channel for kaon and antikaon production from meson scattering is 
$\pi+\pi \rightarrow K+\bar{K}$ which is given by $\simeq 3$ mb \cite{Protopopescu:1973sh}.
If the total electric charge is +1 or -1, there is only one channel: $\pi^++\pi^0\rightarrow K^++\bar{K}^0$ or $\pi^0+\pi^-\rightarrow K^0+K^-$. On the other hand, $\pi^++\pi^-$ and $\pi^0+\pi^0$ can produce either $K^++K^-$ or $K^0+\bar{K}^0$.
We assume that
\begin{eqnarray}
&&\sigma_{\pi^+\pi^-\rightarrow K^+K^-}=\sigma_{\pi^+\pi^-\rightarrow K^0 \bar{K}^0}
\label{pipiKK}\\
&&=\sigma_{\pi^0\pi^0\rightarrow K^+K^-}=\sigma_{\pi^0\pi^0\rightarrow K^0 \bar{K}^0}=1.5~[\rm mb].  \nonumber\\
&&\sigma_{\pi^+\pi^0\rightarrow K^+ \bar K^0}=\sigma_{\pi^-\pi^0\rightarrow K^- K^0}
= 3~[\rm mb]. 
\nonumber
\end{eqnarray}
The annihilation of $K$ and $\bar{K}$ is taken into account by the detailed balance consistently.

$\bullet$ Again, at relativistic energies the strangeness can be produced 
by the excitation and decay of meson-meson strings above the threshold, 
which is taken to be $\sqrt{s_{th}}(mm \ str)=$1.3 GeV.

\subsection{Resonance decays}

The decay of $K^*(\bar{K}^*$) and $\phi$ mesons produces a (anti)kaon and antikaon-kaon
pairs, respectively.
$K^*(\bar{K}^*$) decays into $K (\bar{K})\pi$ with almost 100 \% probability, 
while $\phi$ meson decays into $K^+K^-$ with 49.2 \% and $K_L^0K_S^0$ 
with 34 \%~\cite{Tanabashi:2018oca}.
The total decay widths of $K^*(\bar{K}^*)$ and $\phi$ are respectively about 50 MeV and 4 MeV in vacuum~\cite{Tanabashi:2018oca} (and getting broader in the medium).

\section{Off-shell kaon, antikaon production/absorption}\label{medium}

The properties of (anti)kaons are modified in nuclear matter with respect to the vacuum. The energy or the rest mass of the kaon increases in nuclear matter while that of the antikaon decreases. This effectively shifts the threshold energy for kaon production up and that of antikaon production down.

The modification of the reaction thresholds in the nuclear medium has important 
consequences for the strangeness production: the reduction of the antikaon threshold 
leads to an enhancement of antikaon production in the medium. Moreover, it opens  
sub-threshold, i.e. below the vacuum threshold, production of antikaons. 
This sub-threshold antistrangeness production in $A+A$ reactions - with respect to 
the elementary $N+N$ reactions - has been an exciting experimental discovery and extensively 
studied by transport models in the past (cf. the review for details \cite{Hartnack:2011cn}).
Contrary to antikaons, the kaon threshold in the nuclear medium is shifted to larger energies, i.e. above the vacuum threshold for corresponding elementary $BB$ and $mB$
($B=N,\Delta$) reactions for kaon production, which suppresses the production of kaons 
in $A+A$ collisions when accounting for the in-medium effects compared to the 
calculations without medium effects.

In this Section we present our modelling of the in-medium strangeness production
and propagation which accounts for the modification of the (anti)kaon properties 
in the nuclear environment.
We stress that the antikaon scattering in the medium is evaluated
following the coupled-channel G-matrix approach as described in Section \ref{gmatrix}.

\subsection{Antikaon production in the nuclear medium}

The antikaon production cross section in the medium is evaluated by folding the 
corresponding vacuum production cross section with the in-medium spectral function in Eq.~(\ref{spectralFt}): 
\begin{eqnarray}
\sigma_{\bar{K}}^*(\sqrt{s})=
\int_0^{(\sqrt{s}-m_4)^2} \frac{dm^2}{2\pi}A(m^2)~\sigma_{\bar{K}}(\sqrt{s}-\Delta m_{\bar{K}}),
\label{massshiftbar}
\end{eqnarray}
where $m_4$ is the invariant mass of the final particles different from antikaons
(i.e. for $\bar K$ production via $1+2\to \bar K + X$,  
$m_4$ is the mass of all particles $X$),
$m$ is the off-shell mass of the $\bar K$ meson in the medium defined by
the spectral function $A(m^2)$ according to Eq. (\ref{spectralFt}); 
$m_{\bar K}$ is the vacuum mass of the antikaon in the final state
 and $\Delta m_{\bar{K}}=m-m_{\bar{K}}$ describes the deviation from the vacuum mass.
In Eq. (\ref{massshiftbar}) we have assumed that the mass shift of the antikaon is realized in the scattering cross section by a shift of the incoming 
invariant energy $\sqrt{s}$. An explicit derivation is given in Appendix~\ref{app:cs-medium}.
The mass of a produced antikaon is determined by a Monte Carlo procedure  
on the basis of the probability distribution function 
\begin{eqnarray}
\frac{dP(m^2)}{dm^2}=
\frac{A(m^2)~\sigma_{\bar{K}}(\sqrt{s}-\Delta m_{\bar{K}})}{2\pi \sigma_{\bar{K}}^*(\sqrt{s})}.
\end{eqnarray}

The self-energy of the antikaon, which fully determines the spectral function, explicitly depends on the antikaon three momentum in the rest frame of the nuclear medium.
In the present study we take the velocity of the c.m. frame of the scattering
particles in the medium as the antikaon velocity assuming that the kinetic energy 
of the antikaon is small in the c.m. frame.
In other words, the three-momentum of the antikaon in the nuclear medium - on which self-energy depends - is given by $m_K\gamma\vec{\beta}$ with $m_K$ and $\vec\beta$ being the antikaon mass in vacuum and the velocity of the c.m. frame in the medium (with $\gamma=1/\sqrt{1-\beta^2}$).
Considering that the collision energy of heavy-ions in the present study is close to the threshold energy for antikaon production, this should be a reasonable approximation.

\subsection{Antikaon propagation in the nuclear medium}

The equations of motion for off-shell particles are given by the
Cassing-Juchem off-shell transport equations for testparticles  ~\cite{Cassing:1999mh}
used also for the propagation of antikaons in the early HSD study \cite{laura03}:
\begin{eqnarray}
\frac{dr_i}{dt}&=&\frac{1}{1-C}\frac{1}{2E}\bigg[2p_i+\nabla_p{\rm Re}\Sigma+\frac{M^2-M_0^2}{{\rm Im} \Sigma}\nabla_p{\rm Im}\Sigma\bigg],\nonumber\\
\label{x-update}\\
\frac{dp_i}{dt}&=&\frac{-1}{1-C}\frac{1}{2E}\bigg[\nabla_r{\rm Re}\Sigma+\frac{M^2-M_0^2}{{\rm Im} \Sigma}\nabla_r{\rm Im}\Sigma\bigg],\label{momentum-update}\\
\frac{dE}{dt}&=&\frac{1}{1-C}\frac{1}{2E}\bigg[\partial_t{\rm Re}\Sigma+\frac{M^2-M_0^2}{{\rm Im} \Sigma}\partial_t{\rm Im}\Sigma\bigg],\label{energy-update}
\end{eqnarray}
where $M_0$ is the pole mass in vacuum, $M^2=E^2-p_i^2-{\rm Re}\Sigma$, and
\begin{eqnarray}
C=\frac{1}{2E}\bigg[\frac{\partial}{\partial E}{\rm Re}\Sigma+\frac{M^2-M_0^2}{{\rm Im} \Sigma}\frac{\partial}{\partial E}{\rm Im}\Sigma\bigg].
\end{eqnarray}
However, as shown in Appendix~\ref{app:mass-update}, the time-evolution of the off-shell particle mass is more practical for actual calculations than the time-evolution of the 
energy, and Eq.~(\ref{energy-update}) is substituted by
\begin{eqnarray}
\frac{dM^2}{dt}=\frac{M^2-M_0^2}{{\rm Im} \Sigma}\frac{d{\rm Im}\Sigma}{dt},
\label{mass-update}
\end{eqnarray}
which manifests directly that the off-shell mass $M$ approaches $M_0$ as the imaginary part of the self-energy converges to zero when propagating in the vacuum where the antikaon 
is on-shell, i.e. stable with respect to strong interaction.

\subsection{Kaons in nuclear matter}

In this study we assume that the kaon-nuclear potential  increases linearly with baryon density $\rho$ as
\begin{eqnarray}
V_K=25 ~{\rm MeV} ~(\rho/\rho_0),
\label{VK25}
\end{eqnarray}
where $\rho_0$ is the nuclear saturation density (taken to be $0.16/{\rm fm^3}$).
Since the single-particle energy of the kaon, $\mathcal{E}$, in nuclear matter is approximated in the 
nonrelativistic limit by
\begin{eqnarray}
\mathcal{E}=\sqrt{m_K^2+p^2+{\rm Re}\Sigma}\simeq E_K+\frac{{\rm Re}\Sigma}{2E_K}=E_K+V_K,
\label{hamiltoniank}
\end{eqnarray}
where $m_K$ is the kaon mass in vacuum, ${\rm Re}\Sigma$ is the real part of the kaon self-energy 
and $E_K=\sqrt{m_K^2+p^2}$.
The in-medium kaon mass $m_K^*$ is expressed in terms of $V_K$ as follows:
\begin{eqnarray}
m_K^*=\sqrt{m_K^2+{\rm Re}\Sigma}=\sqrt{m_K^2+2E_K V_K}~~~~~~~~~~\nonumber\\
\simeq m_K\bigg(1+\frac{E_KV_K}{m_K^2}\bigg)\simeq m_K\bigg(1+\frac{25~ {\rm MeV}}{m_K}\frac{\rho}{\rho_0}\bigg).
\label{kpmass}
\end{eqnarray}
%
As in case of antikaon production, the cross section is shifted by the mass difference:
\begin{eqnarray}
\sigma_{NN\rightarrow NYK}(\sqrt{s})\rightarrow \sigma_{NN\rightarrow NYK^*}(\sqrt{s}-\Delta m_K)
\label{massshift}
\end{eqnarray}
where $\Delta m_K=m_K^*-m_K$.
Since $\Delta m_K$ is positive, the mass shift suppresses $K$ production in heavy-ion collisions.

In some scattering channels such as $N+N\rightarrow N+N+K+\bar{K}$ and $\pi +N\rightarrow N+ K+\bar{K}$, the kaon and antikaon are simultaneously produced.
In this case the cross section is expressed as the combination of Eqs.~(\ref{massshiftbar}) and (\ref{massshift}):
\begin{eqnarray}
\sigma_{K\bar{K}}^*(\sqrt{s})=
\int_0^{(\sqrt{s}-m_4)^2} \frac{dm^2}{2\pi}A(m^2)\nonumber\\
\times \sigma_{K\bar{K}}(\sqrt{s}-\Delta m_K-\Delta m_{\bar{K}}).
\label{massshift2}
\end{eqnarray}

The production of kaon and antikaon through the decay of $K^*$ and $\phi$ is also modified in nuclear matter.
Assuming that the transition amplitude does not depend on the mass of the 
daughter particle, as shown in Appendix~\ref{app:cs-medium},
the mass distribution of the antikaon - produced from the decay - follows
\begin{eqnarray}
\frac{dP(m^2)}{dm^2}=
\frac{A(m^2)p^{c.m.}(m^2)}{8\pi^2 M^2\Gamma}=A(m^2)~~~~~~~~~\nonumber\\
\times\frac{\sqrt{\{M^2-(m+m_2)^2\}\{M^2-(m-m_2)^3\}}}{(4\pi)^2 M^3\Gamma},
\end{eqnarray}
where $p^{c.m.}$ is the three-momentum of the antikaon in the c.m. frame and 
$M$ and $m_2$ are, respectively, the masses of the mother particle and the other daughter
particle different from the antikaon.

The position and momentum of the kaon are updated in nuclear matter according to the equations of motion given by
\begin{eqnarray}
\frac{dr_i}{dt}&=&\frac{\partial H}{\partial p_i}=\frac{p_i}{E},\nonumber\\
\frac{dp_i}{dt}&=&-\frac{\partial H}{\partial r_i}=-\nabla V_K(r),
\end{eqnarray}
where $i=1,2,3$. The kaon energy is defined by Eq.~(\ref{hamiltoniank}).


\section{$K$/$\bar{K}$ production in heavy-ion collisions}\label{heavy-ion}

We step now to the investigation of strangeness dynamics in heavy-ion collisions within
the PHSD 4.5 which incorporates the strangeness cross sections and their 
modifications in the medium (defined in Sec.~\ref{medium}) and 
the off-shell propagation of antikaons according to 
Eqs.~(\ref{x-update})-(\ref{mass-update}).
The kaon cross sections and potential depend on the baryon density while
the antikaon properties, defined by self-energies of the G-matrix, depend 
on the  baryon density, the temperature and the three-momentum of the antikaon 
in the rest frame of the medium.

\begin{figure*}[h!]
\centerline{
\includegraphics[width=8.6 cm]{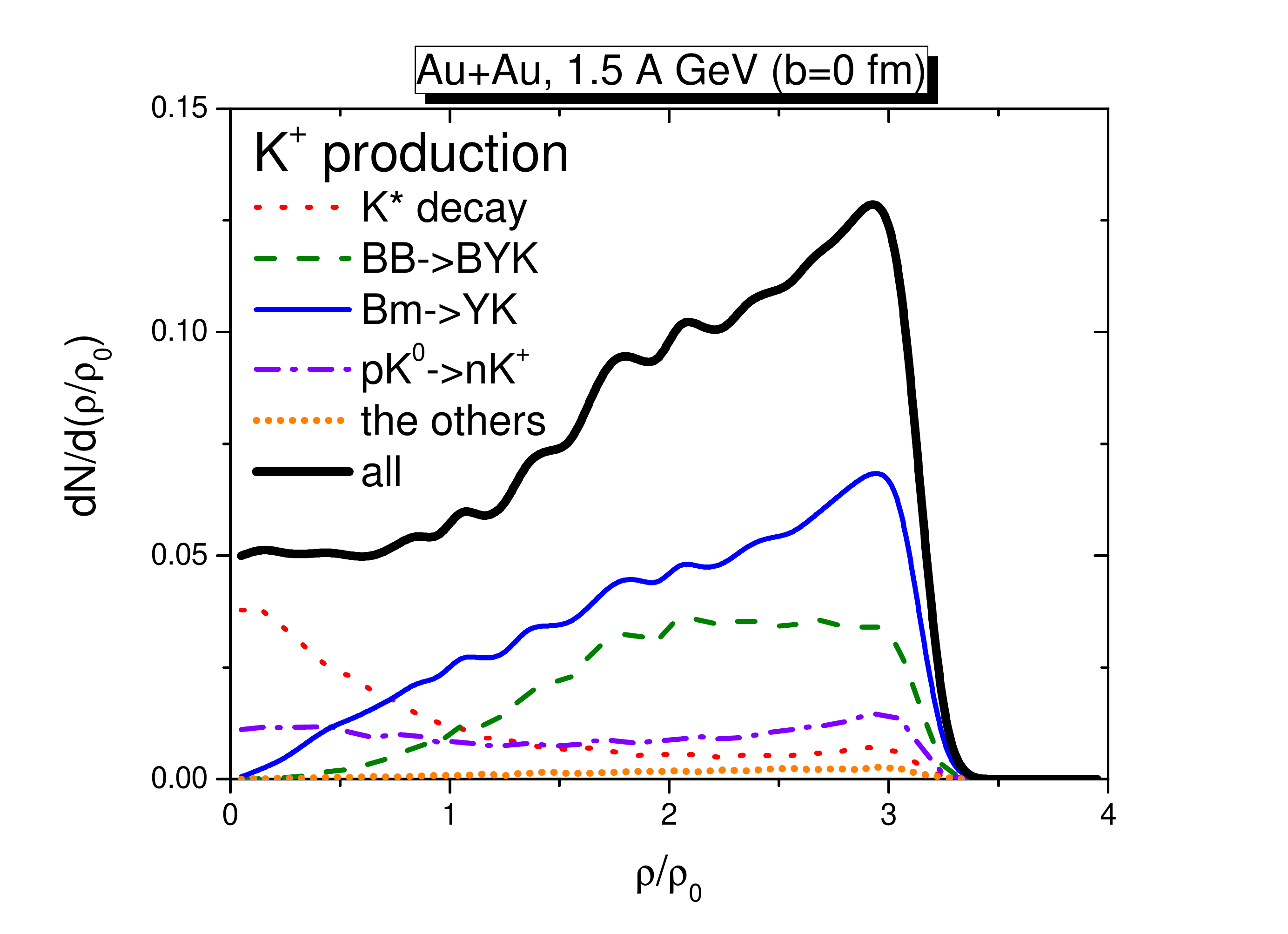}
\includegraphics[width=8.6 cm]{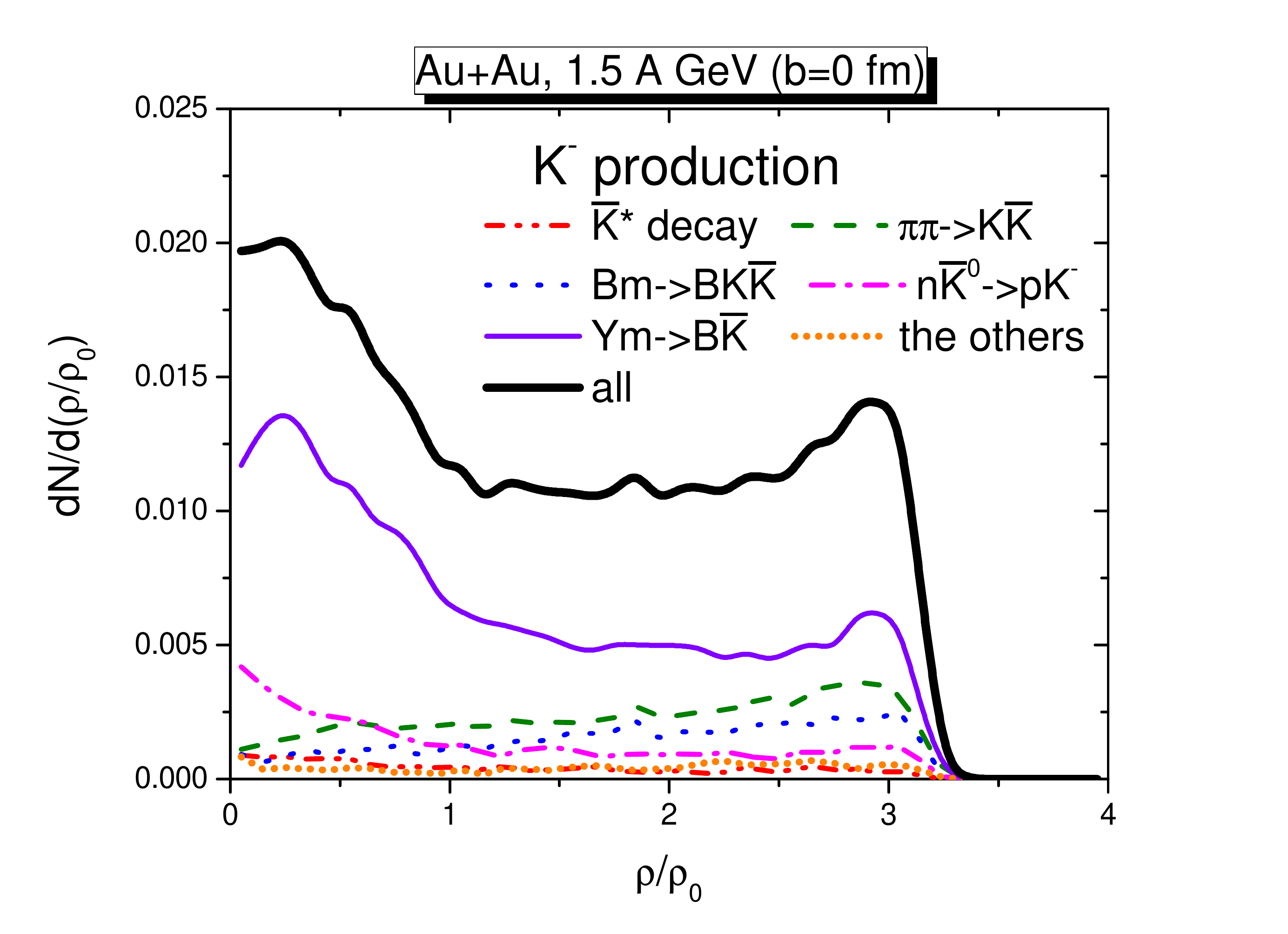}}
\centerline{
\includegraphics[width=8.6 cm]{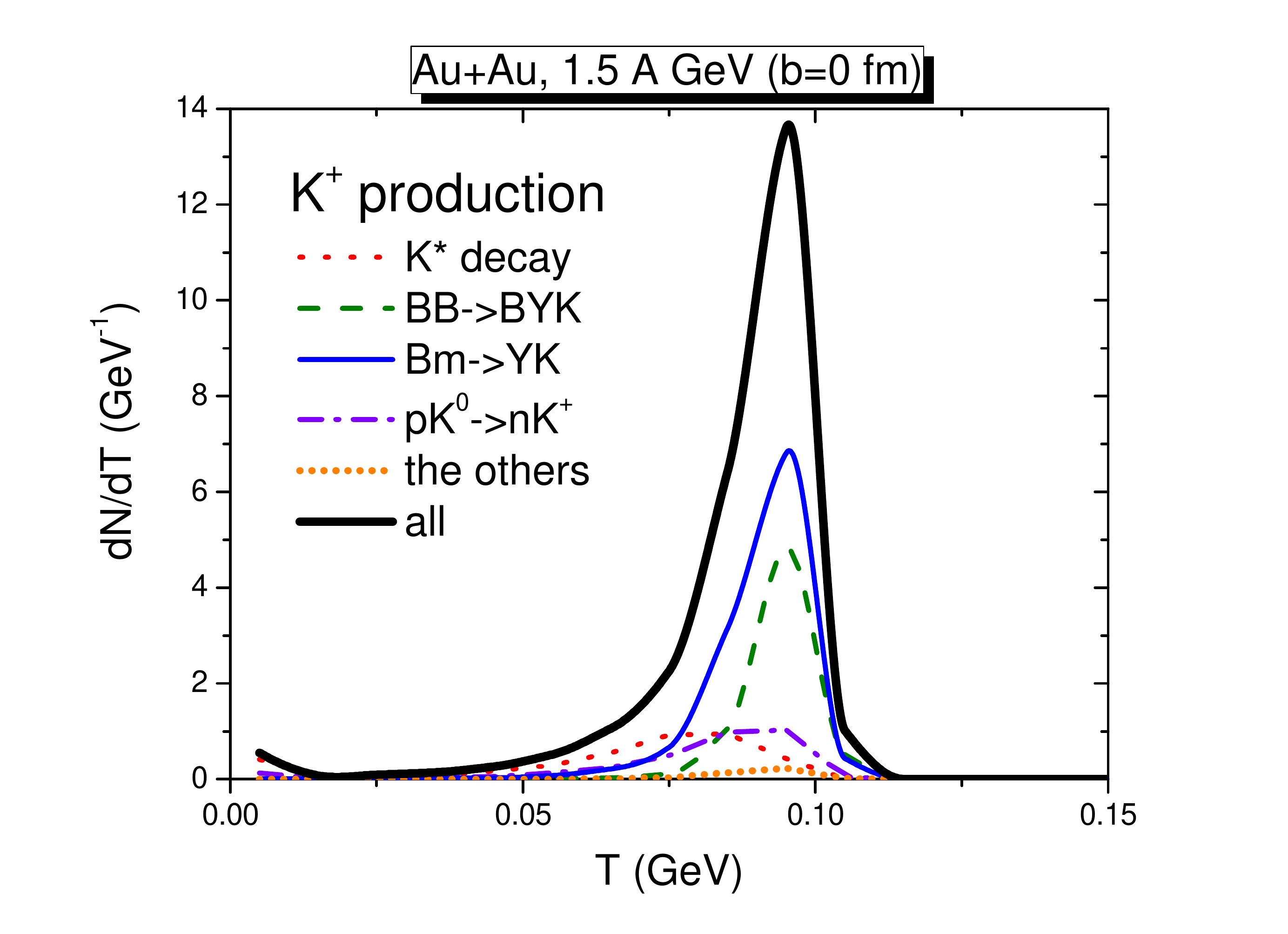}
\includegraphics[width=8.6 cm]{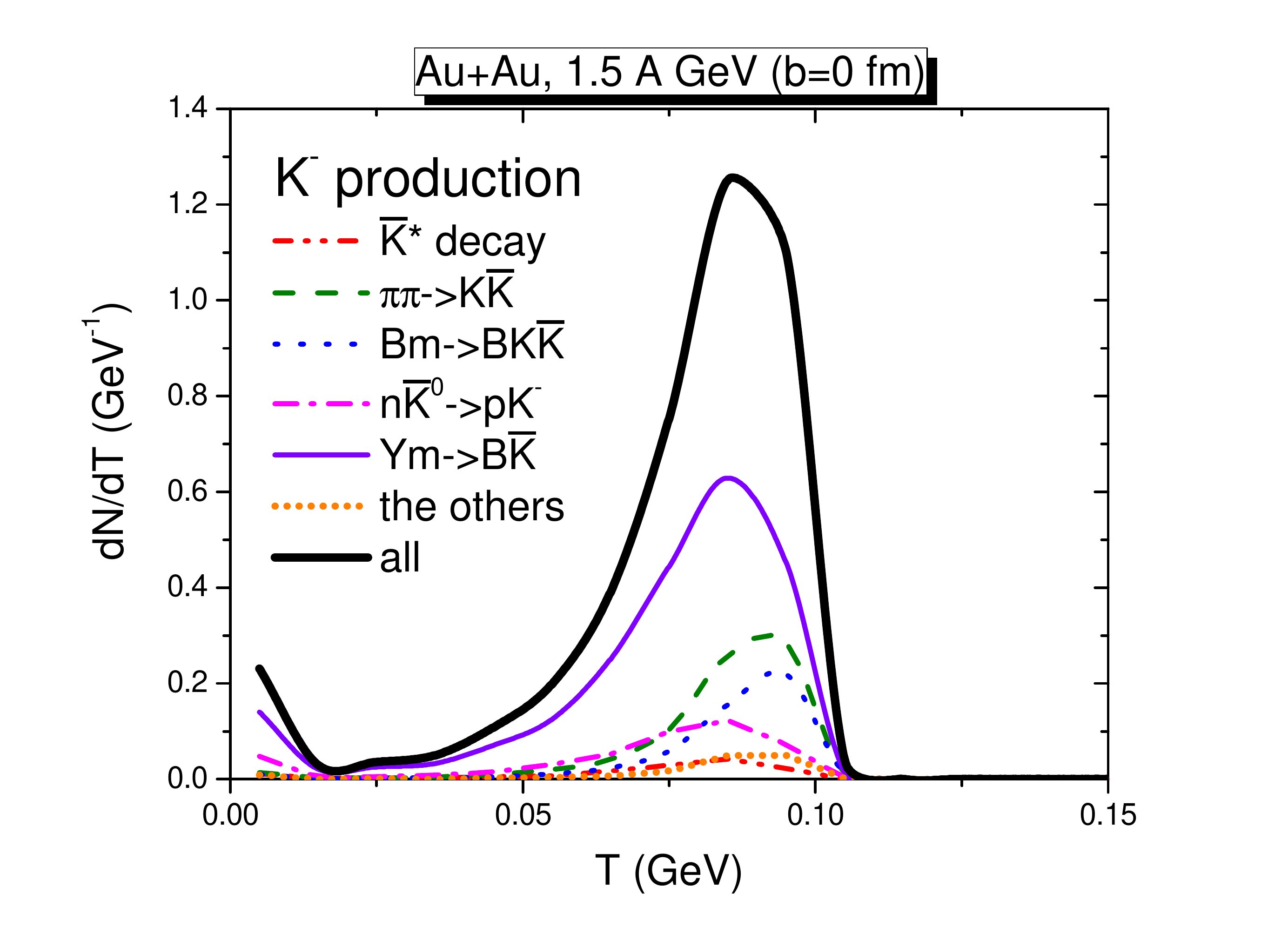}}
\centerline{
\includegraphics[width=8.6 cm]{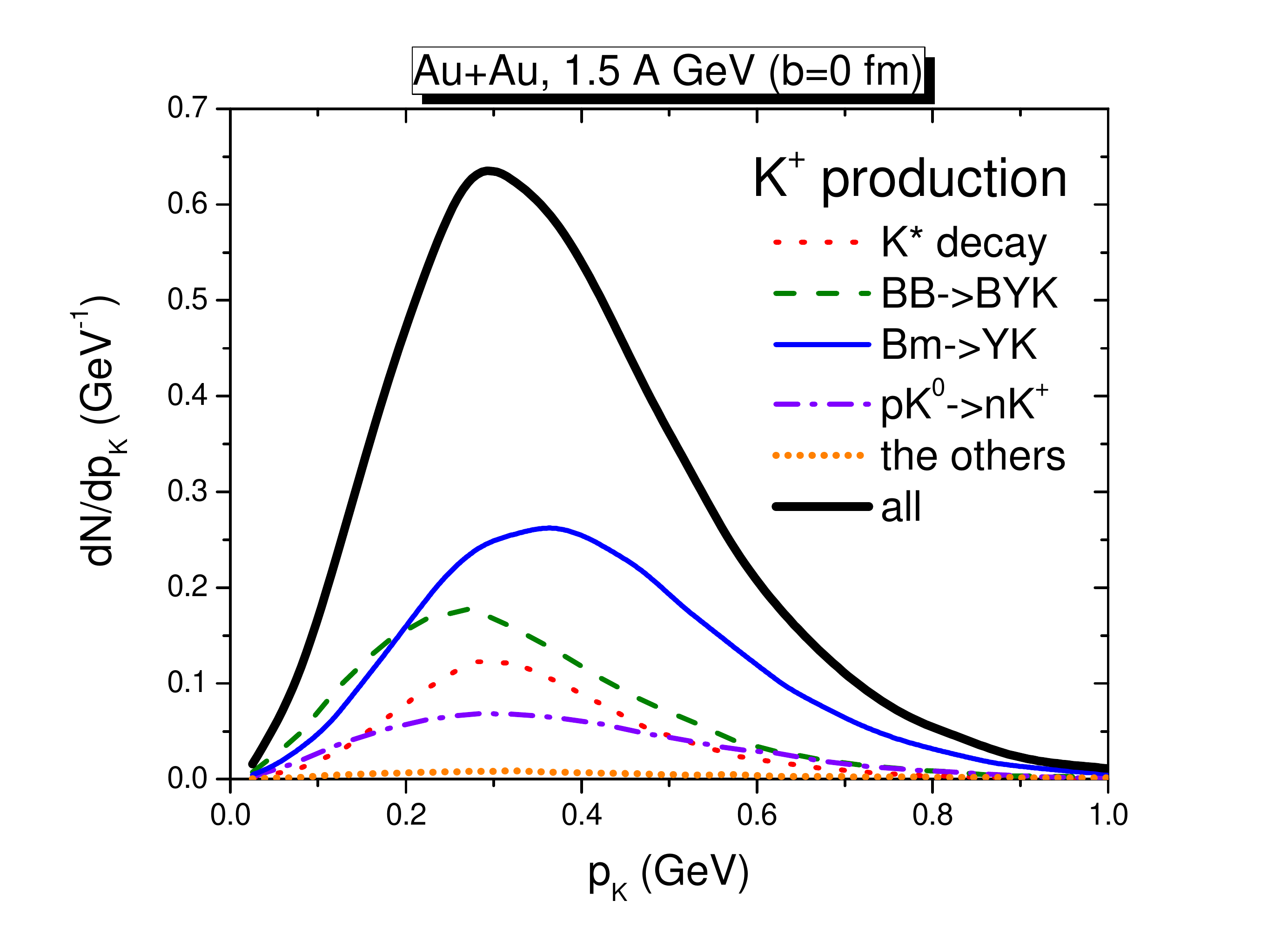}
\includegraphics[width=8.6 cm]{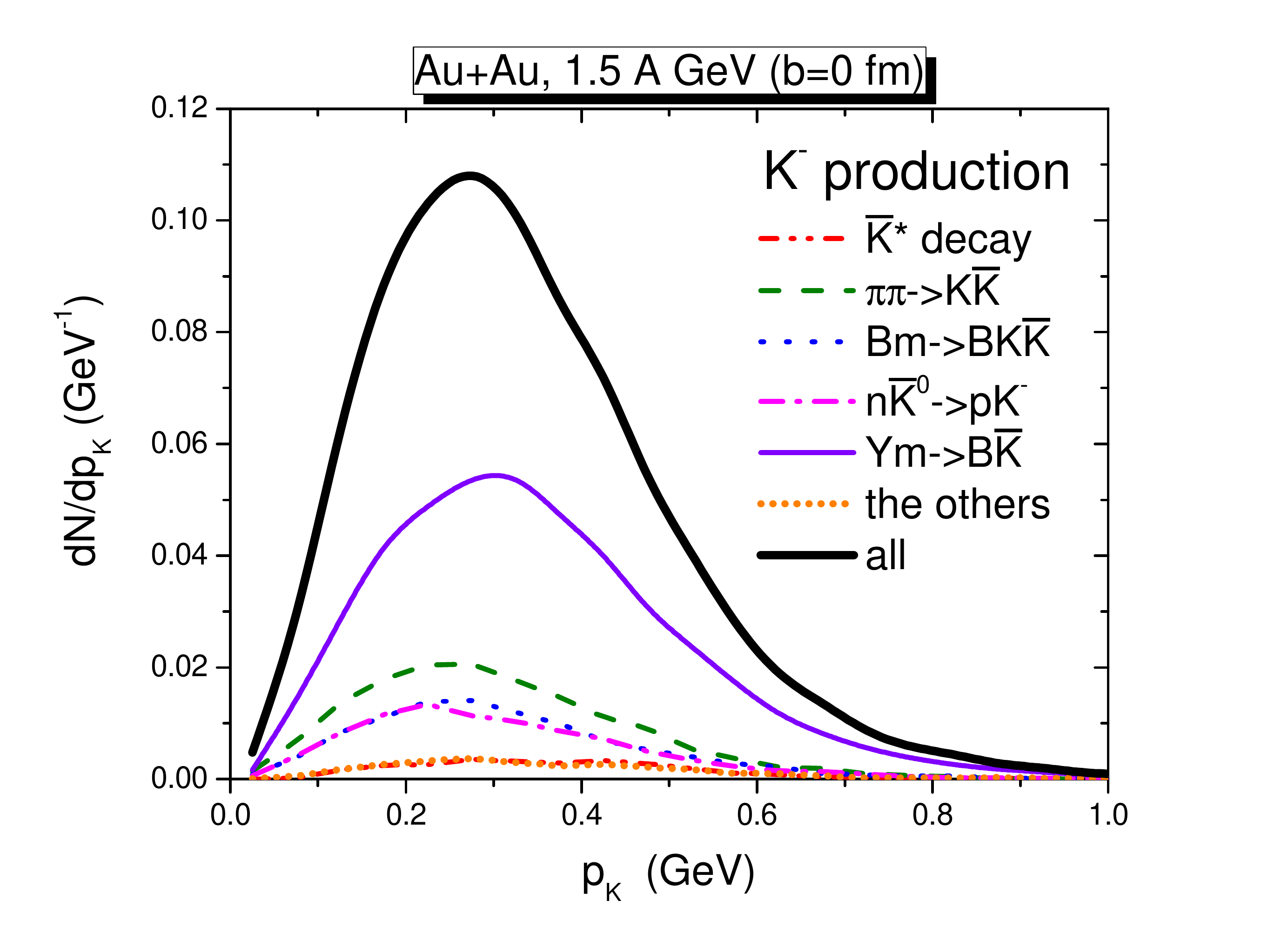}}
\caption{(Color online) The distributions in  baryon density $\rho/\rho_0$ (upper), 
temperature $T$ (middle)  and  momentum $p_K$ in the matter-rest frame (lower) for  $K^+$'s (left)
and  $K^-$'s (right) at their production points in central Au+Au collisions ($b=0$ fm) 
at 1.5 A GeV.
The lines show the individual production channels involving
non-strange baryons $B$, hyperons $Y=(\Lambda,\Sigma)$ and non-strange mesons $m$.
The PHSD calculations include the in-medium effects, i.e.
a repulsive potential for $K^+$ and the self-energy from the G-matrix approach 
for $K^-$.
}
\label{prodk}
\end{figure*}

Figure~\ref{prodk} shows the distributions of baryon density, temperature and momentum 
in the matter-rest frame of $K^+$ and $K^-$ at their production points for individual 
production channels and for their sum (solid black lines) in central Au+Au collisions 
at 1.5 A GeV. 
The PHSD calculations include the in-medium effects, i.e.
a repulsive potential for $K^+$ and the self-energy from the G-matrix approach 
for $K^-$.
The channel decomposition involves 
non-strange baryons $B$, hyperons $Y=(\Lambda,\Sigma)$ and non-strange mesons $m$.
From the upper row of Fig. \ref{prodk} one can see that the density at $K^+$ production 
is larger than at $K^-$ production.
That is due to the fact that the main production channels of $K^+$'s is 
$B+B\rightarrow B+Y+K$ which occurs promptly in heavy-ion collisions while $K^-$'s 
are produced not directly but in two steps, first $B+B\rightarrow B+Y+K$ 
and then $Y+m\rightarrow B+\bar{K}$.
We note that  we consider here central Au+Au collisions at $b=$ 0 fm, where
the pion density is larger than for the minimal bias selection. 
This enhances the contribution from $B+\pi \rightarrow Y+K$ compared to that 
from $B+B\rightarrow B+Y+K$ for the $K^+$ production.
There are many production channels which are not explicitly shown in the figure.
For example, $\pi+\pi\rightarrow K+\bar{K}$ and $B+m\rightarrow B+K+\bar{K}$ 
are only shown for $K^-$ production, though they contribute to both, $K^+$ and $K^-$
production, because their relative contribution to $K^+$ production is small.
The lines 'other' channels in Fig.~\ref{prodk}
for $K^+$ and $K^-$ production include $\phi$ decay, 
$B+B\rightarrow B+B+K+\bar{K}$, and string decay of $B+B$, $B+m$ and $m+m$ 
(if the final state is different from the channels listed in the figure explicitly).
The channels $Y+m\rightarrow \Xi+K$, $\Xi^0+K^0\rightarrow \Xi^- K^+$ contribute 
only to $K^+$ production and $\Xi+K\rightarrow B+\bar{K}$ and $\Xi+m\rightarrow Y+\bar{K}$
only to $K^-$ production are also included in 'the others.'

The strangeness production in heavy-ion collisions at threshold energies 
is dominated by secondary reactions in the hadronic medium, i.e. 
through multi-step processes.
For example, the primary nucleon+nucleon scattering excites either one or both nucleons 
into a $\Delta$, a process which is more favourable than $NN$ to produce strangeness via
$\Delta+N\rightarrow N+Y+K$, because the $\Delta$ is heavier than the nucleon.
Or the $\Delta$ decays into a nucleon and a pion which produces strangeness 
via $B+\pi\rightarrow Y+K$.

\begin{table}
\begin{center}
\begin{tabular}{c|c}
\hline
channel & produced $K^+$ per event \\
\hline
$N+m\rightarrow Y+K$ & 8.79$\times 10^{-2}$\\
$\Delta+m\rightarrow Y+K$ & 2.30$\times 10^{-2}$\\
\hline
$N+N\rightarrow B+Y+K$ & 1.87$\times 10^{-2}$\\
$N+\Delta\rightarrow B+Y+K$ & 3.39$\times 10^{-2}$\\
$\Delta+\Delta\rightarrow B+Y+K$ & 1.22$\times 10^{-2}$\\
\hline
$N+m\rightarrow B+K+\bar{K}$ & 8.79$\times 10^{-4}$\\
$\Delta+m\rightarrow B+K+\bar{K}$ & 8.79$\times 10^{-4}$\\
\hline
\end{tabular}
\end{center}
\caption{The number of produced $K^+$'s (with in-medium effects) per event from several channels 
in central Au+Au collisions ($b=0$ fm) at 1.5 A GeV.}
\label{delta}
\end{table}

In order to quantify the importance of secondary processes involving $\Delta$'s and pions,
we present in Table~\ref{delta} the number of produced $K^+$'s per event in subchannels
of the channels $B+m\rightarrow Y+K$, $B+B\rightarrow B+Y+K$ and 
$B+m\rightarrow B+K+\bar{K}$, depending on whether the $\Delta$ baryon is involved 
in strangeness production or not.
In the case of $B+m\rightarrow Y+K$ about 20 $\%$ of $K^+$ production is induced 
by a $\Delta$ baryon.
On the other hand, $\Delta+N\rightarrow B+Y+K$ is about 80 $\%$ more frequent 
than $N+N\rightarrow B+Y+K$.
However, $\Delta+\Delta\rightarrow B+Y+K$ is kinematically more favourable for $K^+$
production, but it is less frequent since the possibility that two $\Delta$'s  scatter 
is lower.
In the case of $B+m\rightarrow B+K+\bar{K}$ the contribution from $N+m$ and that from 
$\Delta+m$ are similar.
We note that the contribution of each subchannel depends on the collision system and energy
as well as on centrality.
It is worth noting that the contribution from $K^*(\bar{K}^*)$ decay is not negligible for $K^+(K^-)$ production. Thus up to 15 \% for $K^+$ and 3.7 \% for $K^-$ come from
their decays.
Since the decay width of the $K^*(\bar{K}^*)$ is about 50 MeV, their decays 
($K^*\rightarrow K+\pi$ or $\bar{K}^*\rightarrow \bar{K}+\pi$) happen on avarage 4 fm/c after $K^*(\bar{K}^*)$ production.
Therefore the baryon density at $K(\bar{K})$ production through $K^*(\bar{K}^*)$ decays is quite low as follows from  Fig.~\ref{prodk}.
On the other hand, the $\phi$ decay barely contributes to (anti)kaon production: 
0.07 \% for $K^+$ and 0.67 \% for $K^-$.

The middle panels in Fig.~\ref{prodk} show the temperature distributions at the 
$K^+$ and $K^-$ production points.
The temperature is extracted from the local energy density and baryon density 
by using the hadron resonance gas model 
(HRG) which includes all mesons up to 1.5 GeV of mass and all (anti)baryons up to 2 GeV, i.e. in line with the hadronic degrees of freedom of the PHSD. 
A similar HRG model has been used in the past to study the thermal equilibration in infinite hadron-string matter as well as in HICs within 
the BUU \cite{Bratkovskaya:2000qy}, however, here we consider all hadrons "on-shell",
i.e neglect their spectral functions (as realized in most of the statistical models, 
cf. \cite{Cleymans:1992zc}), which might slightly influence the final temperature
\cite{Bratkovskaya:2000qy}. Although the produced nuclear matter is not in complete thermal equilibrium, we calculate local energy density and baryon density in the rest frame of the local cell and map them into those from the HRG model to obtain a corresponding temperature, which is a commonly used method in coarse graining. Indeed, such mapping of non-equilibrium distributions to its equilibrium values can be considered only as a leading order approximation. 
However, we note that the temperature dependence of the antikaon spectral function is relatively
weak compared to its density dependence which washes out the uncertainties related to the extraction of local temperatures within our HRG model.

One can see that the temperature distribution is highly peaked around $T=$ 90-100 MeV for $K^+$ while in the case of $K^-$ it is less sharply peaked around $T=$ 80-90 MeV.
Considering that $K^+$ is earlier produced than $K^-$, this is expected.
In addition, the temperature distribution of $K^*(\bar{K}^*)$ decay is lower than 
for the other channels, since it happens later when the matter is cooling due to
the expansion.

The distributions in three-momentum of the (anti)kaon at its production time 
in the local rest frame of the baryon matter are displayed in the lower panels 
of Fig.~\ref{prodk}.
The figure shows that the three-momentum of $K^+$ is peaked around 300 MeV$/c$ while 
for $K^-$ it is a bit lower.
In general the three-momentum of $K^+$ is slightly larger than that of $K^-$, 
though the distributions are quite similar.
One can also find that the momentum of $K^+$ from $B+m\rightarrow Y+K$ is larger
than that from other channels and $K^-$ from $Y+m\rightarrow B+\bar{K}$ has a larger momentum than that from $\pi+\pi\rightarrow K+\bar{K}$.

\begin{figure}[t!]
\centerline{
\includegraphics[width=8.6 cm]{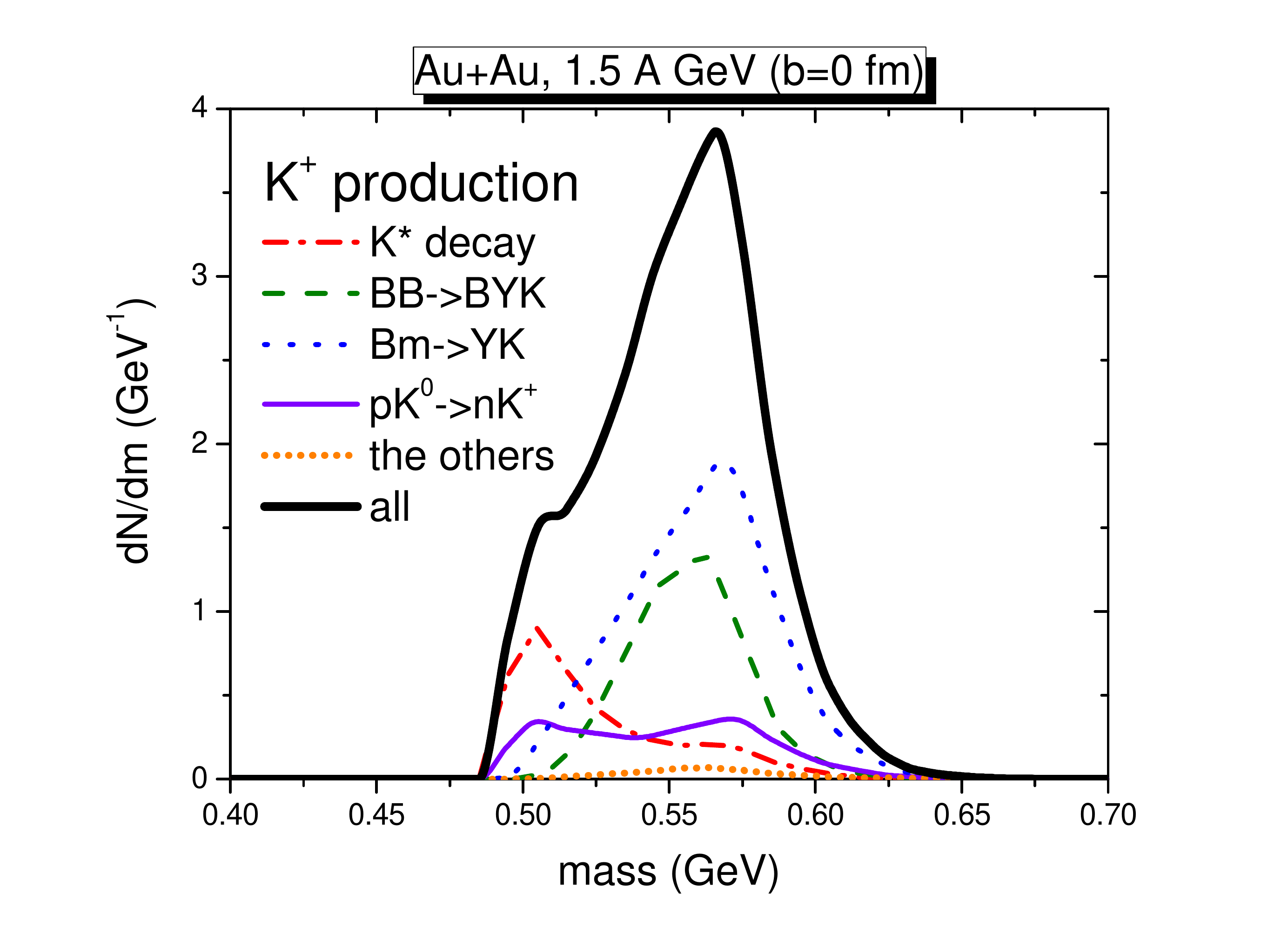}}
\centerline{
\includegraphics[width=8.6 cm]{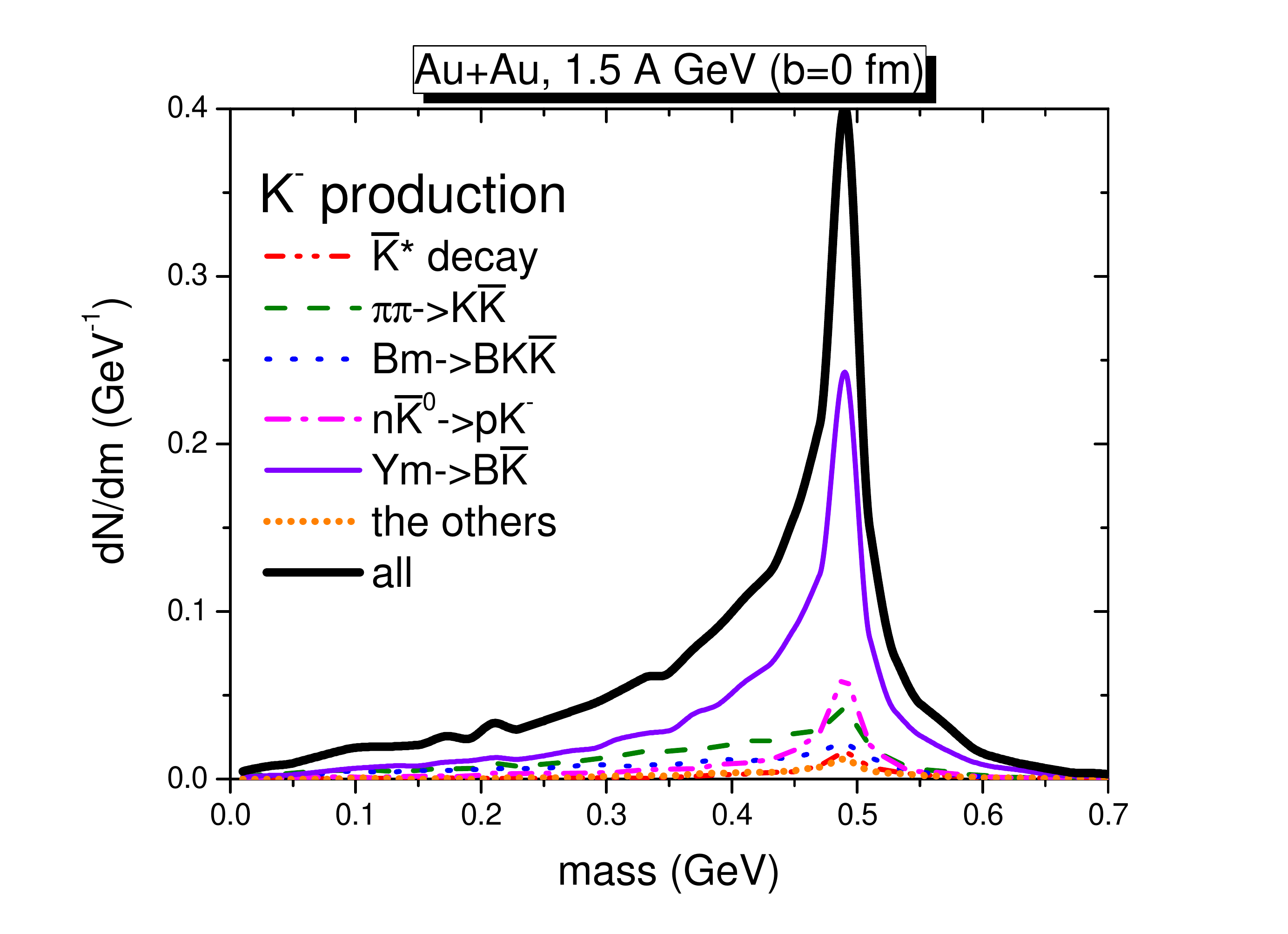}}
\caption{(Color online) The mass distribution of  $K^+$'s (upper) and  $K^-$'s (lower) 
at their production points in central Au+Au collisions ($b=0$ fm) at 1.5 A GeV.
The lines show the individual production channels.
The PHSD calculations  include the in-medium effects for $K^+$ and $K^-$.
}
\label{dist-mass}
\end{figure}

Figure~\ref{dist-mass} shows the mass distributions of $K^+$'s (upper plot) 
and $K^-$'s (lower plot) at their production points in central Au+Au collisions 
($b=0$ fm) at 1.5 A GeV by summing up all production contributions discussed above
with respect to Fig.~\ref{prodk}.
The $K^+$ mass in the medium mainly depends on the baryon density as follows 
from Eq.~(\ref{kpmass}).
Considering that the baryon density at $K^+$ production is peaked around 
$3\rho_0$ (as follows from  Fig.~\ref{prodk}),
it is consistent that the $K^+$ mass is distributed around 560-570 MeV in 
Fig.~\ref{dist-mass}.
We note that the $K^+$ mass from the $K^*$ decay is relatively low because this 
decay happens later at low baryon densities.

On the other hand, the $K^-$ mass depends on baryon density, temperature and 
the three-momentum of the antikaon in the rest frame of the baryonic medium 
through the self-energy of the antikaon.
Figure~\ref{prodk} shows that the $K^-$ is produced in a wide range of baryon densities 
from 0 to $3\rho_0$ and at temperatures between 50 MeV and 100 MeV with 
the three-momentum peaking at 200-400 MeV$/c$.
As follows from Fig.~\ref{self-energy}, the absolute value of the real part of the $K^-$ 
self-energy rapidly decreases with increasing antikaon momentum and is less 
than 20 MeV for $p=$ 300 MeV$/c$, while the imaginary part is not much affected by the 
change of momentum.
Therefore, the pole mass of the $K^-$ spectral function changes only slightly 
and the spectral width broadens as shown in Fig.~\ref{dist-mass}.

\begin{figure}[t!]
\centerline{
\includegraphics[width=8.6 cm]{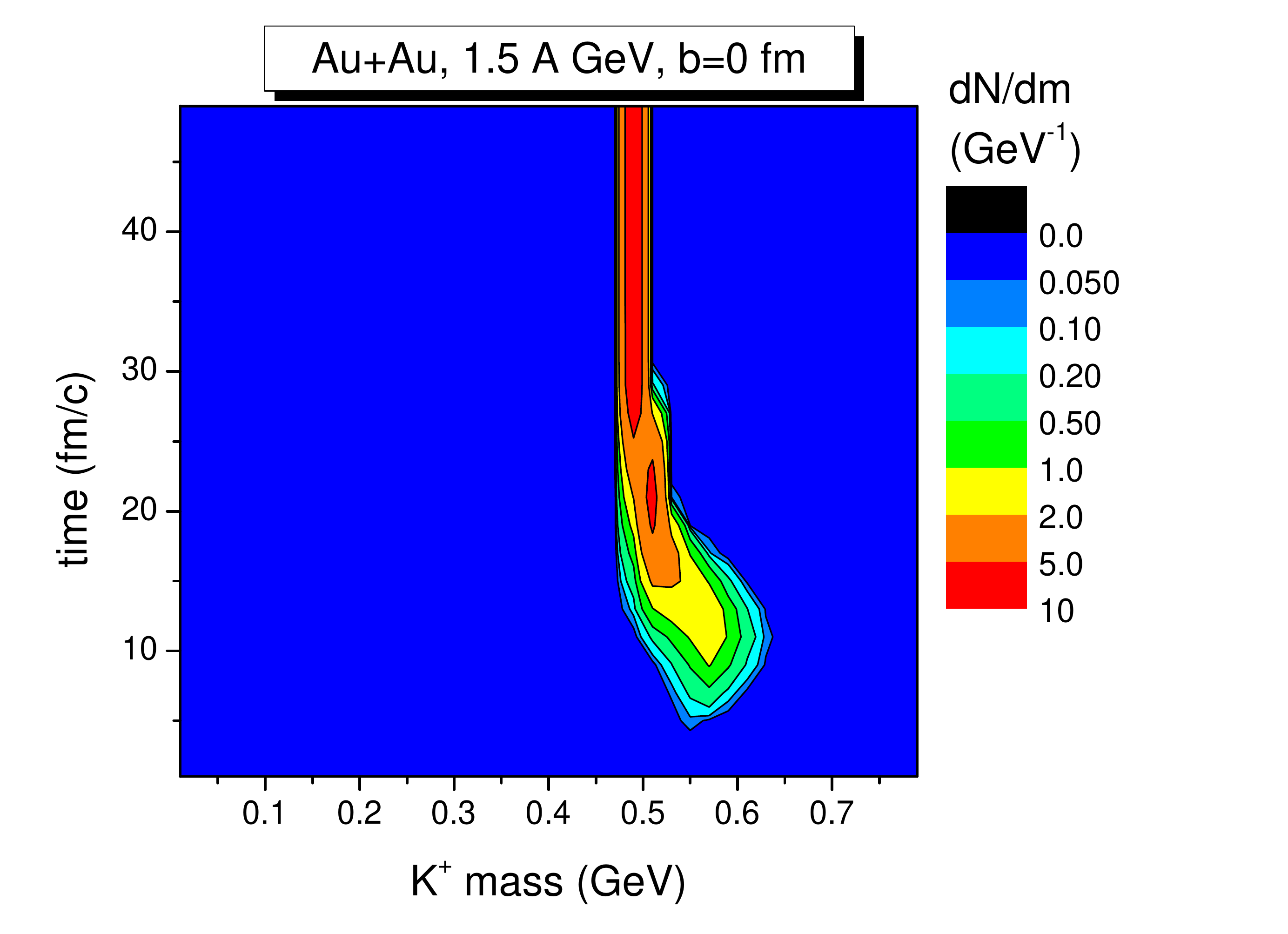}}
\centerline{
\includegraphics[width=8.6 cm]{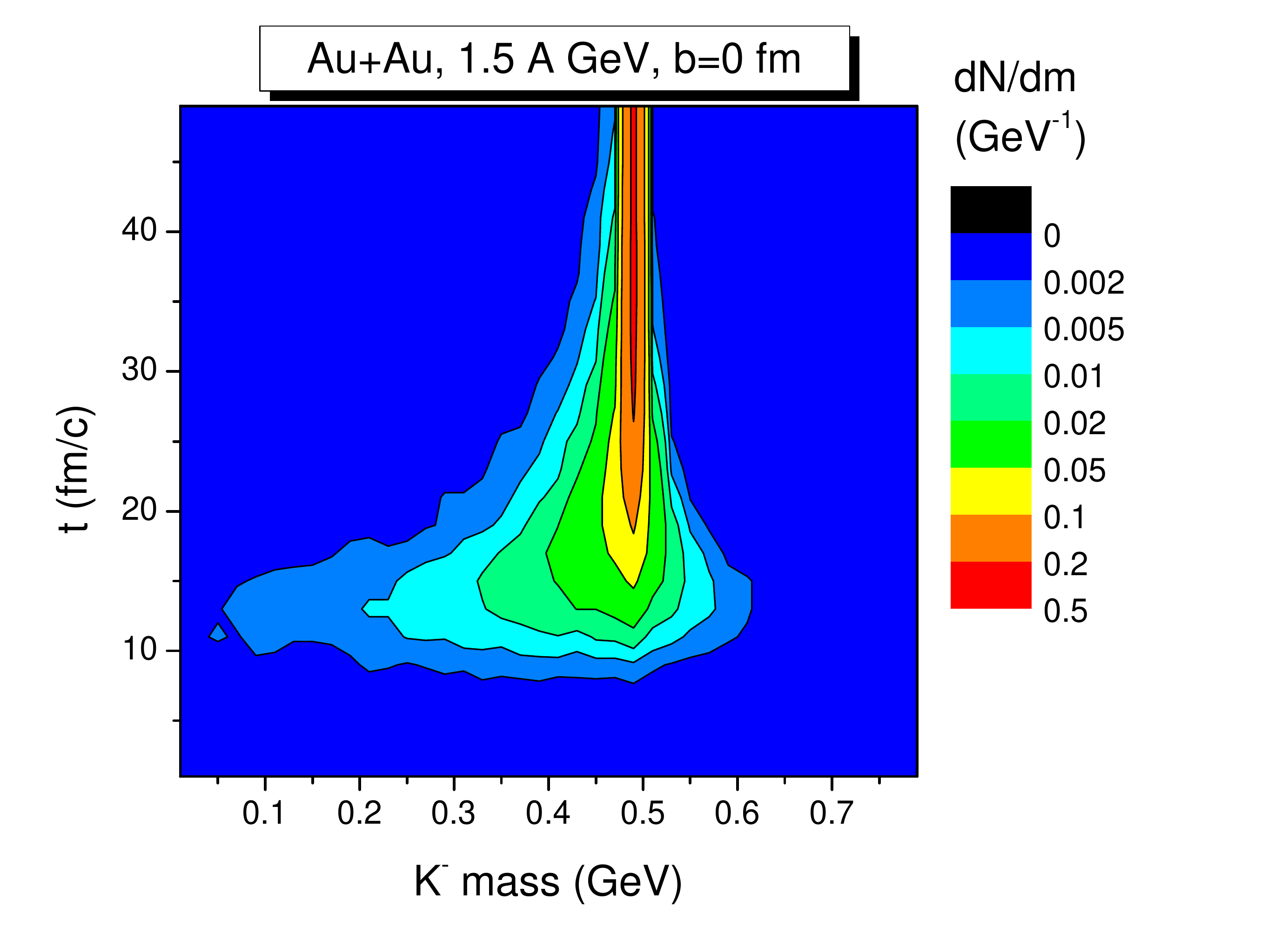}}
\caption{(Color online) The contour plot for the mass distributions of $K^+$'s (upper) 
and $K^-$'s (lower) as a function of time in central Au+Au collisions ($b=0$ fm) 
at 1.5 A GeV.
The PHSD calculations include the in-medium effects for $K^+$ and $K^-$.
}
\label{time-evolution}
\end{figure}

Figure~\ref{time-evolution} shows the contour plot for the mass distributions 
of $K^+$'s (upper plot) and $K^-$'s (lower plot) as a function of time for  
central Au+Au collisions ($b=0$ fm) at 1.5 A GeV.
While at the production times the masses of $K^+$ and $K^-$ are highly off-shell 
they approach dynamically (according to the off-shell propagation) to the 
on-shell masses following the decrease of the baryon density and temperature with time  according to Eqs.~(\ref{mass-update}) and (\ref{kpmass}), respectively.
As displayed in Fig.~\ref{dist-mass}, the $K^+$ mass is initially larger 
than the vacuum mass while the pole position of the $K^-$ mass is close to the 
vacuum mass with a large spreading in spectral width towards lower masses.
The decrease of the baryon density with time leads to a reduction of the off-shell
$K^+$ mass to the vacuum mass and shrinks the $K^-$ spectral function such that 
the width vanishes in vacuum.

\begin{figure}[t!]
\centerline{
\includegraphics[width=8.6 cm]{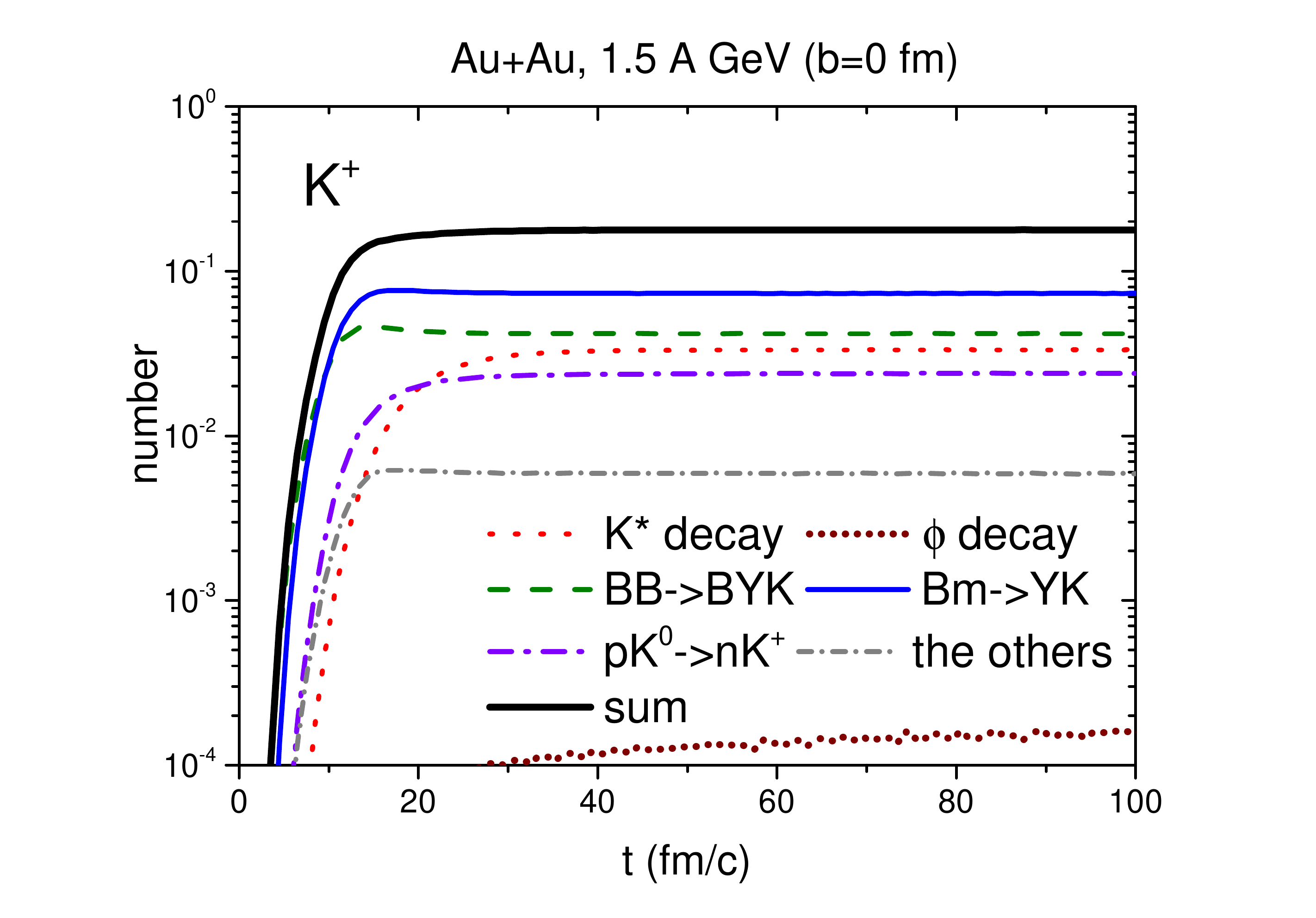}}
\centerline{
\includegraphics[width=8.6 cm]{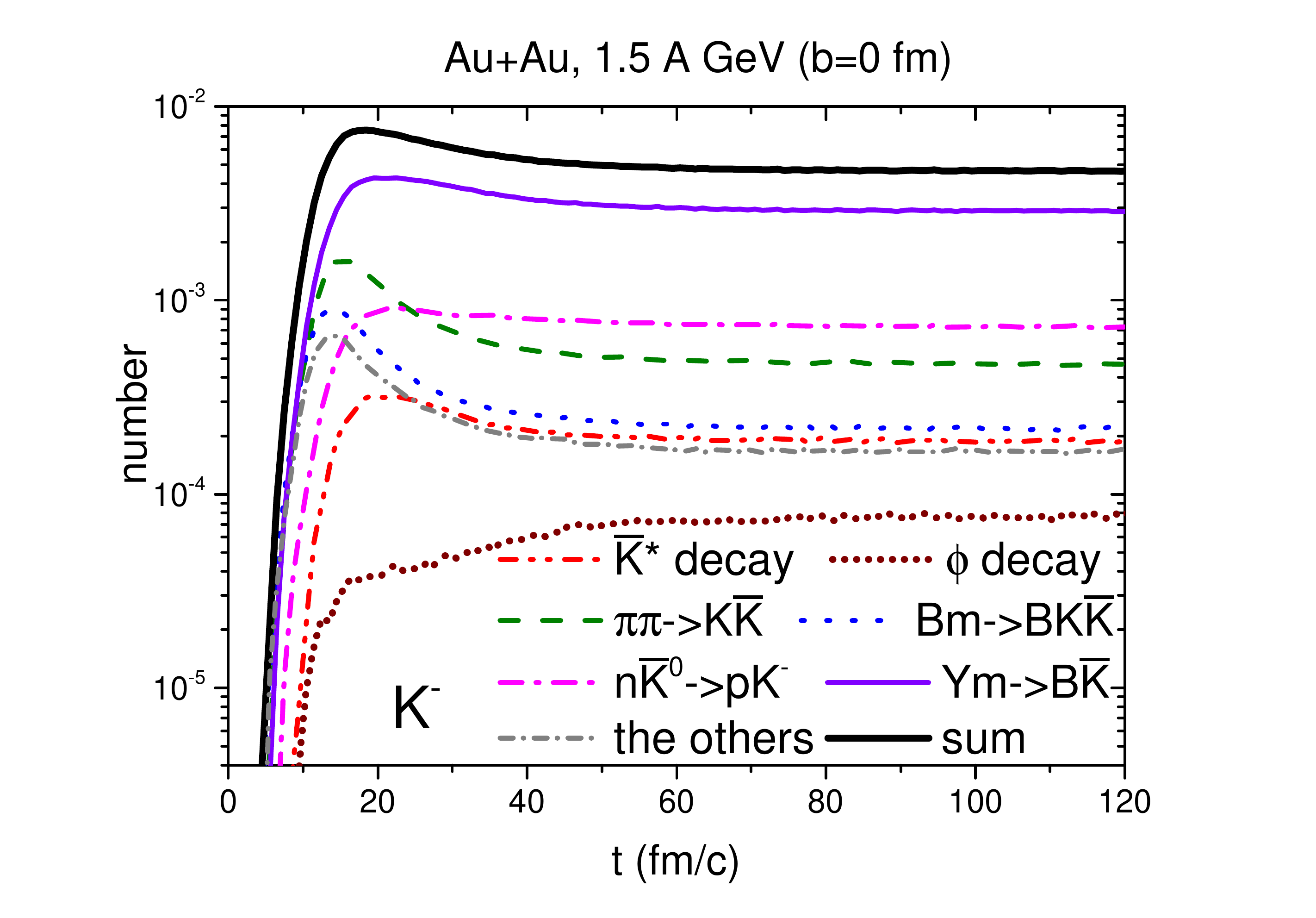}}
\caption{(Color online) The channel decomposition for $K^+$ (upper) and $K^-$ (lower) 
production as a function of time in central Au+Au collisions ($b=0$ fm) at 1.5 A GeV.
The PHSD calculations include the in-medium effects for $K^+$ and $K^-$.
}
\label{sources}
\end{figure}

We show in Fig.~\ref{sources} the channel decomposition of the $K^+$ and $K^-$ 
production as a function of time in central Au+Au collisions ($b=0$ fm) at 1.5 A GeV where
the color lines indicate the dominant production channels.
One can see that the $K^+$ production via the $B+B\rightarrow B+Y+K$ reaction 
takes place early, dominantly due to  initial $N+N$ scattering. 
The channels involving secondary particles - mesons or baryons - come with
some delay, since their production also needs some time.
The $B+m\rightarrow Y+K$ channel involves produced mesons;
the isospin exchange reactions ($p+K^0\rightarrow n+K^+$) can happen only after 
kaons ($K^0$) are produced;  the $K^*$ decays come even with a larger delay
due to their later production.

In the case of $K^-$'s, the reactions $Y+m\rightarrow B+\bar{K}$, 
$\pi+\pi\rightarrow K+\bar{K}$, and $B+m\rightarrow B+K+\bar{K}$ take place 
at a similar time after meson production, dominantly pions from $\Delta$ decays.
After that the isospin exchanges ($n+\bar{K}^0\rightarrow p+K^-$) and then $K^*$ decays 
are involved.
As follows from Fig.~\ref{sources}, the dominant channel for $K^-$ production is
a meson-hyperon reaction which is in line with previous findings 
\cite{Barz:1985xc,Cass97,Hartnack:2011cn}.
We note that many antikaons are absorbed after production because the absorbtion
cross section of antikaons by nucleons is very large (cf.  Fig.~\ref{check}).
That is why the $K^-$ masses  in Fig. \ref{time-evolution} bent down after initial strong rise.

\begin{table}[h!]
\begin{center}
\begin{tabular}{c|c|c}
\hline
channel & produced & survived \\
\hline
total $K^+$ & 2.57$\times 10^{-1}$ & 1.68$\times 10^{-1}$\\
$B+B\rightarrow B+Y+K$ & 6.48$\times 10^{-2}$ & 3.92$\times 10^{-2}$ \\
$B+m\rightarrow Y+K$ & 1.13$\times 10^{-1}$ & 6.90$\times 10^{-2}$ \\
$p+K^0\rightarrow n+K^+$ & 3.15$\times 10^{-2}$ & 2.25$\times 10^{-2}$\\
$K^*\to \pi+K$ decay & 3.89$\times 10^{-2}$ & 3.14$\times 10^{-2}$ \\
\hline
total $K^-$ & 4.10$\times 10^{-2}$ & 4.13$\times 10^{-3}$ \\
$B+m \rightarrow B+K+\bar{K}$ & 4.57$\times 10^{-3}$ & 1.85$\times 10^{-4}$\\
$n+\bar{K}^0\rightarrow p+K^-$ & 4.65$\times 10^{-3}$ & 6.19$\times
10^{-4}$\\
$Y+m\rightarrow B+\bar{K}$ & 2.16$\times 10^{-2}$ & 2.61$\times 10^{-3}$\\
$\pi+\pi\rightarrow K+\bar{K}$ & 7.42$\times 10^{-3}$ & 3.69$\times
10^{-4}$\\
$\bar{K}^*\to \pi +\bar K$ decay & 1.41$\times 10^{-3}$ & 2.0$\times 10^{-4}$\\
\hline
\end{tabular}
\end{center}
\caption{The numbers of produced and survived $K^+$ and $K^-$ mesons per event from 
several channels in central Au+Au collisions ($b=0$ fm) at 1.5 A GeV.
The PHSD calculations are done including the in-medium effects for $K^+$ and $K^-$.
}
\label{survival}
\end{table}

\begin{figure*}[th!]
\centerline{
\includegraphics[width=8.6 cm]{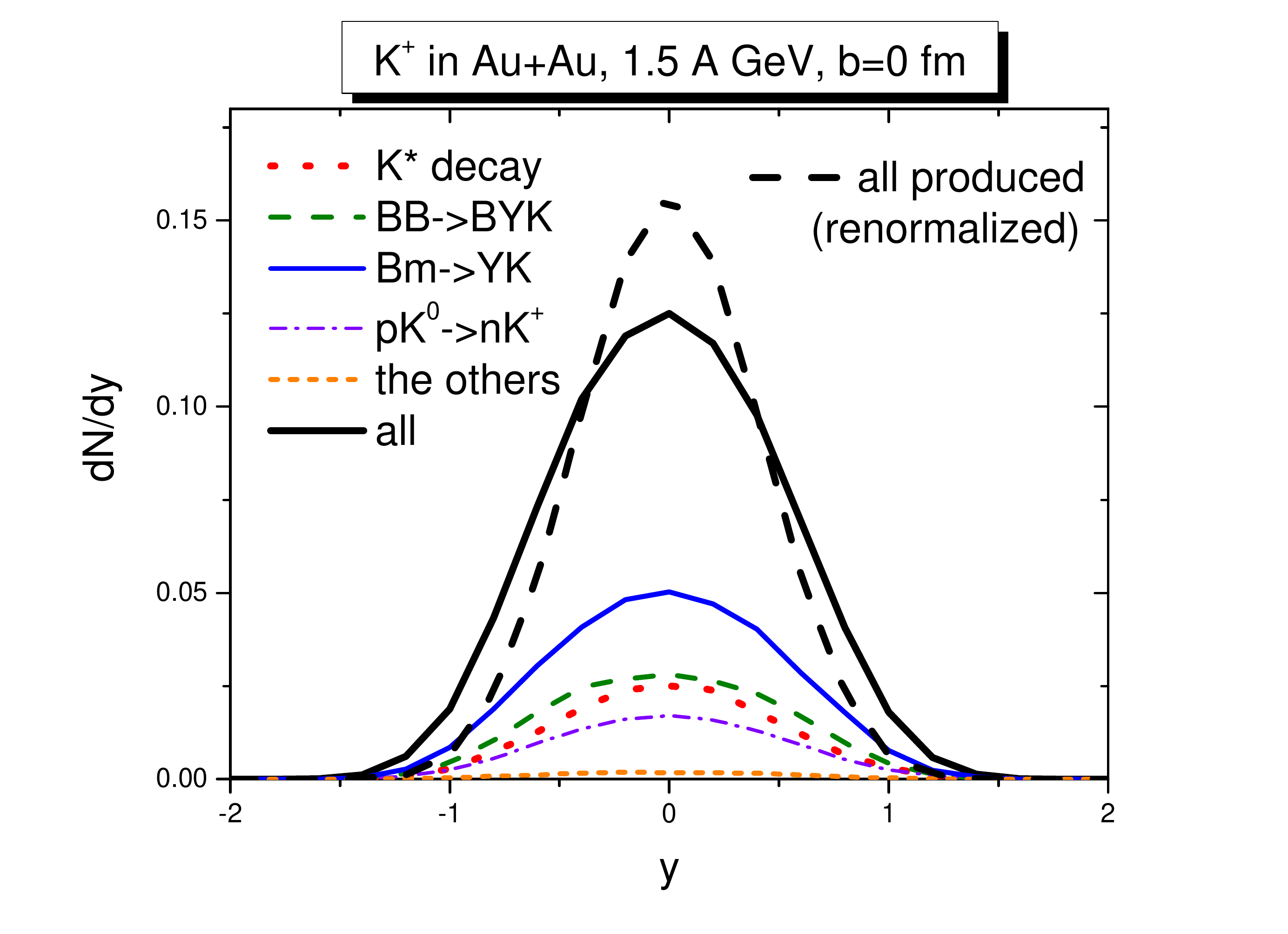}
\includegraphics[width=8.6 cm]{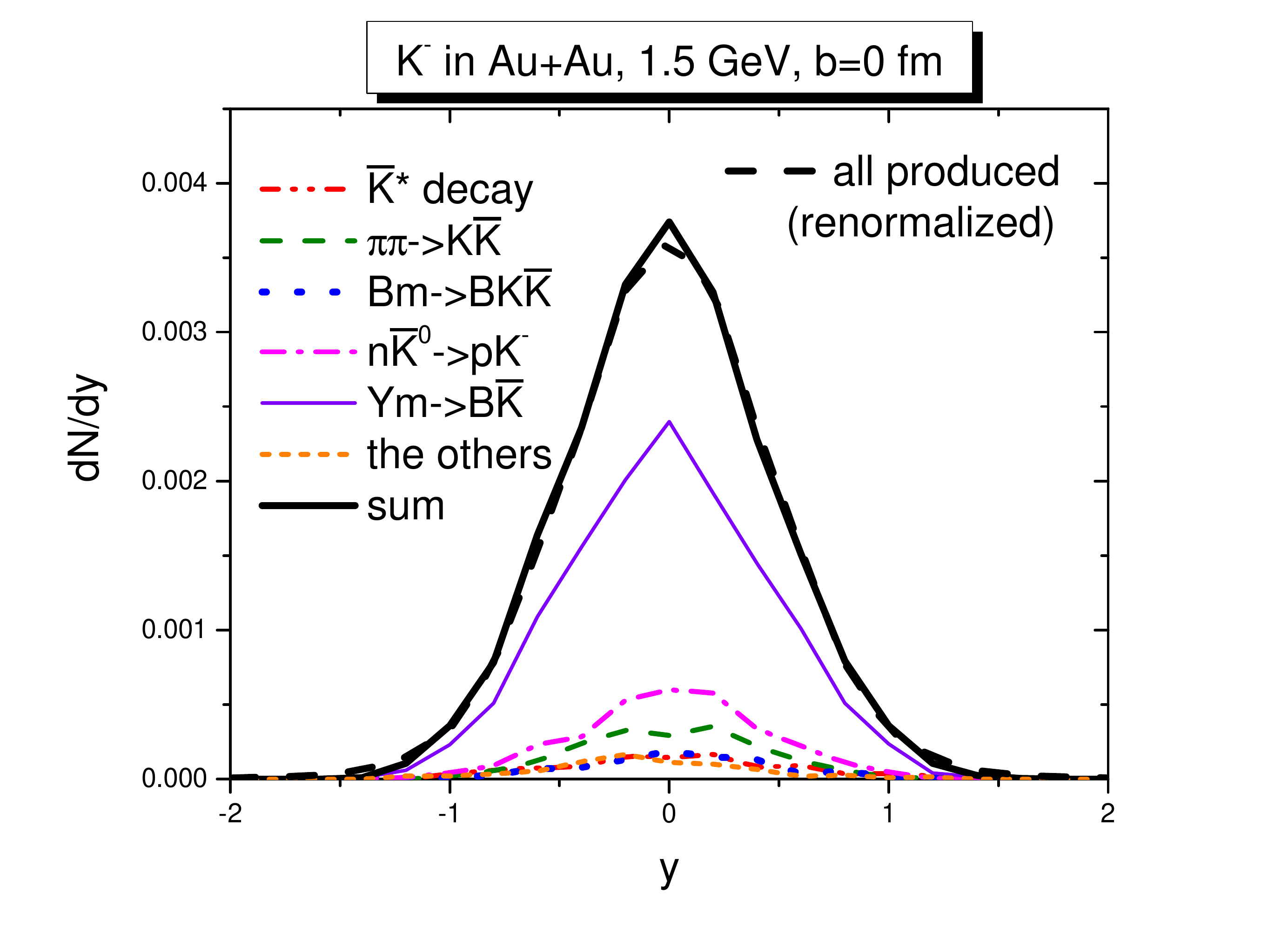}}
\centerline{
\includegraphics[width=8.6 cm]{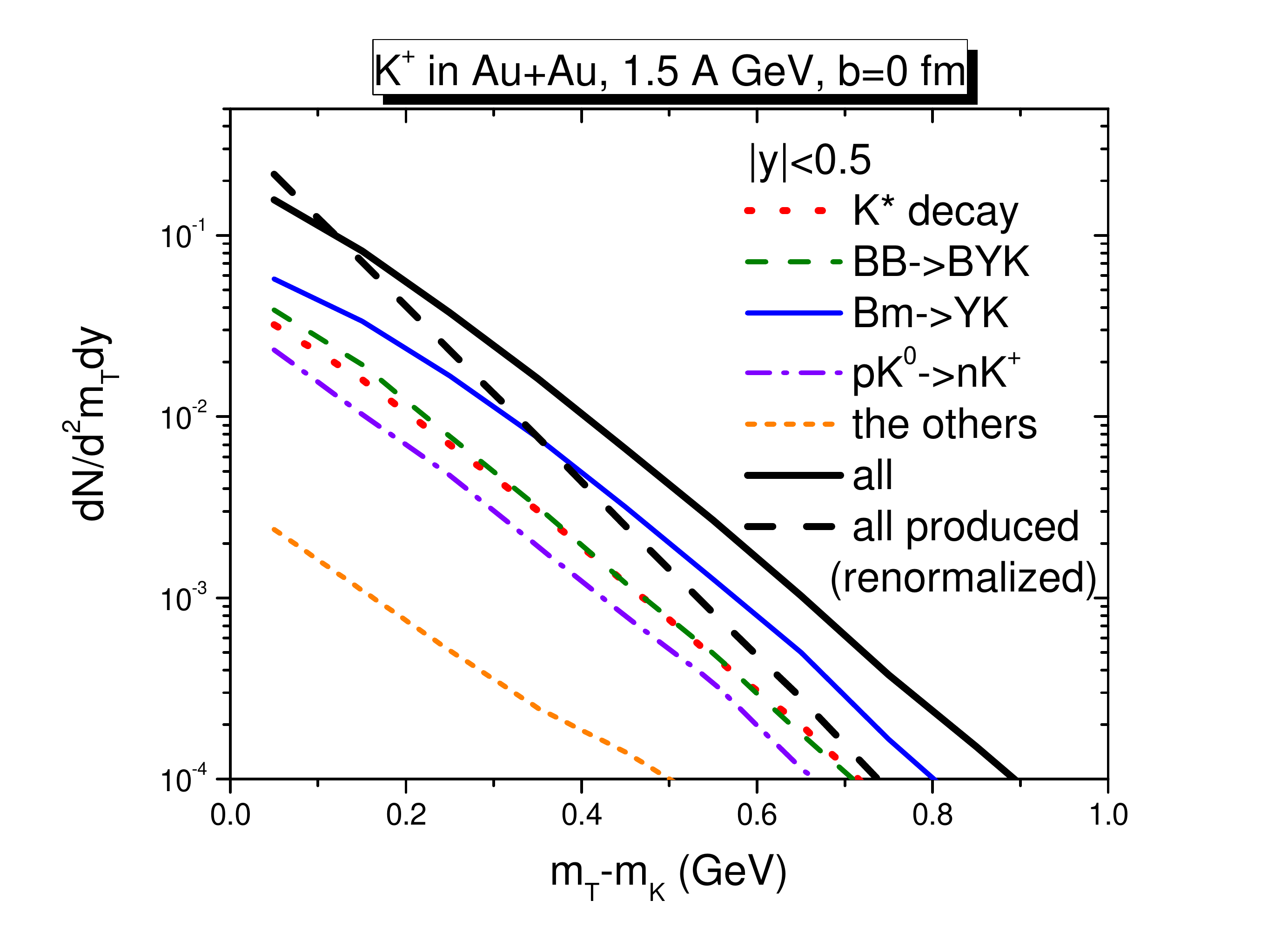}
\includegraphics[width=8.6 cm]{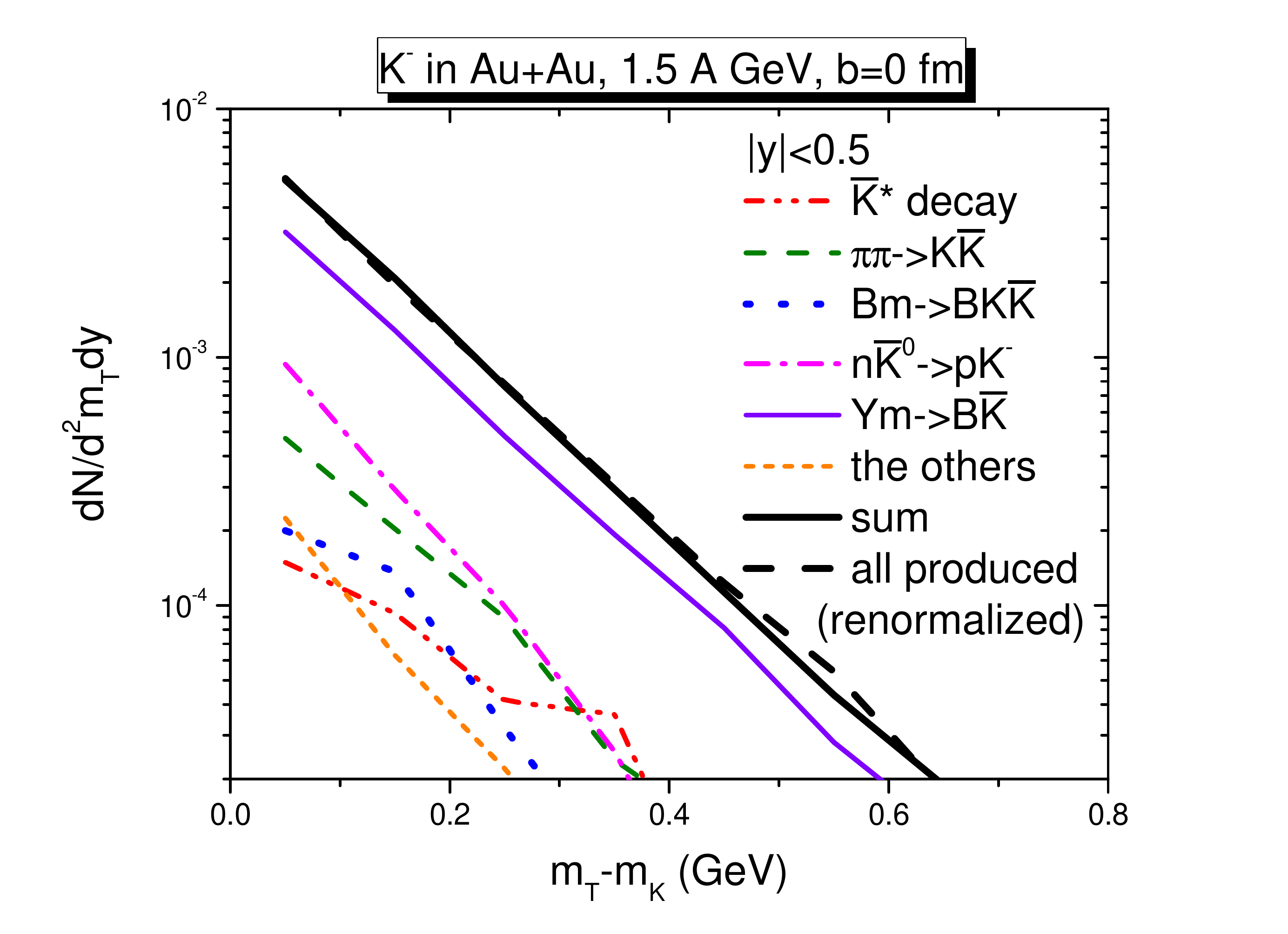}}
\caption{(Color online) The rapidity distributions (upper) and  $m_T$-spectra (lower) 
of $K^+$ (left)  and  $K^-$ (right) mesons at freeze-out (i.e. survived) 
in central Au+Au collisions ($b=0$ fm) at 1.5 A GeV.
The colored lines show the different production channels. The black dashed line
show the rapidity distribution of $K^+$ and $K^-$ at the production point normalized 
to the total number of survived (anti)kaons (solid black lines). 
The PHSD calculations include the in-medium effects for $K^+$ and $K^-$.
}
\label{components}
\end{figure*}

Table~\ref{survival} lists the number of produced and surviving $K^+$'s and $K^-$'s 
for each production channel per event in central Au+Au collisions at 1.5 A GeV.
One can see that about 65 \% of produced $K^+$ mesons survive, while only 10 \% of $K^-$ remain and the other $K^-$ are absorbed by nucleons or switched to $\bar{K}^0$ by isospin exchange.
The survival probability of $K^+$ from $K^*$ decay is relatively large and amounts 
to 81 \%, because the $K^*$ decays happen later (at low baryon density)
than for other production channels.

We note that (anti)kaon production in Table~\ref{survival} as well as in 
Figs.~\ref{prodk} and \ref{dist-mass} does not only include primary production but 
also secondary and third productions, e.g. a $K^+$ turns to a $K^0$ 
by isospin exchange and then returns to $K^+$ by another isospin exchange. 
In this case $K^+$ production is counted twice.


Figure~\ref{components} shows the rapidity distribution and $p_T$-spectra within 
the rapidity window $|y|<0.5$ of $K^+$ and $K^-$ mesons at freeze-out in central 
Au+Au collisions at 1.5 A GeV. 
The integrated yields correspond to the numbers in the right column ('survived') 
of Table~\ref{survival}. 
The black dashed lines show the rapidity distribution of $K^+$ and $K^-$ at 
the production point normalized to the total number of surviving (anti)kaons 
(solid black lines). The colored lines indicate the different production channels
of the surviving $K^+, K^-$  mesons.
One can see that the rapidity distribution of surviving $K^+$ mesons is broader than
that of produced $K^+$ while the rapidity distribution of surviving $K^-$ has
practically the same shape as at production.

The lower panels of Fig.~\ref{components} show the $m_T$-spectra 
of $K^+$ and of $K^-$. Again the colored lines indicate the different 
production channels of surviving $K^+, K^-$  mesons. One can see that the
production channels have slightly different shapes, however, the final shape
is related to the dominant channels: $B+m\to YK$ for  $K^+$ mesons and 
$Y+m\to B+\bar K$  for $K^-$ mesons. 
The final $m_T$-spectrum (black solid line) of  $K^+$ mesons is harder than that at the  production point (black dashed line), while 
the final $K^-$ spectrum becomes slightly softer. 

We note (without showing the results explicitly) that the channel decomposition depends on the system size:
the lighter the colliding system, the larger is the role of primary 
baryon-baryon collisions for $K^+$ production. 
For a light system as C+C, the production of $K^+$
by the $BB\to BYK$ reactions exceeds that by the $Bm\to YK$ reactions
which is dominant channel for Au+Au collisions at the same energy of 1.5 A GeV - cf. Fig.~\ref{components}. 
The $K^-$ mesons are dominantly produced by secondary reactions involving mesons
and hyperons. Such the $Ym\to B\bar K$ channel is dominant for the C+C
collisions similar to the Au+Au collisions.
The total yield is, however,  substantially reduced due to the low abundances 
of produced hyperons and mesons in C+C collisions 
as compared to more heavy systems.

\begin{figure}[th!]
\centerline{
\includegraphics[width=8.6 cm]{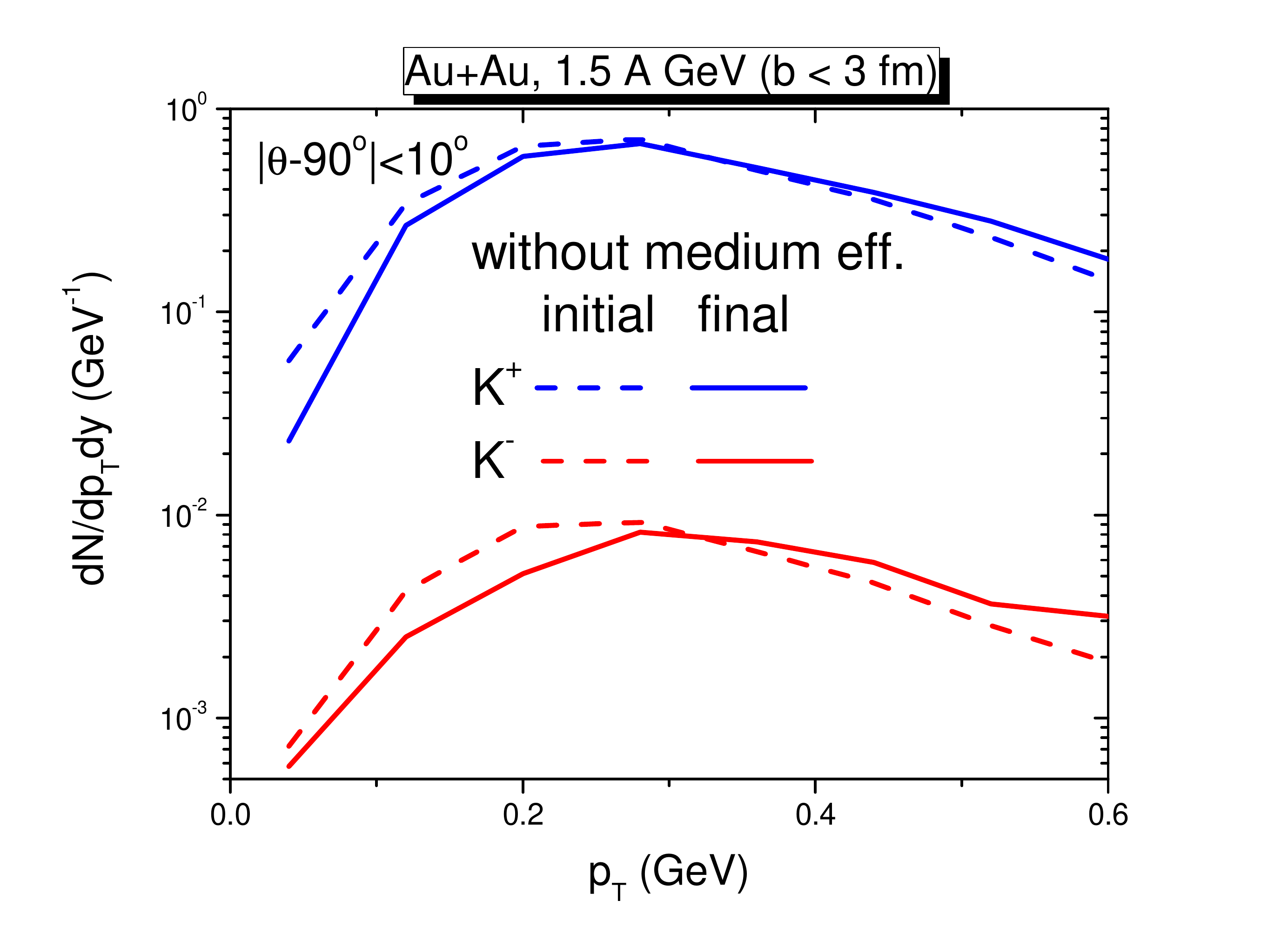}}
\centerline{
\includegraphics[width=8.6 cm]{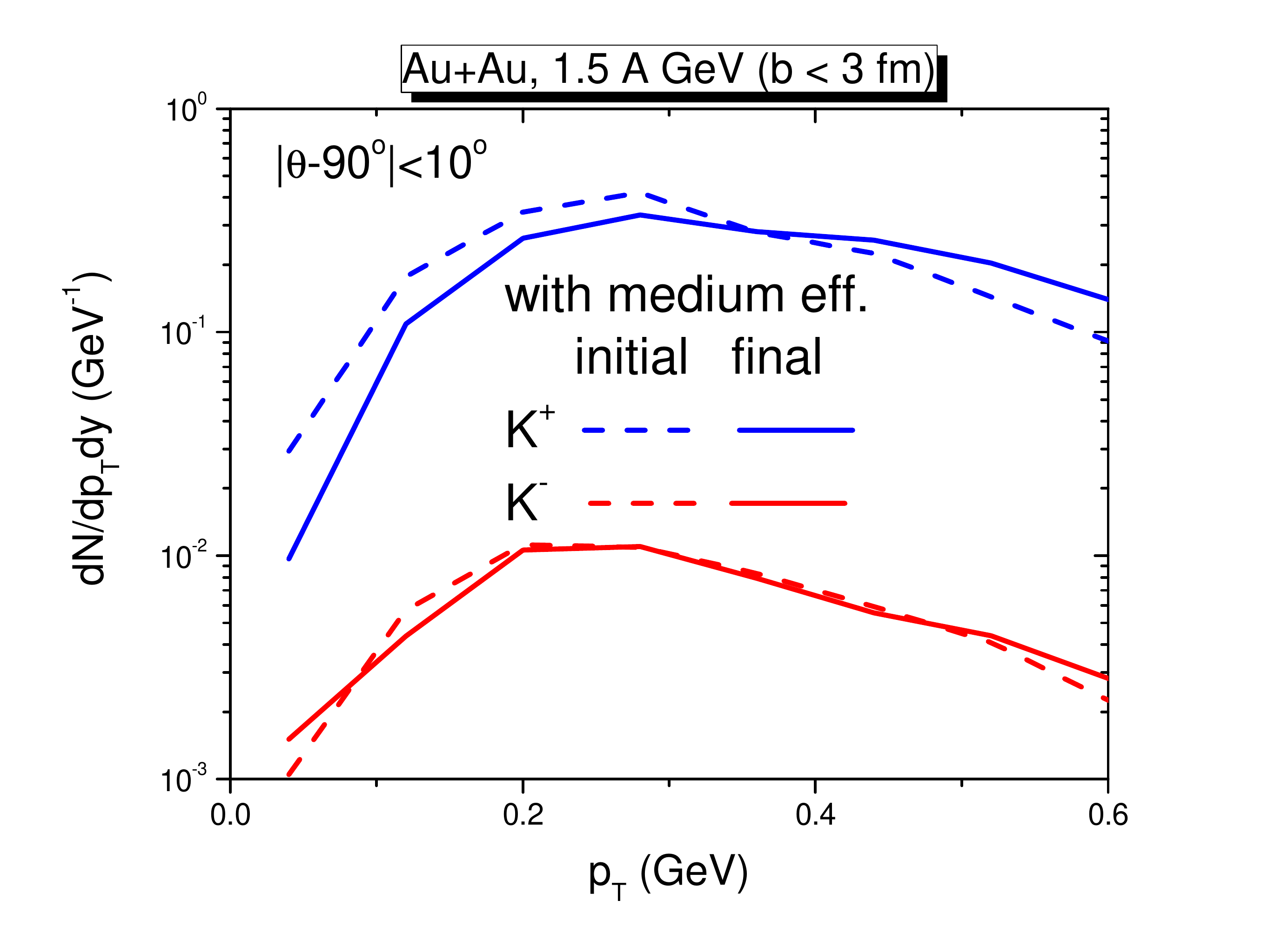}}
\caption{(Color online) Comparison of initial (dashed lines) and final 
(solid lines) $p_T$-spectra of $K^+$ (upper blue lines) and $K^-$ (lower red lines) mesons
without (upper plot) and  with (lower plot) in-medium effects 
(repulsive potential for $K^+$ and the self-energy within the G-matrix approach for $K^-$) in central 
Au+Au collisions ($b<3$ fm) at 1.5 A GeV including the angular cut $|\theta-90^o|<10^o$.
Here the number of produced (anti)kaons is normalized to the number of surviving
(anti)kaons in each plot.}
\label{pt-shift}
\end{figure}

The modification of the final rapidity and $m_T$-distributions of 
the $K^+$ and $K^-$ mesons has two origins  \cite{Hartnack:2011cn} -
the rescattering  in the hadronic medium and the in-medium effects in terms of a
repulsive potential for $K^+$ and  self-energies within the G-matrix approach for $K^-$.
The influence of each  effect is quantified  in Fig.~\ref{pt-shift}:
The upper plot displays the $p_T$- distributions of produced (initial) 
i) and surviving (final) $K^+$ and 
$K^-$ mesons without in-medium effects  in central Au+Au collisions at 1.5 A GeV 
at mid-rapidity (which corresponds to the angular cut $|\theta-90^o|<10^o$).
Since the number of produced (anti)kaons is different than for surviving (anti)kaons, 
the initial distribution is normalized to the number of surviving (anti)kaons for easy comparison.
One can see that both $p_T$- distributions of the produced $K^+$ and $K^-$ are shifted 
to larger transverse momenta due to scattering/absorption in the medium with baryons and mesons. We note that the $K^-$ absorption is stronger for slow $K^-$.
ii) The lower panel of Fig.~\ref{pt-shift} shows the PHSD results 
including the kaon potential and the self-energy within the G-matrix approach for $K^-$.
One sees that the $p_T$- distribution of the surviving $K^+$ is further shifted 
to larger $p_T$ due to the repulsive forces.
On the other hand, the $p_T$- distribution  of surviving $K^-$ is shifted back 
to small $p_T$ because of the attractive  potential related to the real part
of antikaon self-energy and partially due to the broadening of the spectral 
function related to the imaginary part of the self-energy (as will be also discussed 
in Section VII.A). As a result, the final $p_T$- distribution
turns out to be not far from the $p_T$- distribution of the produced $K^-$,
as has been already seen in the $m_T$-spectra of $K^-$
in Fig. \ref{components} (low, right).
We note also  that -as demonstrated in Fig. \ref{components} - the 
shape of the $K^-$ $m_T$- (or $p_T$-) distribution follows the shape of
the dominant production channel  $Y+m\to B+\bar K$ 
(cf. also Fig. \ref{sources}) which involves the secondary particles,
i.e. such processes happen with a time delay when the density of
the system is decreasing due to the expansion. Consequently, the $K^-$ mesons
feel a relatively lower density by propagation in the medium - 
cf. Fig. \ref{prodk}. Then, one should expect larger medium effects for 
$K^-$ for heavy systems where the baryon density and the size of the fireball are larger. This will be demonstrated in the next section.

\section{Comparison with experimental data}\label{compare}

In this Section we compare our results on strangeness production 
with experimental data from the KaoS, FOPI and HADES Collaborations at SIS energies
and demonstrate the influence of different effects on the observables such 
as the in-medium modifications of (anti)kaon properties  
in terms of the kaon-nuclear potential for kaons and the self-energy within the G-matrix approach for antikaons - as well as scattering effects
and the influence of the equation-of-state.

We compare here two scenarios: with and without medium effects in the following sense:
\begin{itemize}
\item  {\it with medium effects}: \\
i) for kaons ($K^+, K^0$) we  include a repulsive density-dependent 
potential which leads to a shift of the kaon mass to a larger value and thus 
to a shift of the production threshold, which leads to a modification 
of kaon production and its interaction cross sections - cf. Section V.C.\\
ii) for antikaons ($K^-, \bar K^0$) the off-shell dynamics includes the
full density and temperature dependent self-energy within the G-matrix approach: 
the modification of the antikaon masses in line with the in-medium spectral function, 
off-shell dispersion relations including complex self-energies, in-medium cross sections for the production and interactions as well as $\bar K$ off-shell propagation 
- cf. Sections II and V.A,B.
\item {\it without medium effects}: \\
i) for kaons ($K^+, K^0$) we employ on-shell dynamics without a kaon potential 
and use  free cross sections - cf. Section IV.\\
ii) for antikaons ($K^-, \bar K^0$) we employ on-shell dynamics using the free production cross sections in the coupled channels - cf. Sections II and IV.
\end{itemize}
Since strangeness is produced always in pairs of $(s,\bar s)$, 
the kaon and antikaon (and hyperon) dynamics is coupled.
The in-medium effects are applied for  kaons and antikaons simultaneously.

\subsection{$y$-distributions}

\begin{figure*}[th!]
\includegraphics[width=5.7 cm]{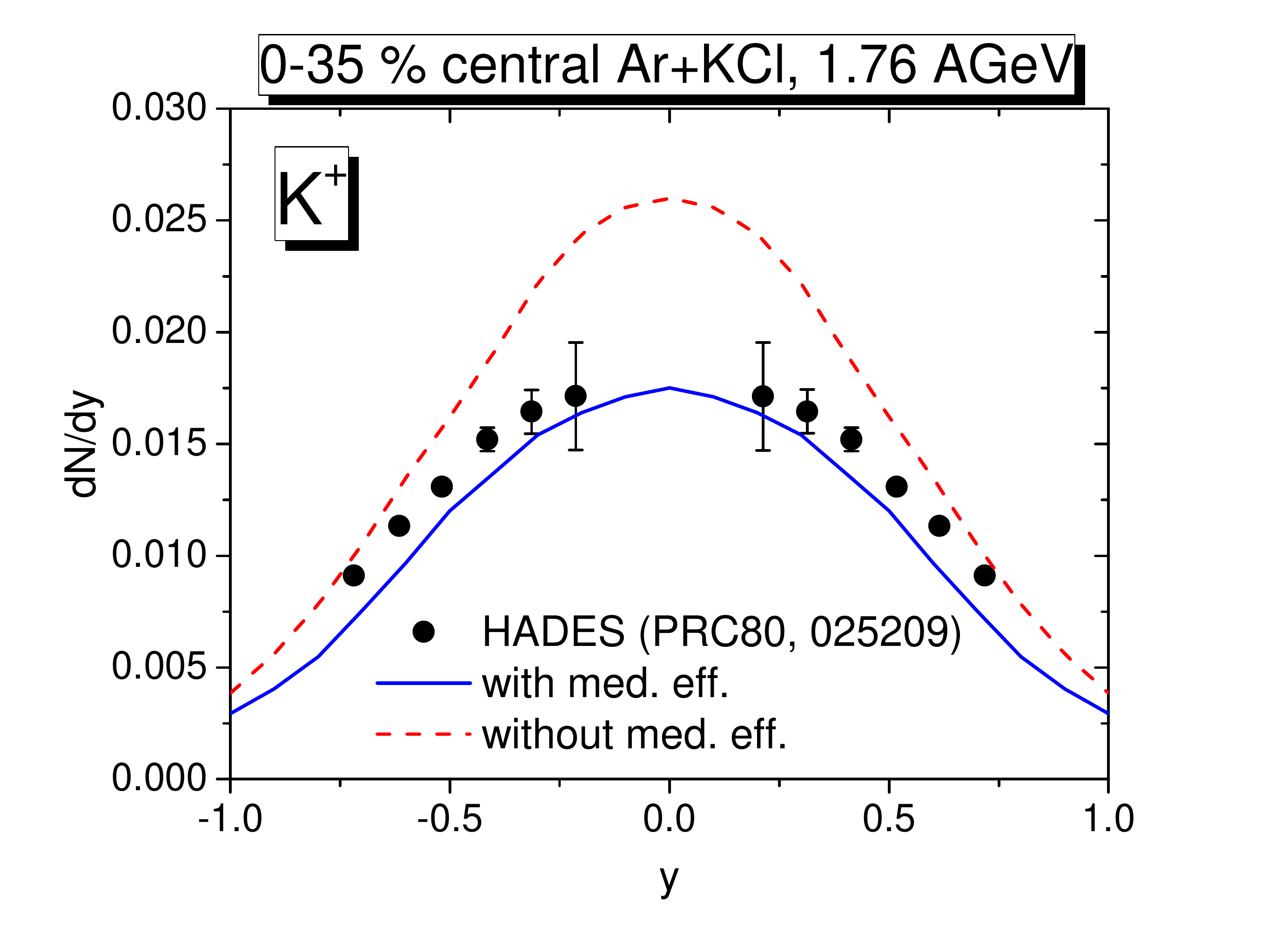}
\includegraphics[width=5.7 cm]{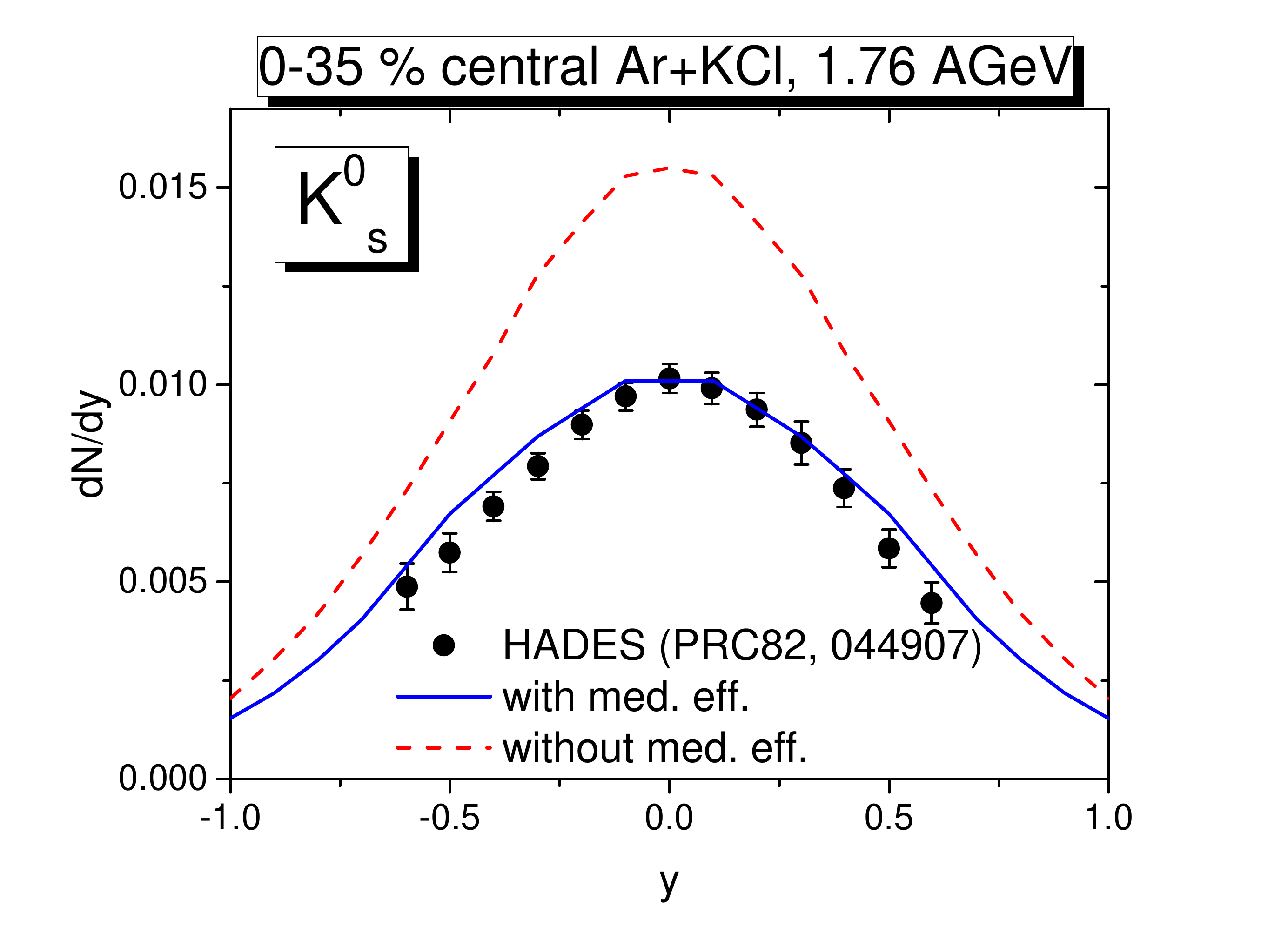}
\includegraphics[width=5.7 cm]{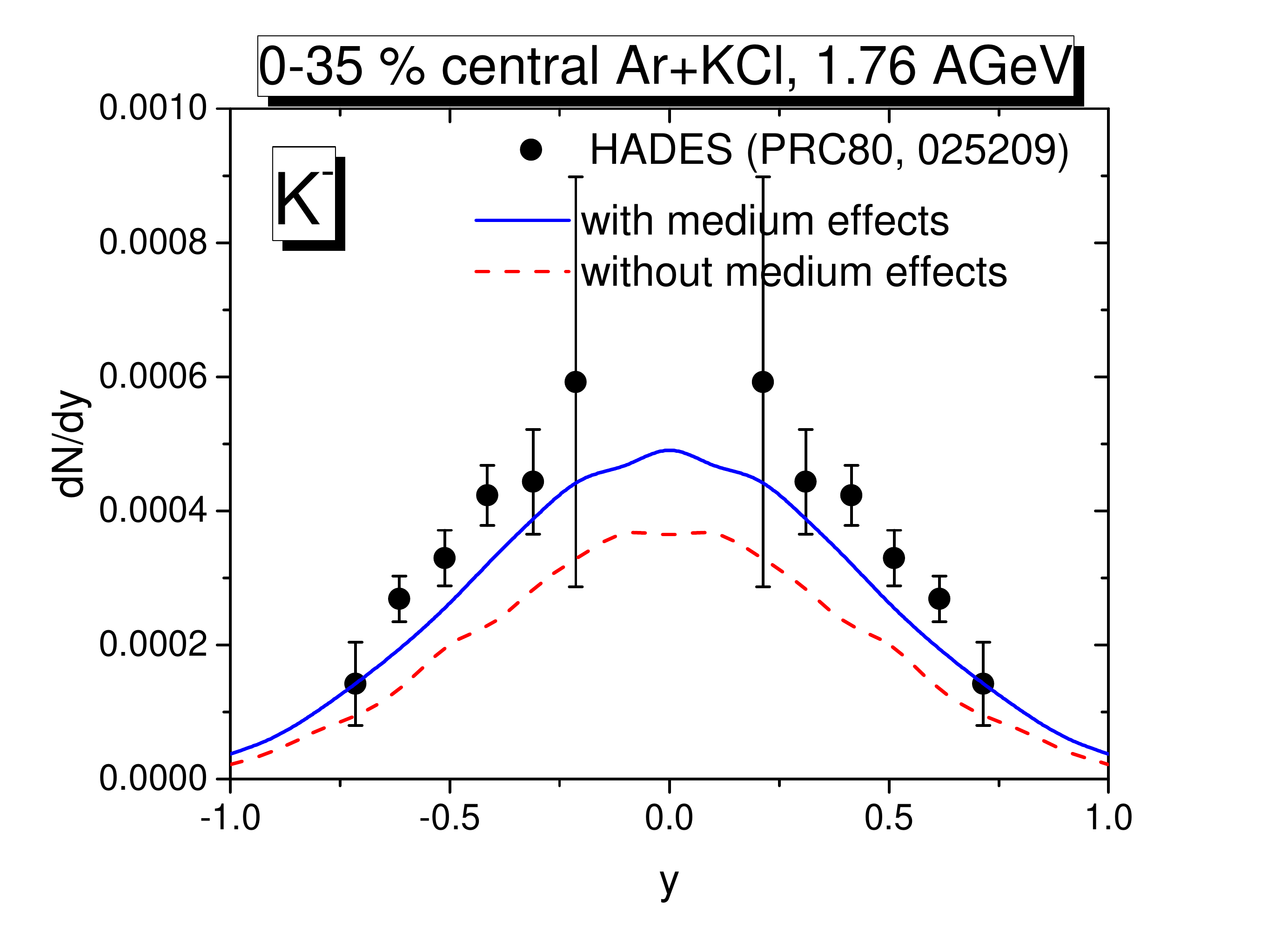} 
\includegraphics[width=5.7 cm]{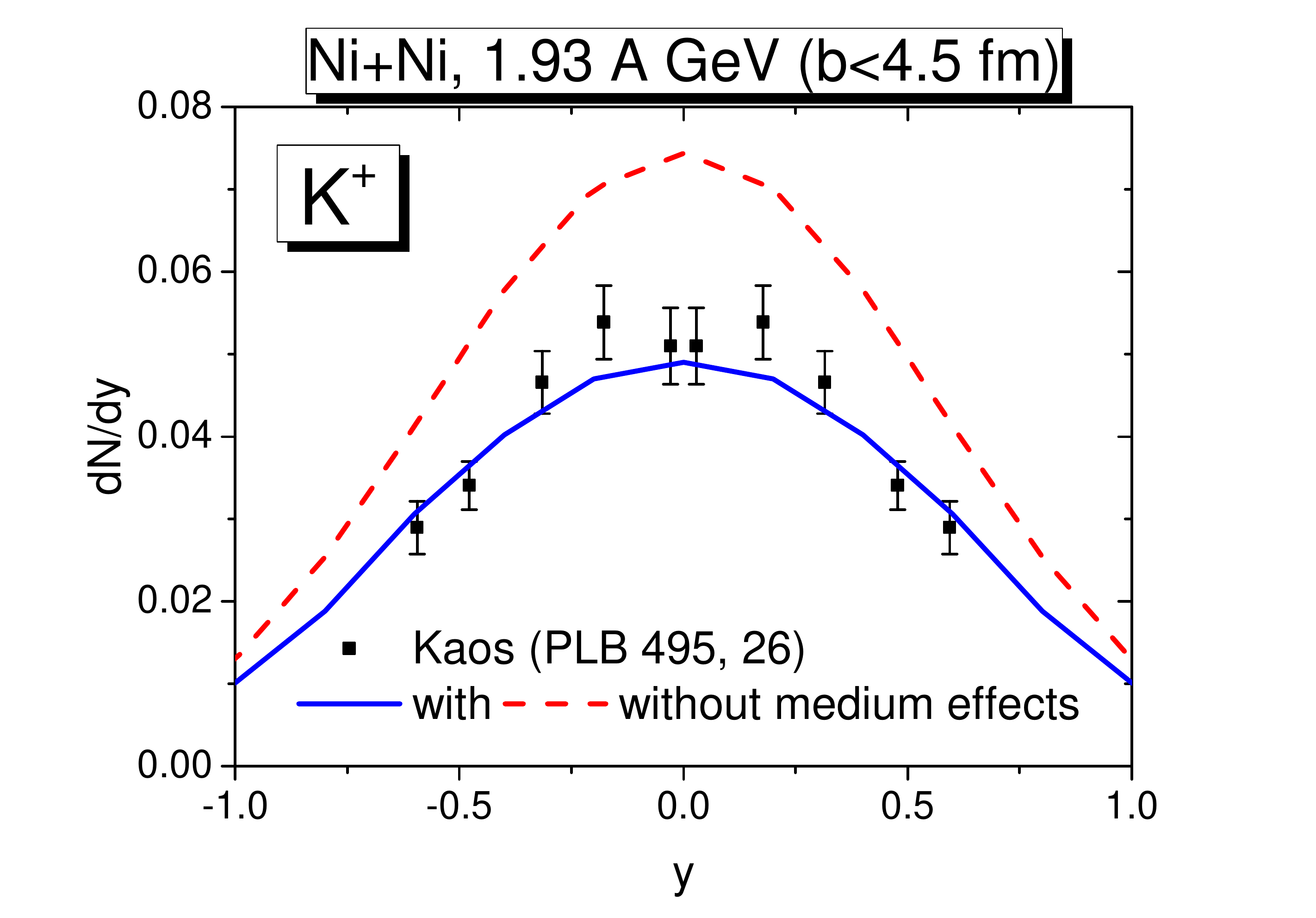}
\includegraphics[width=5.7 cm]{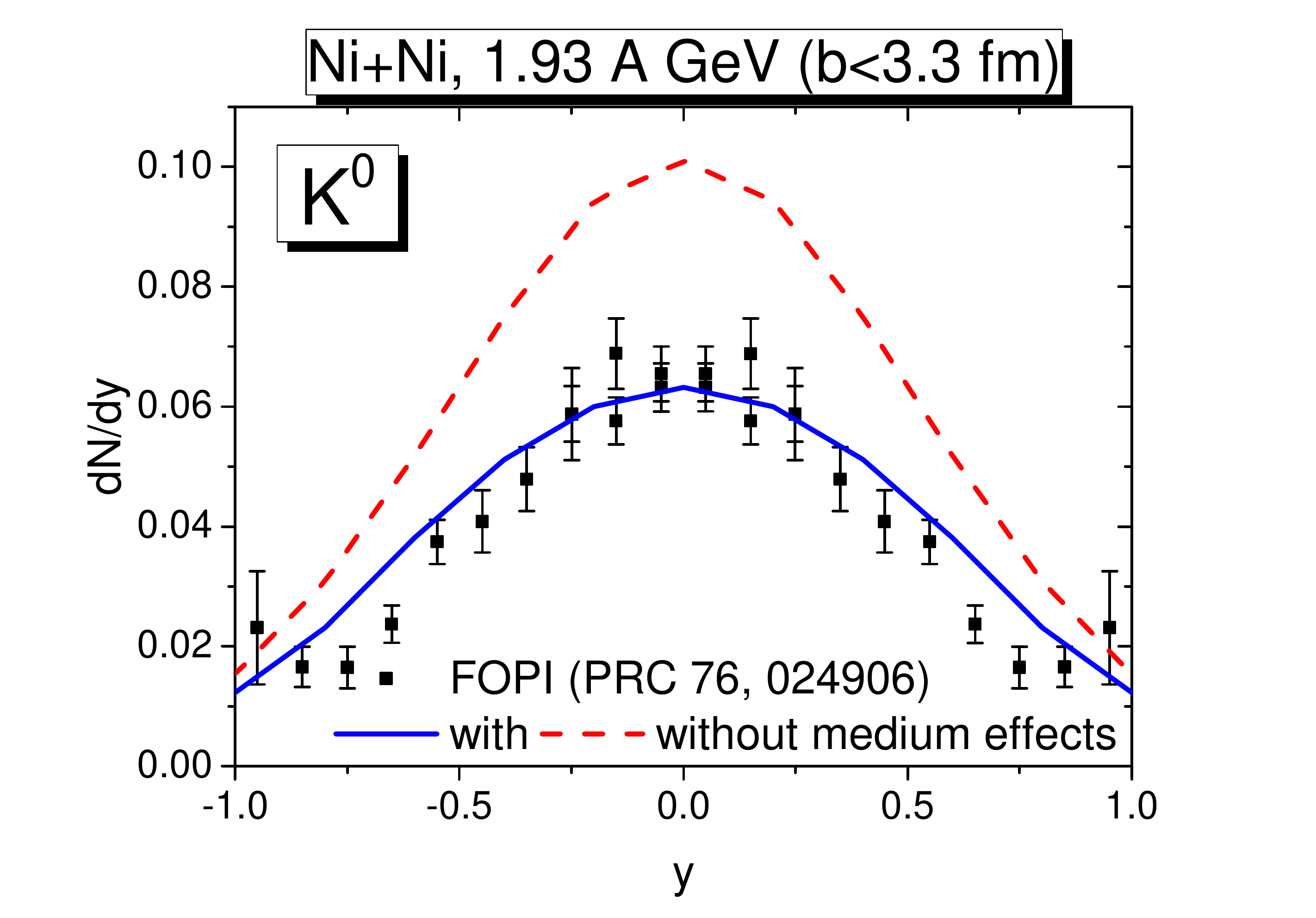}
\includegraphics[width=5.85 cm]{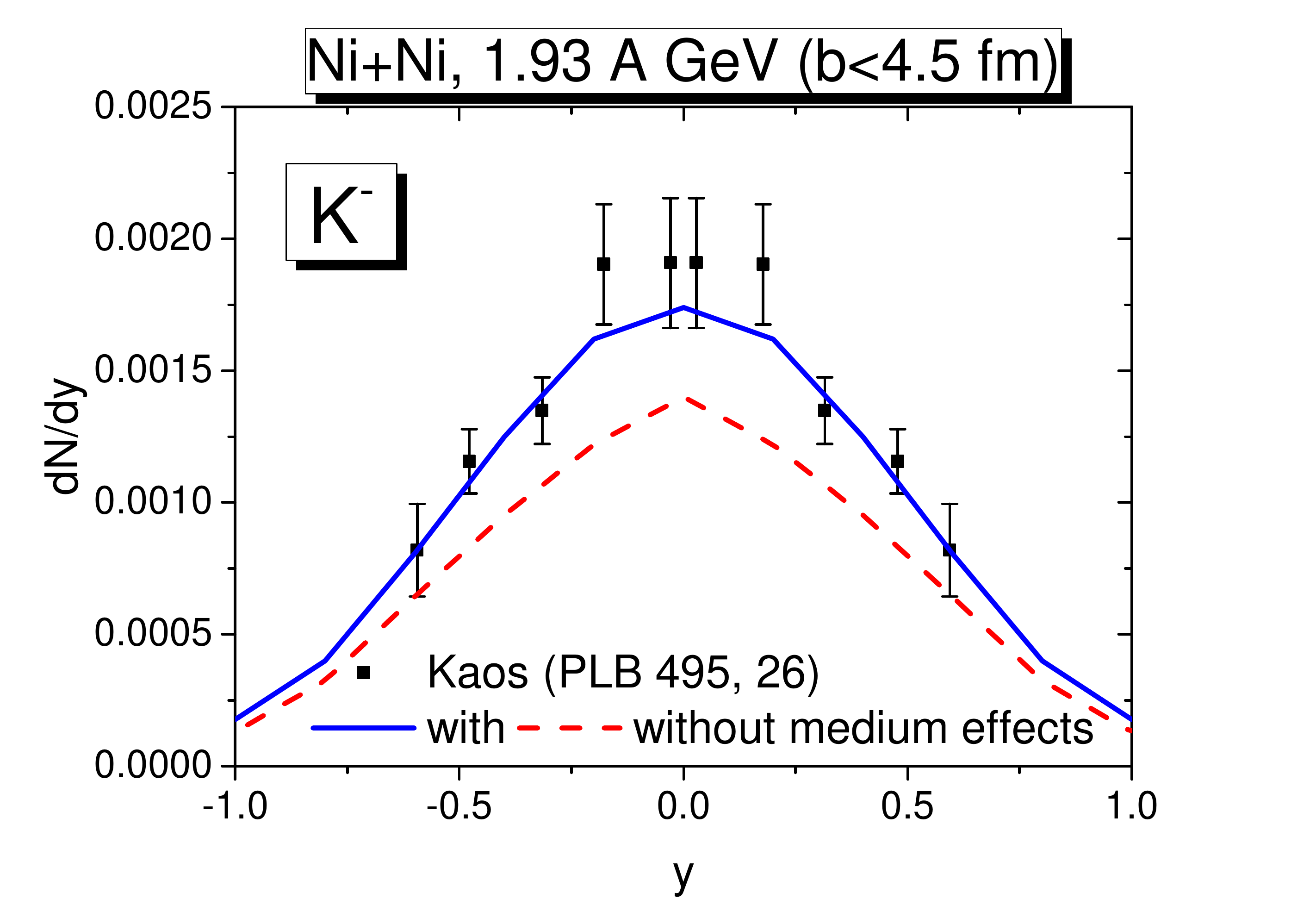} 
\includegraphics[width=5.7 cm]{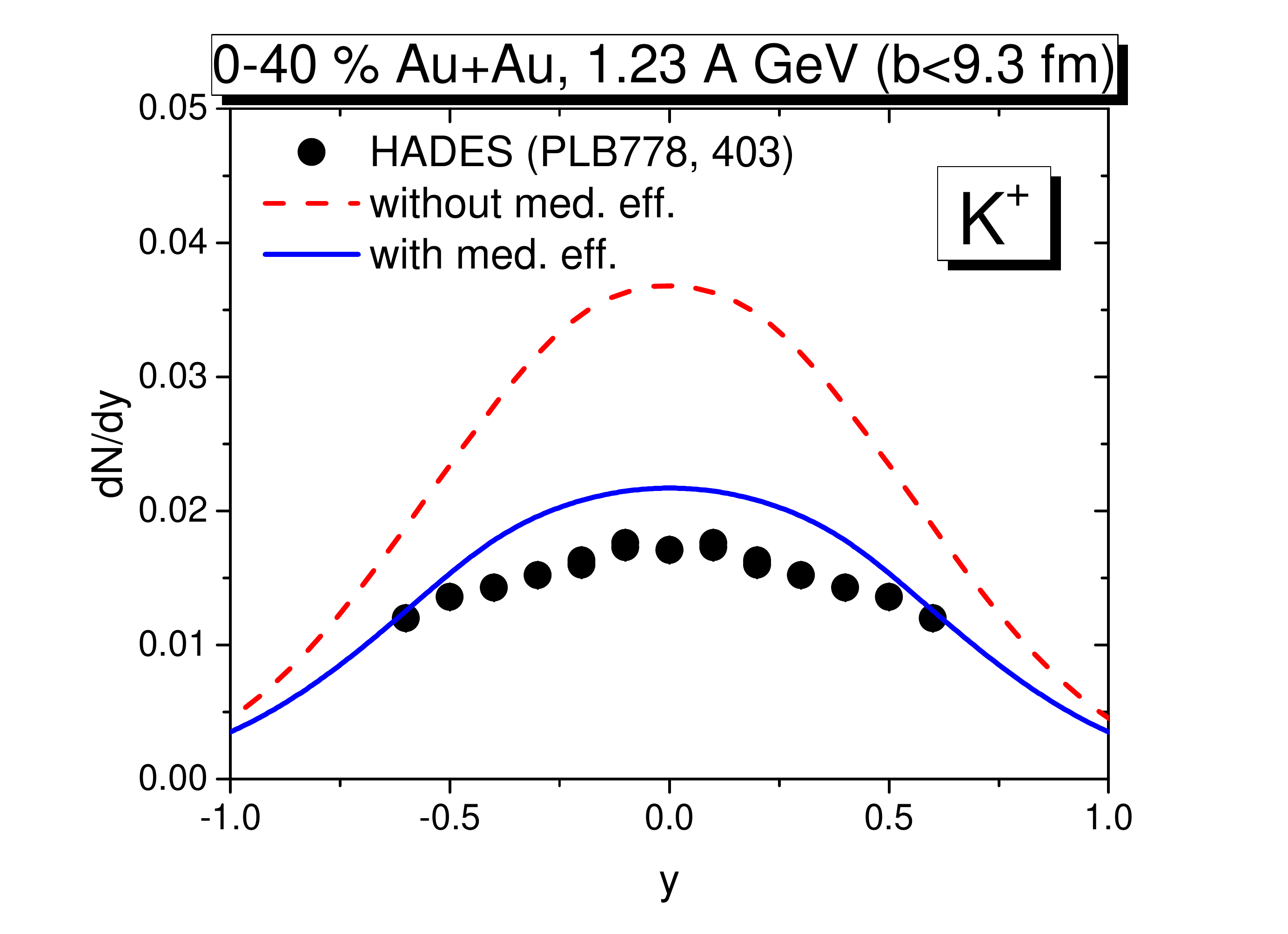}
\includegraphics[width=5.7 cm]{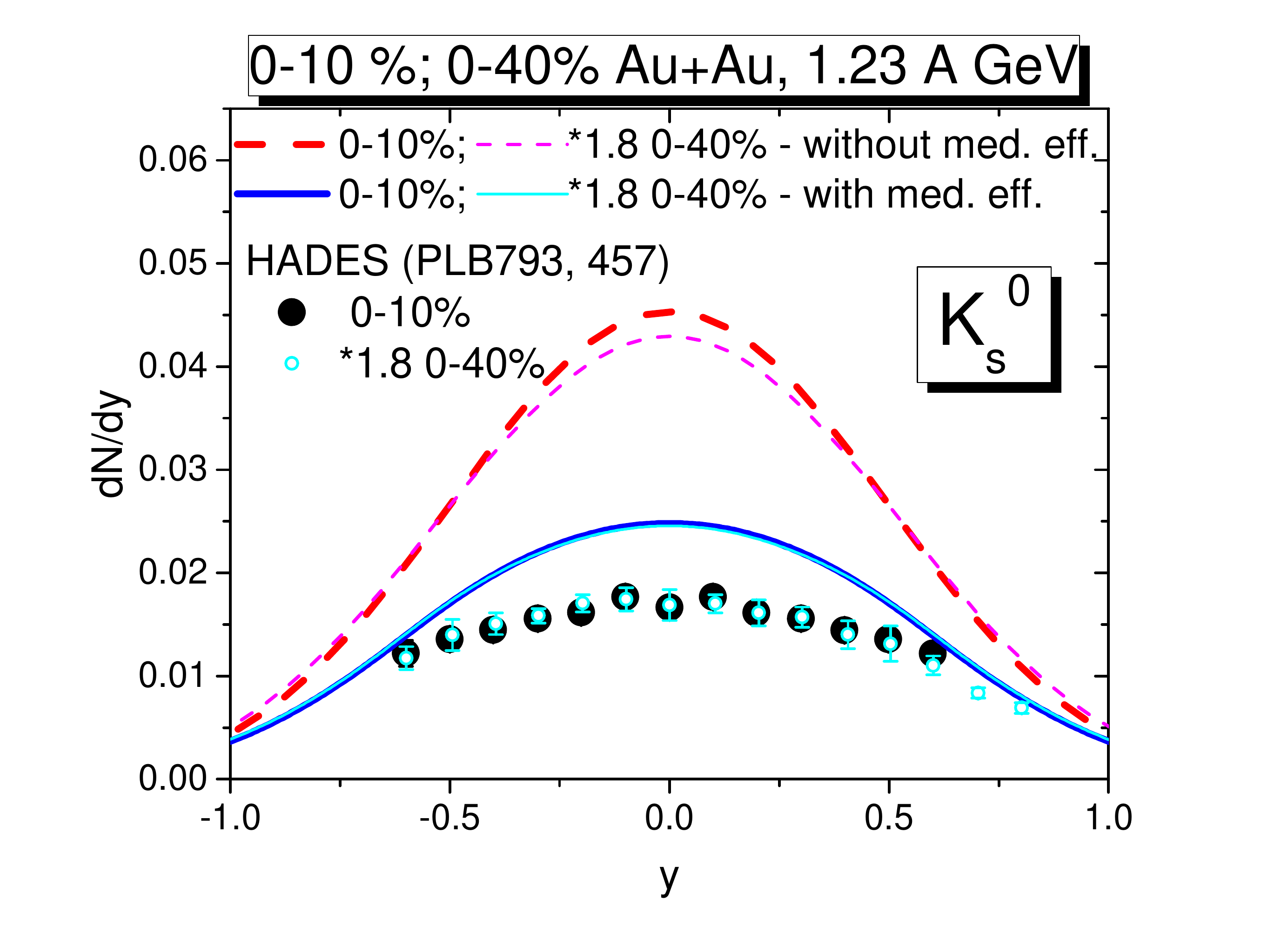}
\includegraphics[width=5.85 cm]{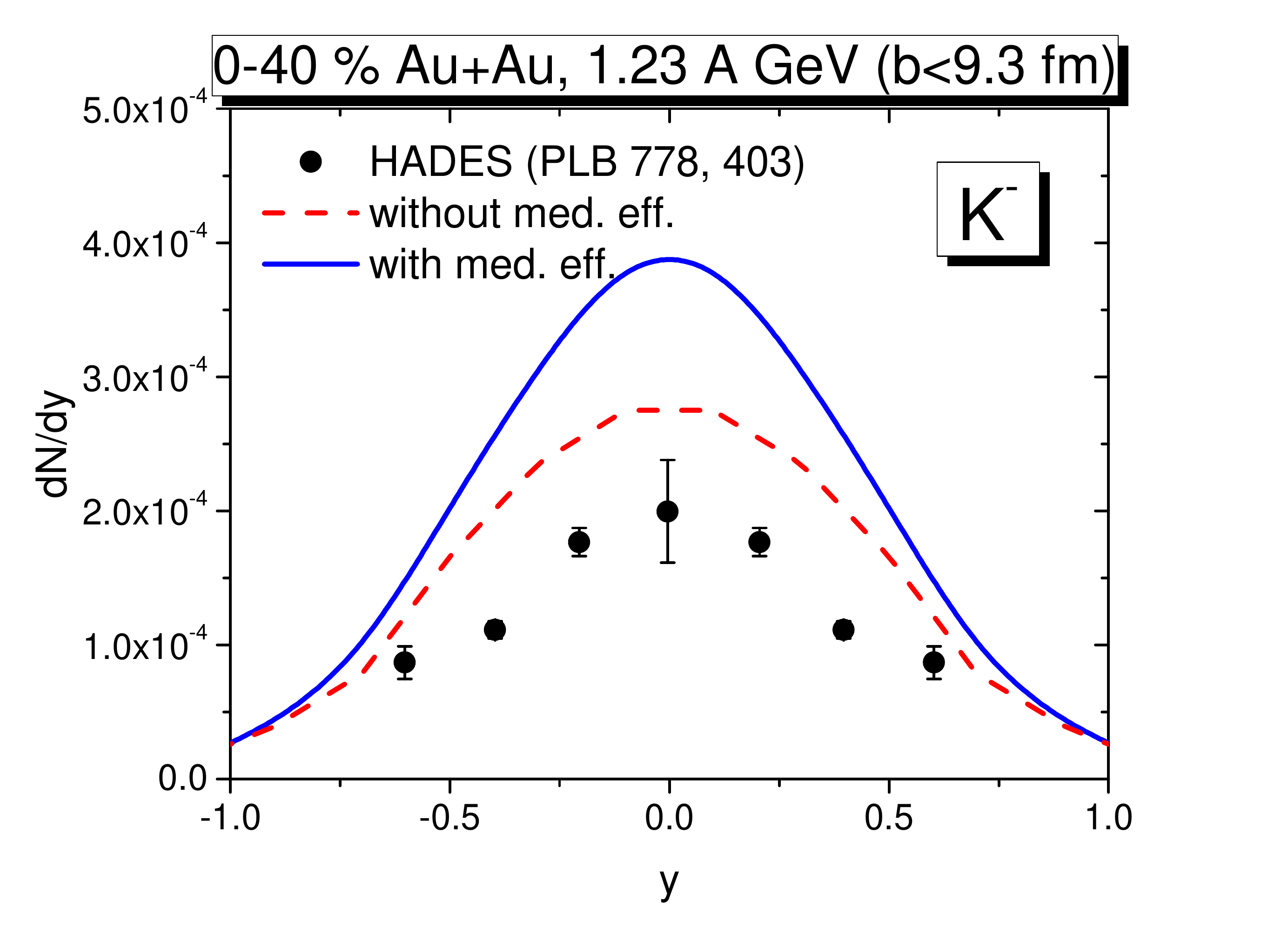}  
\caption{(Color online) 
Upper row:
the PHSD results for the rapidity distributions of  $K^+$ (left), $K^0_s$ (middle) 
and  $K^-$ (right) mesons in 35\% central Ar+KCl collisions at 1.76 A GeV  
in comparison to the experimental data of the HADES Collaboration 
\cite{Agakishiev:2009ar,Agakishiev:2010zw}.
Middle row: 
the PHSD results for the rapidity distributions of $K^+$ (left), $K^0$ (middle) 
and $K^-$ (right)  mesons in central Ni+Ni collisions at 1.93 A GeV in comparison to the experimental data of the KaoS Collaboration \cite{Menzel:2000vv} for $K^+, K^-$ and
the FOPI Collaboration \cite{Merschmeyer:2007zz} for $K^0_s$.
Lower row:
the PHSD results for the rapidity distributions 
of $K^+$ (left) and $K^-$ (right) mesons in 40\% central Au+Au collisions at 1.23 A GeV  
in comparison to experimental data of the HADES Collaboration  \cite{Adamczewski-Musch:2017rtf} and  
of $K^0_s$ mesons (where $K^0_s=(K^0+\bar K^0)/2$) (middle)
in 10\% central and 40\% central (scaled by a factor of 1.8) 
Au+Au collisions at 1.23 A GeV  in comparison to experimental data of the HADES 
Collaboration \cite{Adamczewski-Musch:2018xwg}.
The dashed lines indicate the PHSD results without medium effects for (anti)kaons, while
the solid lines display the results with the medium effects: a repulsive potential
for kaons and the self-energy within the G-matrix approach for antikaons.}
\label{NiNi-y-central}
\end{figure*}

Figure~\ref{NiNi-y-central} shows the rapidity distributions of $K^+$, $K^0$,  
$K^0_s=(K^0+\bar K^0)/2$ and $K^-$ in central Ar+KCl  collisions at 1.76 A GeV 
Ni+Ni collisions at 1.93 A GeV and Au+Au collisions at 1.23 A GeV.
The dashed lines indicate the results without medium effects for (anti)kaons, while
the solid lines denote the results with the medium effects: a repulsive potential 
for kaons and the self-energy within the G-matrix approach for antikaons.
One can see that the kaon potential suppresses $K^+$, $K^0$ (or $K^0_s$) production 
since it increases the kaon mass in the dense medium and thus increases the threshold
for their production. 
Oppositely, the inclusion of in-medium effects within the G-matrix approach for $K^-$ leads 
to an enhancement of their production due to a reduction of the threshold
in the medium. 
The latter is caused dominantly by a broadening of the $K^-$ spectral function 
in the hadronic medium rather than by a shift of the pole mass of the spectral function 
as discussed in the previous Section.
As seen from Fig.~\ref{NiNi-y-central}, the rapidity distributions of 
$K^+$ and $K^0$ with potential are broader than those without potential 
due to the repulsive forces.
On the other hand, the rapidity distribution of $K^-$ becomes slightly narrower 
with the medium effects although the broadening of the antikaon spectral function 
in the medium leads to the increase in mass  for some fraction of antikaons
 - cf. Fig.~\ref{dist-mass}. 
The narrowing of the final rapidity distributions occurs due to the following reasons: 
i) Due to the attractive nature of the antikaon potential, related to the real part of
the antikaon self-energy $Re\Sigma$ as seen in Fig.\ref{self-energy}. 
ii) In addition to $Re\Sigma$, the imaginary part of the antikaon self-energy 
$Im\Sigma$ plays an important role in the off-shell propagation since it is entering 
the equations-of-motion Eqs.~(\ref{x-update},\ref{momentum-update},\ref{energy-update})
and influence the forces. The effect of $Im\Sigma$ can be illustrated in the 
following example: let's assume ${\rm Re}\Sigma \approx 0$, i.e. that the pole mass 
does not change, then Eq.~(\ref{momentum-update}) is approximated by
\begin{eqnarray}
\frac{dp_i}{dt}\approx - \frac{1}{2 E}\frac{M^2-M_0^2}{{\rm Im} \Sigma}\nabla_r{\rm Im}\Sigma,
\end{eqnarray}
where $M_0$ is the $K^-$ mass in vacuum.
Since  ${\rm Im}\Sigma$ is negative in dense nuclear matter, 
$\nabla_r{\rm Im}\Sigma$ is directed outward in heavy-ion collisions.
Therefore $K^-$ mesons lighter than $M_0$ feel an attractive force while heavier 
$K^-$ mesons feel a repulsive force.
As shown in Fig.~\ref{dist-mass}, much more $K^-$ mesons are lighter than $M_0$ 
and thus are attracted inwards or to low rapidity.

One can see from Fig.~\ref{NiNi-y-central} that in-medium effects  
are necessary to explain the experimental data on the rapidity distributions 
of $K^+$, $K^0$ and $K^-$ mesons in Ni+Ni collisions at 1.93 A GeV from the KaoS and FOPI Collaborations as well as $K^+$, $K^0_s$ and $K^-$ mesons in Ar+KCl collisions 
at 1.76 A GeV from the HADES collaboration.
However, there is a tension between the PHSD results on $K^+$, $K^0_s$ and $K^-$ 
meson rapidity distributions in Au+Au Collisions at 1.23 A GeV 
and the HADES data (which are slightly lower in energy) - without as well 
as with medium effects. 
The overestimation for $K^+$ and $K^0_s$ (including in-medium effects) is about 20\%, 
while it is about of factor 2 for  $K^-$  with medium effects.
Thus, within the same model we do not find a consistent description of 
all three experiments. The origin for this contradictory message from the comparison 
of our results to the experimental data of the HADES Collaboration, and to 
the FOPI and the Kaos Collaborations requires  further investigation.

\begin{figure}[th!]
\centerline{
\includegraphics[width=8.6 cm]{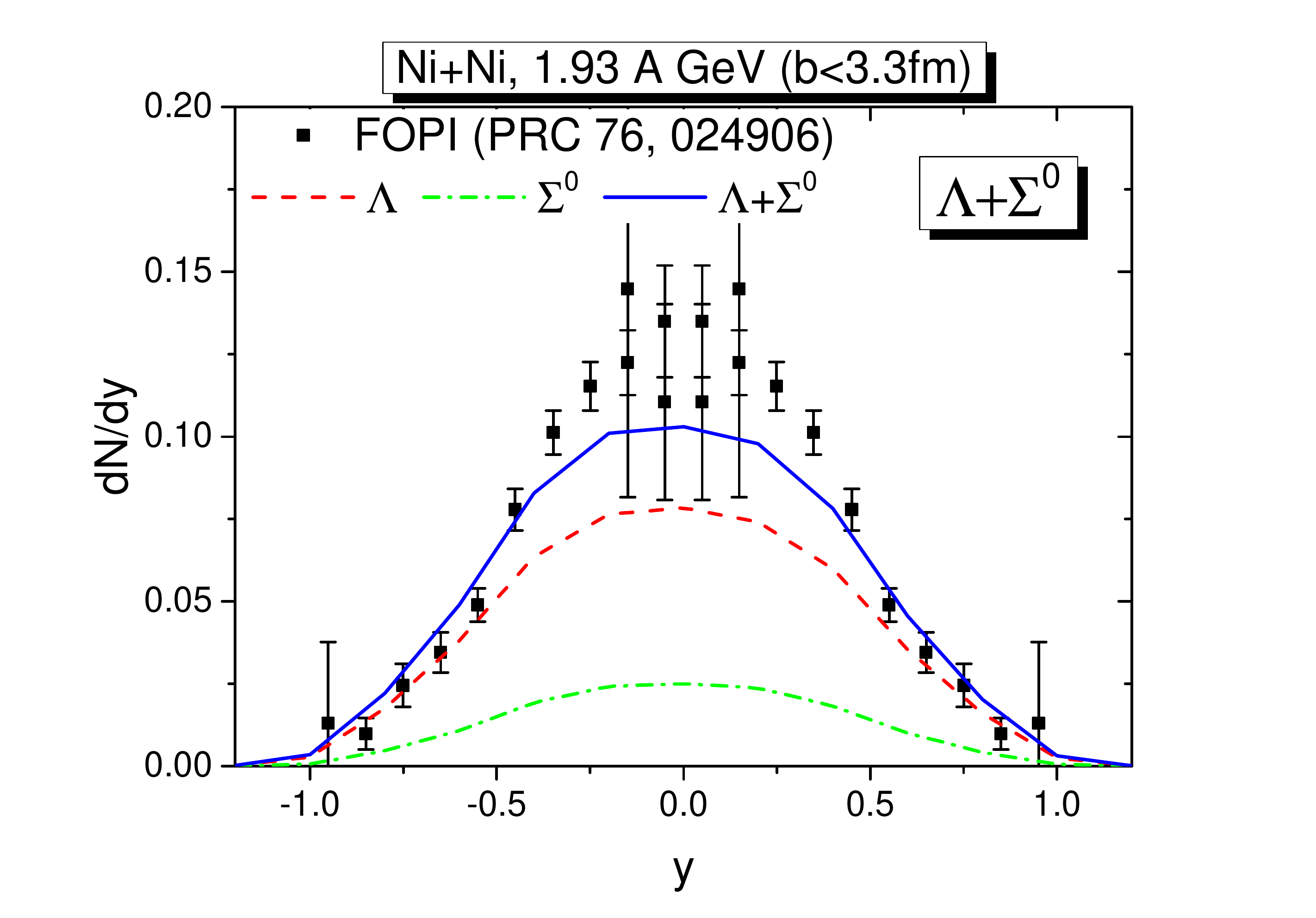}}
\vspace*{-2mm}\centerline{
\includegraphics[width=8.6 cm]{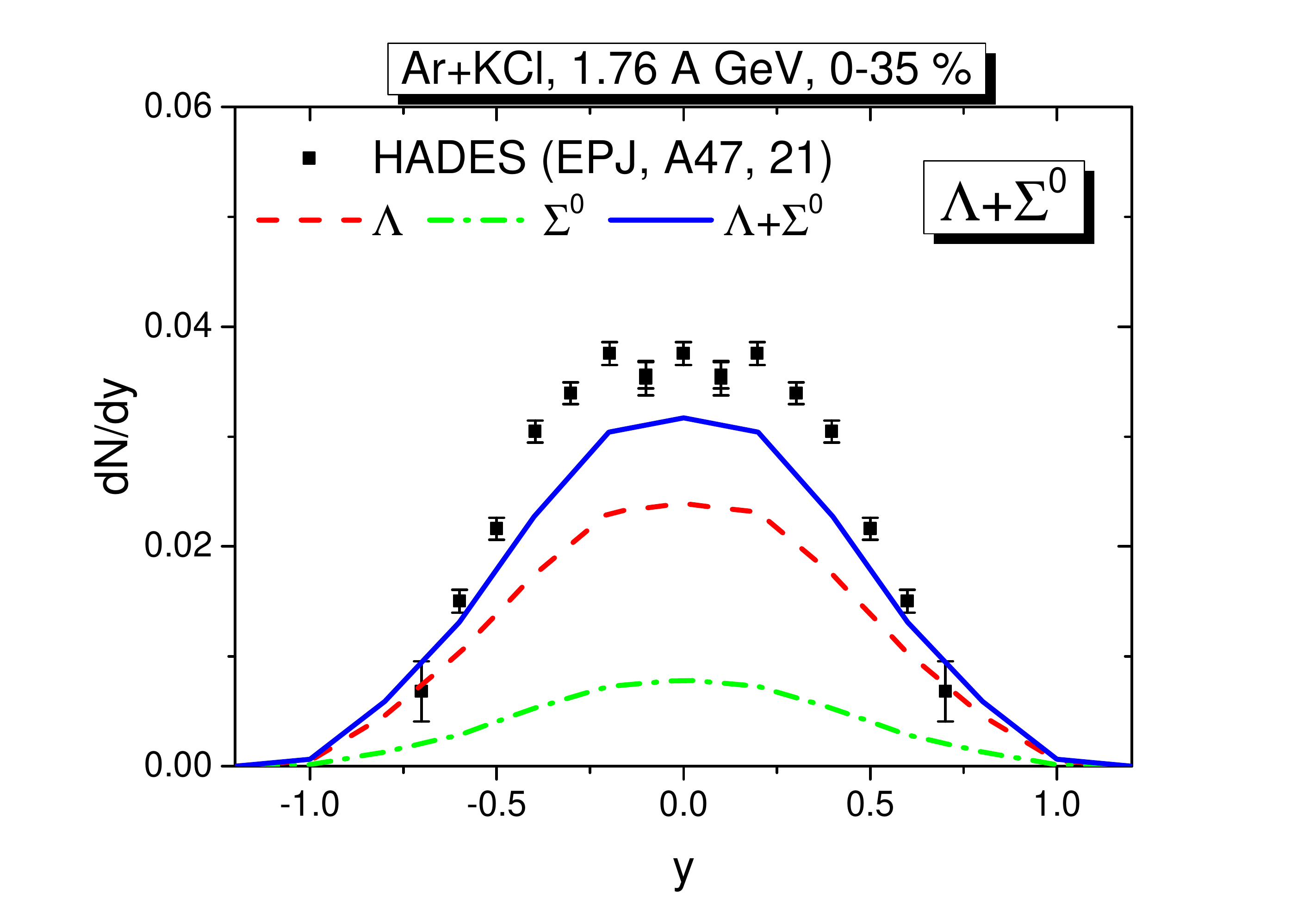}}
\vspace*{-2mm}\centerline{
\includegraphics[width=8.6 cm]{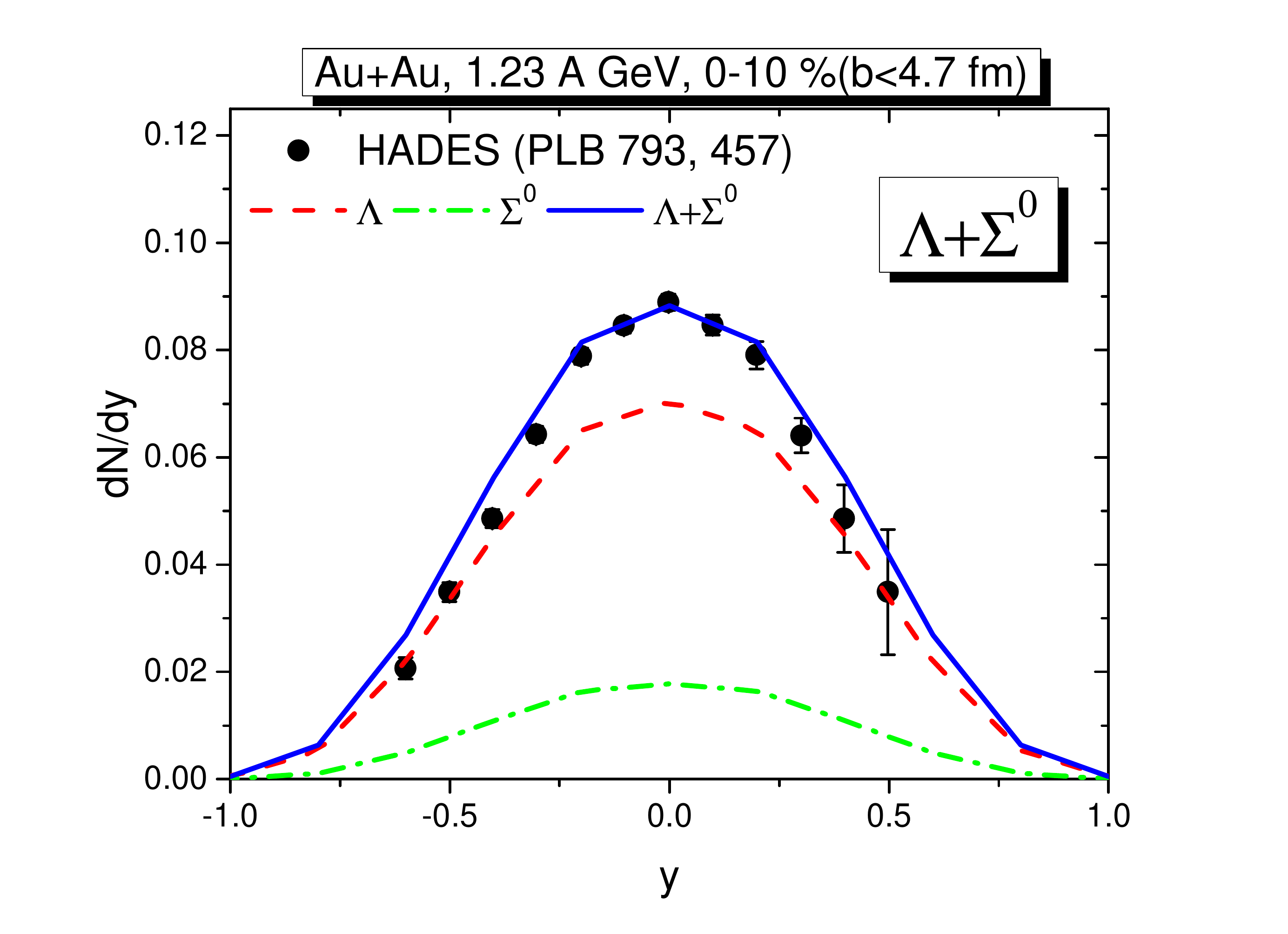}}
\caption{(Color online) The PHSD results for the rapidity distributions 
of neutral hyperons $\Lambda+\Sigma^0$ (blue solid lines)  in central (upper) Ni+Ni collisions at 1.93 A GeV
in comparison to the experimental data from the FOPI Collaboration \cite{Merschmeyer:2007zz},
(midle) in central Ar+KCl collisions at 1.76 GeV and (lower) in central
Au+Au collisions at 1.5 GeV in comparison to the experimental data from the HADES
Collaboration \cite{Agakishiev:2010rs,Adamczewski-Musch:2018xwg}.
The red dashed lines show the contribution from $\Lambda$ and the green dot-dashed lines 
present the contribution from $\Sigma$ hyperons.}
\label{y-NiNi-Y}
\end{figure}

Since strangeness is not present in the initial nuclei,  
strangeness conservation holds strictly , i.e. the number of strange hadrons 
is always equal to that of antistrange hadrons. Therefore the number of produced kaons 
is the same as that of hyperons and antikaons (the production of multi-strange states 
at low energies is negligible in the total strangeness balance).
Thus, the dynamics of hyperons and antikaons is closely linked 
since the $\pi +Y \to \bar K +N$ reaction is the dominant channel for antikaon production.
Since the PHSD results for kaon and antikaon rapidity and the $p_T$-spectra for Ni+Ni 
are in a good agreement with experimental data, we confront our results
for the  hyperon distribution with the experimental data, too.
We recall that in the PHSD the strangeness is strictly conserved.

In the upper part of Fig.~\ref{y-NiNi-Y} we show the rapidity distribution 
of neutral hyperons $\Lambda+\Sigma^0$ in central Ni+Ni collisions at 1.93 A GeV in 
comparison to the experimental data from the FOPI Collaboration~\cite{Merschmeyer:2007zz}.
A small deviation is seen at midrapidity where our results underestimate the FOPI data, and
the theoretical distribution is slightly broader than the data.
A similar disagreement is found in the middle part of Fig.~\ref{y-NiNi-Y} 
which shows the  comparison with the experimental data from the HADES 
Collaboration~\cite{Agakishiev:2010rs} in central Ar+KCl collisions at 1.76 A GeV.
However, the PHSD result for the  $y$- distribution of $\Lambda+\Sigma^0$ agrees 
very well with the HADES data for Au+Au collisions at 1.23 A GeV as seen 
from the lower part of Fig.~\ref{y-NiNi-Y}.

\subsection{$p_T$-distributions}


We note that the rapidity distributions are not measured directly in the experiments.
They are extrapolated from the measured $p_T$-spectra assuming a thermal
distribution. In this respect it is very important to compare the model results
with directly measured quantities such as $p_T$- distributions which we present 
in this subsection.

\begin{figure}[h!]
\centerline{
\includegraphics[width=8.6 cm]{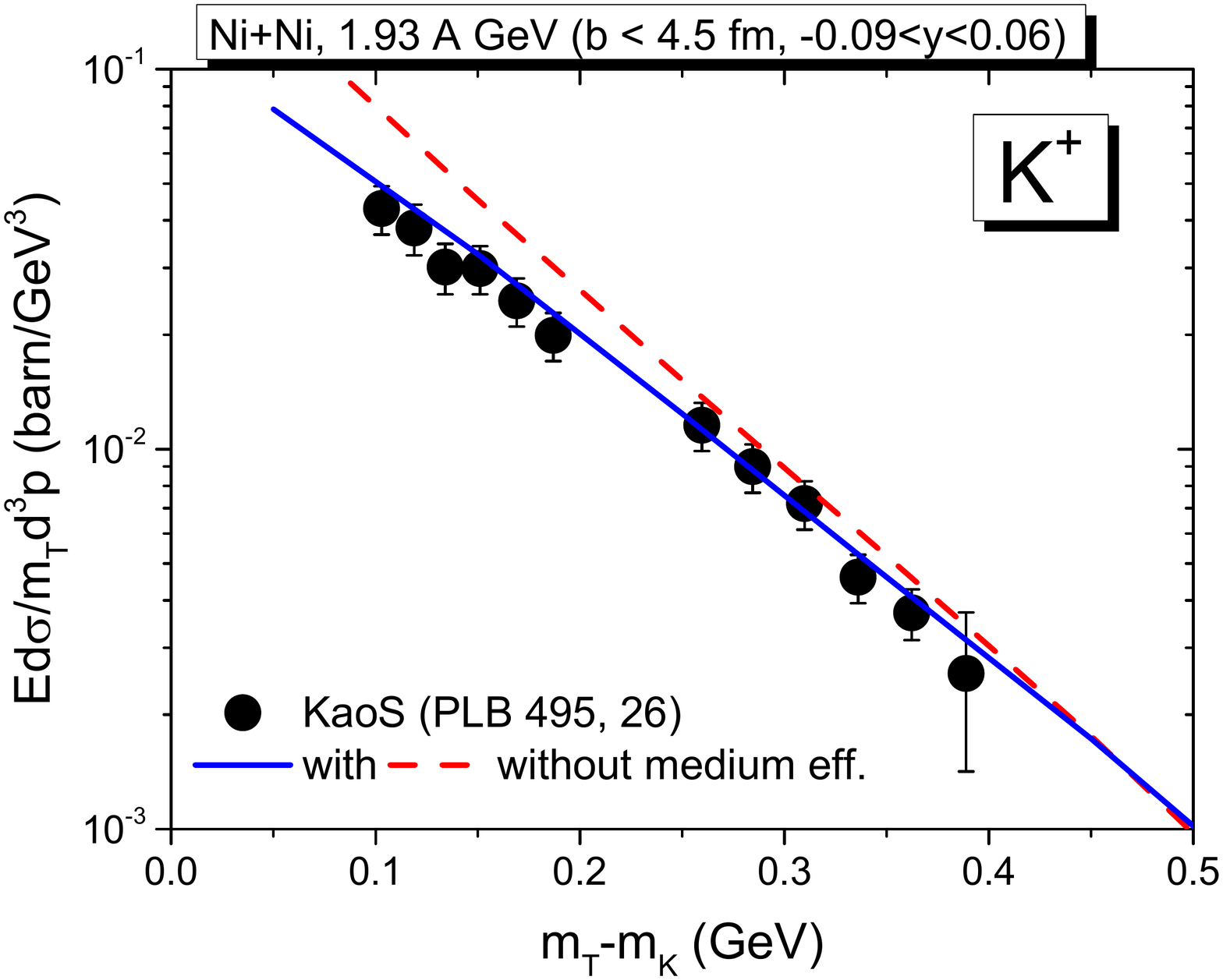}}
\vspace*{-2mm}
\centerline{
\includegraphics[width=8.6 cm]{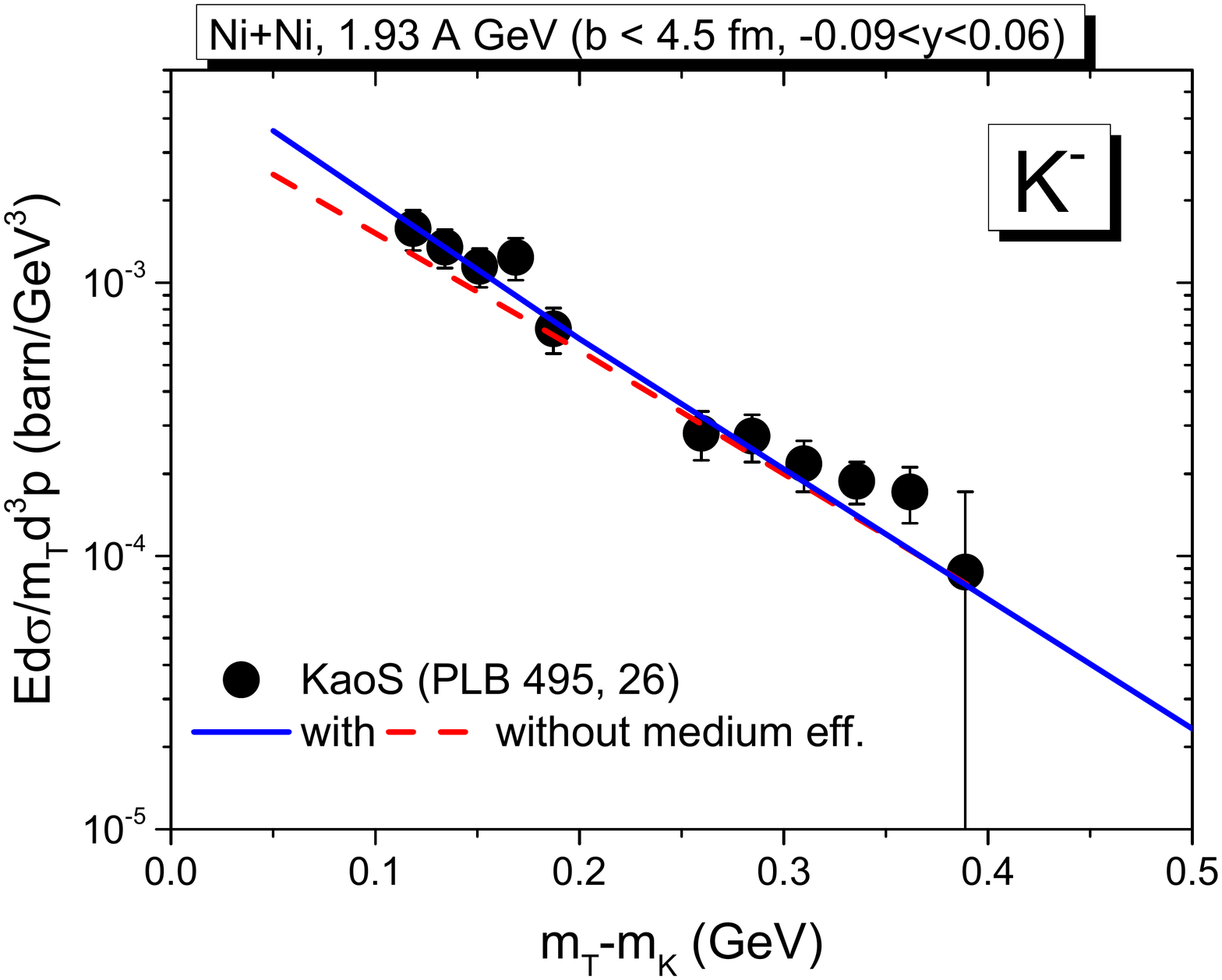}}
\caption{(Color online) The PHSD results for the $m_T$-spectra of  $K^+$ (upper) and  
$K^-$ (lower) 
as a function of the transverse kinetic energy $m_T-m_K$ 
in central Ni+Ni collisions at 1.93 A GeV in comparison to the experimental 
data of the KaoS Collaboration ~\cite{Menzel:2000vv}.
The dashed lines indicate the results without medium effects for (anti)kaons, while
the solid lines display the results with the medium effects: 
a repulsive potential for kaons and the self-energy within the G-matrix approach for antikaons.}
\label{NiNi-pt-central}
\end{figure}

\begin{figure}[h!]
\centerline{\includegraphics[width=8.6 cm]{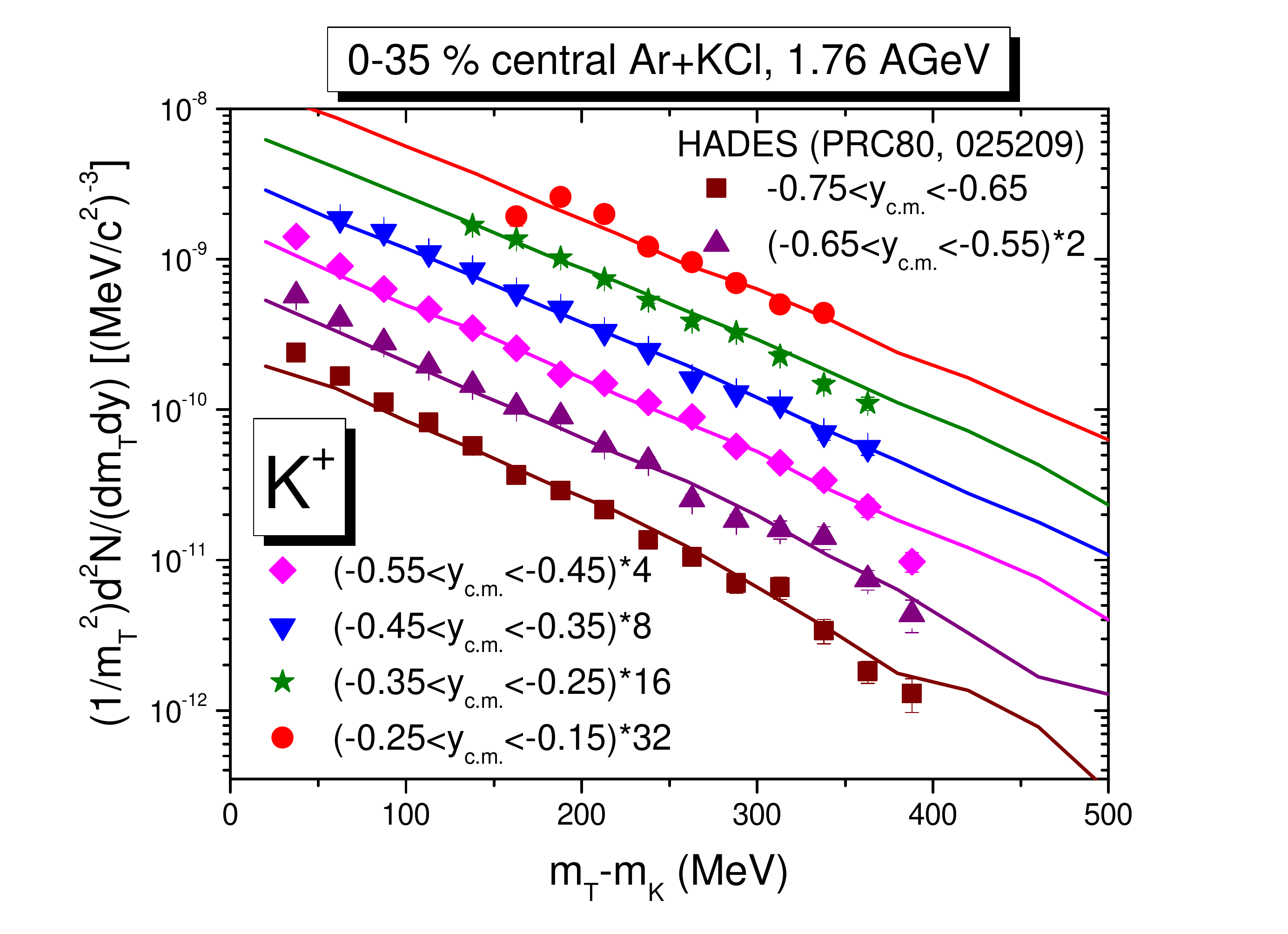}}
\vspace*{-1mm}
\centerline{\includegraphics[width=8.6 cm]{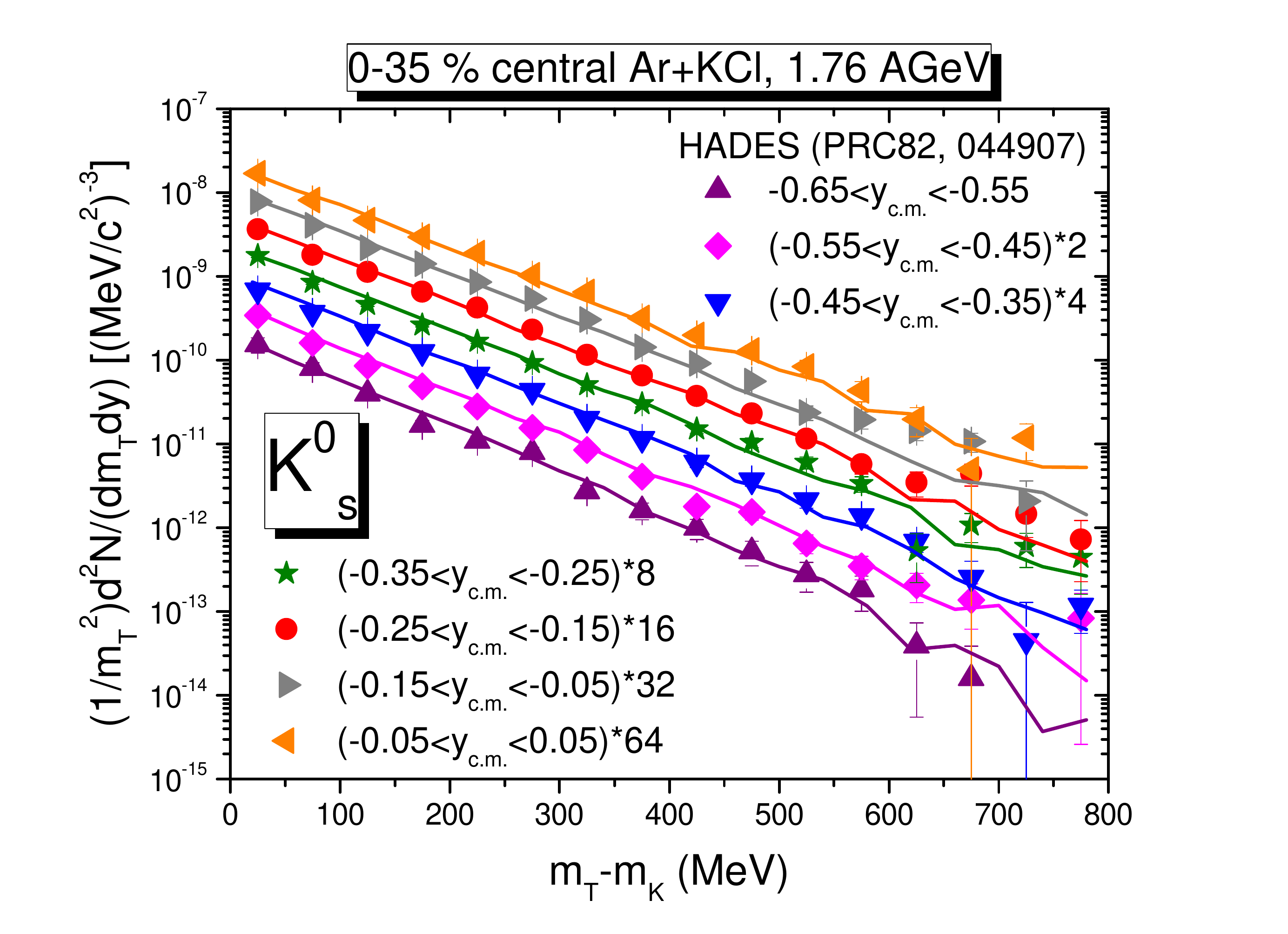}}
\vspace*{-1mm}
\centerline{\includegraphics[width=8.6 cm]{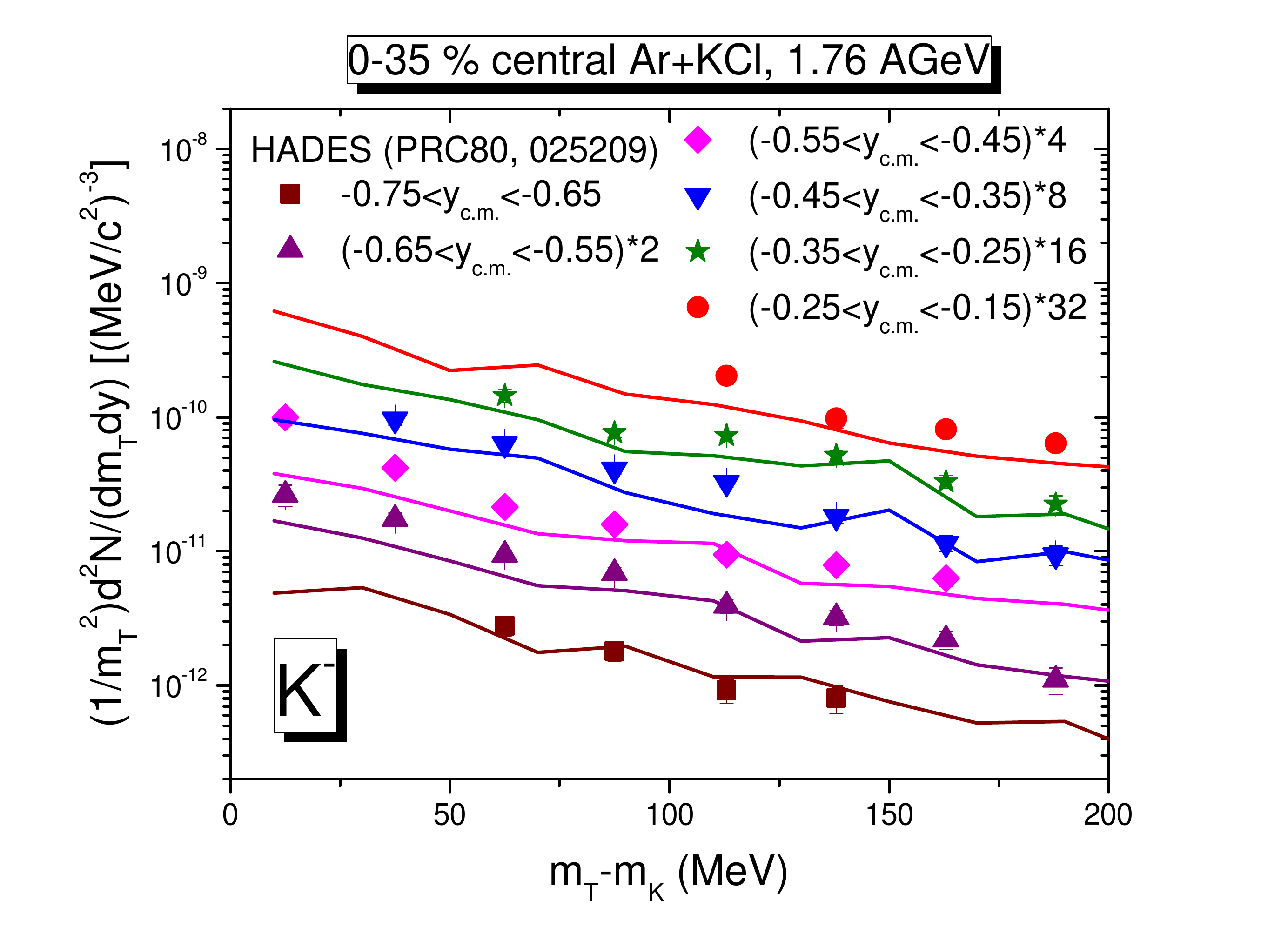}}
\caption{(Color online) The PHSD results (including in-medium effects) 
for the $m_T$-spectra of $K^+$ (upper), $K^0_S$ (midle) and  $K^-$ (lower) 
as a function of the transverse kinetic energy $m_T-m_K$ for different rapidity bins
in 35\% central Ar+KCl collisions at 1.76 A GeV 
in comparison to the experimental data of the 
HADES Collaboration \cite{Agakishiev:2009ar,Agakishiev:2010zw}.
}
\label{HADES-ymt-ArKCl}
\end{figure}

\begin{figure}[h!]
\centerline{\includegraphics[width=8.6 cm]{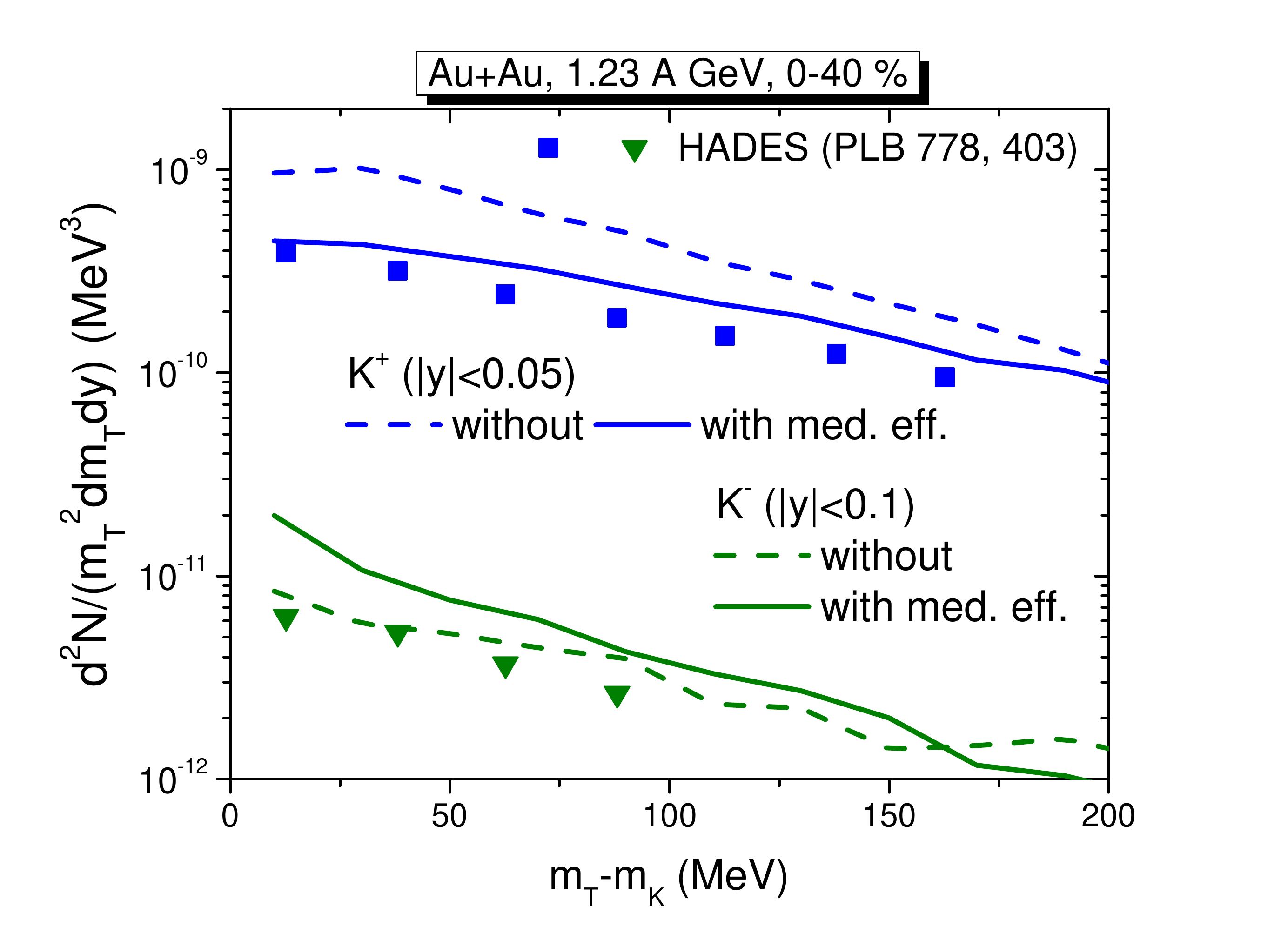}}
\caption{(Color online) The PHSD results for the midrapidity $m_T$-spectra of  
$K^+$ (blue line) and  $K^-$(green line) as a function of the transverse 
kinetic energy $m_T-m_K$ in 40\% central Au+Au collisions at 1.23 A GeV 
in comparison to the experimental data of the 
HADES Collaboration \cite{Adamczewski-Musch:2017rtf}.
The dashed lines indicate the results without medium effects for (anti)kaons, while
the solid lines display the results with the medium effects: 
a repulsive potential for kaons and the self-energy within the G-matrix approach 
for antikaons.}
\label{HADES-mt}
\end{figure}
\begin{figure}[h!]
\centerline{\includegraphics[width=8.6 cm]{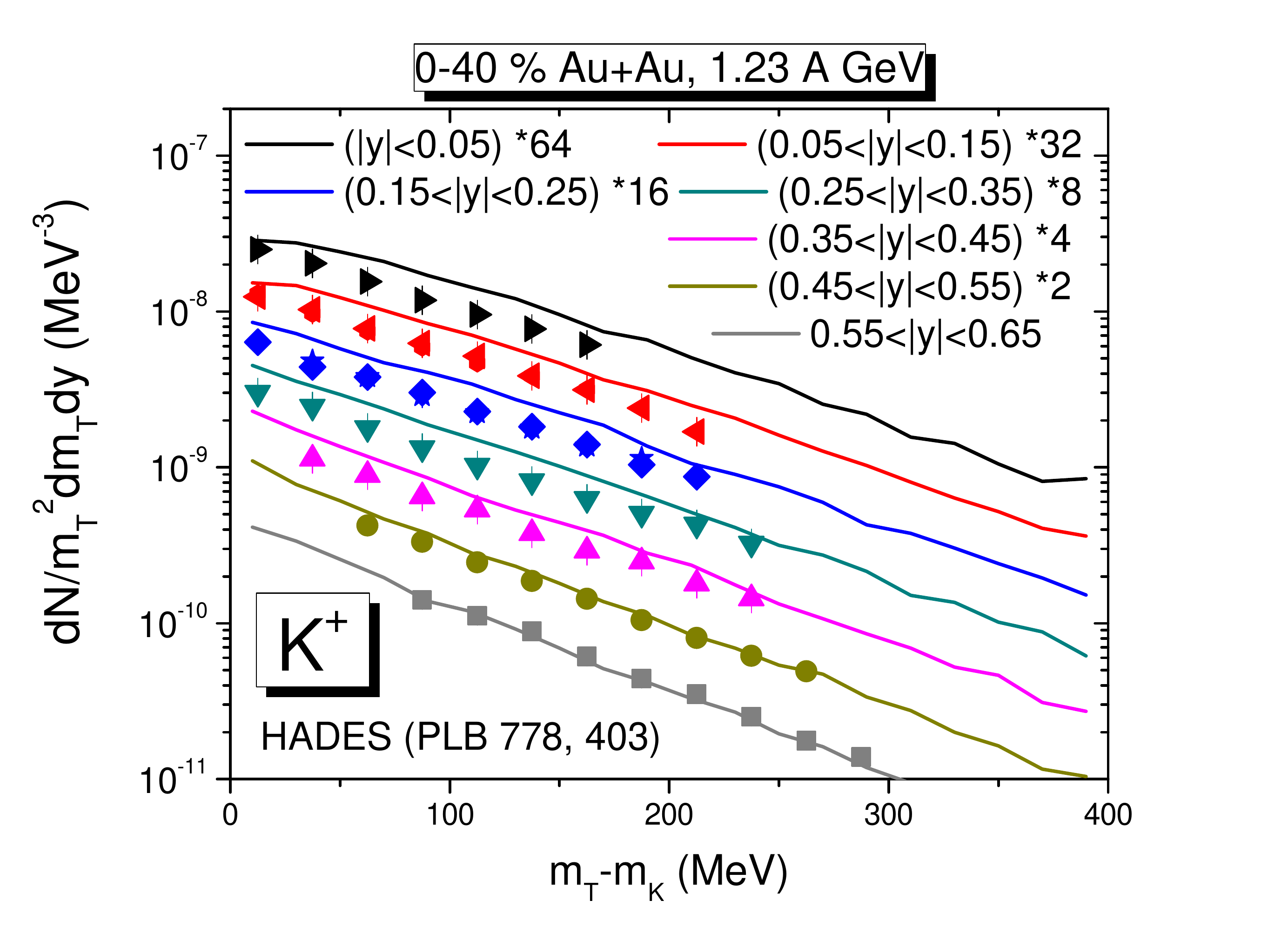}}
\vspace*{-3mm}
\centerline{\includegraphics[width=8.6 cm]{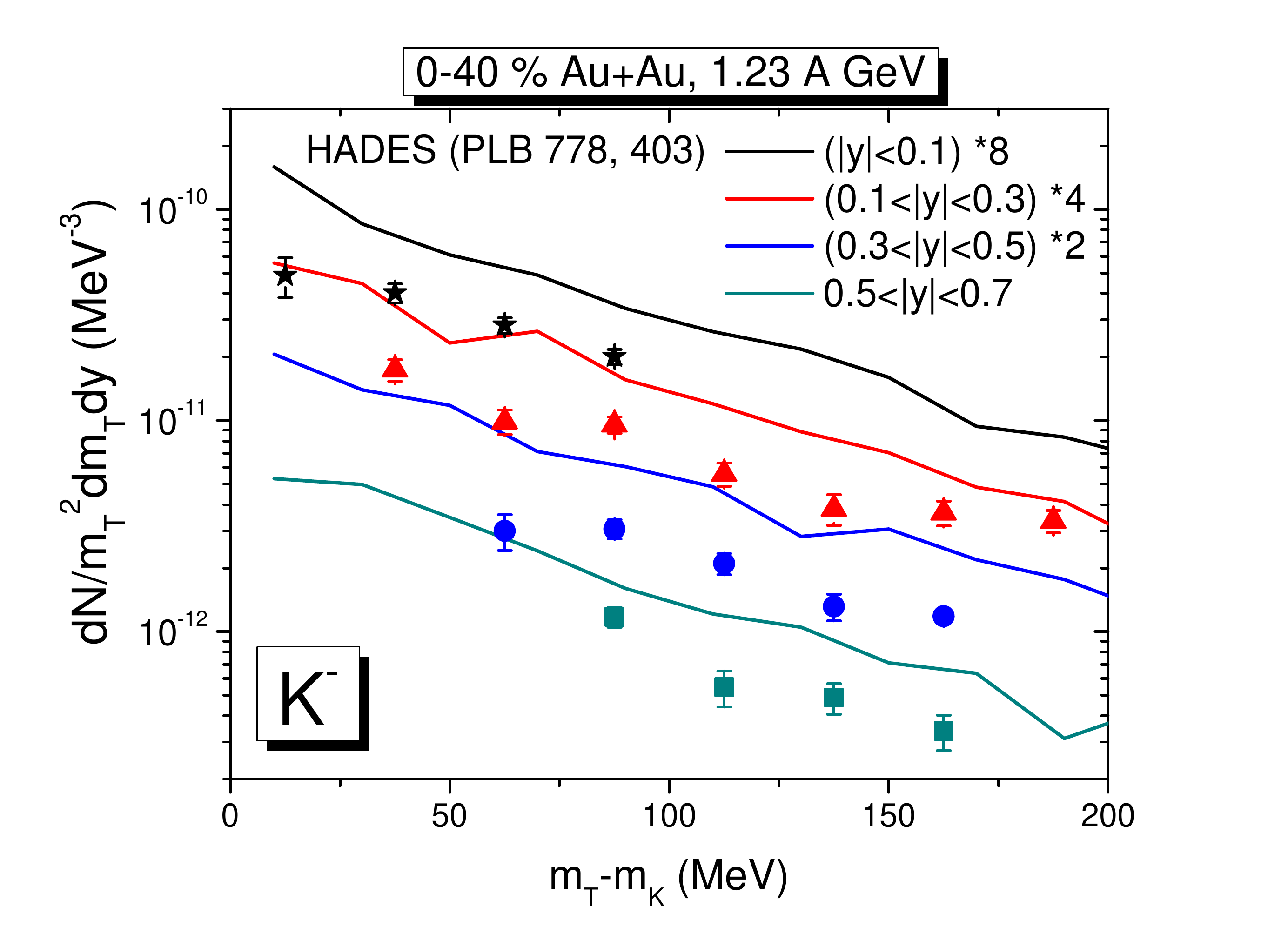}}
\caption{(Color online) The PHSD results (including in-medium effects) 
for the $m_T$-spectra of $K^+$ (upper) and  $K^-$ (lower) as a function of 
the transverse kinetic energy $m_T-m_K$ for different rapidity bins
in 40\% central Au+Au collisions at 1.23 A GeV 
in comparison to the experimental data of the 
HADES Collaboration \cite{Adamczewski-Musch:2017rtf}.
}
\label{HADES-ymt}
\end{figure}

In Fig. \ref{HADES-ymt-ArKCl} we show the PHSD results (including in-medium effects) 
for the $m_T$-spectra of $K^+$ (upper), $K^0_S$ (midle) and  $K^-$ (lower) 
as a function of the transverse kinetic energy $m_T-m_K$ for different rapidity bins
in 35\% central Ar+KCl collisions at 1.76 A GeV 
in comparison to the experimental data of the 
HADES Collaboration \cite{Agakishiev:2009ar,Agakishiev:2010zw}.
One can see that the PHSD agrees with HADES data for Ar+KCl very well for all
rapidity bins.

Figure~\ref{NiNi-pt-central} shows the PHSD results for the $m_T$-spectra 
versus the transverse kinetic energy $m_T-m_K$ of $K^+$ and $K^-$ 
at midrapidity in central Ni+Ni collisions at 1.93 A GeV in comparison to 
the experimental data of the KaoS Collaboration ~\cite{Menzel:2000vv}.
We note that here the invariant $p_T$-spectra are divided by transverse mass $m_T$  
to extract the effective temperature which will be discussed later in this section;
therefore the spectra are not Lorentz-invariant.
As for the rapidity distribution the kaon-nuclear potential suppresses the $K^+$ yield 
and pushes the $K^+$ spectrum to larger transverse momenta.
On the other hand, the antikaon potential enhances the $K^-$ yield and pulls 
the $p_T$-spectrum toward smaller transverse momenta.
Again the kaon potential is necessary to reproduce the experimental data on the $p_T$- 
spectra of $K^+$;  the data for $K^-$ are also in favour of the inclusion of the
self-energy within the G-matrix approach, similar to their yields and rapidity distributions, but, as discussed in the last section, scattering of antikaons
in the medium and the in-medium effect within the G-matrix for antikaons 
partially compensate each other.

\begin{figure*}[th!]
\centerline{
\includegraphics[width=8.6 cm]{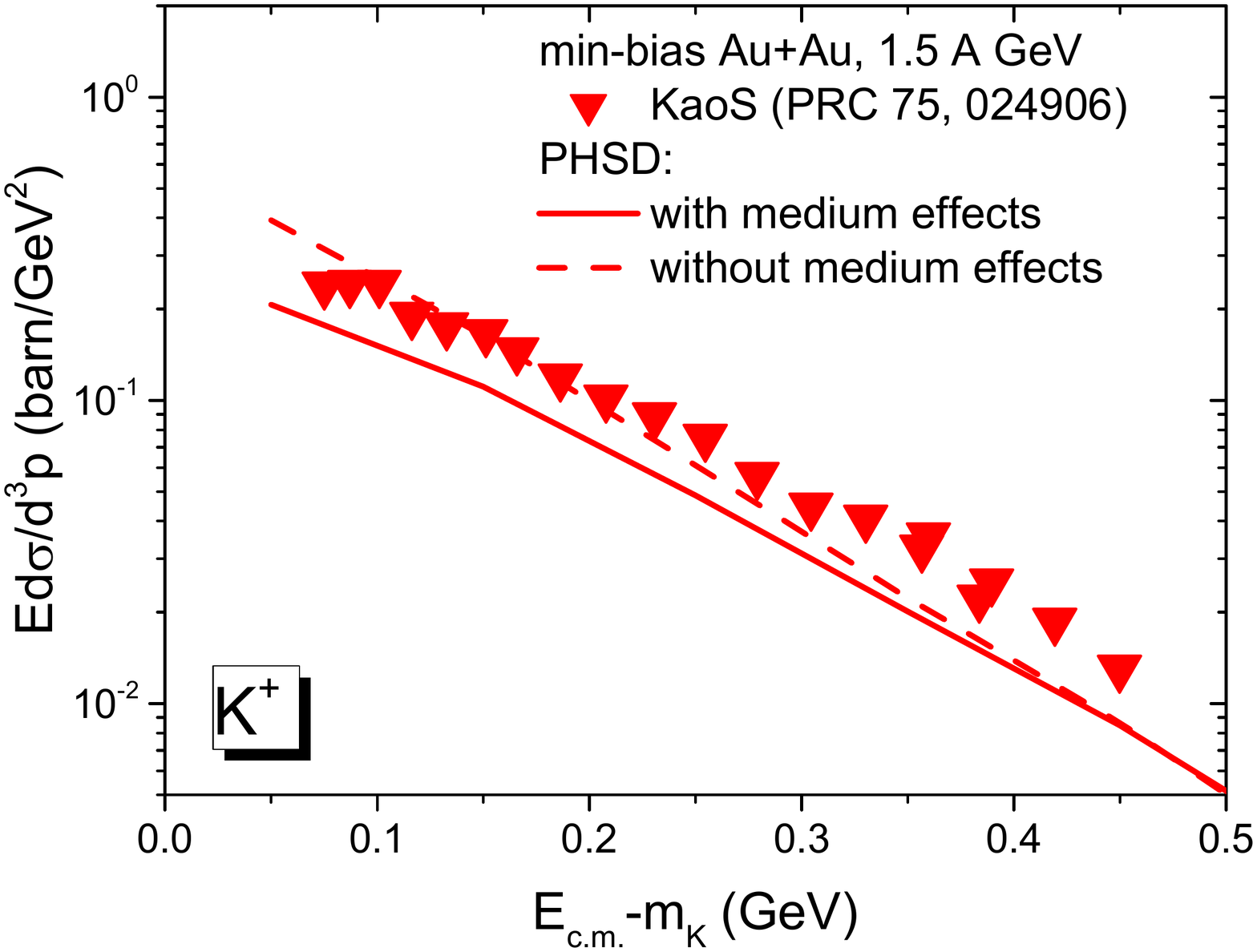}
\includegraphics[width=8.6 cm]{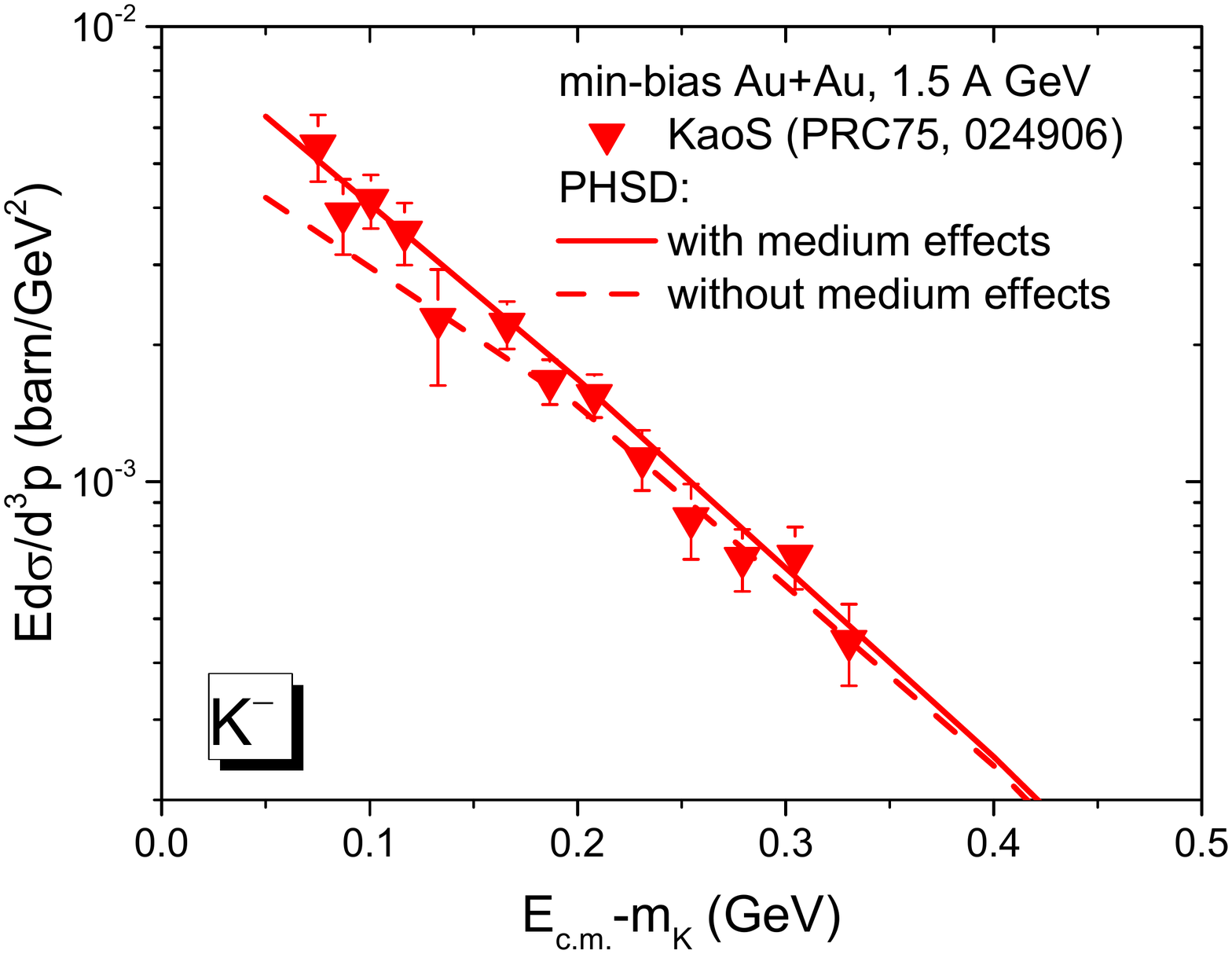}}
\vspace*{-5mm}
\centerline{
\includegraphics[width=8.6 cm]{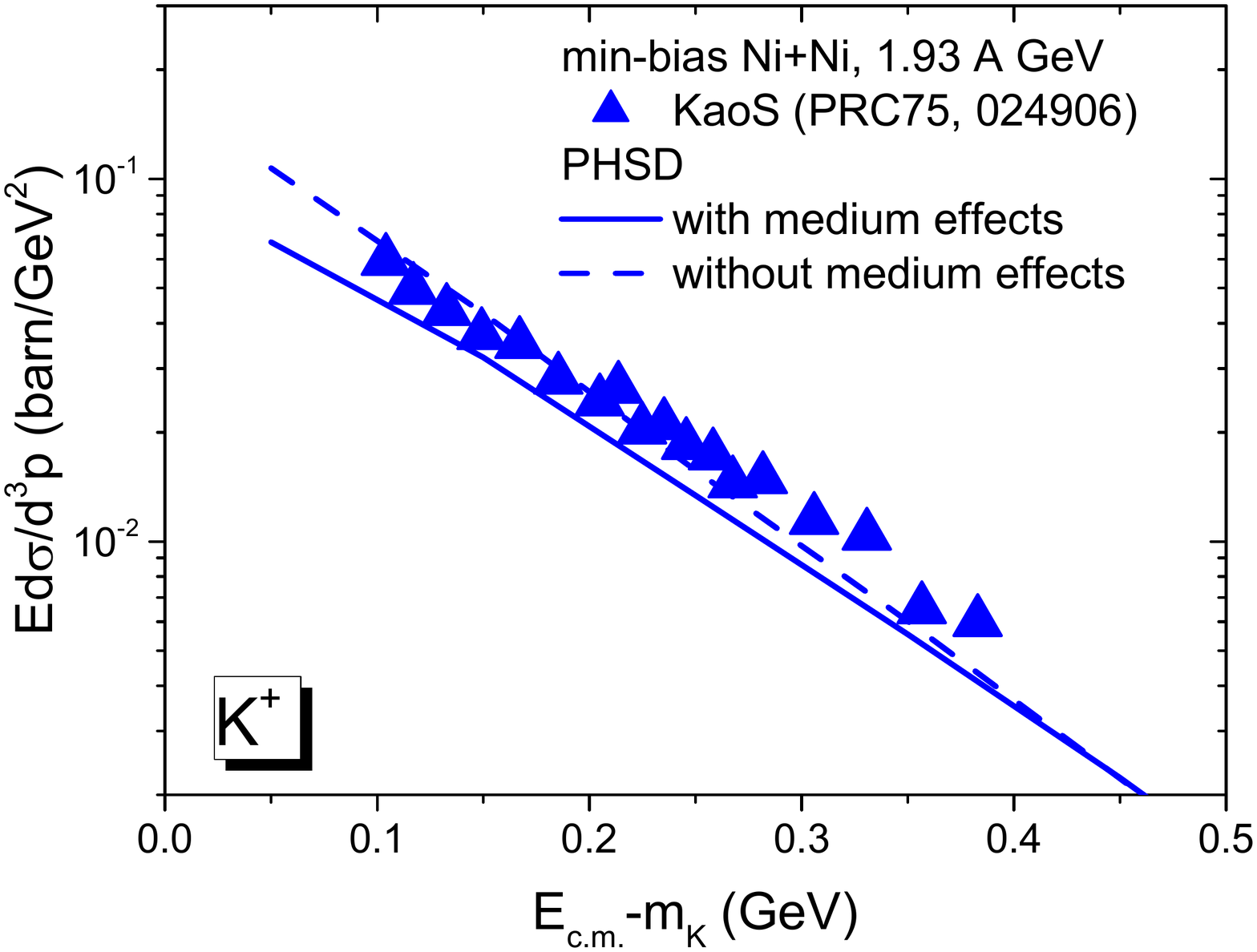}
\includegraphics[width=8.6 cm]{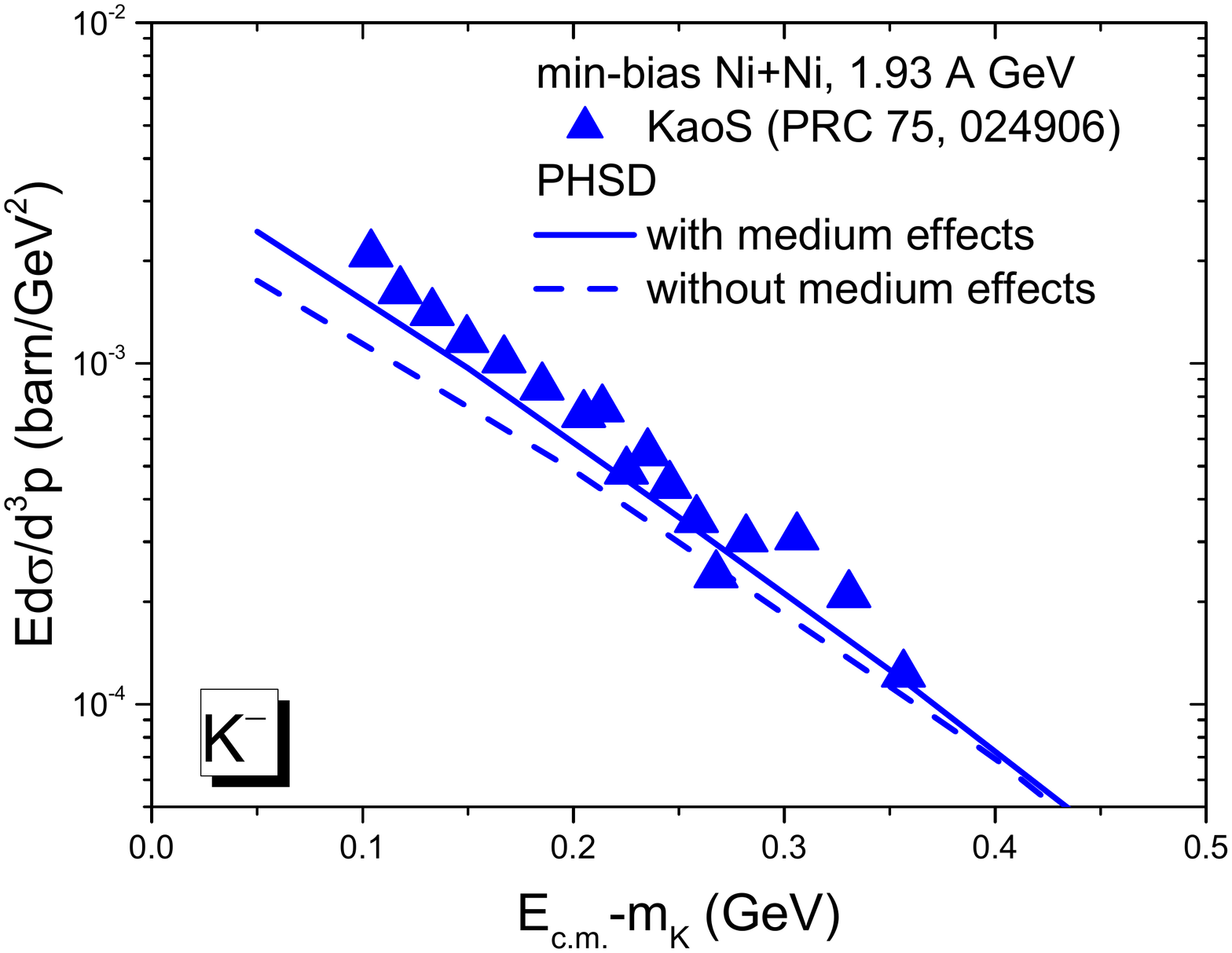}}
\vspace*{-5mm}
\centerline{
\includegraphics[width=8.6 cm]{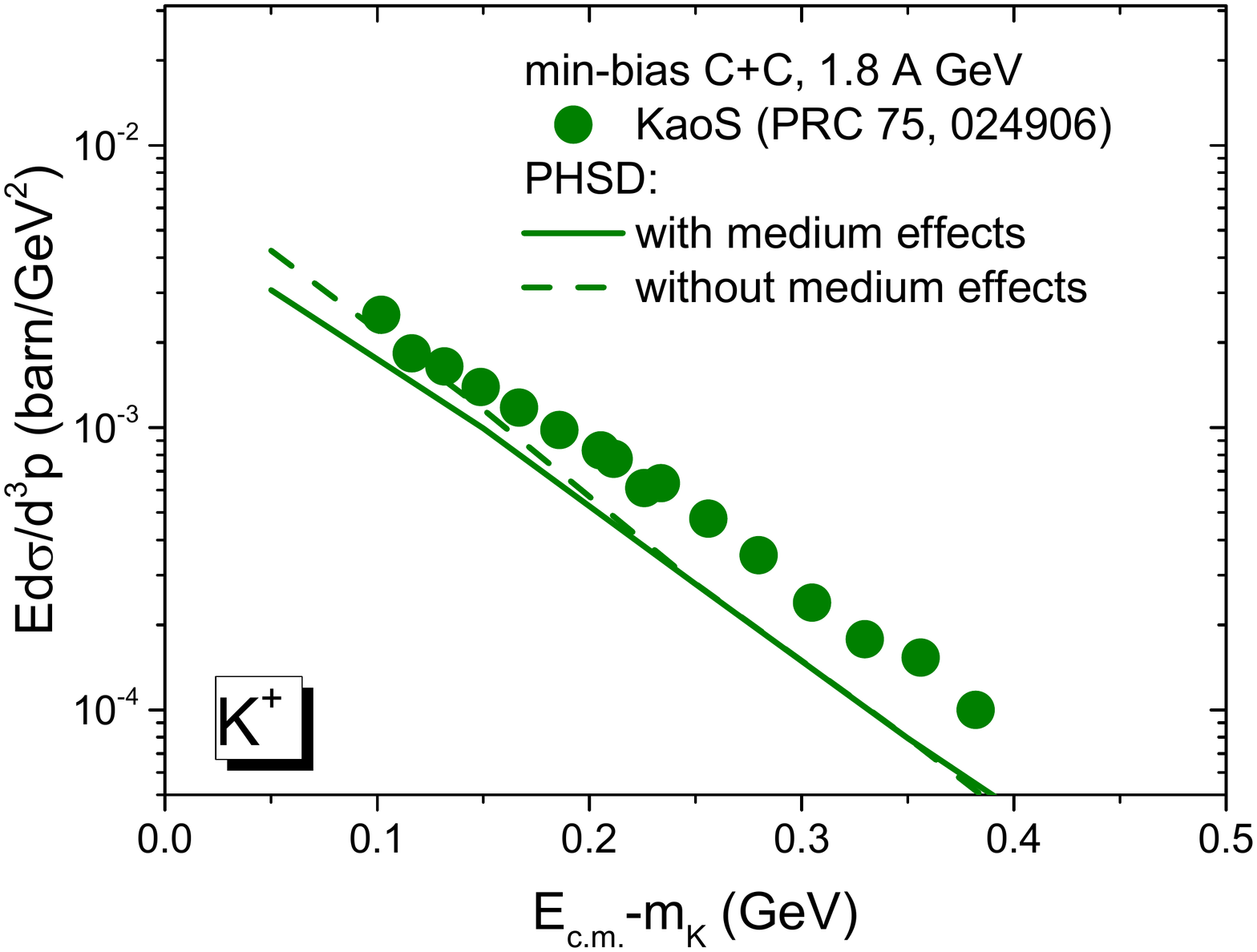}
\includegraphics[width=8.6 cm]{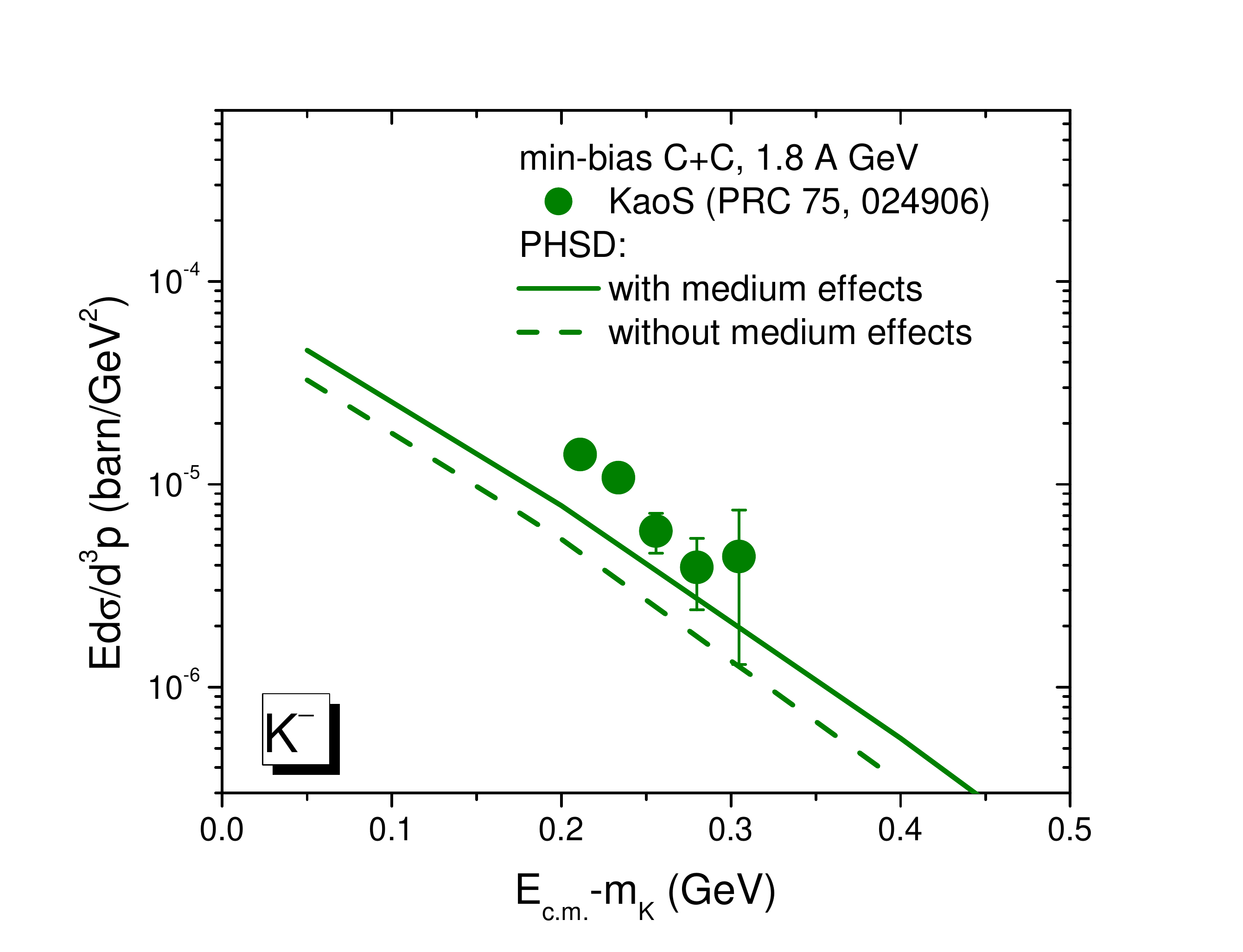}}
\caption{(Color online) The PHSD results for the inclusive invariant cross sections 
at midrapidity as a function of the kinetic energy $E_{c.m.}-m_K$ for $K^+$ (left) and  $K^-$ (right) mesons 
in minimum-bias Au+Au collisions at 1.5 A GeV (upper),  Ni+Ni collisions at 1.93 A GeV (middle)
and C+C collisions at 1.8 A GeV (lower) with (solid lines) and without (dashed lines) medium effects, compared to the experimental data from the KaoS Collaboration~\cite{Forster:2007qk}.
The midrapidity condition is a selection of $\theta_{c.m.} = 90^0 \pm 10^0$.
}
\label{pt-minb}
\end{figure*}

\begin{figure}[th!]
\centerline{
\includegraphics[width=8.6 cm]{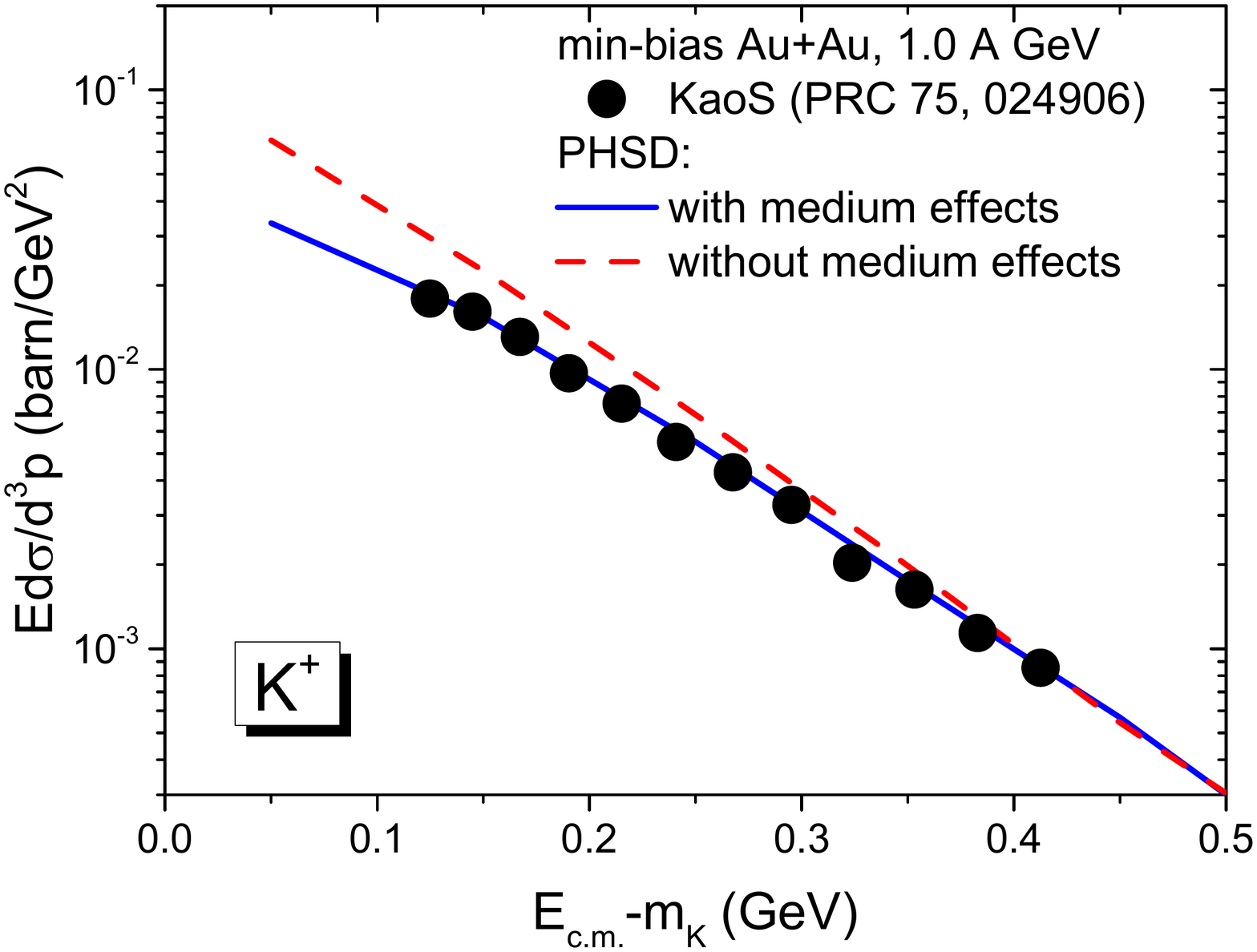}
}
\caption{(Color online) The PHSD results for the inclusive invariant cross section 
at midrapidity as a function of the kinetic energy $E_{c.m.}-m_K$ for $K^+$ mesons 
in minimum-bias Au+Au collisions at 1.0 A GeV  without (red dashed line) 
and with (blue solid line)  medium effects compared to the experimental data from the KaoS Collaboration~\cite{Forster:2007qk}.
The midrapidity condition is a selection of $\theta_{c.m.} = 90^0 \pm 10^0$.
}
\label{pt-minbAu10}
\end{figure}

Now we advance to a heavy system and show in Fig. \ref{HADES-mt}
the PHSD results for the midrapidity $m_T$-spectra of $K^+$ (blue line) 
and  $K^-$(green line) as a function of the transverse 
kinetic energy $m_T-m_K$ in 40\% central Au+Au collisions at 1.23 A GeV 
in comparison to the HADES data \cite{Adamczewski-Musch:2017rtf}.
The HADES data for $K^+$ are overestimated by about 20-30\% 
at large $m_T$ and agree with the data at small $m_T$ for the in-medium scenario 
(solid blue line).
Without medium effects (dashed blue line) we over-predict the data by a factor of 2.  
The discrepancies for the $K^-$ with medium effects are about factor of 2
at low $m_T$ (solid green line) and by about 30\% (within statistical fluctuations) 
at larger $m_T$ for the free scenario (dashed green line). 

It is interesting to compare our results for the $m_T$-distributions
not only to the mid-rapidity data as shown in Fig. \ref{HADES-mt},
but also to the non-central rapidity bins which have been used for the extrapolation 
of the HADES results for $dN/dy$ shown in Fig. \ref{NiNi-y-central}.
In Fig. \ref{HADES-ymt} we display the PHSD $m_T$-spectra of $K^+$ (upper) and  
$K^-$ (lower) (including in-medium effects)  for different rapidity bins
in 40\% central Au+Au collisions at 1.23 A GeV  in comparison to the experimental data 
of the HADES Collaboration \cite{Adamczewski-Musch:2017rtf}.
As one can see, the slope of the $m_T$-spectra of $K^+$ is getting harder 
when coming to midrapidity. The $m_T$-spectra of $K^+$ for rapidity bins $|y|>0.45$
are very well described, however, the agreement is getting worse (up to 30\% overestimation
for large $m_T$) when coming to the midrapidity bin, as discussed 
above for Fig. \ref{HADES-mt}.  
For $K^-$ mesons the slopes of the $m_T$-spectra for each $y$-bin are approximately 
in line with the HADES data, however, the absolute yield for each $y$-bin is 
overestimated by about of a factor of 2.
This is reflected in the $y$- distribution of $K^+$ and $K^-$ of Fig. \ref{NiNi-y-central}, too.


Figure~\ref{pt-minb} shows the inclusive invariant cross sections 
at midrapidity as a function of the kinetic energy $E_{c.m.}-m_K$
for $K^+$ and $K^-$ mesons in minimum-bias 
Au+Au collisions at 1.5 A GeV, Ni+Ni collisions at 1.93 A GeV and C+C collisions 
at 1.8 A GeV with and without medium effects, which are compared to the experimental 
data from the KaoS Collaboration~\cite{Forster:2007qk}.
Contrary to the kaon $p_T$-spectra in central Ni+Ni collisions (shown in 
Fig.~\ref{NiNi-pt-central}), the yields of $K^+$ mesons in minimum-bias are 
slightly underestimated as compared to the experimental data, which implies that 
our $K^+$ production in non-central collisions is smaller than in the experimental data.
However, the slopes of the $K^+$ spectra are close to the experimental data 
with the kaon potential switched on.
As seen from Fig.~\ref{NiNi-pt-central}, the repulsive force from the kaon potential 
hardens the spectra relative to those without kaon potential.
The effects of the kaon potential are stronger in Au+Au collisions than in 
C+C collisions, since the baryon density achieved in Au+Au collisions 
is higher than in C+C collisions.

The same applies for the $K^-$ mesons (right panel of Fig.~\ref{pt-minb}):
the maximal softening of the $p_T$-spectrum of $K^-$ occurs in Au+Au collisions.
The softening of the $p_T$-spectrum is also well visible in minimum-bias
Ni+Ni collisions, however, the effect is smaller than in 
central Ni+Ni collisions in Fig.~\ref{NiNi-pt-central}, since the baryon density 
is larger for central collisions than in minimum-bias Ni+Ni collisions.
For C+C collisions the softening is invisible (with the statistics achieved in the calculations), moreover, the absolute yield is underestimated for 
C+C even with the inclusion of in-medium effects.

Additionally, in Fig. \ref{pt-minbAu10} we show the inclusive invariant cross section 
at midrapidity as a function of the kinetic energy $E_{c.m.}-m_K$ for the $K^+$ mesons
in minimum-bias Au+Au collisions at strongly sub-threshold energy of 1.0 A GeV  
without (red dashed line) and with (blue solid line) medium effects compared to the experimental data from the KaoS Collaboration~\cite{Forster:2007qk}.
One can see that the PHSD describes the $K^+$ spectra very well with 
in-medium effects and substantially overestimates the data without 
the repulsive potential.

\begin{figure}[th!]
\centerline{
\includegraphics[width=8.6 cm]{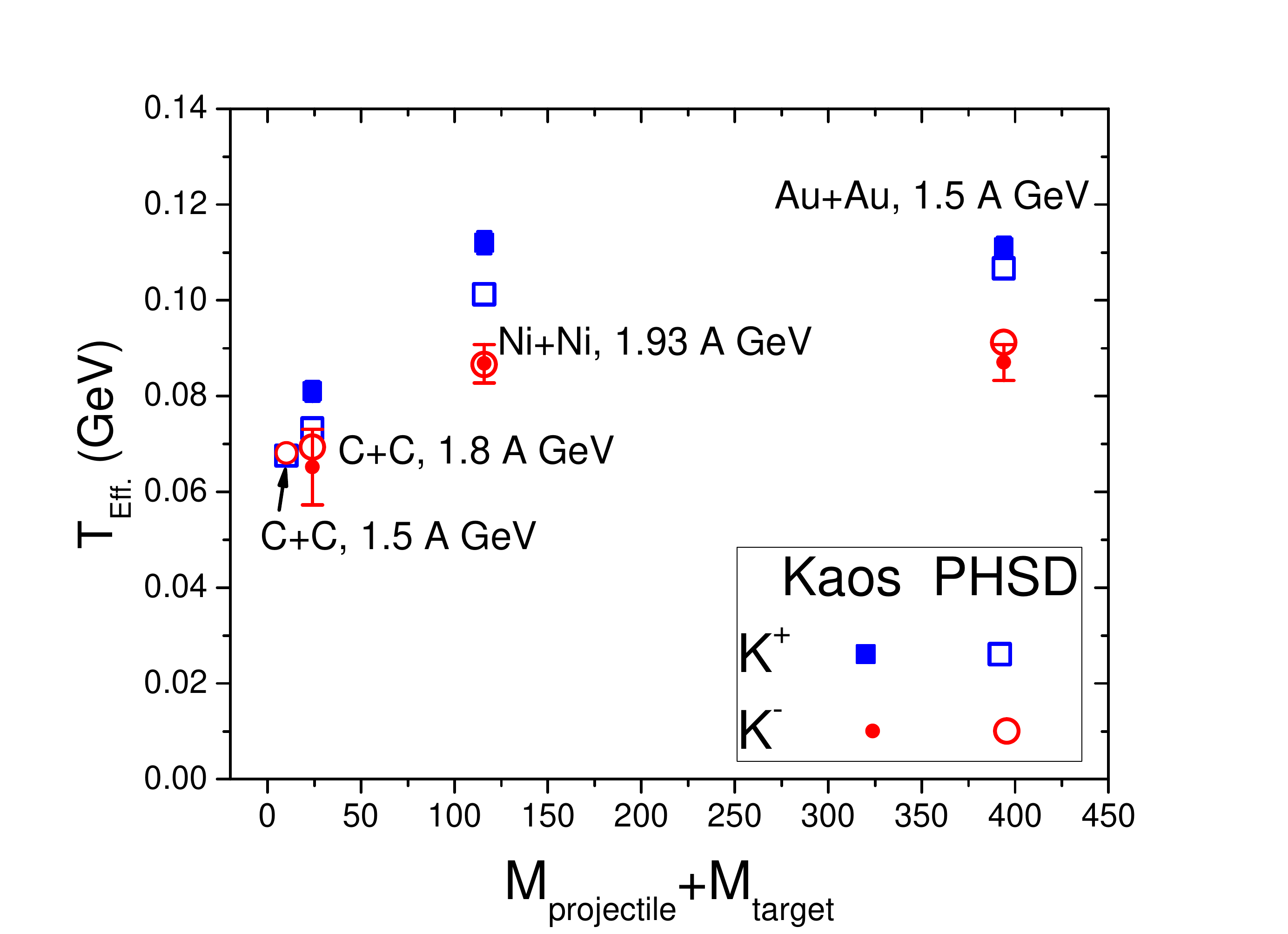}}
\caption{(Color online) The PHSD results for the effective temperatures (or the inverse slopes) of 
$K^+$ and $K^-$mesons  as a function of the mass number of projectile and target nuclei in 
minimum-bias collisions of C+C at 1.5 and 1.8 A GeV, 
Ni+Ni at 1.93 A GeV and Au+Au at 1.5 A GeV
in comparison to the experimental data from the KaoS Collaboration~\cite{Forster:2007qk}. 
We shift the PHSD results for C+C at 1.5 A GeV to a lower mass number
for better visibility.} 
\label{Teff}
\end{figure}

The transverse momentum spectra can be characterized by an effective temperature 
$T_{eff}$ which can be obtained from a fit of the (anti)kaon spectra 
with a thermal distribution function: 
\begin{eqnarray}
E\frac{d\sigma}{d^3p}\sim E\exp\bigg(-\frac{E}{T}\bigg).
\label{temeff}
\end{eqnarray}
Since radial flow, which hardens the spectrum, contributes to the slope, 
$T$ is called an effective temperature $T_{eff}$.
Fig.~\ref{Teff} shows the effective temperatures for $K^+$ and $K^-$ mesons in minimum-bias 
C+C at energies of 1.5 and 1.8 A GeV;
Ni+Ni and Au+Au collisions, respectively, at energies of 1.93 and 1.5 A GeV 
as a function of the mass number of target and projectile nuclei.
The effective temperature depends on the range of transverse momenta considered and our results 
are fitted to the spectra between $0<E-m_K<0.5~{\rm GeV}$.
Because of collective flow the effective temperature in Au+Au collisions 
is higher than that in C+C collisions, although the collision energy is higher 
in the latter case.

One can see that the effective temperature for $K^+$ mesons is higher than for $K^-$ mesons, 
because the kaon potential - which generates a repulsive force - hardens the $K^+$ spectra, while  the inclusion of the self-energy within the G-matrix approach
softens the $K^-$ spectra. This effect is stronger for the heavy systems:
the splitting of the effective temperatures for $K^+$ and $K^-$ is small 
in C+C collisions, where the size and density of nuclear matter is small, and 
the splitting is larger for Au+Au collisions which produce a much larger 
and denser fireball. Moreover, the larger density achieved in Au+Au collisions
compared to  C+C collisions lead to the relative increase of the $T_{eff}$
of $K^+$ and $K^-$ with system size.
Our results are close to the experimental data of the 
KaoS Collaboration~\cite{Forster:2007qk} and qualitatively consistent.

The difference of the effective temperatures of (anti)kaons 
for C+C and Au+Au at 1.5 A GeV is mostly attributed to in-medium effects.
As demonstrated in Fig. \ref{pt-shift} (upper plot), the elastic scattering increases 
the effective temperature for both, kaons and antikaons, considerably 
(since the (anti)kaons are produced near the thresholds, they are not energetic due to
phase space limitation). 
The interaction with the medium is repulsive for kaons and mostly attractive for antikaons, and this leads to the separation of their effective temperatures in the Au+Au reaction.
Such separation is not visible for the light C+C system 
due to the small volume of the medium and lower baryon densities.

We note that the different slopes of the $K^+$ and $K^-$  $m_T$-spectra for Au+Au collisions at 1.23 A GeV have been interpreted by HADES collaboration 
\cite{Agakishiev:2009ar,Adamczewski-Musch:2017rtf}
by the feed-down from $\phi$-meson decay, which substantially softens the spectra 
of $K^-$ mesons compared to the 'thermal' hard component - cf. Fig. 2 
in Ref. \cite{Adamczewski-Musch:2017rtf}.
However, according to the present PHSD calculations, the contribution to the $K^-$ 
abundances for Au+Au collisions at 1.5 GeV from the $\phi$ decay plays a sub-leading 
role - cf. Fig. \ref{sources} and  the different slopes of $K^+$ and $K^-$
is attributed to the in-medium effects (as discussed above).  
This result is in line with the experimental data of the KaoS Collaboration 
which show that the splitting $T_{eff}$ of $K^+$ and $K^-$ spectra increases 
with the system size.
One can emphasize that if the antikaons are dominantly produced by $\phi$ decay 
one would expect a different behaviour:  since the $\phi$-nucleon interaction 
cross section  is small (cf. Ref. \cite{joos}) and the $\phi$-meson  life time 
is long, i.e. the most of the $\phi$ mesons decay to $K\bar K$ outside of the fireball
at almost zero baryon density,  one would expect that the slope of the spectra
of antikaons do not show the prominent dependence on the system size.  
Thus, we stress, that by a measurement of the effective temperature of $K^-$ 
for different system sizes for the same beam energy one can 
shed more light on the origin of the antikaon production and the 
role of the in-medium effects.

We note that our results for the rapidity and $p_T$-spectra  
are consistent with previous findings within the HSD \cite{Cass97,brat97,CB99,laura03} 
and other groups \cite{Hartnack:2011cn,Kolomeitsev:2004np} 
about the necessity to include medium effects for the 
proper description of the experimental data on strangeness production at low energies.

We stress again that the comparison to the HADES data on $m_T$-spectra
for Au+Au collisions at 1.23 A GeV falls out of this systematics, although
the PHSD results with in-medium effects (as well as the results of other 
transport models \cite{Hartnack:2011cn,Kolomeitsev:2004np}) agree well 
with the KaoS and FOPI data even for Au+Au collisions at 1.0 and  
1.5 A GeV. Moreover, the PHSD reproduces  well the KaoS and FOPI data 
for more light system as Ni+Ni at 1.93 A GeV
and the HADES data for Ar+KCl at 1.76 A GeV. 
This tension is presently unsolved and requires further investigations
from theoretical and experimental sides.

In this respect,  we note that in the present calculation the $s-$wave 
and $p-$wave of the antikaon-nucleon 
interaction have been obtained from the leading-order contribution together with the
$\Lambda, \Sigma$ and $\Sigma^*$ pole terms. In order to improve the agreement with
data for large momenta, future work should include the analysis of the effect of next-to-leading-order corrections to the antikaon-nucleon interaction as well as the inclusion of strange baryonic resonances located at energies between 1.89 and 2.35 GeV,  well above the antikaon-nucleon threshold, as  recently discussed in Ref.~\cite{Feijoo:2018den}.

\subsection{Polar distributions}

\begin{figure*}[th!]
\centerline{
\includegraphics[width=8.6 cm]{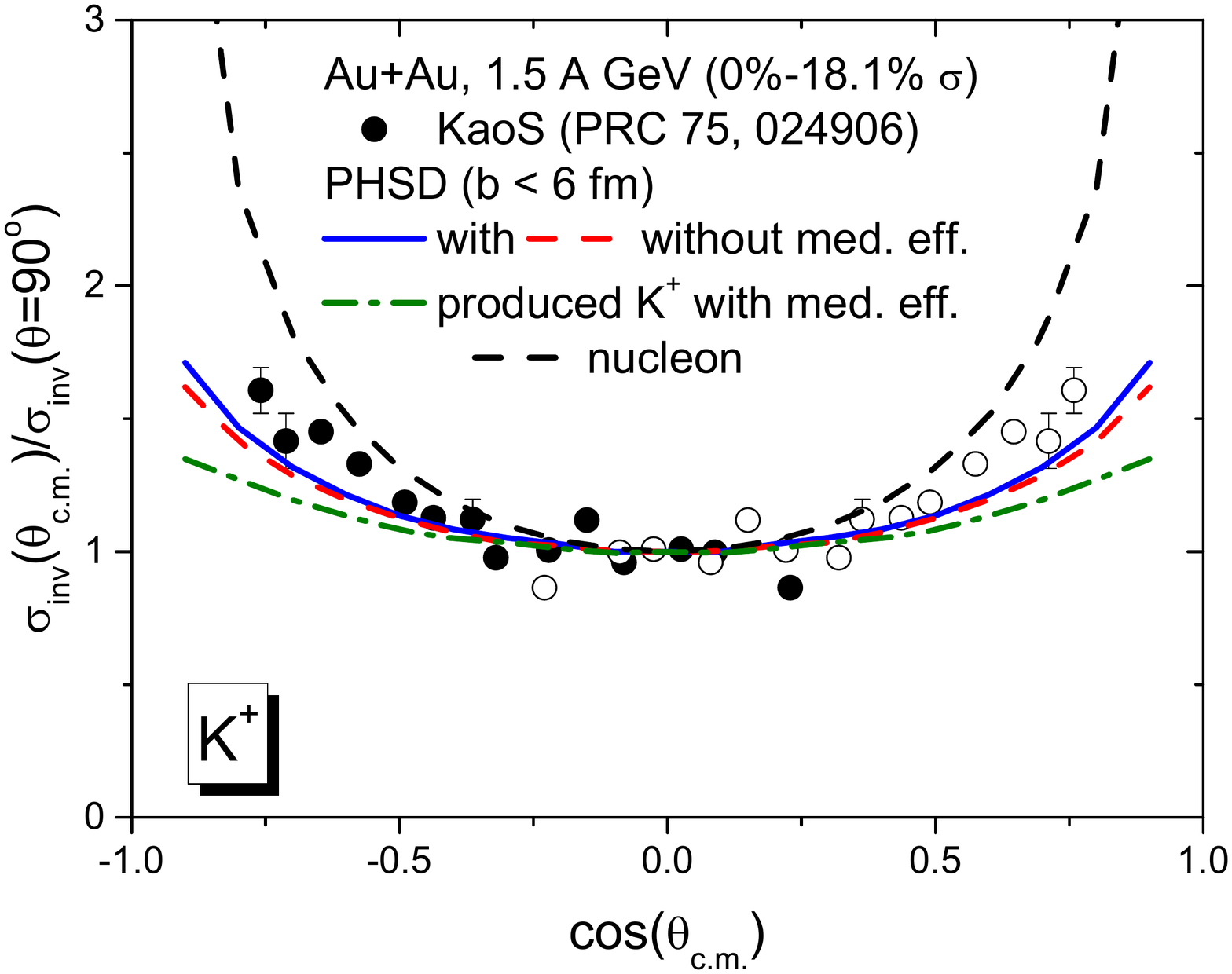}
\includegraphics[width=8.6 cm]{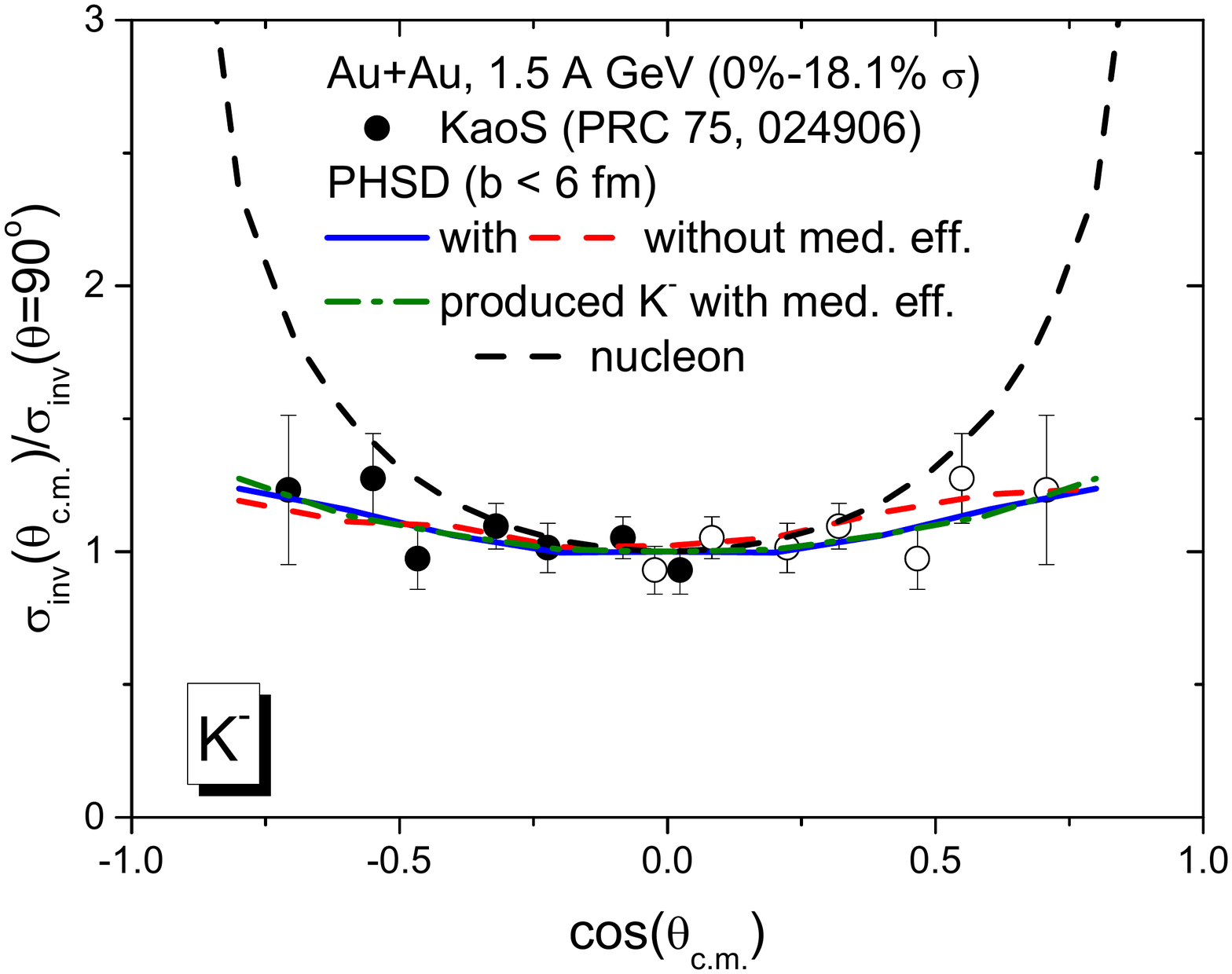}}
\centerline{
\includegraphics[width=8.6 cm]{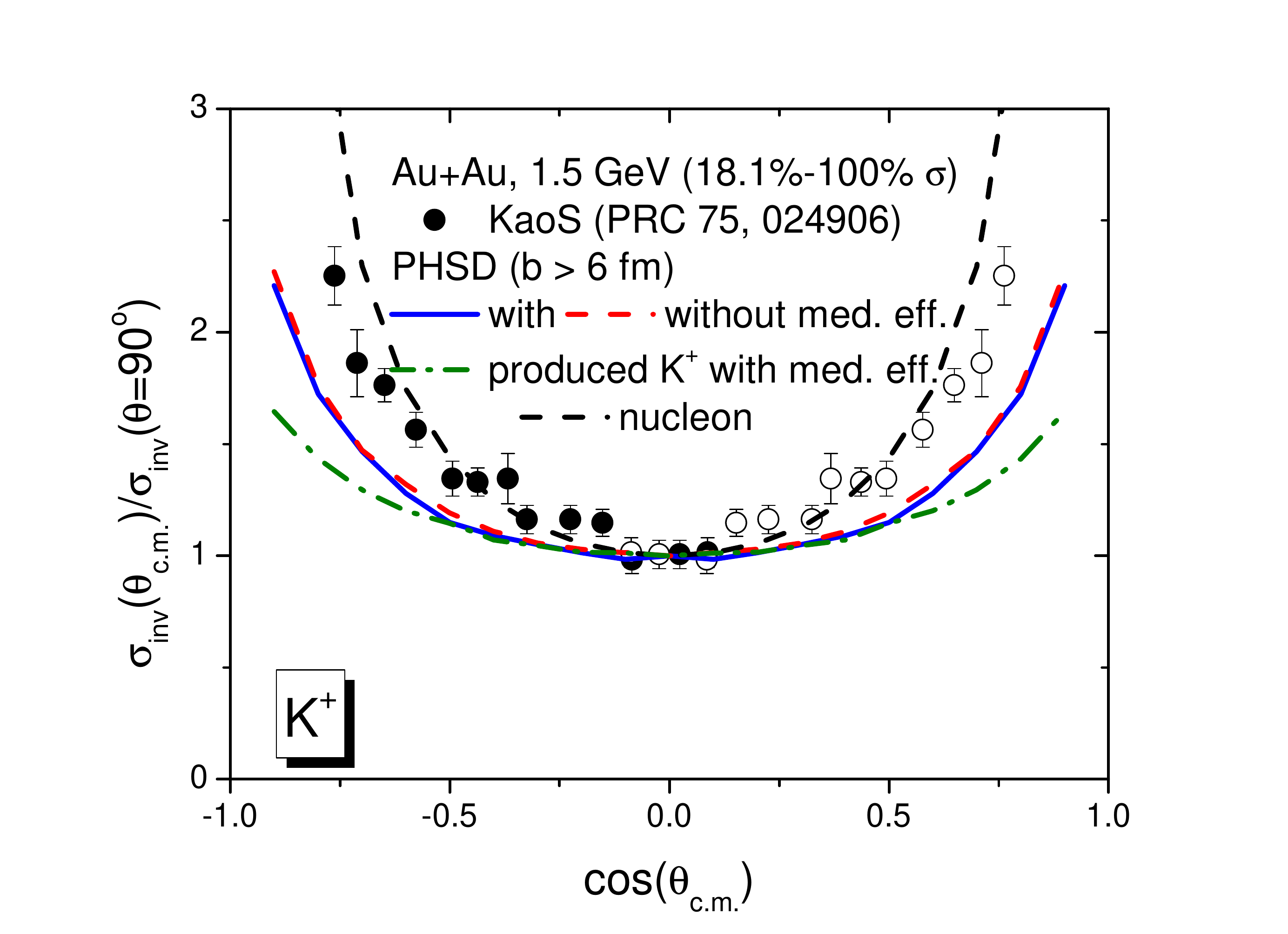}
\includegraphics[width=8.6 cm]{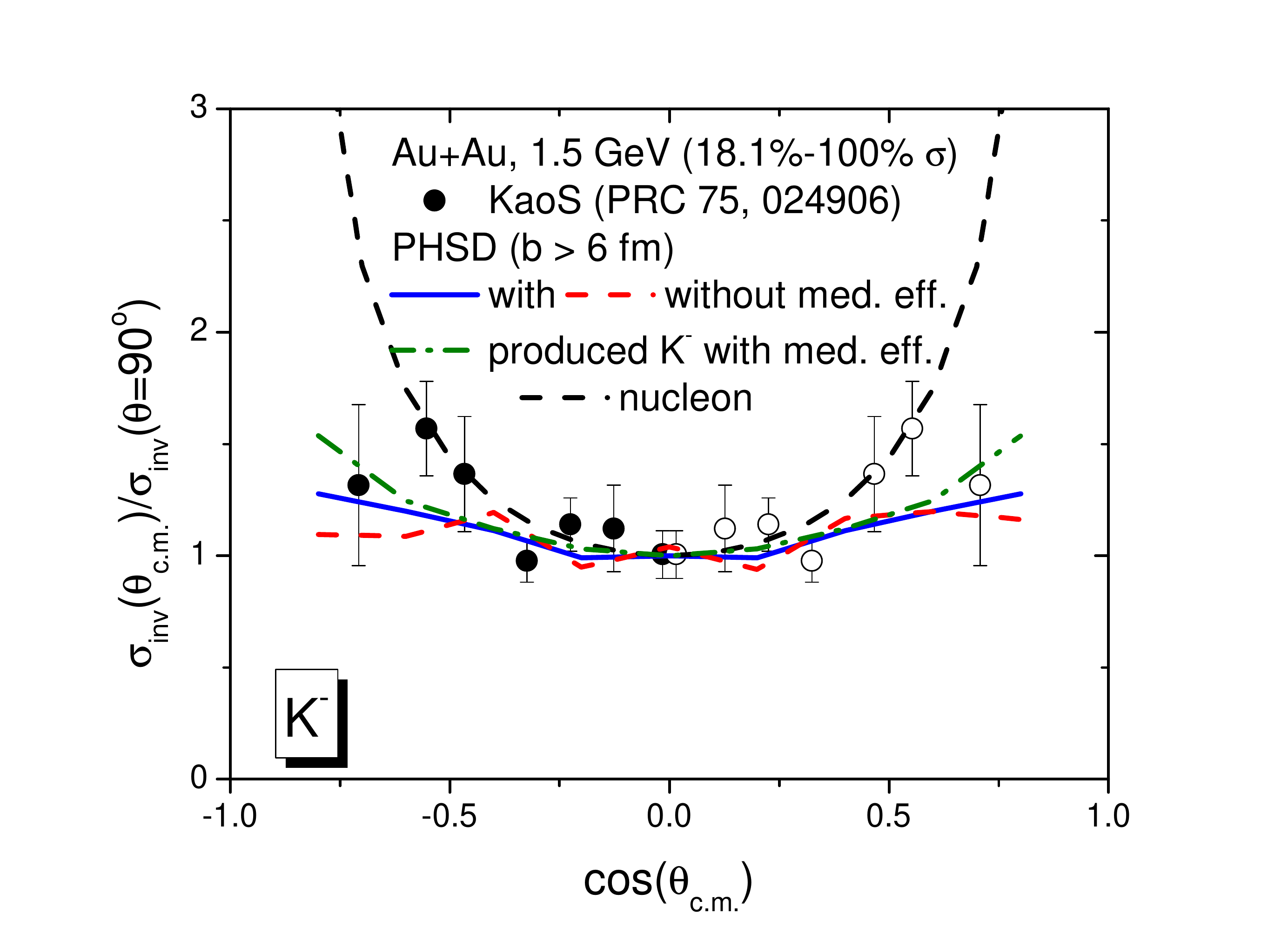}}
\caption{(Color online) The polar angular distributions of  $K^+$ (left) and  $K^-$ (right)
mesons in  central (upper) and noncentral  (lower) Au+Au collisions at 1.5 A GeV 
compared to the experimental data from the KaoS Collaboration~\cite{Forster:2007qk}.
The distributions are normalized to unity at $\cos(\theta_{c.m.})=0$, 
i.e. presented as the ratios  
$\sigma_{inv}(\theta_{c.m.})/\sigma_{inv}(\theta_{c.m.}=90^0)$.
The solid blue and dashed red lines show the (anti)kaon angular distributions 
with and without the medium effects, respectively. The  dot-dashed green lines show their
distribution at the production point (with medium effects). The black dashed lines 
indicate the nucleon angular distributions.
 }
\label{polar-AuAu}
\end{figure*}

\begin{figure*}[th!]
\centerline{
\includegraphics[width=8.6 cm]{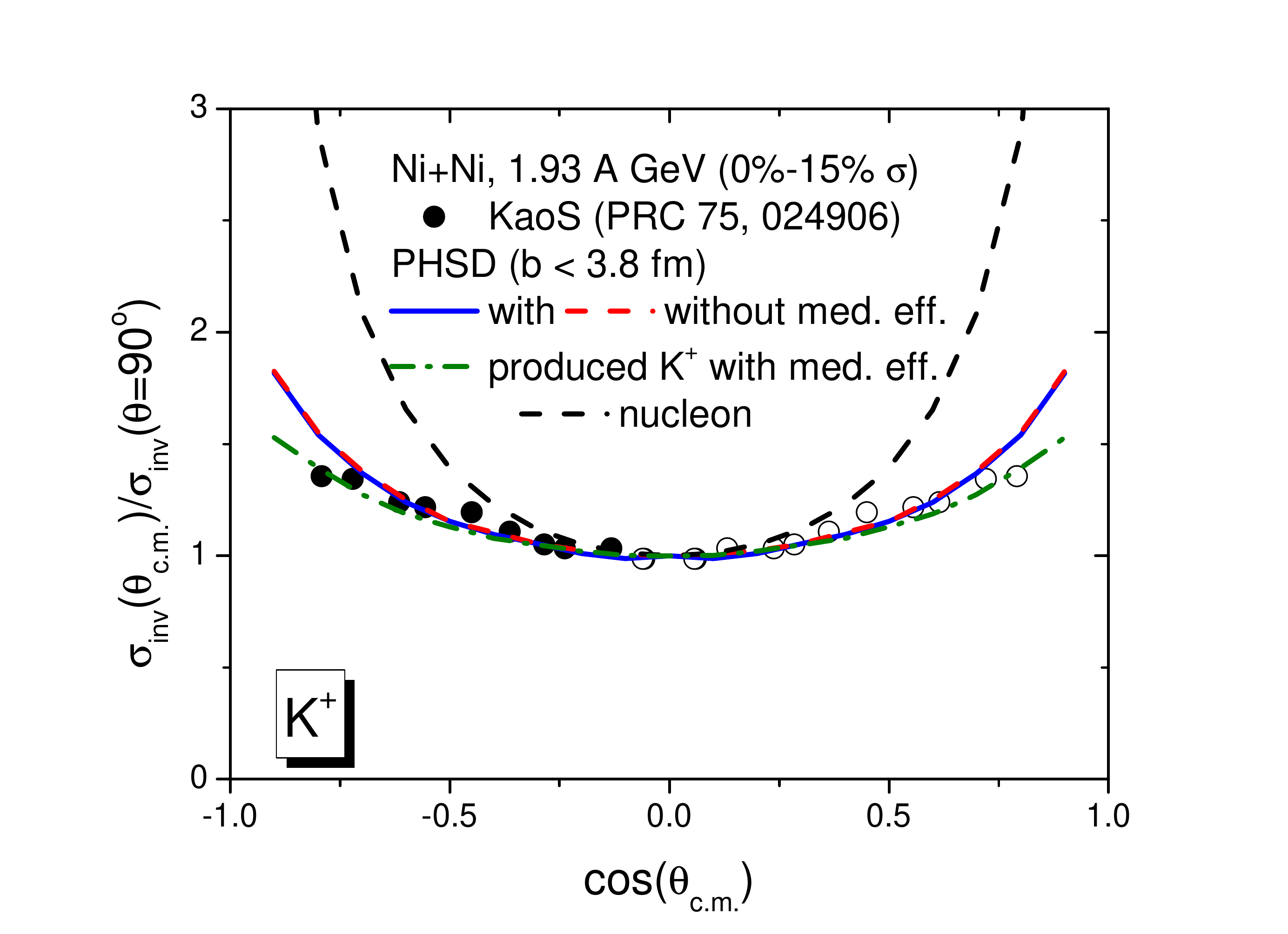}
\includegraphics[width=8.6 cm]{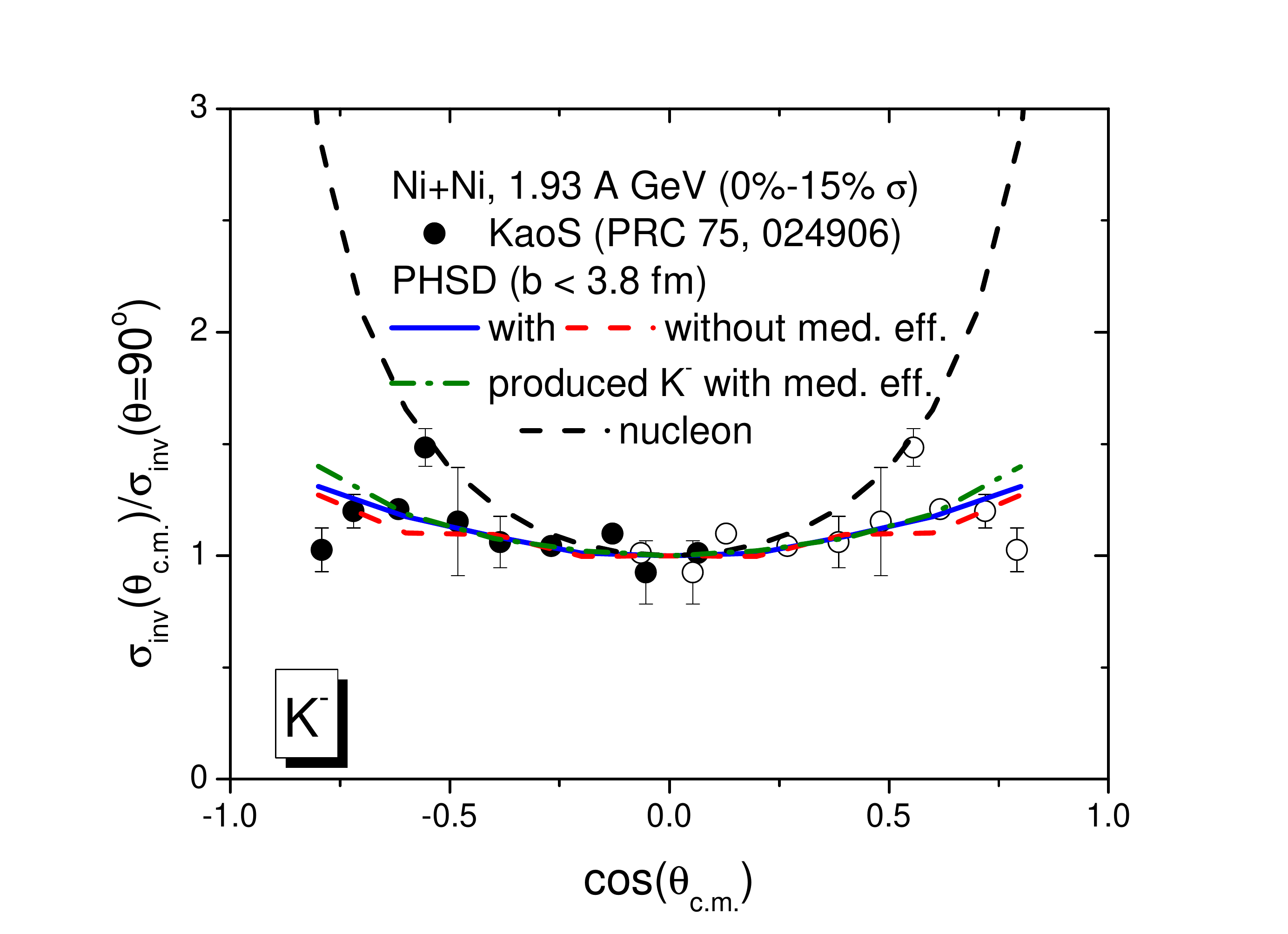}}
\centerline{
\includegraphics[width=8.6 cm]{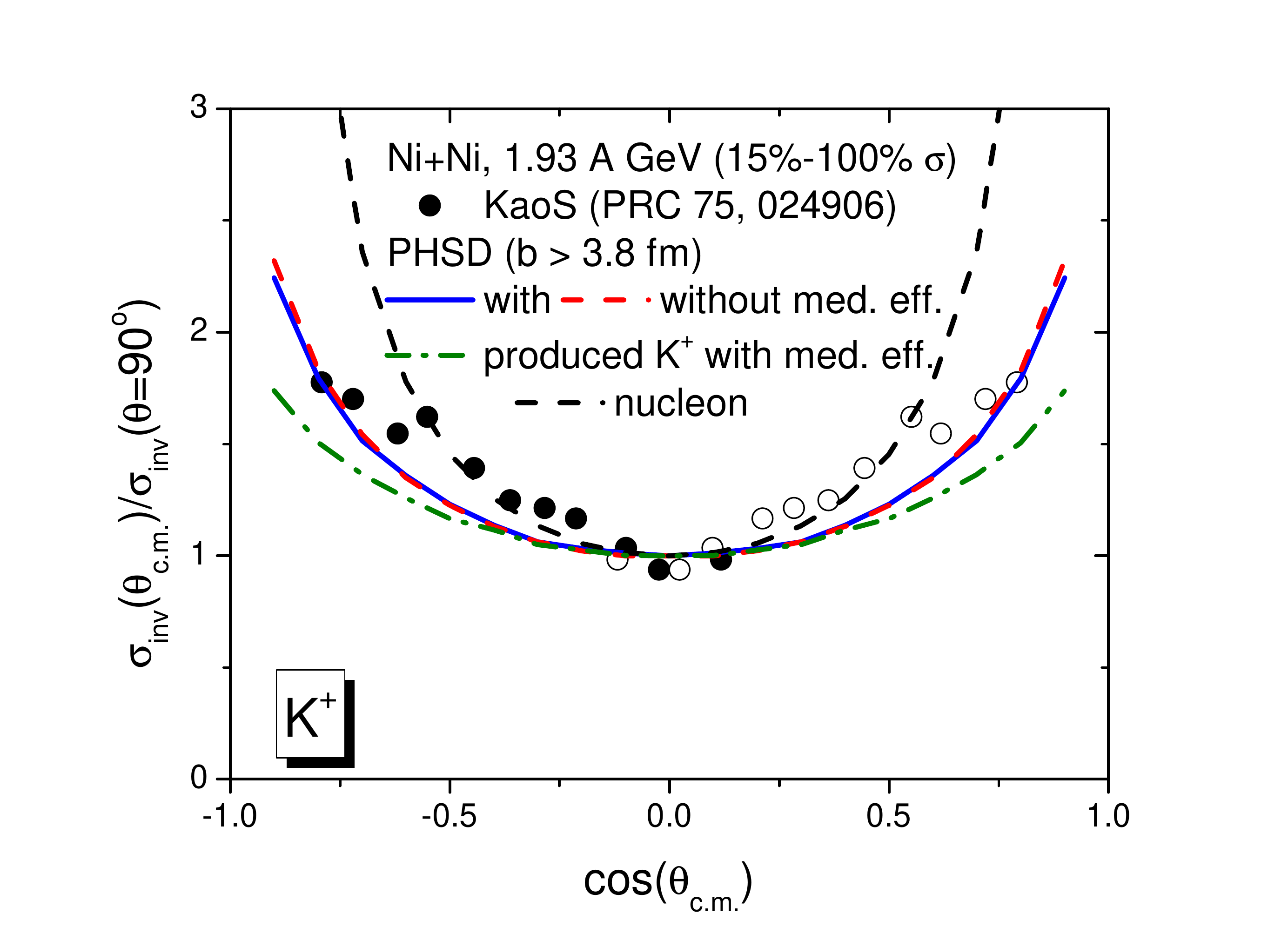}
\includegraphics[width=8.6 cm]{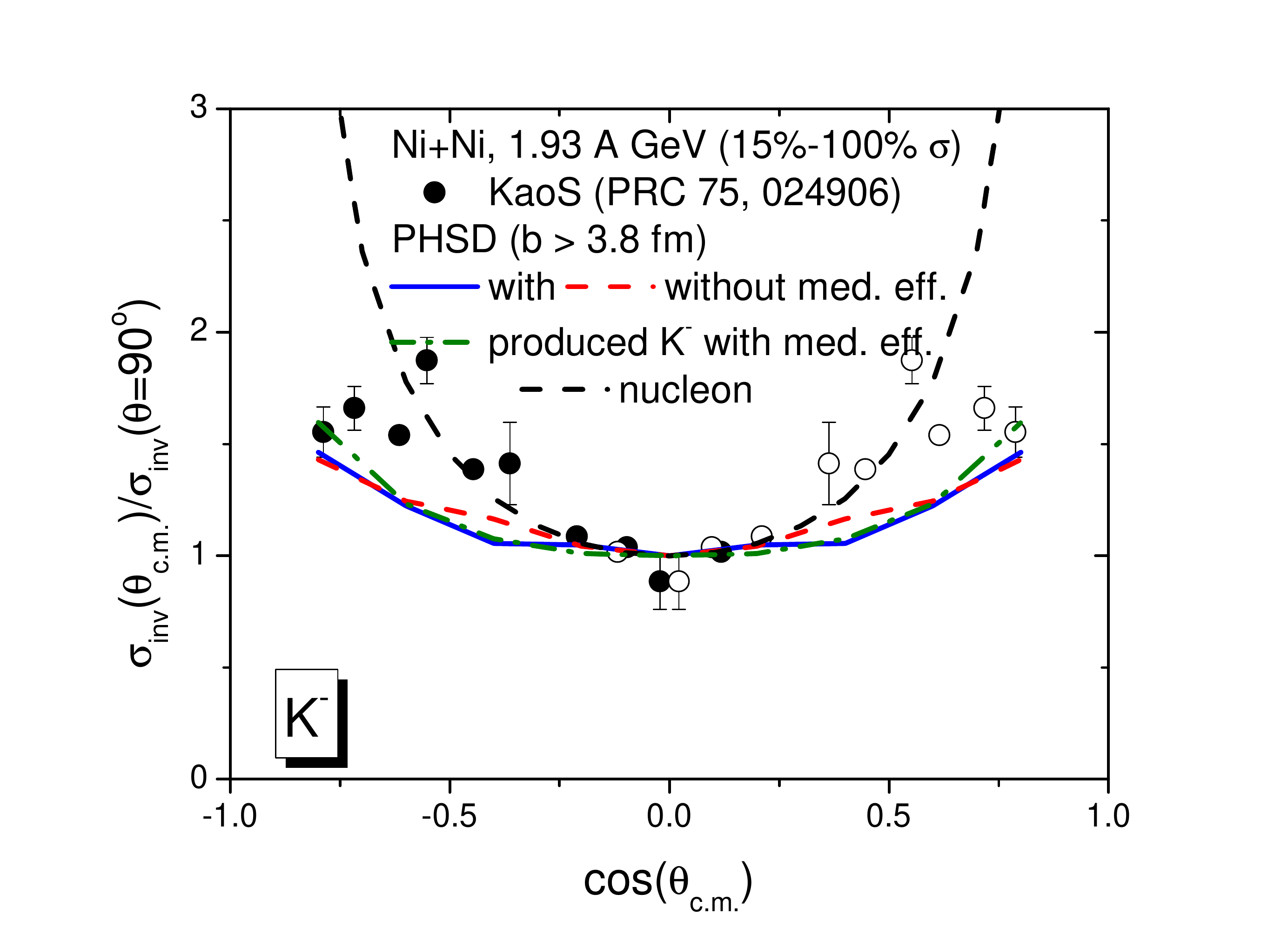}}
\caption{(Color online) The polar angle distributions of  $K^+$ (left) and  $K^-$ (right)
mesons in  central (upper) and  noncentral (lower) Ni+Ni collisions at 1.93 A GeV 
compared to the experimental data from the KaoS Collaboration~\cite{Forster:2007qk}.
The line description is the same as in Fig. \ref{polar-AuAu}.
}
\label{polar-NiNi}
\end{figure*}

Figure~\ref{polar-AuAu} shows the polar angle distributions of $K^+$ and $K^-$ in 
central (upper row: $(0-18.1)\%$ of total reaction cross section $\sigma$) 
and in noncentral (lower row: $(18.1-100)\% \cdot\sigma$) Au+Au collisions 
at 1.5 A GeV in comparison to the experimental data from the 
KaoS Collaboration~\cite{Forster:2007qk}.
The polar angle is defined as the angle between the beam axis and the particle, i.e.
$\cos(\theta_{c.m.})=\pm 1$ corresponds to the beam direction and 
$\cos(\theta_{c.m.})=0$ to midrapidity.
The distributions are normalized to unity at $\cos(\theta_{c.m.})=0$, 
i.e. presented as the ratios  
$\sigma_{inv}(\theta_{c.m.})/\sigma_{inv}(\theta_{c.m.}=90^0)$,  
where $\sigma_{inv}(\theta_{c.m.})$ is the invariant particle production 
cross section measured at a  polar angle $\theta_{c.m.}$.

As seen from Fig.~\ref{polar-AuAu},  the polar distribution of $K^+$ mesons 
in  noncentral collisions is strongly peaked in beam direction, 
while the distribution in central collisions is rather flat. 
This results from a strong correlation of $K^+$ mesons with nucleons
through the production of $K^+$ from nucleon+nucleon or nucleon+pion scattering and the 
interactions of $K^+$ mesons with nucleons.
One can see that the polar distribution of $K^+$ follows that of nucleons which 
is shown by black dashed lines both in central and noncentral Au+Au collisions.

The dot-dashed green lines display the polar distributions of produced $K^+$.
We note that the distribution of produced $K^+$ is only slightly affected 
when including the kaon potential.
For central collisions  the angular distribution of kaons is also rather flat, 
while for noncentral it grows  towards beam direction.
The differences between the dot-dashed green line and the dashed red line 
are caused by scattering and those between the dashed red line and the 
blue solid line is the effect of the kaon potential. Since the kaon potential is repulsive, 
it pushes the polar distributions of $K^+$ mesons towards forward and backward directions.
One can see that the effects of scattering on the polar distribution of $K^+$ mesons
are stronger than those of the kaon potential.

The right two panels of Fig.~\ref{polar-AuAu} show the polar distributions of $K^-$ 
mesons in central and noncentral Au+Au collisions, respectively.
Though the statistics is limited in noncentral collisions, one can see that
the scattering pushes $K^-$ mesons backward and forward as in the case of $K^+$.
However, the effects of the antikaon potential are opposite such that it pulls 
$K^-$ mesons into the middle of the fireball, since
the antikaon feels attraction.

In Fig.~\ref{polar-NiNi} we show the same polar distributions for Ni+Ni 
collisions at 1.93 A GeV. One can see a rather similar behaviour of 
$K^+$ and $K^-$ polar distributions as for Au+Au at 1.5 A GeV, 
although the effects of scattering and the potential are slightly weaker than in Au+Au collisions since the size and density of the produced fireball is smaller.

The minor influence of the in-medium effects on the polar distribution for Ni+Ni and Au+Au collisions might seem to be in conflict with the enhancement of the $K^+$ 
rapidity distribution shown in Fig.~\ref{NiNi-y-central}. However, both results are consistent due to the following reasons:  one has to keep in mind that the polar distribution is normalized to $\sigma_{inv}(\theta=90^0)$, i.e. at $\cos\theta_{c.m.}=0$. For kaons with a moderate transverse momentum in the range of 
$0.2 ~{\rm GeV} < p_T <0.8 ~{\rm GeV}$ (to which most kaons belong to), 
the rapidity $y=0.5$ corresponds to $0.8<\cos\theta_{c.m.}<0.9$.
Therefore, a wide range of $\cos\theta_{c.m.}$ in Fig.~\ref{polar-NiNi} corresponds 
to a narrow range close to midrapidity in Fig.~\ref{NiNi-y-central}.
If $dN/dy$ with and without medium effects would be rescaled 
(i.e. assuming the same number of kaon production with and without medium effects), the shapes of the 
$y-$ distribution near mid-rapidity $(-0.5 < y < 0.5)$ would be similar to each other.
That is why the polar distributions with and without medium effects 
are similar to each other in Figs. \ref{polar-AuAu} and \ref{polar-NiNi}, too.


\subsection{Azimuthal distributions}

The azimuthal distribution of particles is parameterized in a Fourier series as:
\begin{eqnarray}
\frac{dN(p_T,y)}{d\phi}=C[1+2v_1(p_T,y)\cos\phi\nonumber\\
+2v_2(p_T,y)\cos(2\phi)+\cdots],
\end{eqnarray}
where the coefficients $v_1$ and $v_2$ are, respectively, denoted as the directed 
and elliptic flows obtained by
\begin{eqnarray}
v_1(p_T,y)&=&\frac{\int d\phi \frac{dN(p_T,y)}{d\phi}\cos\phi}{\int d\phi \frac{dN(p_T,y)}{d\phi}},\\
v_2(p_T,y)&=&\frac{\int d\phi \frac{dN(p_T,y)}{d\phi}\cos(2\phi)}{\int d\phi \frac{dN(p_T,y)}{d\phi}}.
\end{eqnarray}

\begin{figure}[h!]
\centerline{
\includegraphics[width=8.3 cm]{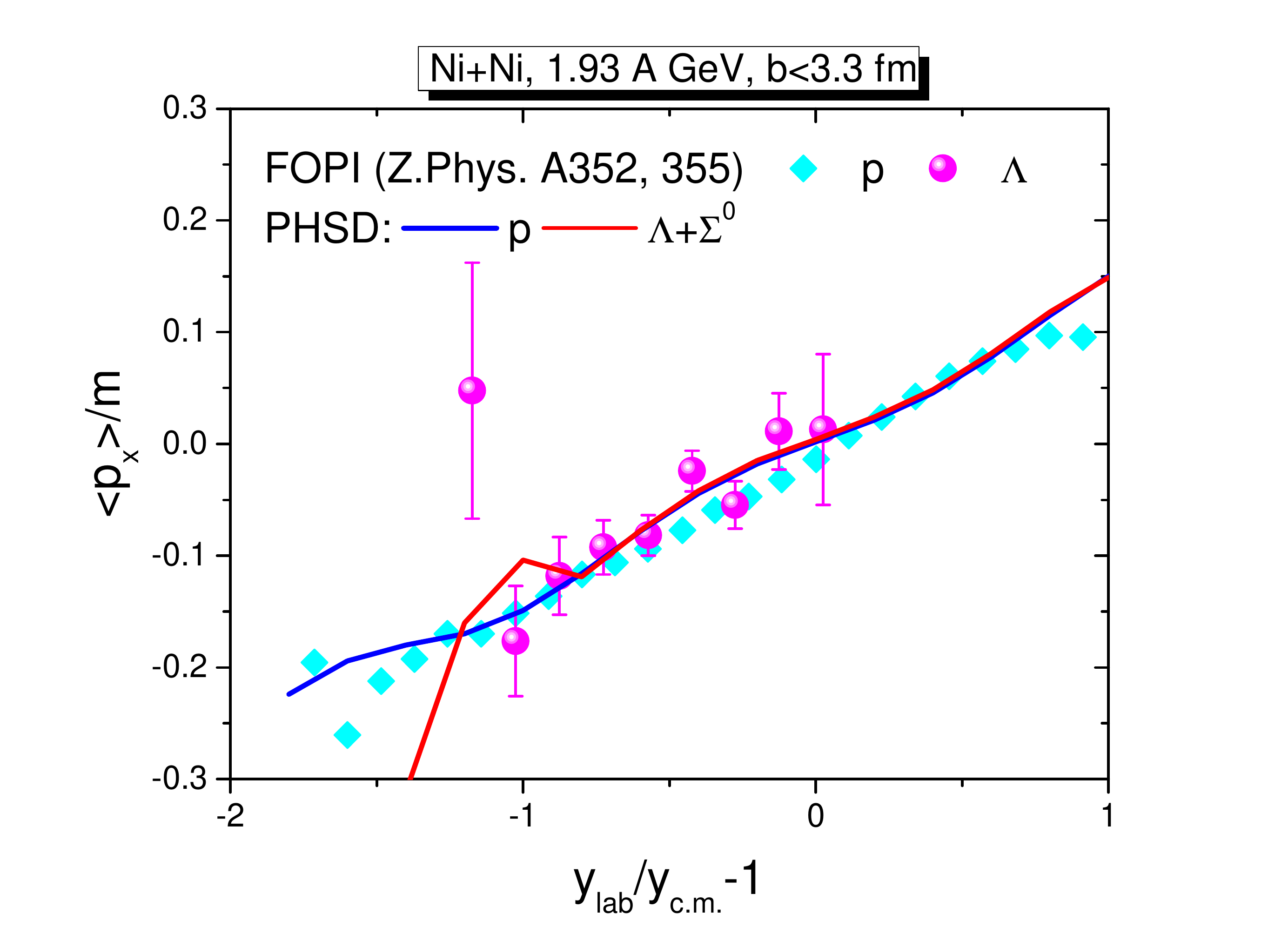}}
\centerline{
\includegraphics[width=8.3 cm]{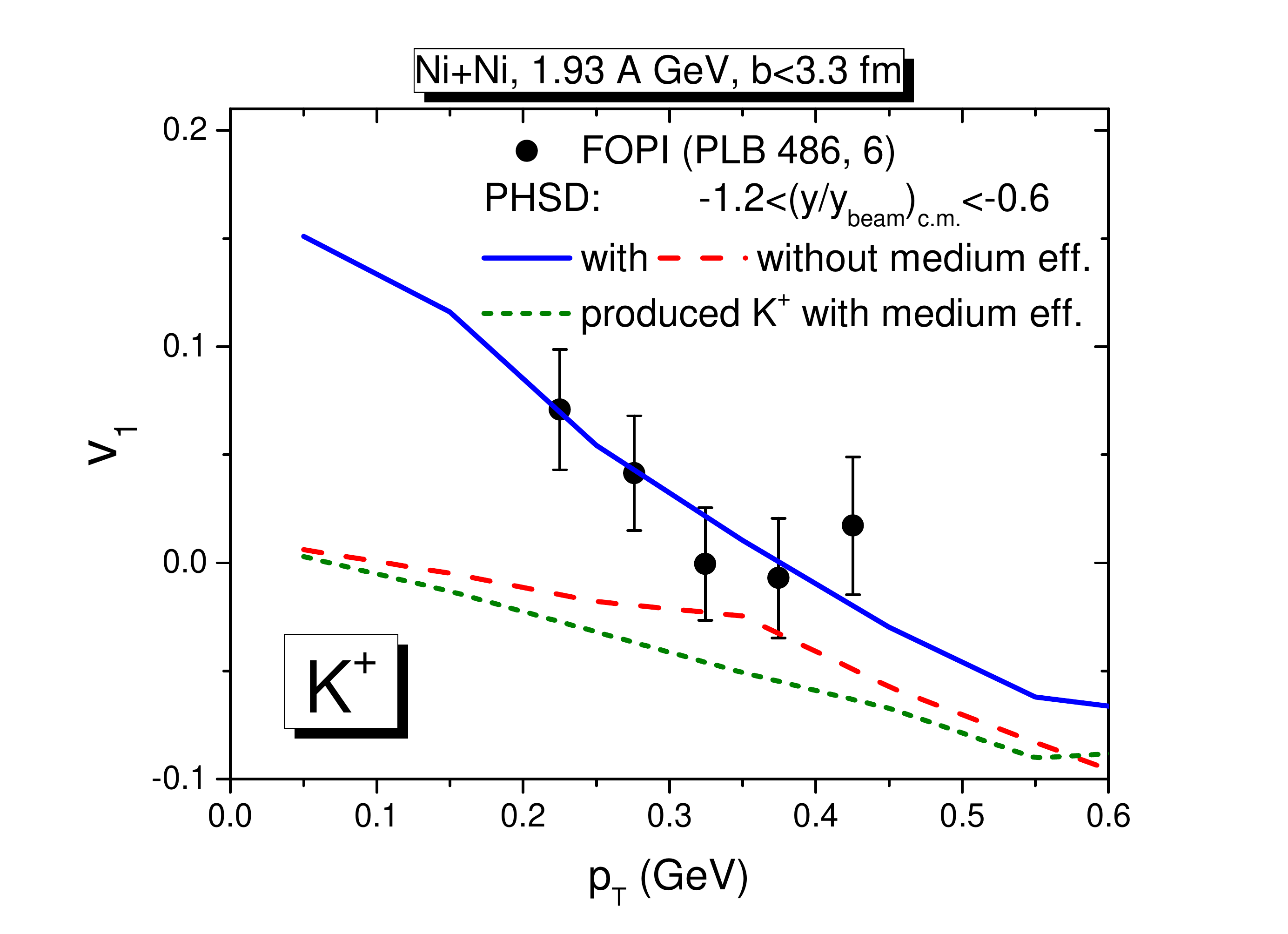}}
\centerline{
\includegraphics[width=8.3 cm]{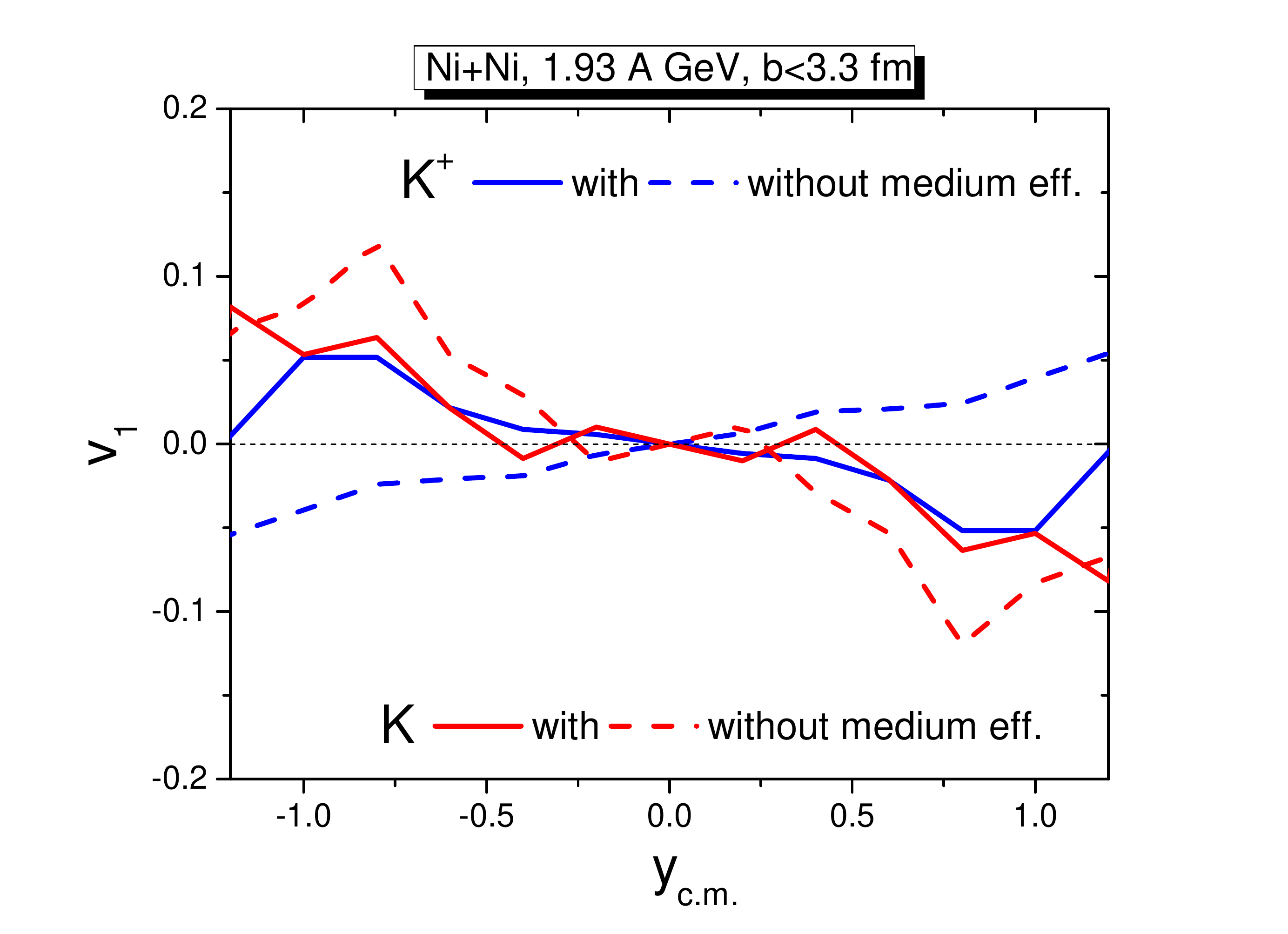}}
\caption{(Color online) (Upper) The $\left<p_x\right>/m$ of $\Lambda$ (red solid line) and 
proton (blue solid line) as a function of normalized rapidity 
and (middle) the $v_1$ of $K^+$ mesons as a function of $p_T$ in central Ni+Ni collisions 
at 1.93 A GeV in comparison to 
the experimental data from the FOPI Collaboration~\cite{Ritman:1995tn,Crochet:2000fz}.
(Lower) The $v_1$ of $K^+$ and $K^-$ with and without medium effects as a function of rapidity in the same collisions.
The blue solid line in the lower plot shows the PHSD results for the $v_1$ 
of final $K^+$ mesons with medium effects, the green dotted line indicates the $v_1$ of kaons
at the production point for the in-medium scenario, while the red dashed line shows
the $v_1$ of kaons without medium effects.
}
\label{v1-NiNi}
\end{figure}

Since the collective flow of kaons is related to the flow of nucleons involved 
in the production and interactions of kaons,
we first show in Fig.~\ref{v1-NiNi} (upper plot) the directed flows of protons 
and $\Lambda$'s (which are the associated partner for the kaon production) 
as a function of the normalized rapidity in central Ni+Ni collisions at 1.93 A GeV.
In our initialization of A+A collisions, the projectile nucleus moving 
in $z$-direction is located at $x=b/2$ and the target nucleus at $x=-b/2$, where 
$b$ is the impact parameter.
One can see that the average $p_x$ of the protons and $\Lambda$'s is antisymmetric 
with respect to $y=0$ and shows a linear increase with rapidity in line with the FOPI data.

The middle panel of Fig.~\ref{v1-NiNi} shows the directed flow $v_1$ of $K^+$ mesons 
as a function of $p_T$ in central Ni+Ni collisions at the same energy.
Considering $y_{\rm lab}/y_{\rm c.m.}-1=(y/y_{\rm beam})_{\rm c.m.}$, 
where $y_{c.m.}$ is the c.m. rapidity of two colliding heavy-ions,
the directed flow of nucleons within $-1.2<(y/y_{\rm beam})_{\rm c.m.}<-0.65$ is  
negative as seen from  the upper panel of Fig.~\ref{v1-NiNi}.
In fact, the directed flow $v_1$ of the produced $K^+$ mesons - which is shown by the dotted line - 
is negative, since they are produced by the scattering of nucleons.
The dashed line represents the $v_1$ of $K^+$ mesons after freeze-out ('survived') without 
kaon potential. The $v_1$ of $K^+$ mesons is only slightly changed due to the interaction
with nucleons which have a negative $v_1$. 
Including the kaon potential, however, the $v_1$ of $K^+$ drastically changes 
and becomes positive. This happens because the nucleons from the target nucleus, 
which are dominantly located in $-x$ direction, push the $K^+$ in $+x$ direction.
Similar phenomena occur at forward rapidity, where the nucleons from the
projectile nucleus push the $K^+$ in $-x$ direction. The effects of the kaon potential 
are stronger for $K^+$ mesons with  small transverse momentum 
and weaker for those with large transverse momentum, since it is harder to change 
the direction of $K^+$ mesons which have large momentum.
One can see that the experimental data are well explained if one
includes the kaon-nuclear potential in the calculations.
The lower panel shows the $v_1$ of $K^+$ and $K^-$ as a function of rapidity.
It demonstrates more clearly the effect of the (anti)kaon potential on the directed flow.
As explained above the repulsive kaon potential pushes the directed flow of $K^+$ away from that of nucleons while the attractive antikaon potential pulls the directed flow of $K^-$ closer.

\begin{figure}[th!]
\centerline{
\includegraphics[width=8.6 cm]{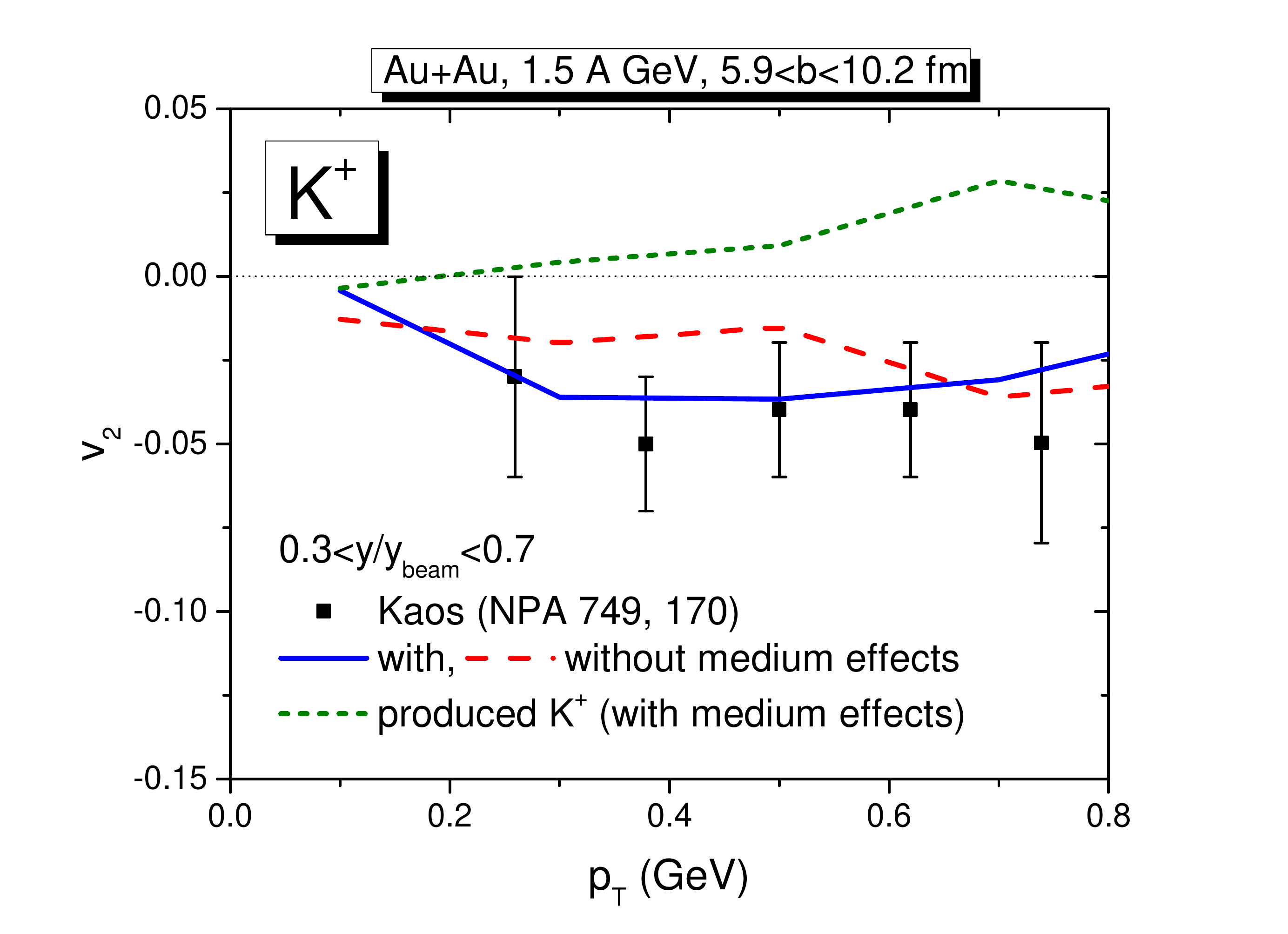}}
\vspace*{-3mm}
\centerline{
\includegraphics[width=8.6 cm]{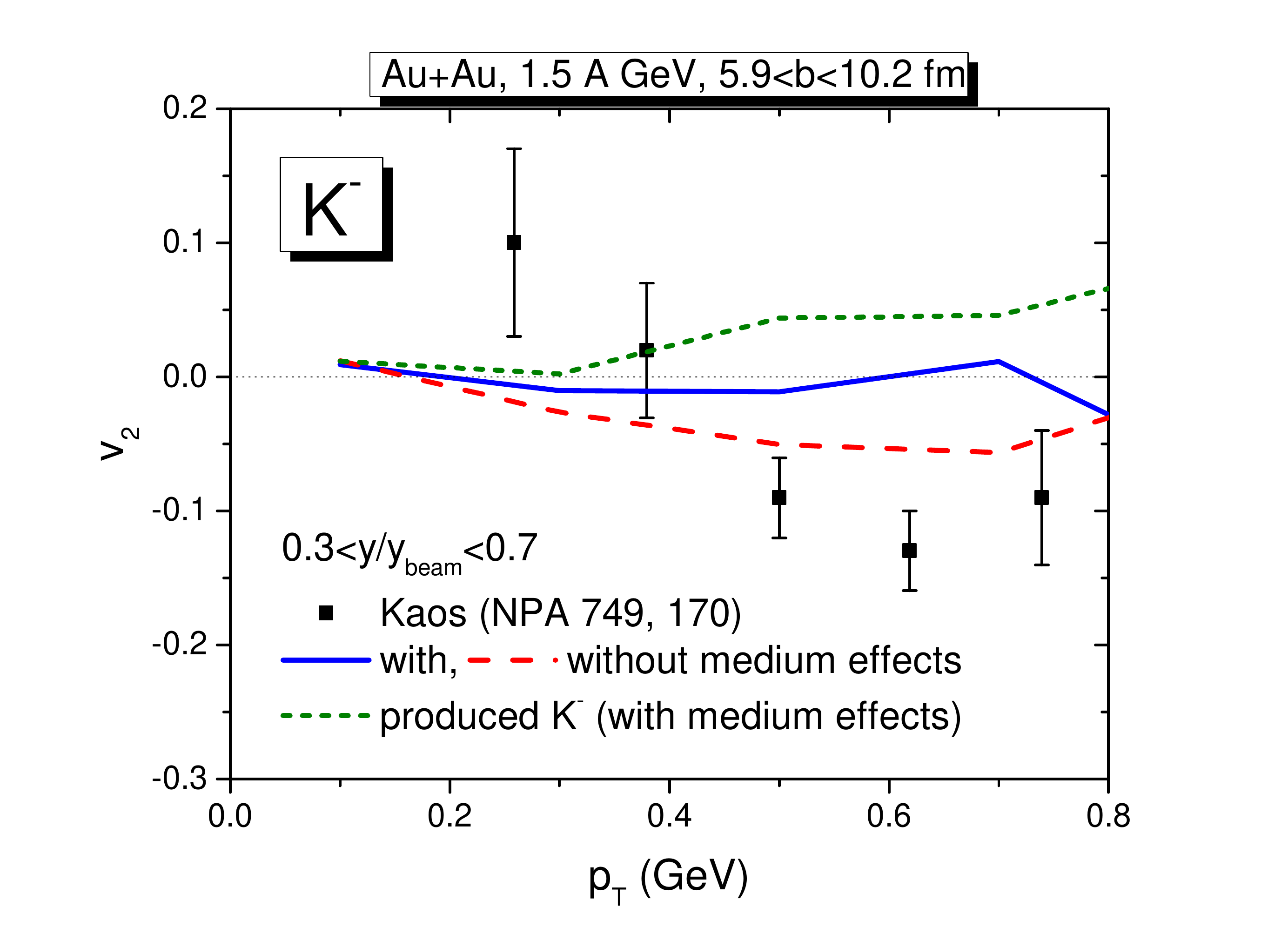}}
\vspace*{-3mm}
\centerline{
\includegraphics[width=8.6 cm]{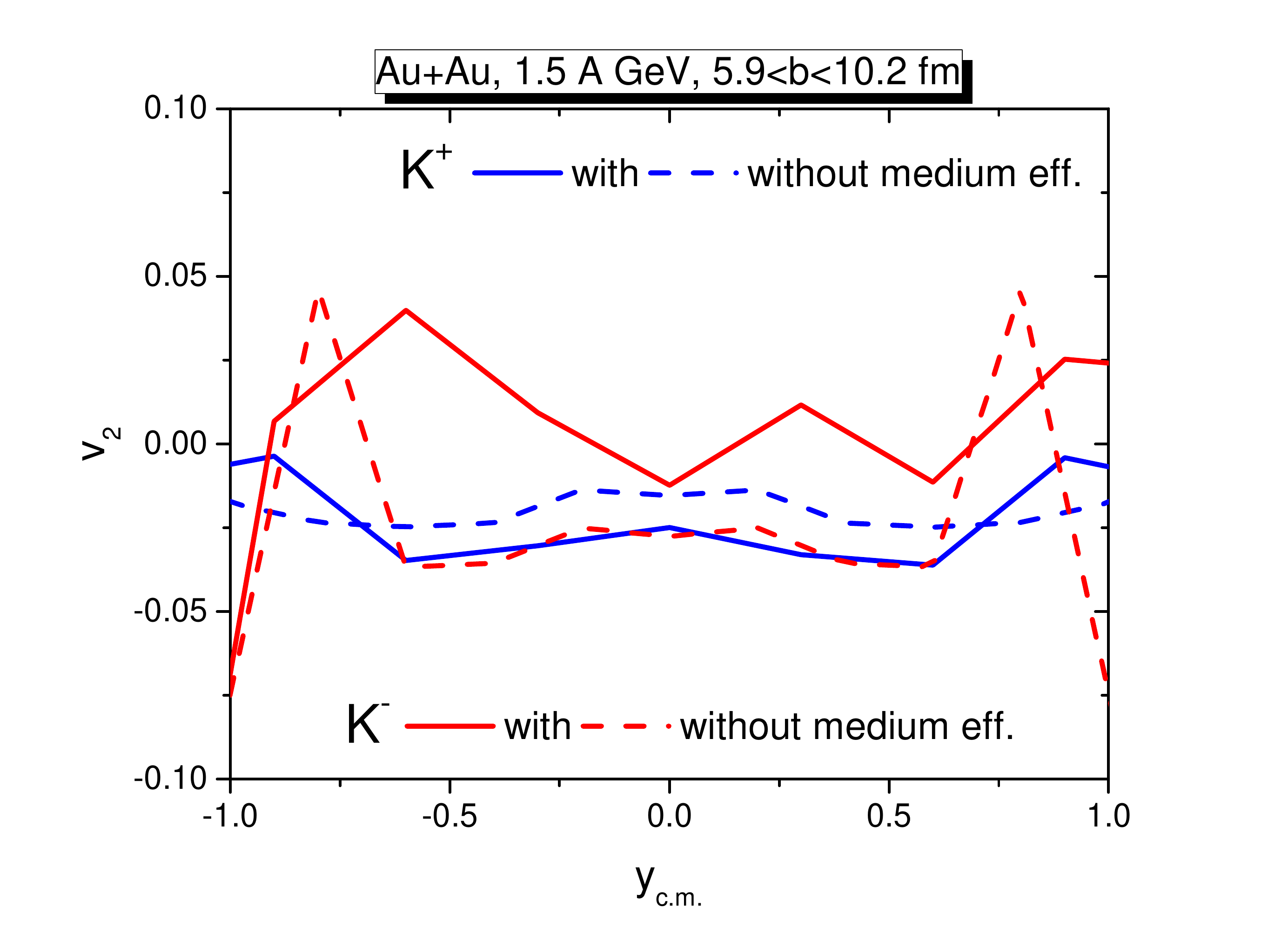}}
\caption{(Color online) Upper and middle plots: 
The flow coefficient $v_2$ of $K^+$ (upper) and  $K^-$ (middle) 
mesons as a function of $p_T$ in semi-central Au+Au collisions at 1.5 A GeV in comparison 
to the experimental data from the KaoS Collaboration~\cite{Ploskon:2005qr}.
The solid blue and dashed red solid lines show the $v_2$ with and without medium effects,
respectively, while the green dotted lines represent $v_2$ at the production point 
for the in-medium scenario.
Lower plot: The $v_2$ of $K^+$ and $K^-$ with and without medium effects as a function of rapidity in the same collisions.}
\label{v2-AuAu}
\end{figure}

\begin{figure}[th!]
\centerline{
\includegraphics[width=8.6 cm]{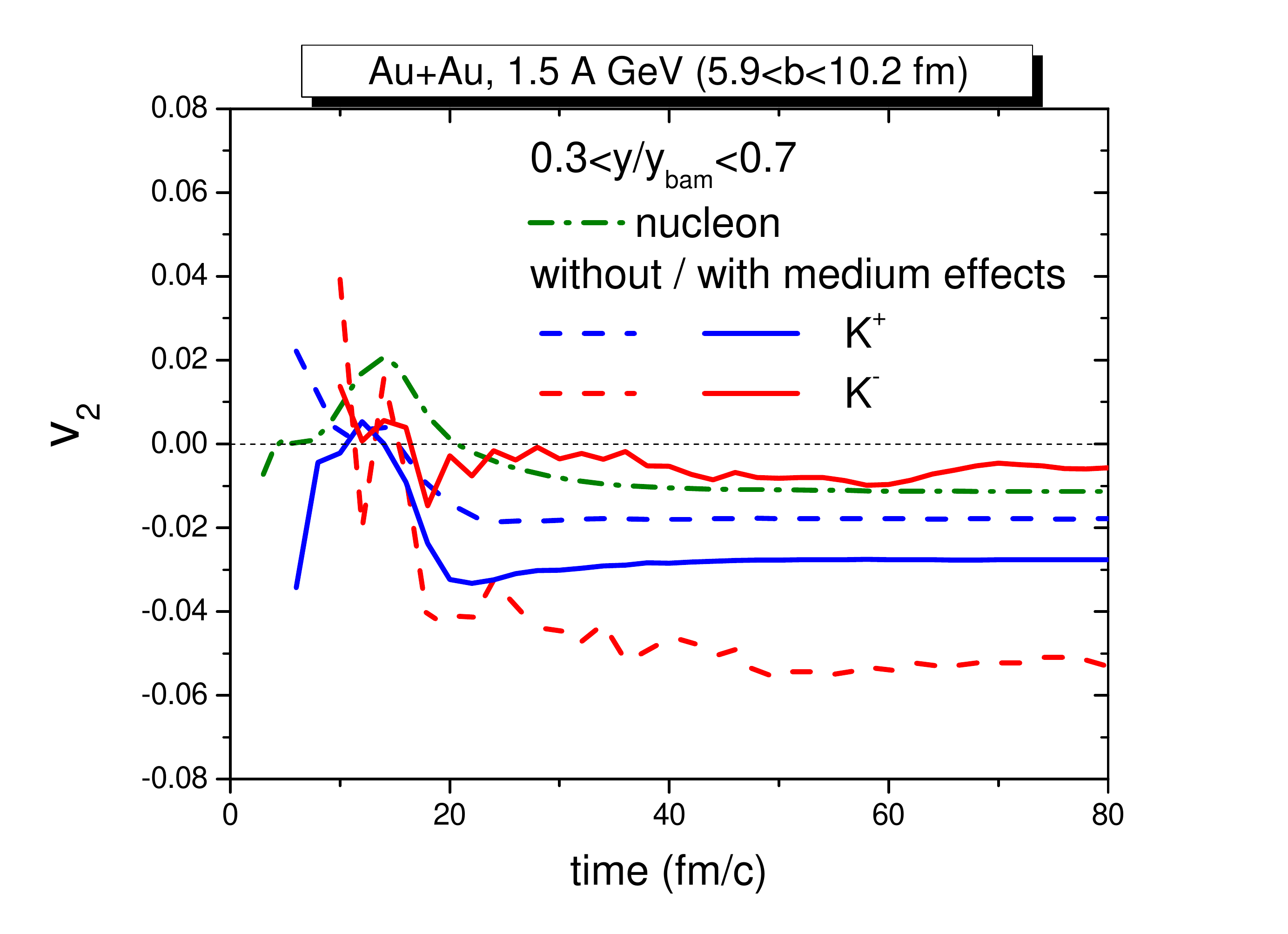}}
\caption{(Color online) the integrated $v_2$ of $K^+$ (blue lines) 
and  $K^-$ (red lines) as a function of time with (solid lines) and without 
(dashed lines) medium effects in the same collisions as in figure~\ref{v2-AuAu}. The green dot-dashed line corresponds to the $v_2$ 
of nucleons.}
\label{v2-time}
\end{figure}

Figure~\ref{v2-AuAu} shows the elliptic flow $v_2$ of $K^+$ (upper plot) and $K^-$ (middle plot)
mesons as a function of $p_T$ and $v_2$ of $K^+$ and $K^-$ as a function of 
rapidity $y$ (lower plot) in semi-central Au+Au collisions at 1.5 A GeV
in comparison to the experimental data from the 
KaoS Collaboration~\cite{Ploskon:2005qr}.
Since $0.3<y/y_{\rm beam}<0.7$ is equivalent to $-0.2<(y/y_{\rm beam})_{c.m.}<0.2$, 
the shown (in the upper and middle plots) $v_2$ corresponds to midrapidity.
For a better understanding of the results we display in Fig.~\ref{v2-time} the integrated $v_2$ of $K^+$,  $K^-$ mesons and 
also nucleons as a function of time in the same rapidity
range as well as the same collision system.
The dotted lines in the upper and middle panels are, respectively, the $v_2$ 
of produced $K^+$ and $K^-$ mesons.
Since (anti)kaons are produced through the scattering of nucleons with other 
nucleons or mesons, their $v_2$ is closely related to the $v_2$ of nucleons.
Considering that the baryon density is peaked around $t=10~{\rm fm/c}$ for 
this centrality/energy and that most of the $K^+$ and $K^-$ mesons are produced before
$t=15~{\rm fm/c}$, one can find that the $v_2$ of nucleons at the production time 
of (anti)kaons is positive.
This explains the positive $v_2$ of the produced $K^+$ and $K^-$ mesons.
After that the $v_2$ of nucleons decreases and changes sign due to the attractive nuclear
potential. Since the produced $K^+$ and $K^-$ interact with the nucleons, 
their $v_2$ follows the $v_2$ of nucleons as shown by the blue and red dashed lines 
in the lower panel, which, respectively, indicate the $v_2$ of $K^+$ and of $K^-$ mesons
considering scattering turned on, but (anti)kaon-nuclear potential turned off.
As a result, the final $v_2$ of $K^+$ and of $K^-$ without the potential turn 
out negative as seen from the upper and middle panels of Fig.~\ref{v2-AuAu}.
This is also well seen on the lower panel of Fig.~\ref{v2-AuAu} which shows that
$v_2$ of $K^+$ and $K^-$ without medium effects is negative. An inclusion of the repulsive kaon potential leads to a further reduction of the $K^+$ elliptic flow, while the in-medium effects for $K^-$ lead to an enhancement of $v_2$, which fluctuates around zero with the presently achieved numerical statistics.

The difference between the dashed line and solid line shows the effects of 
the (anti)kaon-nuclear potential on the $v_2$ of $K^+$ and $K^-$ mesons.
One can see that the kaon potential - which is repulsive - shifts the $v_2$ of $K^+$ 
mesons to a more negative value, while the antikaon potential - which is attractive - moves 
the $v_2$ of $K^-$ mesons towards less negative values.
We note that the kaon-nuclear potential is necessary to reproduce the experimental 
data from the KaoS Collaboration, but antikaon medium effects do not help to explain 
the KaoS experimental data for the $v_2$ of $K^-$ mesons.
\begin{figure}[th!]
\centerline{
\includegraphics[width=8.6 cm]{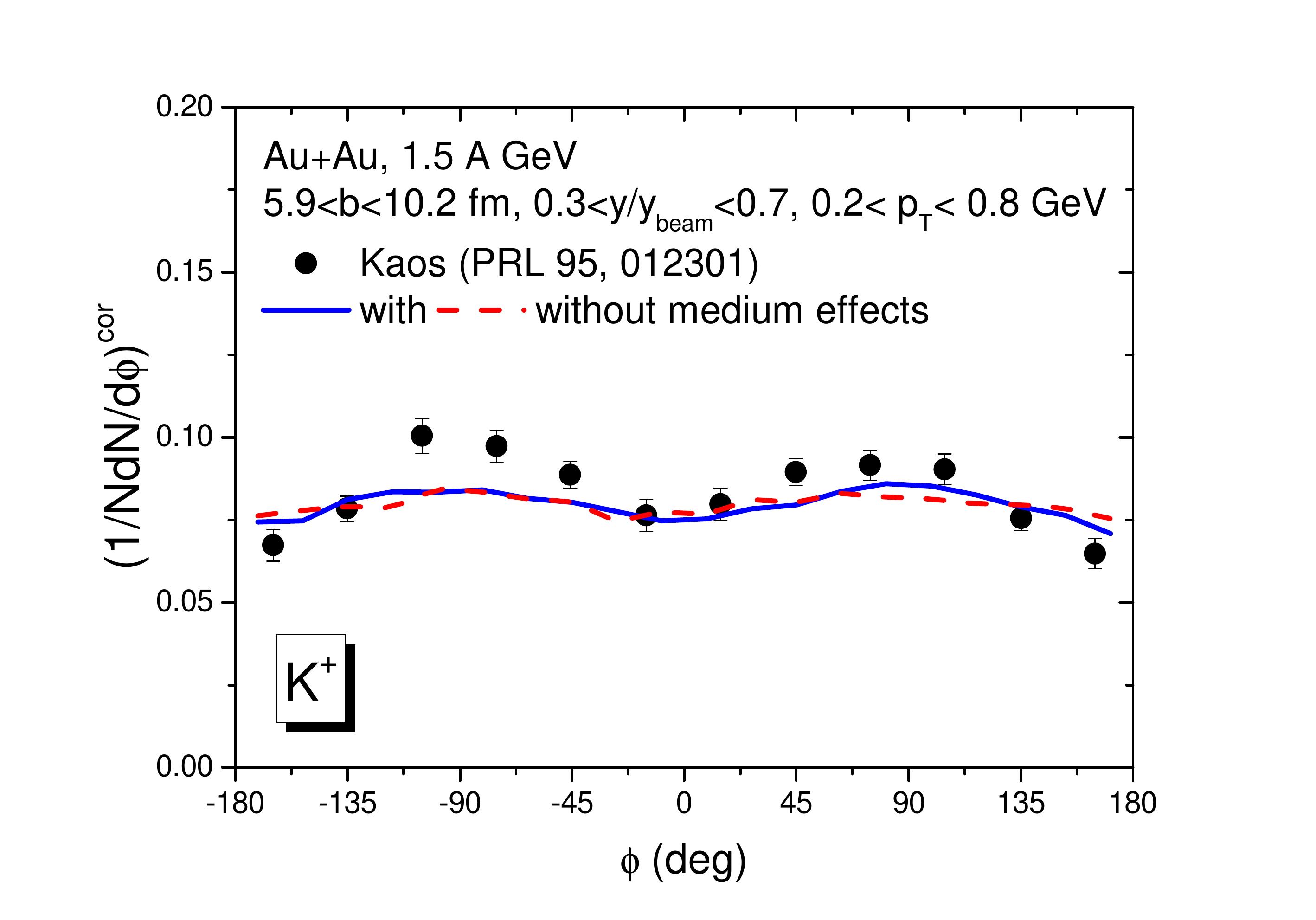}}
\vspace*{-5mm}
\centerline{
\includegraphics[width=8.6 cm]{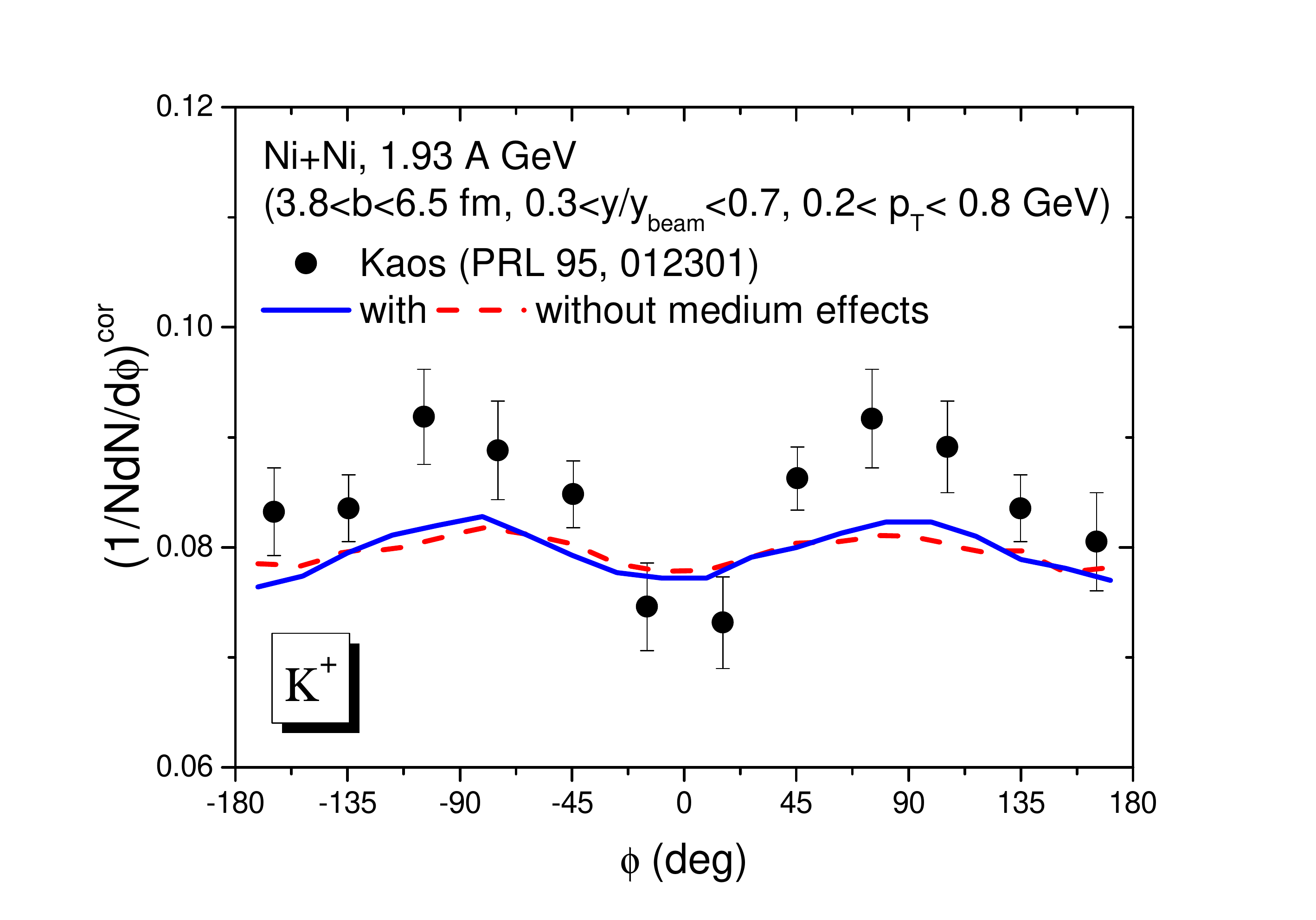}}
\caption{(Color online) The azimuthal angular distributions of $K^+$  
in semi-central Au+Au collisions at 1.5 A GeV (upper) and
in semi-central Ni+Ni collisions at 1.93 A GeV (lower) in comparison 
to the experimental data from the KaoS Collaboration~\cite{Uhlig:2004ue}.
The solid lines show the calculations with medium effects, while the dashed
lines display results  without medium effects.}
\label{dndphi-kp}
\end{figure}

Finally we show in Fig.~\ref{dndphi-kp} the azimuthal angular distribution 
of $K^+$ in semi-central Au+Au collisions at 1.5 A GeV (upper plot) 
and  in semi-central Ni+Ni collisions at 1.93 A GeV (lower plot).
Since the rapidity range is the same as in Fig.~\ref{v2-AuAu}, the contributions 
from $v_1$ are small and the distributions are dominated by $v_2$.
One can see that the $v_2$ of $K^+$ mesons is negative both in Au+Au and Ni+Ni collisions 
and turning on the kaon-nuclear potential enhances the negative flow, which is consistent
with  the results for $v_2$ in Fig.~\ref{v2-AuAu}. 
We note that an increase of the kaon-nuclear potential (Eq. (\ref{VK25})) 
from +25 MeV at $\rho_0$ to a large value (e.g. +35 MeV) leads to some small increase 
of the anisotropy in the azimuthal angular distribution. However, the 
transverse momentum spectra of $K^+$ mesons become harder which is not supported
by the experimental data.
Though the $v_2$ of $K^+$ mesons - with the default kaon-nuclear potential of +25 MeV 
at $\rho_0$ - is smaller than with larger repulsive potential, 
our results qualitatively agree with the experimental 
data from the KaoS Collaboration~\cite{Uhlig:2004ue}. 
However, as seen from Fig.~\ref{dndphi-kp}, the azimuthal angular distribution depends
only very modestly on the inclusion of the medium effects for (anti)kaons.

\subsection{Multiplicity dependence on $A_{part}$ and on the EoS}


The threshold kaon production in heavy-ion collisions is one of the sensitive probes 
for the nuclear equation-of-state~\cite{Aichelin:1986ss,Fuchs,Hartnack:2011cn}.
In the present study  we use a static Skyrme potential \cite{Skyrme} parameterized by
\begin{eqnarray}
U(\rho)=a\bigg(\frac{\rho}{\rho_0}\bigg)+b\bigg(\frac{\rho}{\rho_0}\bigg)^\gamma,
\end{eqnarray}
where $a= -153$ MeV, $b=$ 98.8 MeV, $\gamma=$ 1.63. The
parameters a,b,$\gamma$ are chosen to assure that the energy per nucleon has 
a minimum of $\frac{E}{A}(\rho_0)=-16$ MeV at $\rho_0$. The third condition 
to fix the parameters is the choice of the  
nuclear compressibility $\chi = \frac{1}{V}\frac{d V}{d P}$, where $P$ is the pressure
and $V$ is the volume of the system. Usually  its inverse quantity -  the compression 
modulus $K$ - is quoted ~\cite{Song:2015hua} which is defined as
\begin{eqnarray}
K= -V\frac{d P}{dV} = 9\rho^2 \frac{\partial^2 (E/A)}{\partial \rho^2}\bigg|_{\rho_0}
\end{eqnarray}
A large compression modulus $K$ reflects that the nuclear matter can be hardly 
compressed, thus the equation-of-state is called a 'hard' EoS.
Oppositely, a small $K$ stands for a 'soft' EoS.
The soft EoS enhances the production of kaons in heavy-ion collisions 
for two reasons: i) it allows for the formation of  dense nuclear matter where
nucleons have more chances to collide and produce kaons;
ii) the nucleons loose less compression energy and the 'saved' energy can be used for 
the production of kaons. Our set of parameters  
gives a compression modulus $K\simeq 300$~MeV at  saturation density, 
which is in between the soft and hard equations-of-state, i.e. a 'middle-soft' EoS.
This 'middle-soft' EoS provides an optimal description of the different observables 
at SIS energies and, thus is used as 'default' value in the PHSD 
for low energy heavy-ion collisions.
In order to investigate the sensitivity of the $K^+$, $K^-$ production on the EoS,
we have changed the EoS in PHSD to a softer with $K= 210$ MeV and to a harder
with $K= 380$ MeV. 
We found that the softening of the EoS leads to an increase of the $K^+$ and $K^-$ 
yields by $\sim 13$ \% and $\sim 24$ \%, respectively, in central Ni+Ni collisions 
at 1.93 GeV, while a harder EoS decreases them, respectively, by $\sim 14$ \% 
and $\sim 19$ \%.
We  mention that also momentum dependent Skyrme interaction have been
successfully employed to study the in medium properties of kaon
\cite{Aichelin,Hartnack:2011cn}.

The stiffness of the nuclear equation-of-state affects also the dependence 
of kaon production on the centrality of heavy-ion collisions.
The softer the equation-of-state is, the more kaons are produced in central 
collisions compared to  semi-central and peripheral 
collisions~\cite{Hartnack:2011cn}. 
The centrality dependence might be converted to the dependence on
$A_{\rm part}$ being the number of participants. 
The multiplicity of (anti)kaons in heavy-ion collisions is proportional to 
$M \sim (A_{\rm part})^{\alpha}$ where $\alpha$ is a fitting parameter.
The deviation from a scaling with $A_{part}$ ($\alpha=1$) points towards 'collective production 
effects': e.g. a formation of $K^+$ by scattering of resonances such as
$\Delta N \to N \Lambda K^+$ collisions. The probability that a $\Delta$ interacts with nucleons and produces a kaon (before it disintegrates to a pion and nucleon) 
increases with baryon density, which is larger for the most central collisions.
For pion production the deviation from the linear $A_{part}$ dependence is 
not expected since, in spite of multiple $\Delta$ decay and regeneration by pion-nucleon scattering, the number of final pions is not much affected. 
Indeed, as found by the KaoS Collaboration~\cite{Forster:2007qk}, 
in the case of pion production, the experimental value $\alpha_\pi$ is 
compatible with  unity, 
($\alpha_\pi (Au) = 0.96 \pm 0.05$, \ $\alpha_\pi (Ni) = 1.0 \pm 0.05$).
while for kaons $\alpha_K(Au)=1.34\pm 0.16$,
i.e. the kaon multiplicity rises stronger than linear with $A_{part}$. 
A similar behaviour has been observed for antikaons ($\alpha_{\bar K}(Au)=1.22\pm 0.27$)  
despite the different production thresholds. 
This leads to constant ratio of $K-/K+$ versus centrality.
We note that the PHSD reproduces the KaoS results on the pion 
value of $\alpha_\pi$ rather well.

\begin{figure}[th!]
\centerline{\includegraphics[width=8.6 cm]{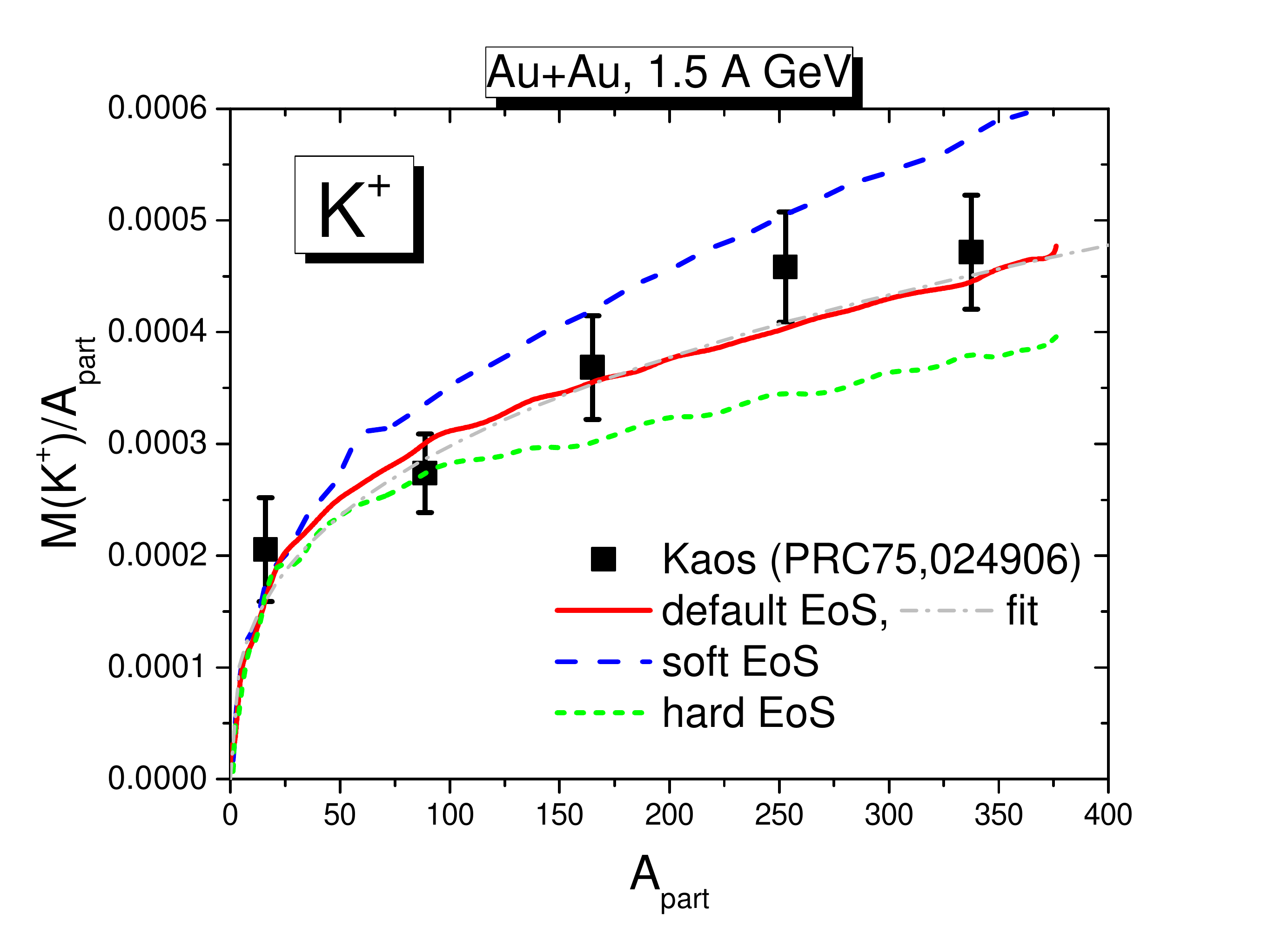}}
\vspace*{-2mm}
\centerline{\includegraphics[width=8.6 cm]{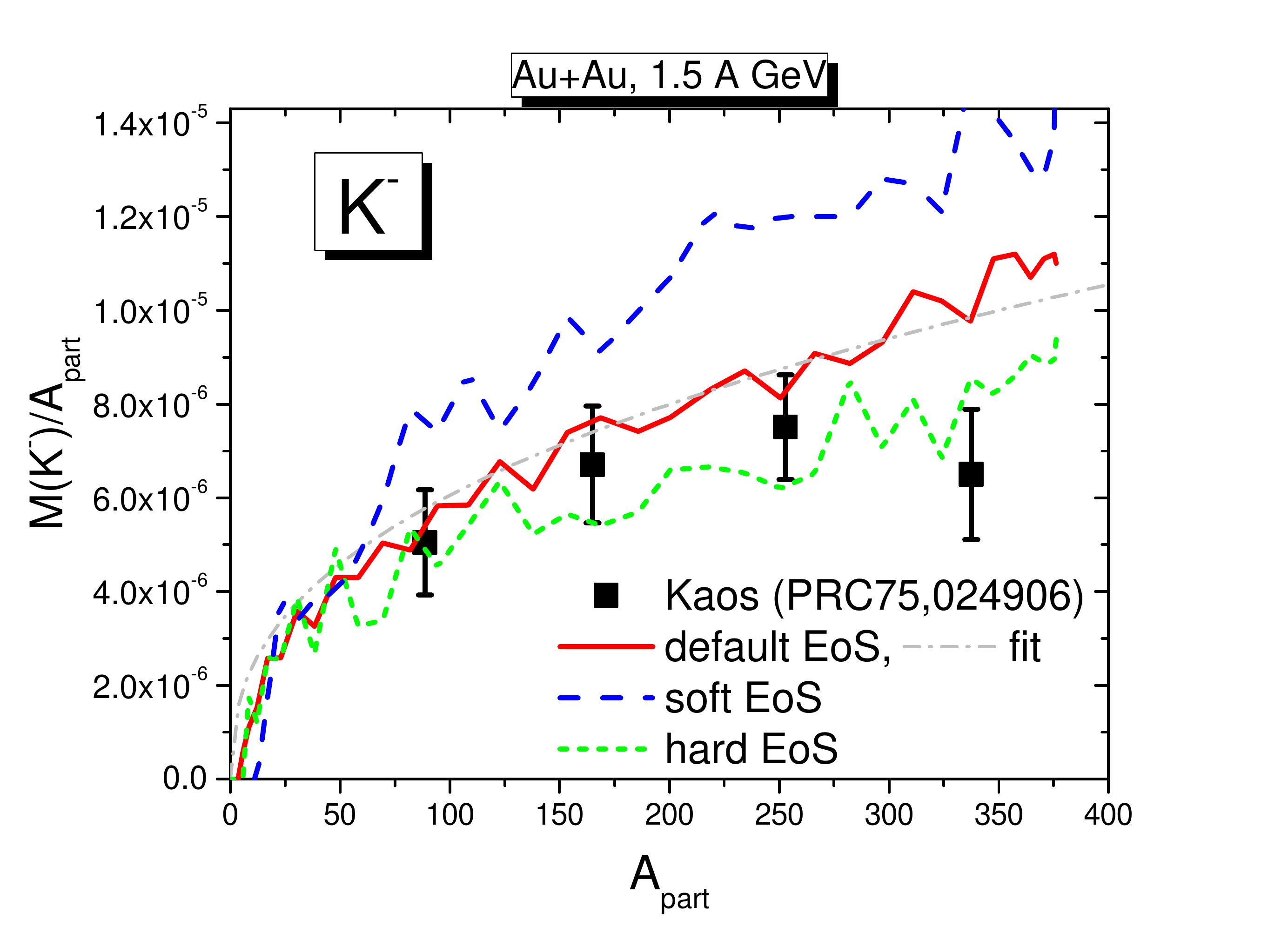}}
\vspace*{-2mm}
\centerline{\includegraphics[width=8.6cm]{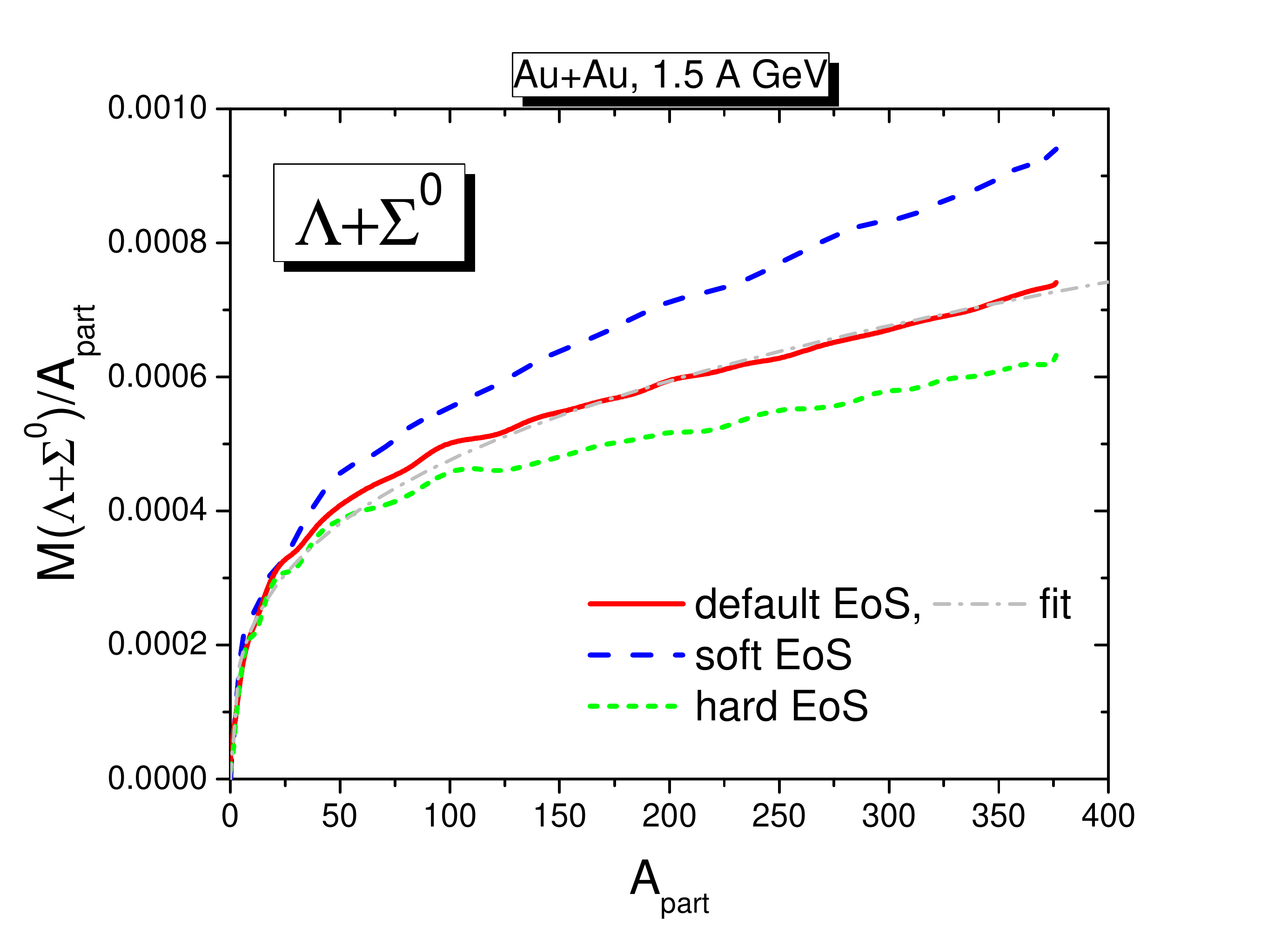}}
\caption{(Color online) Multiplicities per mean number of participants $A_{part}$
of $K^+$ (upper), $K^-$ (middle) and $\Lambda+\Sigma^0$ (lower) 
in Au+Au collisions at 1.5 A GeV.
The solid red lines correspond to the PHSD results (including in-medium effects) 
with default EoS, the blue dashed line and green dotted show the results
 with a hard and soft EoS, respectively.
The grey dot-dashed lines indicate the fit of the PHSD results with the default EoS
(see text). 
The solid squares show the KaoS data for $K^+$ taken from 
Ref. \cite{Forster:2007qk}.}
\label{kaon2pion}
\end{figure}

Figure~\ref{kaon2pion} shows the PHSD results (including the medium effects) 
for the multiplicities per mean number 
of participants $A_{part}$ of $K^+$ (upper), $K^-$ (middle) and 
$\Lambda+\Sigma^0$ (lower) in Au+Au collisions at 1.5 A GeV in comparison 
to the KaoS data for $K^+$ taken from Ref. \cite{Forster:2007qk}.
The solid red lines represent the PHSD results with the default EoS, the blue dashed line  
and green dotted show the results with  a hard and soft EoS, respectively.
One can see a strong sensitivity of the multiplicity of kaons and  $\Lambda+\Sigma^0$
on the compression modulus of the EoS - the hard EoS leads to a strong reduction of the multiplicity in central collisions while the soft EoS to an enhancement of kaons and $\Lambda+\Sigma^0$.
We note that we obtain the best description of the experimental data with the default 
 'middle-soft' EoS in the PHSD.

The grey dot-dashed lines indicate a fit of the PHSD results (with default EoS)
for the multiplicities as $M \sim (A_{\rm part})^{\alpha}$: we find  for kaons
$\alpha_K^{PHSD}(\rm{Au}) \simeq 1.34$ in agreement with the KaoS data. Moreover,
for $\Lambda+\Sigma^0$ -- $\alpha_{\Lambda+\Sigma^0}^{PHSD}(Au) \simeq 1.32$ which
is almost the same as for kaons.  For the antikaons $\alpha_K^{PHSD}(\rm{Au})\simeq 1.4$
which is a bit larger than for kaons, but still in the range of experimental 
errorbars \cite{Forster:2007qk}. 

We stress that the change of the equation-of-state leads 
to a modification of the whole dynamics of the heavy-ion collisions at low energies 
which is seen in the 'bulk' observables such as proton and pion
rapidity and $p_T$-spectra, angular distributions, flow harmonics $v_1, v_2$ etc.
The strange hadrons are reflecting these changes since their production mechanisms
are tightly linked to the dynamics of non-strange baryons and pions, especially 
at subthreshold energies considered in this study.
Thus, in order to pin down a robust information on the nuclear EoS from 
low energy experimental data, one has to analyse of all observables - for strange
and for non-strange hadrons - simultaneously.

\begin{figure}[t!]
\centerline{
\includegraphics[width=8.6 cm]{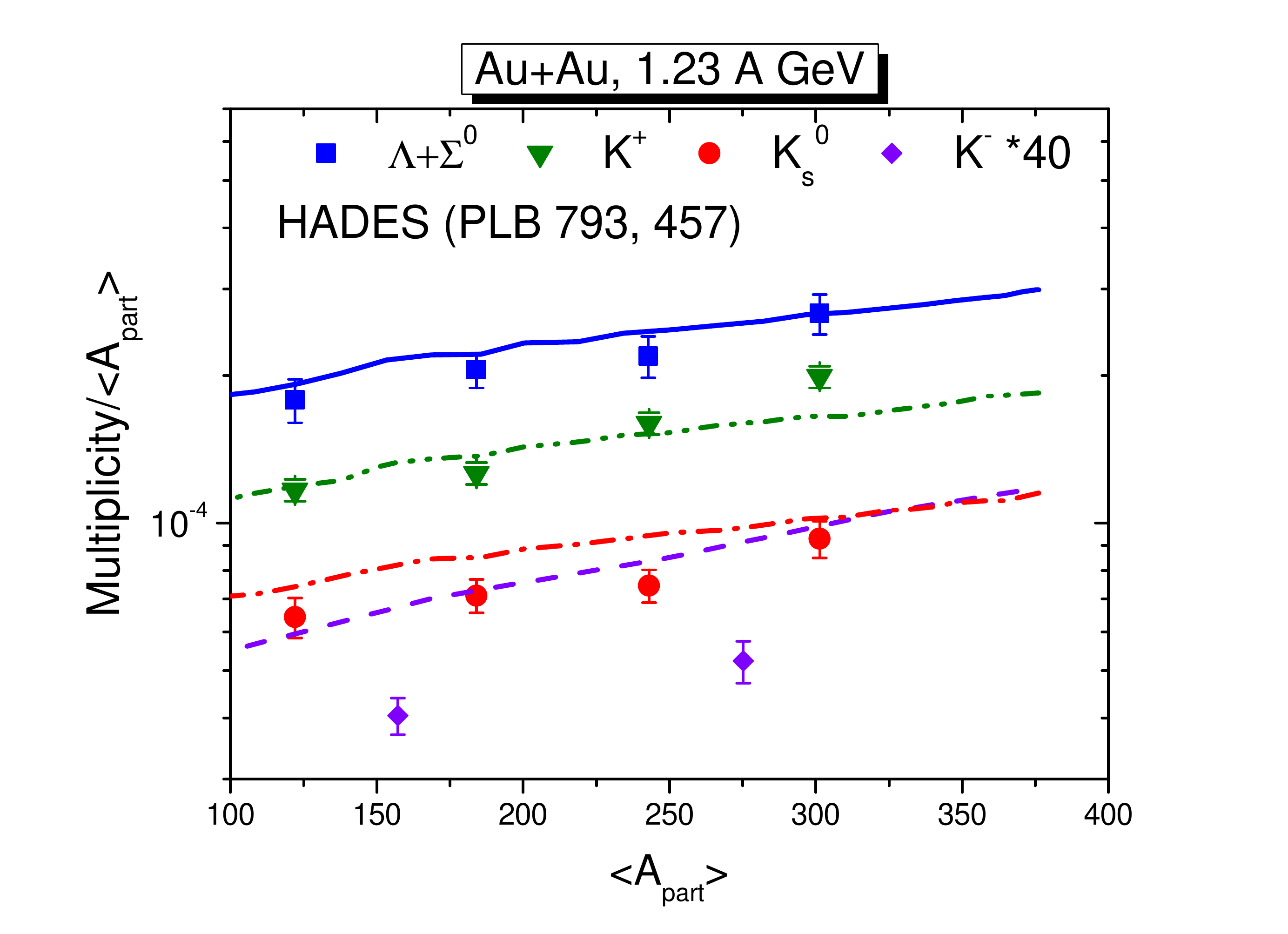}}
\caption{(Color online) The PHSD results for the $A_{part}$ dependence of the
multiplicity over averaged $A_{part}$ of $\Lambda+\Sigma^0$ (blue solid line), 
$K^+$ (green dot-dot-dashed line), $K^0_s=(K^0+\bar K^0)/2$ (red dot-dashed line) and 
$K^-$ (violet dashed line, multiplied by 40)  in comparison to the experimental data from
the HADES Collaboration \cite{Adamczewski-Musch:2018xwg}.  }
\label{NpartAll}
\end{figure}

As follows from Fig.~\ref{kaon2pion} the PHSD reproduces well the 
 $A_{part}$ dependence $K^+$ for the Au+Au at 1.5 A GeV. Now we step down in energy
and come to sub-threshold $K^+$ production in Au+Au collisions at 1.23 A GeV.
In Fig. \ref{NpartAll} we show the PHSD results for the $A_{part}$ dependence of the
multiplicity over averaged $A_{part}$ of $\Lambda+\Sigma^0$ (blue solid line), 
$K^+$ (green dot-dot-dashed line), $K^0_s=(K^0+\bar K^0)/2$ (red dot-dashed line) and 
$K^-$ (violet dashed line, multiplied by 40)  in comparison to the experimental data
from the HADES Collaboration \cite{Adamczewski-Musch:2018xwg}. 
One can see that the $\Lambda+\Sigma^0$ and $K^+$ data are reasonable well described 
by PHSD. However, as expected,  we see the same tension of the PHSD results 
for the $K^0_s$ and $K^-$ with the HADES data  as in Figs. \ref{NiNi-y-central}
and \ref{HADES-mt}: the $K^0_S$ data are overestimated by 20\% and $K^-$ data are
overestimated by a factor of 2.

By fitting the experimental data with $M\simeq <A_{part}>^\alpha$ the HADES Collaboration extracted  $\alpha_{HADES} =1.45 \pm 0.06$ for $K+, K^-, K^0_s$ 
and $\Lambda+\Sigma^0$  \cite{Adamczewski-Musch:2018xwg}.
On the other hand in Ref. \cite{Adamczewski-Musch:2018xwg} the comparison of the
HADES results to the model predictions (by HSD, IQMD (with and without medium effects) 
as well as by UrQMD) has been shown. A similar tension in multiplicity 
of $K^0_s$ between all three models and experimental data has been reported. 
We note that in Ref. \cite{Adamczewski-Musch:2018xwg} the comparison has been done  
using the HSD model with G-matrix from  Ref. \cite{laura03}.
There the HSD results for $K^0_s$ and $\Lambda+\Sigma^0$  have been fitted 
with $\alpha=1.35 \pm 0.02$ (cf. Table II in Ref. \cite{Adamczewski-Musch:2018xwg}). 
Our new results for $K^0_s$ and $\Lambda+\Sigma^0$ are consistent with this finding, too.
Moreover, the $K^+$ and $K^-$ yields can be fitted with the same $\alpha$ which is
in agreement with KaoS results within errorbars.

\subsection{Isospin ratios of $K^+$ over $K^0$}

Finally we check the isospin dependence of the $K^+$ to $K^0$ production.
As advocated in Refs. \cite{DiToro06,Li:2008gp} the inclusive  $K^+/K^0$ 
ratio at subthreshold energies for kaon production is sensitive to the 
symmetry energy of the nucleon-nucleon potential and thus can be used 
to obtain experimental information on this issue.
On the other hand, the isospin decomposition of strangeness production is defined
to a large extend by hadronic inelastic and charge exchange 
reactions  with baryons and mesons. Their cross sections
are not well known experimentally and thus the transport calculations have 
to rely on phenomenological models. In this respect the experimental data 
on the $K^+$ to $K^0$ production in collisions of nuclei of different sizes
provide an opportunity to access the production channels, too.

Experimentally the isospin dependence of $K^+$ and $K^0$ production
at near sub-threshold energies has been investigated by the FOPI Collaboration in Ref. 
\cite{Lopez:2007rh} which measured the ratios of $K^+$ to $K^0$ mesons for 
Ru+Ru to Zr+Zr collisions at the kinetic energy 1.528 A GeV (i.e.
at $\sqrt{s}=2.527$ GeV, while the threshold for kaon production in $pp$ 
collisions is $\sqrt{s_0}=2.546$ GeV).

\begin{table}[h!]
\begin{tabular}{c|c|c}
\hline
\phantom{a}\vspace*{1mm}
Ratio & \  PHSD \  & FOPI \\
\hline
\vspace*{1mm}
$K^+_{Ru}/K^+_{Zr}$ &  1.09 & $1.06 \pm 0.07(stat.) \pm 0.09(syst.)$\\
\hline 
\phantom{a}\vspace*{1mm}
$K^0_{Ru}/K^0_{Zr}$  & 0.96 & $0.94 \pm 0.12(stat.) \pm 0.06(syst.)$\\
\hline 
\phantom{a}\vspace*{1mm}
${(K^+/K^0)_{Ru}\over(K^+/K^0)_{Zr}}$ & 1.14 & $1.13 \pm 0.16(stat.) \pm 0.12(syst.)$\\
\hline
\end{tabular}
\caption{Comparison of the inclusive ratios of $K^+$ and $K^0$ mesons for Ru+Ru to Zr+Zr 
collisions at the kinetic energy 1.528 A GeV in the PHSD versus the FOPI data 
\cite{Lopez:2007rh}.}
\label{TabAsEn}
\end{table}

In Table \ref{TabAsEn} we show the comparison of the inclusive ratios of 
$K^+$ and $K^0$ mesons for Ru+Ru to Zr+Zr collisions -
$K^+_{Ru}/K^+_{Zr}$, $K^0_{Ru}/K^0_{Zr}$, ${(K^+/K^0)_{Ru}\over(K^+/K^0)_{Zr}}$
at the kinetic energy 1.528 A GeV from the PHSD versus the FOPI data \cite{Lopez:2007rh}.
One can see that the PHSD results are in a good agreement
with the measured ratios as well as the model calculations from Ref. \cite{DiToro06}. 
We note, however, that the symmetry energy is not implemented in the potential
used in the PHSD version for this study. Thus, we attribute this agreement  
to the consequence of the isospin dependence of strangeness production and 
rescattering.

\section{Summary}\label{summary}

In this study we have investigated  strangeness production in heavy ion collisions 
at(sub-)threshold energies between 1.0  and 2.0 A GeV within the  
microscopic Parton-Hadron-String Dynamics (PHSD 4.5)  transport approach,
extended for the incorporation of the in-medium effects for
strangeness production in terms of the state-of-the-art  coupled-channel  
G-matrix approach for the modification of antikaon properties in a dense and hot medium.

The PHSD is an off-shell transport approach based on the first order gradient 
expansion (in phase-space representation) of  the Kadanoff-Baym equations 
which allow for a propagation of particles with dynamical spectral functions,
i.e. for a microscopic description of the strongly interacting system.
It includes also the possibility to implement detailed balance on the level of  
$m\leftrightarrow n$ reactions (although its technical implementation
is restricted to selected channels of interest due to the limitation 
in computational resources). For the present study we have 
incorporated the detailed balance on $2\leftrightarrow 3$ level for
the dominant channels for strangeness production/absorbtion by baryons ($B=N, \Delta$)
and pions: $B+B \leftrightarrow N+Y+K$ and  $B+\pi \leftrightarrow N+N+\bar K$,  
as well as  the non-resonant reactions $N+N \leftrightarrow N+N+\pi$ and
$\pi+N \leftrightarrow N+\pi+\pi$ to have a better control on pion dynamics. 
 
The modification of the strangeness properties in the hot and dens medium, which 
is created in the heavy-ion collisions, has been incorporated 
in the PHSD in the following way:

The medium effects for antikaons are determined by a  self-consistent coupled-channel 
G-matrix approach \cite{Cabrera:2014lca}, based
on a chiral Lagrangian, which provides the spectral function of
antikaons as well as the scattering amplitudes (i.e. in-medium cross sections) 
as a function of baryon density, temperature and the three-momentum of the antikaon 
with respect to the matter at rest. This novel G-matrix approach \cite{Cabrera:2014lca} 
incorporates the full self-consistency in $s$- and $p$-waves in the $\bar K N$ interaction at finite density and 
temperature based on a SU(3) chiral Lagrangian. It differs therefore from  the previous work of Ref.~\cite{laura03} 
where a self-consistent unitary approach based on the J\"ulich meson-exchange model \cite{Tolos:2000fj,Tolos:2002ud}  was used.

We point out that the novel G-matrix approach provides a more shallow
dependence of the real part of the complex antikaon self-energy $Re\Sigma$
(and, respectively, $\bar KN$ potential) on baryon density than that from 
Refs.~ \cite{Tolos:2000fj,Tolos:2002ud}, which has been used in the past 
in the off-shell HSD \cite{laura03} for  antikaon dynamics at SIS energies.

Nuclear medium effects on the production and interactions of kaons are implemented 
by a repulsive kaon potential of $V_K= +25$ MeV $(\rho/\rho_0)$. \\

We have studied the effects of the (anti)kaon medium modifications on different 
observables in heavy-ion collisions at SIS energies measured by the 
KaoS, FOPI and HADES Collaborations.
Our study confirms the sensitivity of antikaon observables to the
details of the in-medium models: the in-medium effects on observables obtained
with the novel G-matrix are slightly less pronounced as compared to the previous
G-matrix calculations which provided a much stronger attraction of antikaons in nuclear matter.

Our findings can be summarized as follows:\\
$\bullet$ The kaon-nuclear potential increases the threshold energy for the
kaon production in $BB$ and $mB$ reactions in a hot and dense medium. 
Consequently, the $K^+, K^0$ production in A+A collisions is suppressed
when accounting for the potential compared to the case without potential.
We note, that in A+A collisions at (sub-)threshold energies the kaons are 
dominantly produced by the reactions with $\Delta$'s 
(as $N+\Delta \to Y + K + N$) and mesons (as $\pi +N\to Y+K$). That 
enhances kaon production substantially relative to the production
by $N+N$ reactions only. Moreover, production of kaons below the $N+N$ threshold
is possible only in A+A collisions due to the Fermi-motion of nucleons and 
by secondary nucleon reactions with pions.

$\bullet$ The self-consistent coupled-channel unitarized scheme 
based on a SU(3) chiral Lagrangian  leads to a
broadening of the $\bar K$ spectral function without drastic changes of the pole mass. 
Moreover, the strength of the attractive $\bar K N$ potential decreases 
with increasing three-momentum of $\bar K$ mesons in nuclear matter. 
The $\bar K$ production in A+A collisions is enhanced by the broad spectral width 
(which increases with density) which leads to a reduction of the production threshold.
The antikaons are dominantly produced by the secondary hyperon-meson reactions
$Y+m \to B+\bar K$ which are delayed in time. Consequently, the baryon density 
at the antikaon production point is lower compared to that at the kaon production point, 
that reduces the in-medium effects. 
 
$\bullet$ The kaon-nuclear potential hardens the $p_T$-spectra of 
$K^+, K^0$ mesons while the in-medium effects in terms of self-energies within 
the G-matrix approach soften the $K^-$ spectra. 
These effects are stronger in larger systems such as Au+Au collisions, 
where a higher density is reached and therefore the density gradient (which is proportional to the force) is larger
as compared to smaller systems like C+C collisions.

$\bullet$ For kaons, both collisions with baryons and mesons as well as   the repulsive 
interaction with the nuclear medium 
lead to an increase of the effective temperature $T_{eff}$ of the $p_T$-spectrum 
compared to that at the kaon production.
For antikaons the hardening of the $p_T$-spectra due to collisions and 
the softening due to the in-medium effects counterbalance partially.

$\bullet$ The repulsive kaon potential widens the rapidity distribution of 
$K^+$ mesons, while an attractive antikaon interaction shrinks that of $K^-$ mesons.
It is also visible in the enhancement of the polar angular distribution of $K^+$ mesons
in forward and backward directions, although the effect of $K^+$ rescattering 
in the medium plays an even larger role than the $KN$ potential, especially for 
heavy systems such as Au+Au. The polar angular distribution of $K^-$ mesons
stays rather flat due to the competition of two effects: the scattering
pushes $K^-$ mesons backward and forward similar to $K^+$, however, 
the antikaon potential attracts them to the middle zone of the colliding system;
this push is stronger for central collisions and heavy systems due to the
larger density.

$\bullet$ The collective flow coefficients for $K^+$ and $K^-$  are strongly affected by the flow of nucleons, since (anti)kaons are produced by 
nucleon scattering and secondary interactions with nucleons.
Considering only scattering and neglecting the kaon-nuclear potential, 
the directed flow $v_1$ of $K^+$ mesons follows that of nucleons such that 
the $v_1$ of $K^+$ mesons is positive at forward rapidity and negative at
backward rapidity.
However, the repulsive kaon potential pushes the $K^+$ mesons in the opposite direction 
to the nucleons and, as a result, the $v_1$ of $K^+$ mesons for low transverse momenta increases strongly and becomes positive, while the $v_1$ of $K^+$ mesons with 
large transverse momenta stays negative and finally changes only slightly. \\
The elliptic flow $v_2$ of $K^+$ and $K^-$ mesons are also affected by 
the flow of nucleons. In non-central heavy-ion collisions, the nucleons gain 
a negative elliptic flow due to the geometric shape of the medium and
attractive Skyrme potential.
The $v_2$ of the produced $K^+$ mesons is initially positive, following that of nucleons, 
and becomes negative due to the scattering and repulsive interactions with nucleons.
This is also seen in the azimuthal angular distribution of kaons.
On the other hand, the antikaon-nuclear potential acts in opposite direction:
while the $v_2$ of the initially produced antikaons is also positive (as for $K^+$) 
and becomes also negative due to the rescattering in the matter, the attractive
antikaon potential pushes the $v_2$ towards less negative values such that 
the final $v_2$ of antikaons is close to zero in the present PHSD calculations.

$\bullet$ Moreover, we have investigated the sensitivity of strangeness production
to the nuclear equation-of-state. 
In line with the previous findings \cite{Aichelin:1986ss,Fuchs,Hartnack:2011cn}
we have observed an enhancement of (anti)kaon production by a softening of the EoS.
However, we stress that solid conclusions on the EoS can be obtained only
by considering a variety of observables (including the non-strange
hadrons) since the dynamics of strangeness is tightly bound to the dynamics
of the 'bulk' matter in heavy-ion collisions and the density of pions.

$\bullet$ Finally, our study with the improved description for the
modifications of the (anti)kaon properties in the hot and dense medium (within 
the novel G-matrix approach) as well as an improved general dynamics of the medium itself 
(within the PHSD) confirms the previous findings by different groups 
(summarized in the review \cite{Hartnack:2011cn}) on the observation
of the visible medium effects  in heavy-ion collisions at SIS energies.

\section*{Acknowledgements}

The authors acknowledge valuable discussions with Wolfgang Cassing and his suggestions.
We are grateful to Daniel Cabrera for providing us the G-matrix calculations.
We are also thankful for inspiring discussions with Christoph Blume, Laura Fabbietti,  
Yvonne Leifels, Manuel Lorenz, Pierre Moreau and Iori Vassiliev.  
Furthermore, we acknowledge support by the Deutsche Forschungsgemeinschaft 
(DFG, German Research Foundation): grant BR 4000/7-1,  by the Russian Science Foundation grant 19-42-04101
and   by the GSI-IN2P3 agreement under contract number 13-70.
 L.T. acknowledges support from the Deutsche Forschungsgemeinschaft (DFG, German research Foundation) under the Project Nr. 383452331 (Heisenberg Programme),  Nr. 411563442 (Hot Heavy Mesons). Her research has been also supported  by the Spanish Ministerio de Econom\'ia y Competitividad under contract FPA2016-81114-P and Ministerio de Ciencia e Innovaci\'on  under contract 
PID2019-110165GB-I00. 
This project has, furthermore, received funding from the European Union’s Horizon 2020 research and innovation program under grant agreement No 824093 (STRONG-2020). 
Also we thank the COST Action THOR, CA15213 and CRC-TR 211 'Strong-interaction matter under extreme conditions'- project Nr.315477589 - TRR 211.
The computational resources have been provided by the LOEWE-Center for Scientific Computing and the "Green Cube" at GSI, Darmstadt.



\hfil\break
\appendix
\bigskip

\section{Cross sections for (anti)kaon production}\label{cross-sections}

In this Appendix we collect the parametrizations of the cross sections
for $K, \bar K$ production by $BB, mB$ and $mm$ reactions in the vacuum
used in the PHSD.
The theoretical origin of these parametrizations as well as the comparison to
the available experimental data have been discussed in the original publications
(and the references therein)
\cite{Cass97,brat97,CB99,laura03,Mishra:2004te,Hartnack:2011cn,Kolomeitsev:2004np} 
and are not repeated here.
The in-medium modifications of (anti)kaon production cross sections are explained in Section \ref{medium}.

\subsection{$N+N\rightarrow N+Y+K$}

The reaction cross sections for the channels $N+N\rightarrow N+Y+K$ can be 
approximated by~\cite{Sibirtsev:1995xb},
\begin{eqnarray}
\sigma_{pp\rightarrow p\Lambda K^+}&=&0.732\bigg(1-\frac{s_{01}}{s}\bigg)^{1.8}\bigg(\frac{s_{01}}{s}\bigg)^{1.5}~[\rm mb],\nonumber\\
\sigma_{pp\rightarrow p\Sigma^+ K^0}&=&0.339\bigg(1-\frac{s_{02}}{s}\bigg)^{2.25}\bigg(\frac{s_{02}}{s}\bigg)^{1.35}~[\rm mb], \nonumber\\
\sigma_{pp\rightarrow p\Sigma^0 K^+}&=&0.275\bigg(1-\frac{s_{02}}{s}\bigg)^{1.98}\bigg(\frac{s_{02}}{s}\bigg)~[\rm mb],
\label{Alex0}
\end{eqnarray}
where $\sqrt{s_{01}}=m_\Lambda+m_p+m_K$ and $\sqrt{s_{02}}=m_\Sigma+m_p+m_K$.
For simplicity, an isospin-averaged cross section is introduced by multiplying 
with a factor of 1.5 the cross sections of  Eq.~(\ref{Alex0}):
\begin{eqnarray}
\sigma_{NN\rightarrow N\Lambda K}&=&\frac{3}{2}\sigma_{pp\rightarrow p\Lambda K^+},
\label{Alex1}\\
\sigma_{NN\rightarrow N\Sigma K}&=&\frac{3}{2}\bigg(\sigma_{pp\rightarrow p\Sigma^+ K^0}+\sigma_{pp\rightarrow p\Sigma^0 K^+}\bigg).
\nonumber
\end{eqnarray}
The  different isospin channels are then evenly weighted, for example,
\begin{eqnarray}
\sigma_{pn\rightarrow n\Lambda K^+}=\sigma_{pn\rightarrow p\Lambda K^0}=\frac{1}{2}\sigma_{NN\rightarrow N\Lambda K},\nonumber\\
\sigma_{pn\rightarrow n\Sigma^0 K^+}=\sigma_{pn\rightarrow n\Sigma^+ K^0}=\sigma_{pn\rightarrow p\Sigma^- K^+}\nonumber\\
=\sigma_{pn\rightarrow p\Sigma^0 K^0}=\frac{1}{4}\sigma_{NN\rightarrow N\Sigma K}.
\label{Alex2}
\end{eqnarray}
The $\Delta$ baryon is assumed to have the same cross section for strangeness 
production as a nucleon:
\begin{eqnarray}
\sigma_{N\Delta\rightarrow N\Lambda K}=\sigma_{\Delta\Delta\rightarrow N\Lambda K}=\sigma_{NN\rightarrow N\Lambda K},
\label{Alex4}
\end{eqnarray}
where the production thresholds are modified accordingly depending on $\Delta$ mass.

\subsection{$N+N\rightarrow N+N+K+\bar{K}$}

As the center of mass energy $\sqrt{s}$ increases, a strange meson pair can be produced in the channel $N+N\rightarrow N+N+K+\bar{K}$ with a cross section given by~\cite{Sibirtsev:1996rh}:
\begin{eqnarray}
\sigma_{NN\rightarrow NNK\bar{K}}=1.5\bigg(1-\frac{s_{0}}{s}\bigg)^{3.17}\bigg(\frac{s_{0}}{s}\bigg)^{1.96}~[\rm mb],
\label{Alex5}
\end{eqnarray}
where $\sqrt{s_{0}}=2m_N+m_K+m_{\bar{K}}$.
We assume the same form of the cross section if the initial nucleon is replaced by a $\Delta$ 
 baryons, while baryons in the final state are always $n$ or $p$.

\subsection{$m+N\rightarrow K+Y,~K+\bar{K}+N$}

The cross section for $\Lambda$ production ($\pi N\rightarrow K \Lambda$) is given by~\cite{Huang:1994xq}
\begin{eqnarray}
\sigma_{\pi^-p\rightarrow \Lambda K^0}=\frac{0.007665(\sqrt{s}-\sqrt{s_o})^{0.1341}}{(\sqrt{s}-1.72)^2+0.007826} ~[\rm mb],
\label{Cugnon1}
\end{eqnarray}
where $\sqrt{s_0}$ is the threshold energy. Using isospin relations~\cite{Huang:1994xq} we find
\begin{eqnarray}
\sigma_{\pi^0 p\rightarrow \Lambda K^+}&=&\sigma_{\pi^0 n\rightarrow \Lambda K^0}=\frac{1}{2}\sigma_{\pi^- p\rightarrow \Lambda K^0},\\
\sigma_{\pi^+ n\rightarrow \Lambda K^+}&=&\sigma_{\pi^- p\rightarrow \Lambda K^0}.
\label{Cugnon2}
\end{eqnarray}
In heavy-ion collisions  many baryons are in resonance states. The contribution of resonances to  the $\Lambda$ production 
through $\pi+\Delta$ scattering is included by substituting $p$ by a $\Delta^+$ 
and $n$ by a $\Delta^0$ and assuming
\begin{eqnarray}
\sigma_{\pi^- \Delta^{++}\rightarrow \Lambda K^+}=\sigma_{\pi^+ \Delta^-\rightarrow \Lambda K^0}=\sigma_{\pi^- p\rightarrow \Lambda K^0}.
\label{Cugnon3}
\end{eqnarray}

The cross sections for $\Sigma$ baryon production ($\pi N\rightarrow K \Sigma$) are given by~\cite{Tsushima:1994rj,Tsushima:1994pv}
\begin{eqnarray}
\sigma_{\pi^+p\rightarrow \Sigma^+ K^+}&=&\frac{0.03591(\sqrt{s}-\sqrt{s_o})^{0.9541}}{(\sqrt{s}-1.89)^2+0.01548}\label{sigmai} \\
  & +& \frac{0.1594(\sqrt{s}-\sqrt{s_o})^{0.01056}}{(\sqrt{s}-3.0)^2+0.9412}~[\rm mb],
  \nonumber\\
\sigma_{\pi^-p\rightarrow \Sigma^- K^+}&=&\frac{0.009803(\sqrt{s}-\sqrt{s_o})^{0.6021}}{(\sqrt{s}-1.742)^2+0.006583} \label{sigmai1}\\
  & + &\frac{0.006521(\sqrt{s}-\sqrt{s_o})^{1.4728}}{(\sqrt{s}-1.94)^2+0.006248}~[\rm mb],
  \nonumber\\
\sigma_{\pi^+n\rightarrow \Sigma^0 K^+}&=&\frac{0.05014(\sqrt{s}-\sqrt{s_o})^{1.2878}}{(\sqrt{s}-1.73)^2+0.006455}~[\rm mb], \,\,\, ~~~~~  \label{sigmai2}\\
\sigma_{\pi^0n\rightarrow \Sigma^- K^+}&=&\sigma_{\pi^+n\rightarrow \Sigma^0 K^+}, 
  \label{sigmai3}\\
\sigma_{\pi^0p\rightarrow \Sigma^0 K^+}&=&\frac{0.003978(\sqrt{s}-\sqrt{s_o})^{0.5848}}{(\sqrt{s}-1.74)^2+0.00667} \label{Huang1}\\
 & +& \frac{0.04709(\sqrt{s}-\sqrt{s_o})^{2.165}}{(\sqrt{s}-1.905)^2+0.006358}~[\rm mb],
 \nonumber
\end{eqnarray}
and isospin symmetry is assumed. Consequently we find
\begin{eqnarray}
\sigma_{\pi^-n\rightarrow \Sigma^0 K^0}&=&\sigma_{\pi^+p\rightarrow \Sigma^+ K^+},\nonumber\\
\sigma_{\pi^+n\rightarrow \Sigma^+ K^0}&=&\sigma_{\pi^-p\rightarrow \Sigma^- K^+},\nonumber\\
\sigma_{\pi^-p\rightarrow \Sigma^0 K^0}&=&\sigma_{\pi^0p\rightarrow \Sigma^+ K^0}=\sigma_{\pi^+n\rightarrow \Sigma^0 K^+},\nonumber\\
\sigma_{\pi^0n\rightarrow \Sigma^0 K^0}&=&\sigma_{\pi^0p\rightarrow \Sigma^0 K^+},
\label{Huang2}
\end{eqnarray}
As for the $\Lambda$ production, an initial $p$ can be substituted by a $\Delta^+$ and an initial $n$ by a $\Delta^0$. $\Delta^{++}$ and $\Delta^-$ are excluded for $\Sigma$ production.

If the collision energy $\sqrt{s}$ exceeds 1.7 GeV, the cross section for $\Sigma+K$ production in Eqs.~(\ref{sigmai}) to (\ref{Huang2}) is modified to
\begin{eqnarray}
\sigma_{\pi N\rightarrow \Sigma K}(\sqrt{s})= A \sqrt{s}-\sigma_0-\sigma_{\pi N\rightarrow \Lambda K}\nonumber\\
-\sigma_{\pi N\rightarrow N K\bar{K}}~[\rm mb],
\label{above17}
\end{eqnarray}
where $A=1.0$ (${\rm mb/GeV}$) and
\begin{eqnarray}
\sigma_0=1.7~{\rm mb}-\sigma_{\pi N\rightarrow \Sigma K}~~~~~~~~~~~~~~~~~~~~~~~~~~\nonumber\\
-\sigma_{\pi N\rightarrow \Lambda K}
-\sigma_{\pi N\rightarrow N K\bar{K}}\bigg|_{\sqrt{s}=1.7~{\rm GeV}}.
\end{eqnarray}
This means that the total cross section for strangeness production in baryon+meson scattering is a linear function of $\sqrt{s}$ above 1.7 GeV and smoothly connected at $\sqrt{s}=1.7$ GeV.

The cross section $\sigma_{\pi N\rightarrow N K\bar{K}}$ in Eq.~(\ref{above17}) 
is given by~\cite{Sibirtsev:1996rh}
\begin{eqnarray}
\sigma_{\pi^-p\rightarrow p K^0 K^-}=1.121\bigg(1-\frac{s_{0}}{s}\bigg)^{1.86}\bigg(\frac{s_{01}}{s}\bigg)^{2} ~[\rm mb],\nonumber\\
\label{Ko}
\end{eqnarray}
where $\sqrt{s_{0}}=m_N+m_K+m_{\bar{K}}$, and the other channels with various isospin combinations are related to $\sigma_0$ of Eq.~(\ref{Ko}) as 
follows~\cite{Sibirtsev:1996rh}:
\begin{eqnarray}
2\sigma(\pi^-p\rightarrow p K^0 K^-)=2\sigma(\pi^- n\rightarrow nK^0K^-)\nonumber\\
=2\sigma(\pi^+ p\rightarrow pK^+\bar{K}^0)=2\sigma(\pi^+ n\rightarrow nK^+\bar{K}^0)\nonumber\\
=\sigma(\pi^+ n\rightarrow pK^+K^-)=\sigma(\pi^+ n\rightarrow pK^0\bar{K}^0)\nonumber\\
=\sigma(\pi^0 p\rightarrow nK^+\bar{K}^0)=4\sigma(\pi^0 p\rightarrow pK^+K^-)\nonumber\\
=4\sigma(\pi^0 p\rightarrow pK^0\bar{K}^0)=\sigma(\pi^0 n\rightarrow pK^0K^-)\nonumber\\
=4\sigma(\pi^0 n\rightarrow nK^+K^-)=4\sigma(\pi^0 n\rightarrow pK^0\bar{K}^0)\nonumber\\
=\sigma(\pi^- p\rightarrow nK^+K^-)=\sigma(\pi^- p\rightarrow pK^0\bar{K}^0).
\end{eqnarray}
Again an initial $p$ can substituted by a $\Delta^+$ and an initial $n$ by a $\Delta^0$ assuming the same form of the cross section, while in the final state  only $p$ or $n$ are admitted.

The inverse reaction $K+Y\rightarrow \pi+N$ is realized by detailed balance and $N+K+\bar{K}\rightarrow N+\pi$ by using an equation similar to Eq.~(\ref{Alex-reverse}), including 
a $\Delta$ baryon in the final state.
For example,
\begin{eqnarray}
\sigma_{\Lambda K^+\rightarrow \pi^+ \Delta^0}=2\sigma_{\Lambda K^+\rightarrow \pi^0 \Delta^+}=\sigma_{\Lambda K^+\rightarrow \pi^-\Delta^{++}}\nonumber\\
=\bigg(\frac{p_{\Delta}}{p_{\Lambda}}\bigg)^2\sigma_{\pi^+ \Delta^0\rightarrow \Lambda K^+}=\bigg(\frac{p_{N}}{p_{\Lambda}}\bigg)^2\sigma_{\pi^+ n\rightarrow \Lambda K^+},
\end{eqnarray}
where $p_i$ is the three-momentum of particle $i$ in the c.m. frame.

\subsection{$\bar{K}+B \rightarrow \pi+Y$}

The reactions $\bar{K}+N \rightarrow \pi+Y$ are modeled (also in the vacuum)
according to the G-matrix as indicated by Eqs. (\ref{channel1})-(\ref{channel2}).
However, the present coupled channels of G-matrix do not incorporate reactions with
baryonic resonances which are abundant at SIS energies. 
For such reactions (with $B=\Delta, N(1440), N(1535)$) we use the parametrizations 
presented below. 

The cross section for $\bar{K}$ absorption by baryons is equal to the total cross section 
of $\bar{K}B$ scattering subtracted by the $\bar{K}N$ elastic scattering cross section\cite{Efremov:1994ie}. 
The total cross section of $\bar{K}B$ scattering \cite{Efremov:1994ie} is
\begin{eqnarray}
\sigma_{\bar K+B\to \pi+Y}^{tot} =&&\nonumber\\
22.6p_K^{-1.14}~[\rm mb],&&p_K<0.7 ~{\rm GeV/c},\nonumber\\
47.54p_K^{-0.94}~[\rm mb],&&0.7 ~{\rm GeV/c}\leq p_K<1.1 ~{\rm GeV/c},\nonumber\\
69.87p_K^{-3.1}~[\rm mb],&&1.1 ~{\rm GeV/c}\leq p_K<1.3 ~{\rm GeV/c},\nonumber\\
32.4p_K^{-0.17}~[\rm mb],&&1.3 ~{\rm GeV/c}\leq p_K<10 ~{\rm GeV/c},\nonumber\\
22~~[\rm mb],&&10~ {\rm GeV/c}\leq p_K,
\label{crok1}
\end{eqnarray}
while the $\bar{K}N$ elastic scattering cross section is
\begin{eqnarray}
\sigma_{\bar K+B\to \pi+Y}^{elast} =&&\nonumber\\
10.58p_K^{-0.98}~[\rm mb],&& 0.03~{\rm GeV/c}\leq p_K<0.7 ~{\rm GeV/c},\nonumber\\
23 p_K^{1.2}~[\rm mb],&&0.7 ~{\rm GeV/c}\leq p_K<1 ~{\rm GeV/c},\nonumber\\
23 p_K^{-2.6}~[\rm mb],&&1~{\rm GeV/c}\leq p_K<1.5 ~{\rm GeV/c},\nonumber\\
9.56 p_K^{-0.44}~[\rm mb],&&1.5~{\rm GeV/c}\leq p_K<20 ~{\rm GeV/c},\nonumber\\
2.56~~[\rm mb],&&20~ {\rm GeV/c}\leq p_K,
\label{crok2}
\end{eqnarray}
where $p_K$ is the momentum of the $\bar{K}$ in the $B$ rest frame.
The $\bar{K}$ absorption cross section is weighted by 0.3 for $Y=\Lambda$ and 
by 0.2 for $Y=\Sigma$.

\subsection{Elastic and isospin-exchange scattering of kaons}

Finally $K+N$ elastic cross section and the cross section for isospin exchange are parameterized in the units of mb as below
\begin{eqnarray}
\sigma_{elastic}^{K^+ p}&=&\sigma_{elastic}^{K^0 n} \label{Kelast0}\\
&=& 12.3+1.88p_K-2.32p_K^2~~({\rm p_K<2.3 ~GeV/c}),\nonumber\\
&=& 5~[\rm mb]~~~({\rm p_K>2.3 ~GeV/c}).\nonumber
\end{eqnarray} 
Here $p_K$ is the three momentum of kaon (GeV/c) in center-of-mass frame,
\begin{eqnarray}
\sigma_{elastic}^{K^+ n}&=&\sigma_{elastic}^{K^0 p} \label{Kelast}\\
&=&10.51 p_K^{0.82}~~~({\rm p_K<0.6 ~GeV/c}),\nonumber\\
&=&-18.93+76.68 p_K-66.65 p_K^2 +17.74 p_K^3\nonumber\\
&&~~~~~~~~~~~~~~~~({\rm 0.6 ~GeV/c <p_K<1.5 ~GeV/c}),\nonumber\\
&=&6~[\rm mb] ~~~({\rm 1.5 ~GeV/c <p_K}),\nonumber
\end{eqnarray}
and
\begin{eqnarray}
\sigma^{K^+ n\leftrightarrow K^0 p}&=&5.63+4.996 p_K-4.519 p_K^2  \label{Kelast1}\\
&&~~~~~~~~~~~~~~~~~~~({\rm p_K<1.5 ~GeV/c}),\nonumber\\
&=&6.502/p_K^{1.932} ~~~({\rm 1.5 ~GeV/c <p_K}). \nonumber
\end{eqnarray}
The above cross sections are important for the generation of collective flows and isospin diffusion of $K^+$ and $K^0$ in heavy-ion collisions.

\section{3-to-2 interactions}\label{app1}

The scattering cross section for 2-to-3 processes can be expressed as
\begin{eqnarray}
\sigma_{2\rightarrow 3}&=&\frac{1}{4E_1E_2v_{\rm rel}}\int \frac{d^3p_1^\prime}{(2\pi)^32E_1^\prime}\int \frac{d^3p_2^\prime}{(2\pi)^32E_2^\prime}\int \frac{d^3p_3^\prime}{(2\pi)^32E_3^\prime}\nonumber\\
&\times& \overline{|M|}^2(2\pi)^4\delta^{(4)}(p_1+p_2-p_1^\prime-p_2^\prime-p_3^\prime),
\end{eqnarray}
where $v_{\rm rel}$ is the relative velocity of the incident particles and 
$\overline{|M|^2}$ is the transition amplitude squared averaged over initial states:
\begin{eqnarray}
\overline{|M|}^2=\frac{|M|^2}{D_1D_2}
\end{eqnarray}
with $D_1$, $D_2$ denoting the spin-flavor degeneracy of the initial 2 particle states.
Assuming that the transition amplitude does not depend on scattering angle but only on the collision energy $\sqrt{s}$, the cross section can be rewritten as
\begin{eqnarray}
\sigma_{2\rightarrow 3}&=&\frac{1}{4E_1E_2v_{\rm rel}}\frac{|M|^2}{D_1D_2}(PS)_3\nonumber\\
&=&\frac{1}{4p_1^{c.m.}\sqrt{s}}\frac{|M|^2}{D_1D_2}(PS)_3 ,
\label{sigma3}
\end{eqnarray}
where $p_1^{c.m.}=|\vec{p_1}|$ in the c.m. frame and
\begin{eqnarray}
(PS)_3=\int \frac{d^3p_1^\prime}{(2\pi)^32E_1^\prime}\int \frac{d^3p_2^\prime}{(2\pi)^32E_2^\prime}\int \frac{d^3p_3^\prime}{(2\pi)^32E_3^\prime}\nonumber\\
\times (2\pi)^4\delta^{(4)}(p_1+p_2-p_1^\prime-p_2^\prime-p_3^\prime).
\end{eqnarray}
The three-body phase-space integral 
$(PS)_3$ is simplified in the c.m. frame of $p_2^\prime+p_3^\prime$ as follows:
\begin{eqnarray}
(PS)_3=\int \frac{d^3p_1^\prime}{(2\pi)^32E_1^\prime}\int \frac{d^3p_2^\prime}{(2\pi)^24E_2^\prime E_3^\prime}\nonumber\\
\times \delta(E_1+E_2-E_1^\prime-E_2^\prime-E_3^\prime)
\nonumber\\=\frac{1}{16\pi^3\sqrt{s}}\int^{\sqrt{s}-m_1^\prime}_{m_2^\prime+m_3^\prime}dM_{23}p_1^{\prime~c.m.}p_2^* \ ,
\end{eqnarray}
where $M_{23}=\sqrt{(p_2+p_3)^2}$ and $p_1^{\prime~c.m.}=|p_1^\prime|$ in the c.m. frame of $p_1+p_2$ and $p_2^*=|p_2^\prime|$ in the c.m. frame of $p_2^\prime+p_3^\prime$, i.e.
\begin{eqnarray}
p_1^{\prime~c.m.}=\frac{\sqrt{\{s-(m_1^\prime+M_{23})^2\}\{s-(m_1^\prime-M_{23})^2\}}}{2\sqrt{s}},\nonumber\\
p_2^*=\frac{\sqrt{\{M_{23}^2-(m_2^\prime+m_3^\prime)^2\}\{M_{23}^2-(m_2^\prime-m_3^\prime)^2\}}}{2M_{23}}.
\end{eqnarray}

A cross section for 3-to-2 processes cannot be defined but a Lorentz invariant interaction rate for 3-to-2 processes is given by \cite{Cassing:2001ds,Xu:2004mz,Song:2012at}
\begin{eqnarray}
\frac{\Delta N^{3\rightarrow 2}}{\Delta t\Delta V}=\frac{1}{D_1^\prime D_2^\prime D_3^\prime}\int \frac{d^3p_1}{(2\pi)^32E_1}\nonumber\\
\times \int \frac{d^3p_2}{(2\pi)^32E_2}\int \frac{d^3p_1^\prime}{(2\pi)^32E_1^\prime}f_1(p_1^\prime)\nonumber\\
\times \int \frac{d^3p_2^\prime}{(2\pi)^32E_2^\prime}f_2(p_2^\prime)\int \frac{d^3p_3^\prime}{(2\pi)^32E_3^\prime}f_3(p_3^\prime)\nonumber\\
\times |M|^2(2\pi)^4\delta^{(4)}(p_1+p_2-p_1^\prime-p_2^\prime-p_3^\prime),
\end{eqnarray}
where $f_i(p_i^\prime)$ is a particle distribution function including its degeneracy factor $D_i^\prime$.
Keeping in mind that $|M|^2$ does not depend on the scattering angle in the c.m. frame of $p_1+p_2$, the rate may be written as
\begin{eqnarray}
\frac{\Delta N^{3\rightarrow 2}}{\Delta t\Delta V}=\frac{1}{D_1^\prime D_2^\prime D_3^\prime}\frac{p_1^{c.m.}}{4\pi\sqrt{s}}|M|^2\int \frac{d^3p_1^\prime}{(2\pi)^32E_1^\prime}f_1(p_1^\prime)\nonumber\\
\times \int \frac{d^3p_2^\prime}{(2\pi)^32E_2^\prime}f_2(p_2^\prime)\int \frac{d^3p_3^\prime}{(2\pi)^32E_3^\prime}f_3(p_3^\prime).
\end{eqnarray}

Expressing the distribution functions by~\cite{Xu:2004mz}
\begin{eqnarray}
f_i=(2\pi)^3\frac{\Delta N_i}{\Delta p_i^3\Delta V},
\end{eqnarray}
the interaction rate turns to
\begin{eqnarray}
\frac{\Delta N^{3\rightarrow 2}}{\Delta t\Delta V}=\frac{1}{D_1^\prime D_2^\prime D_3^\prime}\frac{p_1^{c.m.}}{4\pi\sqrt{s}}|M|^2
\frac{\Delta N_1^\prime \Delta N_2^\prime \Delta N_3^\prime}{8E_1^\prime E_2^\prime E_3^\prime(\Delta V)^3}.
\end{eqnarray}
Now substituting $|M|^2$ from Eq.~(\ref{sigma3}) one obtains,
\begin{eqnarray}
\frac{\Delta N^{3\rightarrow 2}}{\Delta t}=\frac{D_1D_2}{D_1^\prime D_2^\prime D_3^\prime}\frac{(p_1^{c.m.})^2}{\pi(PS)_3}\sigma_{2\rightarrow 3}\nonumber\\
\times\frac{\Delta N_1^\prime \Delta N_2^\prime \Delta N_3^\prime}{8E_1^\prime E_2^\prime E_3^\prime(\Delta V)^2},
\label{stochastic}
\end{eqnarray}
where $\Delta t$, $\Delta V$ and $\Delta N$ correspond to the simulation time interval, volume of the grid cell, and particle number in the grid cell, respectively. We note that Eq.~(\ref{stochastic}) is equivalent to the expression in Ref.~\cite{Seifert:2017oyb}.

The transition probability for the transition $NYK\rightarrow N'N^{''}$ in the volume $\Delta V$ during the time interval  $\Delta t$ is given by
\begin{eqnarray}
P_{NYK\rightarrow N'N^{''}}&=&\frac{D_{N'}D_{N^{''}}}{D_N D_Y D_K}\frac{(p_{N'}^{c.m.})^2}{\pi(PS)_3}\sigma_{N'N^{''}\rightarrow NYK}\nonumber\\
&\times&\frac{\Delta N_N \Delta N_Y \Delta N_K}{8E_N E_Y E_K(\Delta V)^2}\Delta t,
\label{Alex-reverse}
\end{eqnarray}
with the three-body phase-space integral 
%
\begin{eqnarray}
(PS)_3&=&\int \frac{d^3p_N}{(2\pi)^32E_N}\int \frac{d^3p_Y}{(2\pi)^32E_Y}\int \frac{d^3p_K}{(2\pi)^32E_K}   \hfill \nonumber\\
&\times&(2\pi)^4\delta^{(4)}(p_{N'}+p_{N''}-p_N-p_Y-p_K). \hfill \,\,\,~~~~~~ \label{PS3}
\end{eqnarray}
The degeneracy factors need additional specification: for example,
($\Sigma^0~K^+~n$), ($\Sigma^+~K^0~n$), ($\Sigma^0~K^0~p$), or ($\Sigma^-~K^+~p$) can produce ($p~n$), ($p~\Delta^0$), ($n~\Delta^+$), ($\Delta^+~\Delta^-$), or ($\Delta^{++}~\Delta^-$), and the ratio of the degeneracy factors in Eq.~(\ref{Alex-reverse}) is 1/4, 1/2, and 1/2 for the final states of $(N~N)$, $(N~\Delta)$ and $(\Delta~\Delta)$, respectively, which are distinguished by $p_{N'}^{c.m.}$.
As another example, $(\Lambda~K^+~p)$ to $(p~p)$ has an additional degeneracy factor of 1/2 because two protons in the final state are indistinguishable.
The mass of the $\Delta$ baryon is randomly sampled from its spectral function, and if the total mass of the final states is above $\sqrt{s}$, the transition does not take place. This Monte Carlo sampling is effectively equivalent to the integral of the $\Delta$ spectral function  over invariant mass.

\section{cross section for $\bar{K}$ production in the medium}\label{app:cs-medium}

The phase space for the scattering of on-shell particles is given by

\begin{eqnarray}
\int \frac{d^3p_3}{(2\pi)^3 2E_3}\frac{d^3p_4}{(2\pi)^3 2E_4}(2\pi)^4\delta^{(4)}(p_1+p_2-p_3-p_4).  ~~~\label{Int1} 
\end{eqnarray}
The phase space for $p_3$ can be changed into a more explicitly covariant form
\begin{eqnarray}
\int \frac{d^3p_3}{(2\pi)^3 2E_3}\rightarrow\int \frac{d^4p_3}{(2\pi)^4}2\pi\delta^+(p_3^2-m_3^3)
\label{covariant}
\end{eqnarray}
where $\delta^+(x-x_0)$ is nonzero only for positive $x_0$.
Comparing the normalization condition of Eq.~(\ref{normalization}) and Eq.~(\ref{covariant}), one can find that the spectral function
$A(\omega,{\bf p}_3)$ corresponds to $2\pi\delta^+(p_3^2-m_3^2)$ with the same normalization condition,
\begin{eqnarray}
2\int_0^{\infty}\frac{d\omega}{2\pi}\omega2\pi\delta^+(p_3^2-m_3^2)=\int_0^{\infty}
d\omega 2 \omega \delta^+(\omega^2-E_3^2)    \nonumber\\
=\int_0^{\infty}d\omega\frac{\omega}{E_3}\delta(\omega-E_3)=1,~~~~~
\end{eqnarray}
where $E=\sqrt{m_3^2+p_3^2}$.
Therefore, the covariant form of Eq.~(\ref{covariant}) can be modified for off-shell 
particles into
\begin{eqnarray}
\int \frac{d^3p_3}{(2\pi)^3 2E_3}\rightarrow\int \frac{d^4p_3}{(2\pi)^4}A(p_3^0,{\bf p}_3)\nonumber\\
=\int \frac{dp_3^0}{2\pi}A(p_3^0,{\bf p}_3)\int\frac{d^3p_3}{(2\pi)^3}\nonumber\\
=\int_0^{(\sqrt{s}-m_4)^2} \frac{dm_3^2}{2\pi}A(m_3^2)\int\frac{d^3p_3}{(2\pi)^32p_3^0},
\end{eqnarray}
where $m_3^2=(p_3^0)^2-{\bf p}_3^2$ and $m_4$ is the mass of the particle other than particle $3$ in the final state. The cross section turns to
\begin{eqnarray}
\sigma(\sqrt{s};M_3)\rightarrow \int_0^{(\sqrt{s}-m_4)^2} \frac{dm_3^2}{2\pi}A(m_3^2)~\sigma(\sqrt{s};m_3),
\label{semi-final}
\end{eqnarray}
where $M_3$ and $m_3$ are on-shell and off-shell masses,  respectively.
Applying the same assumption of Eq.~(\ref{massshift}) to Eq.~({\ref{semi-final}), the cross section for off-shell particle production reads
\begin{eqnarray}
\int_0^{(\sqrt{s}-m_4)^2} \frac{dm_3^2}{2\pi}A(m_3^2)~\sigma(\sqrt{s}-m_3+M_3).
\end{eqnarray}

The same spectral function can be applied to the decay of a $\phi$ or $K^*$ meson.
Assuming that the transition amplitude does not depend on the mass of the
daughter particles, the decay width is calculated as
\begin{eqnarray}
\Gamma\sim\frac{1}{M}\int_0^{(M-m_2)^2} \frac{dm_1^2}{2\pi}A(m_1^2)\int \frac{d^3p_1}{(2\pi)^3 2E_1}\nonumber\\
\times \int\frac{d^3p_2}{(2\pi)^3 2E_2}(2\pi)^4\delta^{(4)}(p-p_1-p_2)\nonumber\\
=\frac{1}{4\pi M^2}\int_0^{(M-m_2)^2} \frac{dm_1^2}{2\pi}A(m_1^2)p_1^{c.m.}(m_1^2),
\end{eqnarray}
where $M$ is the mass of the mother particle and $p_1^{c.m.}$ is the three-momentum of 
the off-shell particle in the c.m. frame.
Therefore $m_1$ - produced through the decay - has the mass differential distribution 
\begin{eqnarray}
\frac{dP(m_1^2)}{dm_1^2}=
\frac{A(m_1^2)p_1^{c.m.}(m_1^2)}{8\pi^2 M^2\Gamma}=A(m_1^2)~~~~~~~~~\nonumber\\
\times\frac{\sqrt{\{M^2-(m_1+m_2)^2\}\{M^2-(m_1-m_2)^3\}}}{(4\pi)^2 M^3\Gamma}.
\end{eqnarray}

The probability is proportional to $p_1^{c.m.}$ because the phase space increases 
with $p_1^{c.m.}$.

\section{Time-evolution of the off-shell particle mass}\label{app:mass-update}

As shown in Eq.~(\ref{energy-update}) the energy of the off-shell particle is updated as
\begin{eqnarray}
\frac{dE}{dt}=\frac{1}{2E}\bigg[\partial_t{\rm Re}\Sigma+\frac{M^2-M_0^2}{{\rm Im} \Sigma}\partial_t{\rm Im}\Sigma\bigg],
\label{E-update}
\end{eqnarray}
where $C$ is neglected for simplicity.
Since the partial derivatives are expressed in terms of the total derivatives as
\begin{eqnarray}
\frac{\partial {\rm Re}\Sigma}{\partial t}&=&\frac{d {\rm Re}\Sigma}{d t}-{\bf v}\cdot \nabla {\rm Re}\Sigma,\\
\frac{\partial {\rm Im}\Sigma}{\partial t}&=&\frac{d \Gamma}{d t}-{\bf v}\cdot \nabla {\rm Im}\Sigma,
\end{eqnarray}
Eq.~(\ref{E-update}) turns to
\begin{eqnarray}
\frac{dE}{dt}=\frac{1}{2E}\bigg\{ \frac{d {\rm Re}\Sigma}{d t}+\frac{M^2-M_0^2}{{\rm Im}\Sigma}\frac{d{\rm Im}\Sigma}{d t}\bigg\}\nonumber\\
-\frac{{\bf v}}{2E}\cdot \bigg\{\nabla {\rm Re}\Sigma+\frac{M^2-M_0^2}{{\rm Im}\Sigma}\nabla{\rm Im}\Sigma\bigg\}\nonumber\\
=\frac{1}{2E}\bigg\{ \frac{d {\rm Re}\Sigma}{d t}+\frac{M^2-M_0^2}{{\rm Im}\Sigma}\frac{d{\rm Im}\Sigma}{d t}\bigg\}+{\bf v}\cdot \frac{d{\bf p}}{dt},
\label{E-update2}
\end{eqnarray}
where Eq.~(\ref{momentum-update}) has been substituted in the second equation.
Multiplying by $2E$ on both sides,
\begin{eqnarray}
2E\frac{dE}{dt}-2{\bf p}\cdot \frac{d{\bf p}}{dt}-\frac{d {\rm Re}\Sigma}{d t}=\frac{M^2-M_0^2}{{\rm Im}\Sigma}\frac{d{\rm Im}\Sigma}{d t},
\label{E-update3}
\end{eqnarray}
and using
\begin{eqnarray}
2E\frac{dE}{dt} = \frac{dE^2}{dt},~~~~~2{\bf p}\cdot \frac{d{\bf p}}{dt} = \frac{d{\bf p}^2}{dt},
\label{approx}
\end{eqnarray}
the left-hand-side of Eq.~(\ref{E-update3}) is simplified to
\begin{eqnarray}
\frac{d(E^2-{\bf p}^2-{\rm Re}\Sigma)}{dt}\equiv\frac{dM^2}{d t}=\frac{M^2-M_0^2}{{\rm Im}\Sigma}\frac{d{\rm Im}\Sigma}{d t},
\label{m2-update}
\end{eqnarray}
where $M^2=E^2-{\bf p}^2-{\rm Re}\Sigma$, which is equivalent to Eq.~(\ref{mass-update}).


\end{document}